\documentclass[a4paper,12pt,openany,twoside]{mythesis}
\usepackage{amsmath}
\usepackage{amssymb}
\usepackage{bm}
\usepackage[T1]{fontenc}
\usepackage[latin1]{inputenc}
\usepackage{url}
\usepackage{makeidx}
\usepackage{graphicx}
\usepackage{hyperref}

\newcommand{\op}{\`}
\newcommand{\be}{\begin{equation}}
\newcommand{\ee}{\end{equation}}
\newcommand{\ba}{\begin{eqnarray}}
\newcommand{\ea}{\end{eqnarray}}
\newcommand{\bs}{\begin{subequations}}
\newcommand{\es}{\end{subequations}}
\newcommand{\bc}{\begin{center}}
\newcommand{\ec}{\end{center}}
\newcommand{\ve}{\varepsilon}
\newcommand{\dth}{\bar{\theta}}
\newcommand{\dq}{\bar{q}}
\newcommand{\dy}{\bar{y}}
\newcommand{\da}{\bar{a}}
\newcommand{\de}{\bar{\epsilon}}
\newcommand{\dH}{\bar{H}}
\newcommand{\dalp}{\bar{\alpha}}
\newcommand{\wteta}{{\widetilde{\theta}}}
\newcommand{\teta}{{\widetilde{\eta}}}
\newcommand{\oteta}{{\overline{\theta}}}
\newcommand{\sV}{{\text{\tiny $\phi V$}}}
\newcommand{\T}{{\text{\tiny $T$}}}
\newcommand{\tV}{{\text{\tiny $TV$}}}
\newcommand{\V}{{\text{\tiny $V$}}}

\newcommand{\eff}{{\text{\tiny eff}}}
\newcommand{\narx}{} %also \newcommand{\narx}{arXiv:}
\newcommand{\arx}[1]{[#1]} %also \newcommand{\arx}[1]{[arXiv:#1]}
\newcommand{\case}[2]{{\textstyle\frac{#1}{#2}}}
\renewcommand{\H}{{\text{\tiny $H$}}}
\renewcommand{\S}{{\text{\tiny $\phi$}}}

\makeindex
\newenvironment{abstract}{%
  \thispagestyle{empty} \normalsize
  \begin{center}%
  {\bfseries Abstract}
  \end{center}%
  \quotation}%

\begin{document}

\thispagestyle{empty}

\headsep=1cm \textheight=23cm \hoffset=-.8in
\evensidemargin=75pt \oddsidemargin=100pt 
\voffset=-1cm
\footskip=1.5cm

\begin{center}
{\large \sc Parma University -- Italy} \\
\bigskip

{\large \sc Department of Physics} \\
\vspace{1.0cm}
\begin{figure}[h]
\centering\includegraphics[width=2.5cm,height=2.5cm]{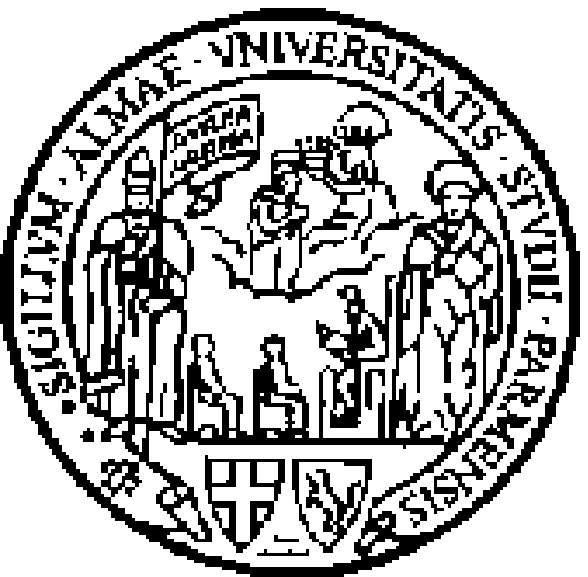}
\end{figure}
\vspace{1cm}
%--------

\vspace{2cm}
{\LARGE \bf BRANEWORLD COSMOLOGY AND\\\bigskip NONCOMMUTATIVE INFLATION}\vspace{1cm}\\
Thesis for the award of the degree of\\
Doctor of Philosophy

\vspace{2cm}
{\Large \bf Gianluca Calcagni}

\vspace{5cm}
{\large March 2005}

\end{center}

\clearpage
\thispagestyle{empty}

\frontmatter
\sloppy % lower linebreaking standard (\fussy is the opposite)
\oddsidemargin=110pt\evensidemargin=70pt

%%%%%%%%%%%%%%%%%%%%%%%%%%%%%%%%%%%%%%%%%%%%%%%%%%%%%%%%%%%%%%%%%%%%%%%%%%%%%%%%%%%%%%%%%%%%%%%%%%%%%%%%%%%%%%%%%%%%%%%%%%%%%%%%%%%%%%%%%%%%%%%%%%%%%%%%%%%%%%%%%%%%%%%%%%%%%%%%%%%%%%%%%%%%%%%%%%%%%%%%%%%%%%%%%%%%%%%%%%%%%%%%%%%%%%%%%%%%%%%%

\thispagestyle{empty}
\begin{abstract}
In this work we develop the patch formalism, an approach providing a very simple and compact description of braneworld-motivated cosmologies with nonstandard effective Friedmann equations. In particular, the Hubble parameter is assumed to depend on some power of the brane energy density, $H^2 \propto \rho^q$. The high-energy limit of Randall-Sundrum ($q=2$) and Gauss-Bonnet ($q=2/3$) braneworlds are considered, during an accelerating era triggered by a single ordinary or tachyonic scalar field. We present a slow-roll formalism generalizing the four-dimensional one; full towers of slow-roll parameters are constructed and the dynamics of the inflaton field explored in detail. The inflationary attractor condition, exact cosmological solutions, and perturbation spectra are provided. Using the latest results from WMAP and other experiments for estimates of cosmological observables, it is shown that future data and missions can in principle discriminate between standard four-dimensional and braneworld scenarios. The issue of non-Gaussianity is also studied within nonlinear perturbation theory.

The introduction of a fundamental energy scale reinforces these results. Several classes of noncommutative inflationary models are considered within an extended version of patch cosmological braneworlds, starting from a maximally invariant generalization of the action for scalar and tensor perturbations to a noncommutative brane embedded in a commutative bulk. Slow-roll expressions and consistency relations for the cosmological observables are provided, both in the ultraviolet and infrared region of the spectrum. The effects of noncommutativity are analyzed in a number of ways and energy regimes. 

Finally, we establish dual relations between inflationary, cyclic/ekpyrotic and phantom cosmologies, as well as between scalar-driven and tachyon-driven cosmologies. The exact dualities relating the four-dimensional spectra are broken in favour of their braneworld counterparts. The dual solutions display new interesting features because of the modification of the effective Friedmann equation on the brane.
\vspace{1cm}

\end{abstract}

\newpage
\thispagestyle{empty}
$\vphantom{dunno how to skip the page}$

\newpage
\thispagestyle{empty}

\begin{center}
\Large{\bf Acknowledgments}
\end{center}

It is a pleasure to thank Robert Brandenberger, Claudia de Rham, James Gregory, Andrew Liddle, Sabino Matarrese, Yun Soo Myung, Antonio Padilla, Max Pietroni, Toni Riotto, M. Sami, Shinji Tsujikawa, and Filippo Vernizzi for the many e-mail and face-to-face useful discussions we had. Shinji Tsujikawa deserves double acknowledgment for the fruitful, stimulating collaboration we had. Actually that was my first team work, which is not unimportant. Also, he run the codes which produced the figures in Sec. \ref{likeli} and is co-author of part of the material presented in Chapter \ref{noncom}.

Thanks also to the organizers of Gran Sasso Summer School 2002, Cargèse Summer School 2004, Strings 2004, and DESY Theory Workshop 2004 for their kind hospitality, without which I couldn't have travelled so much and see so many people during my PhD. In this respect I cannot forget the INFN group in Parma, represented by the exquisite understanding of Marisa Bonini.

Last but not least, I am grateful to my advisor Luca Griguolo for his invaluable suggestions and advice during these three years, especially as regards the topics in Chapter \ref{noncom}.

Finally, a tender thought for Arianna, who gave me moral support from the very beginning with her closeness, love, and patience.

\vspace{3cm}

\begin{center}
\Large{\bf Note to the arXiv version}
\end{center}
All the 31 figures have been omitted in this version (hep-ph/0503044). The interested reader can refer to my original papers or else contact me for a full pdf copy of my thesis.

\newpage
\thispagestyle{empty}
$\vphantom{dunno how to skip the page}$
\newpage

\setcounter{page}{5}
\tableofcontents
\mainmatter
   
\chapter{Introduction} \label{intro}

\begin{quote}
\textsl{This to attain, whether heav'n move or earth,\\ Imports not, if thou reckon right; the rest\\
From man or angel the great Architect\\ Did wisely to conceal, and not divulge\\
His secrets to be scanned by them who aught\\ Rather admire; or if they list to try\\
Conjecture, he his fabric of the heav'ns\\ Hath left to their disputes, perhaps to move\\
His laughter at their quaint opinions wide\\ Hereafter, when they come to model heav'n\\
And calculate the stars \dots}\\ --- Milton, \textit{Paradise lost}, VIII, 70-80
\end{quote}\vspace{1cm}

Imagine to be a two-dimensional man, like a paper silhouette cut in a sheet, living on the surface of a table. Imagine that the table is your world and all you see and touch and speak with lives in this two-dimensional universe. Things might look quite boring relative to our common 3D vision, but just for some more moments we are sheet-guys who actually do not know what the third dimension is. 

Now imagine that, after a life accustomed to flatness, a mad scientist claims the existence of a third, amazing, unprecedented ``extra dimension,'' transverse to the table surface. In the beginning people do not believe him, wondering: `Why should we complicate our world with things we cannot see?' Some colleague of the scientist's even shows that his proposal goes against current observations.
	
Despite all this skepticism, later on the theory is modified, its context changes and widens, and the underlying philosophy enriches with important consequences involving nothing less than our attitude to the interpretation of natural phenomena. People begin projecting experiments and models -- and sub-models, and scenarios within scenarios -- in order to find whether this extra dimension (an object they really cannot figure out, since their minds think in 2D only) produces some visible effect on the table or not. However, this new enthusiasm triggered in the field is not rewarded by a direct, unequivocal check of the idea.

This analogy, borrowed from \emph{Flatland} by Reverend Abbott (indeed farsighted a work, being written 120 years ago!), does not differ much from the fair tale of the real world and the 21st-century science. Recently a number of developments in string theory have given new insights to our comprehension of the high-energy physics and the fundamental behaviour of Nature. Although the mathematical structure of the string lore is so elegant that many theorists believe to be on the right path, its complexity and interpretative difficulties make concrete (i.e. liable to experimental pressure) predictions hard to formulate. Nevertheless, important applications have been proposed and almost fully constructed, particularly in cosmology.

One of the basic statements of these theories is that particles are actually microscopic vibrating ``strings'' which interact in a higher-dimensional spacetime; that is to say, four dimensions are not enough for the physics at the quantum scale. Beside strings, other extended objects (the ``D-branes'') appear in the full spectrum and play special roles of interest.

%%%%%%%%%%%%%%%%%%%%%%%%%%%%%%%%%%%%%%%%%%%%%%%%%%%%%%%%%%%%%%%%%%%%%%%%%%%%%%%%%%%%%%%%%%%%%%%%%%%%%%%%%%%%%%%%%%%%%%%%%%%%%%%%%%%%%%%%%%%%%%%%%%%%%%%%%%%%%%%%%%%%%%%%%%%%%%%%%%%%%%%%%%%%%%%%%%%%%%%%%%%%%%%%%%%%%%%%%%%%%%%%%%%%%%%%%%%%%%%%

\section{Brane worlds and cosmological principle}

The idea that the world we live in has more dimensions than we can see dates back to the 1920s with the works of Theodor Kaluza \cite{kal21} and Oskar Klein \cite{kle26}; however, in order not to violate results coming from gravitational and collider experiments, extra dimensions should be compactified and very tiny (of order of the Planck scale) and so almost unobservable. During recent years new models, exploiting many of the mentioned stringy ingredients, have been explored \cite{HW1,HW2,LOSW,LOW1,LOW2,ADD98,AADD,ADD99,ADDK,RSa,RSb,CIM,kok02}\footnote{See \cite{KMST,StST1,StST2,TW,oht1,oht2,CHNOW} for examples of compactifications on a hyperbolic manifold.} which require dimensions with compactification scale close to the limit of modern measurements of gravity (around millimeter) or even noncompact dimensions. In these cases, the visible universe is confined into a four-dimensional variety (a \emph{brane}) embedded in a larger spacetime, thus called \emph{braneworld}.\footnote{Brane universes in a multidimensional target spacetime were first considered in \cite{aka82,RuS,vis85}.}

Besides regaining classical gravity at low energies, these models have many interesting consequences, such as the mass hierarchy problem solution and the concrete possibility to test and bound the theory by means of astrophysical (supernovae) and cosmological observations (cosmic microwave background temperature fluctuations, large-scale structures) and accelerators (high-energy processes available at future LHCs). However, the problems opened by this new trend of research are far from being fully solved.

In the typical cosmological framework, the background metric on the brane is the Friedmann-Robertson-Walker (FRW) metric [with signature $(+,-,-,-)$]
\be\label{frwmet}
ds^2_{\rm FRW} = dt^2-a^2(t)\, dx_i dx^i\,,
\ee
where $a(t)$ is the scale factor on the 4D variety and Latin indices denote spatial coordinates. The FRW metric is the realization of the \emph{cosmological principle} of the standard big bang model, stating that ``the Universe does not possess any privileged direction or point; it is therefore homogeneous and isotropic, at least with good approximation.''\footnote{Due to the presence of small anisotropies in the microwave relic and the gravitational clustering of matter in the large-scale structures, the cosmological principle gives only an approximated description of the world.} If the Universe follows an homogeneous evolution, one can define a measure of time such that identical physical properties in different places imply synchronized local clocks. This is the reason why $t$ is called synchronous time. The spatial comoving coordinates $x^i$ are glued to the elements of the continuous fluid we assume to represent the (brane) universe, so that a coordinate spatial label $\mathbf{x}$ corresponds to the fluid element passing on that point at the time $t$. Physical distances are given by comoving distances times the scale factor $a(t)$.

The Einstein equations are modified in accordance with the gravity model permeating the whole spacetime. This in turn produces the basic FRW equations for the cosmological evolution. For comprehensive reviews on brane worlds, see \cite{rub01,mar03,BVD,csa04}. In particular, the five-dimensional Randall-Sundrum type 2 model \cite{RSb,BDL,BDEL,CGKT,CGS,FTW,MWBH,CLL} and its Gauss-Bonnet generalization \cite{KKL1,KKL2,NO1,neu00,NOO1,CD,LNO,CNW,NOO2,dav03,GW,LN,GrP,deh04,DLMS,MS,TSM,SST} have received much attention since their birth because of their rich structure in a relatively simple conceptual framework. One of the many aspects of these models is a possibility for the cosmological evolution to change substantially from the standard four-dimensional case, since the extra noncompact dimension can communicate via gravitational interaction with the matter confined in the brane. Gravity is free to propagate in the anti de Sitter (AdS) bulk, which is assumed to be empty in the simplest scenarios.

%%%%%%%%%%%%%%%%%%%%%%%%%%%%%%%%%%%%%%%%%%%%%%%%%%%%%%%%%%%%%%%%%%%%%%%%%%%%%%%%%%%%%%%%%%%%%%%%%%%%%%%%%%%%%%%%%%%%%%%%%%%%%%%%%%%%%%%%%%%%%%%%%%%%%%%%%%%%%%%%%%%%%%%%%%%%%%%%%%%%%%%%%%%%%%%%%%%%%%%%%%%%%%%%%%%%%%%%%%%%%%%%%%%%%%%%%%%%%%%%

\section{Inflation}

According to modern data, the large-scale structure of the Universe, as well as the anisotropies of the cosmic microwave background (CMB), can be explained by an early stage of accelerated expansion (inflation) driven by an effective cosmological constant \cite{gut81,sat81,AlS,lin82,lin83}. This mechanism is triggered by the dynamics of a scalar field (generically dubbed ``inflaton'') rolling down its potential, and may also provide an explanation for the present phase of acceleration; in the most famous version of inflation, the rolling is slow enough to justify the adoption of the \emph{slow-roll} (SR) formalism. 

Inflation was originally devised for solving a number of problems afflicting the hot big bang model, in particular the flatness or entropy problem (Why the Universe is flat? Why does it have so high an entropy?), the horizon problem (Why distant, causally disconnected regions are in thermal equilibrium?), and the monopole problem (Where are the topological defects emerging from the cosmological phase transitions?). However, the reasons of its success rely on an aspect which is much more than a back bonus. In fact, an immediate consequence of this scenario is that cosmological large-scale structures were originated by the exponential dilatation of quantum fluctuations of the inflaton up to macroscopic scales \cite{haw82,sta82,GP,BaSTu}. The study of microcosm allows us to investigate macrocosm in some sense, so cosmological observations are complementary to those with ground-based accelerators. Moreover, they clarify the composition and geometrical structure of the Universe, as well as those primordial elementary processes constituting the basis of our visible world. 

In particular, it is possible to combine the available observables in relations, called \emph{consistency equations}, that are characteristic of the inflationary paradigm, and verify them through CMB and sky-survey data. These relations do not depend on the form of the inflationary potential but do depend on either the type of scalar field (ordinary or tachyonic) on the brane and the details of the high-energy geometrical model. The consistency equations are a typical result from inflation that other theories of structure formation are not able to reproduce, and reflect the common physical origin of scalar and tensor perturbations; this scalar-tensor entanglement is even more pronounced in the braneworld framework.

Recently, due to many progresses made in understanding the vacuum structure of string theory (in particular, see \cite{sen1,sen2,sen3,sen4,HKM,GS,KMM1,KMM2,sen5,sen6,sen7,sen8}), the eventuality that the scalar field is tachyonic has been explored. With ``tachyon'' we mean a causal scalar field with an effective Dirac-Born-Infeld (DBI) action emerging from the low coupling limit of non-perturbative string theory. 

We leave the reader to textbooks such as \cite{CoL,LiL} for an introduction to standard cosmology. See also \cite{lid97} for a review of SR inflation.

%%%%%%%%%%%%%%%%%%%%%%%%%%%%%%%%%%%%%%%%%%%%%%%%%%%%%%%%%%%%%%%%%%%%%%%%%%%%%%%%%%%%%%%%%%%%%%%%%%%%%%%%%%%%%%%%%%%%%%%%%%%%%%%%%%%%%%%%%%%%%%%%%%%%%%%%%%%%%%%%%%%%%%%%%%%%%%%%%%%%%%%%%%%%%%%%%%%%%%%%%%%%%%%%%%%%%%%%%%%%%%%%%%%%%%%%%%%%%%%%

\section{Not only extra dimensions}

In addition to the brane conjecture, one can insert other exotic ingredients, borrowed from string and M theory, that may give rise to characteristic predictions, although at the price of increasing the number and complexity of concurring models. For instance, the introduction of a stringy spacetime uncertainty relation and the associated noncommutative scale leads to modifications of perturbation spectra at large scales, can generate a blue-tilted spectrum, and modifies the observationally allowed regions in the parameter space. Since the uncertainty relation is saturated when a perturbation with a particular wavelength is generated, the standard evolution of commutative fluctuations is altered and large-scale modes are damped. This might partially explain the low-multipole suppression of the CMB spectrum detected by recent data.

%%%%%%%%%%%%%%%%%%%%%%%%%%%%%%%%%%%%%%%%%%%%%%%%%%%%%%%%%%%%%%%%%%%%%%%%%%%%%%%%%%%%%%%%%%%%%%%%%%%%%%%%%%%%%%%%%%%%%%%%%%%%%%%%%%%%%%%%%%%%%%%%%%%%%%%%%%%%%%%%%%%%%%%%%%%%%%%%%%%%%%%%%%%%%%%%%%%%%%%%%%%%%%%%%%%%%%%%%%%%%%%%%%%%%%%%%%%%%%%%

\section{Observations}

Early-Universe observations have come to the golden age. The first-year results of the \textit{Wilkinson Microwave Anisotropy Probe} (WMAP) \cite{ben03,kom03,spe03,pei03} provided high-precision cosmological data sets from which astroparticle models can be tested. The observations strongly support the inflationary paradigm based on general relativity as a backbone of high-energy physics. In particular, nearly scale-invariant and adiabatic density perturbations generated in single-field inflation exhibit an excellent agreement with the observed CMB anisotropies \cite{bri03,BLM,KKMR,teg04}. Together with the upcoming high-precision data by the Planck satellite \cite{planc}, it will be possible to discriminate between a host of inflationary models from observations.

%%%%%%%%%%%%%%%%%%%%%%%%%%%%%%%%%%%%%%%%%%%%%%%%%%%%%%%%%%%%%%%%%%%%%%%%%%%%%%%%%%%%%%%%%%%%%%%%%%%%%%%%%%%%%%%%%%%%%%%%%%%%%%%%%%%%%%%%%%%%%%%%%%%%%%%%%%%%%%%%%%%%%%%%%%%%%%%%%%%%%%%%%%%%%%%%%%%%%%%%%%%%%%%%%%%%%%%%%%%%%%%%%%%%%%%%%%%%%%%%

\section[Big bang singularity, phantoms, and cosmological symmetries]{Big bang singularity, phantoms, and\\ cosmological symmetries}

Recently a lot of attention has been devoted to the symmetries of the cosmological dynamics. Transformations of the Einstein equations, encoded in the Friedmann relation coupled to the equations of motion for the matter content in the Universe, link standard inflationary cosmologies to other possible phases. These are either contracting periods, ideally embedded in some motivated high-energy pre-inflationary framework, or superaccelerating cosmologies, $\ddot{a}/a>H^2$, dominated by a matter component (called \emph{phantom}) with and effective equation of state $p<-\rho$. Such scenarios are of particular interest from both a theoretical and observational point of view, since the first one is intertwined with the big bang problem and the resolution of the initial singularity, while phantoms might explain modern data on the late-time evolution of the Universe. Conversely, bouncing events can leave their imprint on the large-scale perturbation spectra, while a phantom component can arise in a stringy or supersymmetric setup.

%%%%%%%%%%%%%%%%%%%%%%%%%%%%%%%%%%%%%%%%%%%%%%%%%%%%%%%%%%%%%%%%%%%%%%%%%%%%%%%%%%%%%%%%%%%%%%%%%%%%%%%%%%%%%%%%%%%%%%%%%%%%%%%%%%%%%%%%%%%%%%%%%%%%%%%%%%%%%%%%%%%%%%%%%%%%%%%%%%%%%%%%%%%%%%%%%%%%%%%%%%%%%%%%%%%%%%%%%%%%%%%%%%%%%%%%%%%%%%%%

\section{Plan of the thesis}

The material of this work is arranged as follows.
\begin{description}
\item[Chapter \ref{patch}] We introduce the \emph{patch} notation, an approach providing a very simple and compact description of braneworld-motivated cosmologies with nonstandard effective Friedmann equations. The particular cases of Randall-Sundrum and Gauss-Bonnet braneworlds are considered. We present a slow-roll formalism generalizing the four-dimensional case; full towers of parameters involving either the inflaton potential or the Hubble parameter are constructed, and the dynamics of standard and tachyonic fields are considered in detail. The inflationary attractor condition and exact cosmological solutions are provided. Through all these features, scalar-driven and tachyon-driven accelerating eras are compared. The original contribution is based on \cite{cal3}.
\item[Chapter \ref{obs}] This chapter is devoted to cosmological braneworld spectra and the observational imprint on early-Universe structures. An important aspect is the emerging of a set of consistency relations involving some of the most relevant observables, that is the amplitudes and indices of the perturbation spectra generated by quantum fluctuations stretched outside the Hubble horizon during the accelerated expansion. It is shown that, while the degeneracy between 4D and high-energy regimes can come from suitable values of the cosmological observables, exact functional matching between consistency expressions is discarded. Also, it turns out that CMB experiments of this and next generation might be able to discriminate between the standard four-dimensional lore and braneworld scenarios. The original contribution is based on \cite{cal3,cal2,cal5,cal6}.
\item[Chapter \ref{noncom}] We consider several classes of noncommutative inflationary models within an extended version of patch cosmological braneworlds, starting from a maximally invariant generalization of the action for scalar and tensor perturbations to a noncommutative brane embedded in a commutative bulk. Slow-roll expressions and consistency relations for the cosmological observables are provided, both in the ultraviolet and infrared region of the spectrum. The effects of noncommutativity are then analyzed in a number of ways and energy regimes. The original contribution is based on \cite{cal5,cal4,CT}.
\item[Chapter \ref{dlts}] In this chapter we address further theoretical issues and establish a triality between inflationary, cyclic/ekpyrotic, and phantom cosmologies in different patches. The exact dualities relating the four-dimensional spectra are broken in favour of their braneworld counterparts; the dual solutions display new interesting features because of the modification of the effective Friedmann equation on the brane. We then give some qualitative remarks on phantomlike cosmologies without phantom matter. The original contribution is based on \cite{cal9}.
\item[Chapter \ref{concl}] Discussion, conclusions, and future trends.
\item[Appendix \ref{appA}] A couple of examples of exact Randall-Sundrum solutions with late-time constant SR parameters is given. The original contribution is based on \cite{cal1}.
\item[Appendix \ref{appB}] A digression on CMB non-Gaussianities in the braneworld context. We calculate the bispectrum of single-field braneworld inflation, triggered by either an ordinary scalar field or a cosmological tachyon, by means of a gradient expansion of large-scale nonlinear perturbations coupled to stochastic dynamics. The resulting effect is identical to that for single-field 4D standard inflation, the nonlinearity parameter being proportional to the scalar spectral index in the limit of collapsing momentum. If the slow-roll approximation is assumed, braneworld and tachyon non-Gaussianities are subdominant with respect to the post-inflationary contribution. However, bulk physics may considerably strengthen the nonlinear signatures. These features do not change significantly when considered in a noncommutative framework. The original contribution is based on \cite{cal10}.
\end{description}

% version:  December 15, 2004

\newpage
\thispagestyle{empty}
$\vphantom{dunno how to skip the page}$
\newpage

\chapter{Patch cosmology} \label{patch}

\begin{quote}
\textsl{Yet it is possible to see peril in the finding of ultimate perfection. It is clear that the ultimate pattern contains its own fixity. In such perfection, all things move toward death.}\\ --- Frank Herbert, \textit{Dune}
\end{quote}\vspace{1cm}

%%%%%%%%%%%%%%%%%%%%%%%%%%%%%%%%%%%%%%%%%%%%%%%%%%%%%%%%%%%%%%%%%%%%%%%%%%%%%%%%%%%%%%%%%%%%%%%%%%%%%%%%%%%%%%%%%%%%%%%%%%%%%%%%%%%%%%%%%%%%%%%%%%%%%%%%%%%%%%%%%%%%%%%%%%%%%%%%%%%%%%%%%%%%%%%%%%%%%%%%%%%%%%%%%%%%%%%%%%%%%%%%%%%%%%%%%%%%%%%%

\section{Motivations} \label{motiv}

How to deal with the physics of extra dimensions? Are there sensible cosmological setups involving branes and other exotic ingredients? The answer to both questions is not unique, since there are many interconnected approaches by which to treat the braneworld \cite{KaS}. Here we will adopt the point of view of a cosmological observer living on a brane, which is a convenient place to put ourselves in if we want to predict what phenomena can be observed in the sky.

In particular, we shall consider the patch formulation of brane cosmology, in which the effective Hubble parameter $H\equiv \dot{a}/a$ experienced by an observer on the brane is assumed to depend on some power of the brane energy density $\rho$, $H^2 \propto \rho^q$. The advantages of this approach are several. First, it provides a concise and versatile formalism to explore different cosmological models and determine their main features, such as exact classes of solutions, the inflationary attractor, and the inflationary imprint on the structure formation of the early Universe. Second, it allows to treat standard and tachyon cosmologies on the same ground, since the equations of the latter are given by a particular limit of braneworld equations.

As the quotation hints, the patch formalism is far from being the ultimate or even the best framework in which to study the cosmology of extra dimensions. Nonetheless, from one side the advantages in using it are compared with the precision of modern observations; from the other side, the small effort in its formulation is well repaid by the insight into a great number of physical properties.

This model has been used to describe the post-inflationary evolution and in this case has been dubbed ``Cardassian cosmology'' \cite{car1,car2,car3,car4,car5,car6,car7,car8,car9,car10,car11,car12,car13,car14,car15}. Here we will take a rather different perspective, and ask how a period of nonstandard expansion can modify the usual early-Universe picture. Apart by the author, commutative patch cosmology was considered, e.g., in \cite{SST,KM,KM2,KKLM}.

%%%%%%%%%%%%%%%%%%%%%%%%%%%%%%%%%%%%%%%%%%%%%%%%%%%%%%%%%%%%%%%%%%%%%%%%%%%%%%%%%%%%%%%%%%%%%%%%%%%%%%%%%%%%%%%%%%%%%%%%

\subsection{Gauss-Bonnet braneworld and energy patches}

In braneworld scenarios the visible universe is confined into a (3+1)-dimensional variety (a brane) embedded in a larger noncompact spacetime (the bulk). This setup is motivated by M theory as a low-energy product of a dimensionally reduced 11D supergravity to a 10D string theory, down to a 5D effective gravity \cite{HW1,HW2,LOSW,LOW1,LOW2,KK,ArDD} (see also \cite{MO}). The resulting 11D manifold is $\text{AdS}_5 \times X_{CY}$, where the brane is located at the fixed point $y=y_b$ of the $Z_2$ symmetry in the 5D anti de Sitter bulk and the other six dimensions are compactified on a Calabi-Yau 3-fold $X_{CY}$. The 5D gravitational coupling is related to the 11D one by $\kappa_5^2\equiv 8\pi/m_5^3=\kappa_{11}^2/V_{CY}$, where $V_{CY}$ is the internal volume of the Calabi-Yau space and $\kappa_{11}^2 \equiv M_s^{-9}$ encodes the fundamental string mass.\footnote{We use the natural units $c=1$, $\hbar=1$.} For the case of branes in a 6D bulk, see, e.g., \cite{BCCF,cua03,WM}.

One of the first problems one has to deal with when constructing such models is how to stabilize the extra dimension. This can be achieved in a number of ways; in the Randall-Sundrum (RS) example, Goldberger and Wise have provided a mechanism according to which a 5D massive scalar is put into the bulk with a potential of the same order of the brane tension $\lambda$ \cite{GW1,GW2,GW3,GW4,GW5}. If the energy density $\rho$ on the brane is smaller than the characteristic energy of the scalar potential, $\rho/V \sim \rho/\lambda \ll 1$, then the radion is stabilized and one gets the standard Friedmann equation $H^2 \propto \rho$. On the contrary, if the brane energy density is comparable with the stabilization potential, $\rho/\lambda \gtrsim 1$, the bulk backreacts because it feels the presence of the brane matter, the minimum of the potential is shifted, and the well-known quadratic corrections to the Friedmann equation arise \cite{BDL,BDEL}:
\be
H^2 = \frac{\kappa_4^2}{6 \lambda} \rho (2 \lambda+\rho)+\frac{\cal E}{a^4}\,,
\ee
where $\kappa_4^2 \equiv 8\pi/m^2_4$ includes the four-dimensional Planck mass $m_4 \approx 10^{19}\,\text{GeV}$, and ${\cal E}=$ const is the \emph{dark radiation} term which is the time-time component of the five-dimensional Weyl tensor projected on the brane. Gravity experiments impose the bulk curvature scale to be $\lesssim 1\,\text{mm}$, that is $m_5 \gtrsim 10^8$ GeV and $\lambda^{1/4}\gtrsim 10^3$ GeV. We will neglect the dark radiation term since during inflation it is strongly suppressed (see below).

The RS model can be viewed as an intermediate scenario between a ``pure Gauss-Bonnet'' high-energy regime, $H^2 \propto \rho^{2/3}$, and the standard 4D (low-energy) evolution, $H^2 \propto \rho$. The five-dimensional bulk action for the Gauss-Bonnet (GB) braneworld is
\ba
S &=&\frac{1}{2\kappa_5^2} \int d^5x\sqrt{-g_5}\left[R-2\Lambda_5+\alpha\left(R^2-4R_{\mu\nu}R^{\mu\nu}+R_{\mu\nu\rho\sigma}R^{\mu\nu\rho\sigma}\right)\right]\nonumber\\
&&\qquad\qquad+S_\partial+S_\text{\tiny matter}\,.
\ea
Here, $\kappa_5$ is the five-dimensional gravitational coupling, $g_5$ is the determinant of the 5D metric, $R$ is the 5D Ricci scalar, $\Lambda_5<0$ is the bulk cosmological constant, and $\alpha=1/(8g_s^2)>0$ is the Gauss-Bonnet coupling, where $g_s$ is the string energy scale. The action includes a pure geometrical boundary term $S_\partial$ and the matter contribution which is confined on the brane. The gravitational part of the action is a natural generalization in five dimensions of the Einstein-Hilbert action (see \cite{CD} for a general discussion), since it is the only extension giving a second-order symmetric divergence-free tensor and field equations that are second-order in the metric \cite{lov71}. From a fundamental physics point of view, it comes from $\alpha'$-leading-order quantum corrections to gravity in the heterotic string effective action \cite{GrS}. In particular, in the Gauss-Bonnet theory graviton interactions are ghost free and spacetime perturbations are wavelike.

A cosmological solution of the theory can be found via a 5D warped metric such that its projection on the 3-brane is FRW-like, Eq. (\ref{frwmet}). Assuming a perfect fluid\footnote{A 4D perfect fluid with energy density $\rho$ and pressure $p$ is isotropic in its local rest frame, its energy tensor being diagonal and proportional to the pressure $p$ in its spatial projection: $T^{00}=\rho$, $T^{i\!j} = -p\,\delta^{i\!j}$. In the case of a confined fluid, the 5D energy-momentum tensor is $T^{\mu\nu}\propto\delta(y_b)\,{\rm diag}(\rho,-p,-p,-p,0)^{\mu\nu}$, where $y_b$ is the brane position along the extra direction $y$.} matter and a $Z_2$ symmetry across the brane, the effective Friedmann equation on the brane is \cite{CD,dav03,GW}
\begin{subequations} \label{gabo}
\be 
H^2=\frac{c_+ + c_- -2}{8\alpha}\,,
\ee
where $H$ is the Hubble parameter and, defining $\sqrt{\alpha/2}\,\kappa_5^2 \equiv \delta_0^{-1}$,
\be
c_\pm = \left[\sqrt{\left(1+4\alpha\Lambda_5/3\right)^{3/2}+\left(\delta/\delta_0\right)^2} \pm \delta/\delta_0\right]^{2/3};
\ee
\end{subequations}
$\delta$ is the matter energy density which we will assume to be decomposed into a matter contribution plus the brane tension $\lambda$: $\delta=\rho+\lambda$. Expanding Eq. (\ref{gabo}) to quadratic order in $\delta$ one recovers the Friedmann equation of the Randall-Sundrum type 2 scenario with vanishing 4D cosmological constant, provided
\[\kappa_5^4=\frac{6\kappa_4^2}{\lambda}\left(1+\frac{4}{3}\Lambda_5\alpha\right)\,,\qquad \lambda=\frac{3}{2\alpha\kappa_4^2}\left[1-\left(\frac{\lambda \kappa_5^4}{6 \kappa_4^2}\right)^{1/2}\right]\,.\] We can now recognize three main energy regimes resulting in particular limits of the Friedmann equation:
\begin{enumerate}
	\item $\delta/\delta_0 \gg 1\,:$ In this pure Gauss-Bonnet high-energy regime, we have a nonstandard cosmology
	      \be \label{fe1}
	      H^2= \left(\frac{\kappa_5^2}{16\alpha}\right)^{2/3}\rho^{2/3}\,;
	      \ee
	\item $\lambda/\delta_0 \ll \delta/\delta_0 \ll 1\,:$ When the energy density is far below the 5D or string scale but
	      $\rho \gg \lambda$, we have a Hubble parameter
	      \be \label{fe2}
	      H^2 = \frac{\kappa_4^2}{6 \lambda} \rho^2\,;
	      \ee
	\item $\rho/\delta_0 \ll \delta/\delta_0 \ll 1\,:$ The standard four-dimensional scenario is recovered when the brane
	      grows stiff with respect to its matter content, $\rho \ll \lambda$:
	      \be \label{fe3}
        H^2 = \frac{\kappa_4^2}{3} \rho \,.
        \ee
\end{enumerate}
The Friedmann equation (\ref{gabo}) and its energy approximations are plotted in Fig. \ref{fig1}.
\begin{figure}[ht]
\bc
\ec
\caption{\it\label{fig1} The Hubble parameter as a function of the energy density in the Gauss-Bonnet scenario and its energy approximations. The solid line is the full Gauss-Bonnet cosmology [Eq. (\ref{gabo})], the lower dashed line is the high-energy Gauss-Bonnet regime [Eq. (\ref{fe1})], and the upper dashed line is the full Randall-Sundrum regime.}
\end{figure}

Equations (\ref{fe1})--(\ref{fe3}) are considerably simpler than the full Gauss-Bonnet equation (\ref{gabo}), and in many practical cases one of the three regimes is assumed. Therefore it can be useful to study a cosmological patch, that is a region of time and energy in which 
\be \label{FRW}
H^2=\beta_q^2 \rho^q\,,
\ee
where $q$ is constant and $\beta_q>0$ is a constant factor with energy dimension $[\beta_q]= E^{1-2q}$. Then, $q=1$ in the pure 4D (radion-stabilized) regime, $q=2$ in the high-energy limit of the RS braneworld, and $q=2/3$ in the high-energy limit of the GB scenario.

The resulting dynamical equations can be applied to any case of interest, let it be a particular limit of either the Gauss-Bonnet braneworld or other scenarios where nonconventional physics modifies the cosmological evolution. 

The parameter $q$, which describes the effective degrees of freedom from gravity, could live in a nonstandard range of values because of the introduction of nonperturbative stringy effects or, just to mention some possibilities, for the presence of a complicated geometrical framework with either compact and noncompact extra dimensions, multiple and/or folding branes configurations, higher-derivative gravities (\cite{ABF1,ABF2,CDCT} and references therein), and so on. In the latter example \cite{ABF1}, one can consider a class of 4D gravitational actions like
\be\label{higher}
S_g=\int d^4x \sqrt{g} f(R)\,,
\ee
where $f(R)$ is an arbitrary function of the Ricci scalar. It turns out that one can construct suitable expressions for $f(R)$ and get Eq. (\ref{FRW}) in the appropriate limit. The case 
\be\label{negative}
f(R)=R -(\sinh R)^{-1}\,,
\ee
is of particular interest, since in the limit of small curvature $R$ (late times, low energy) one gets $H^2 \approx \rho^{-1}$ and can explain the present (super?)acceleration of the Universe. To the author's knowledge, so far this is the only concrete example of cosmologies with negative $q$. We will come back to this point in Sec. \ref{qbounce}.

When considering a five-dimensional braneworld, bulk moduli modify the Friedmann equation on the brane. In general, to a given orbifolded 5D spacetime and matter source confined on the brane there will correspond a set of junction conditions determining the matter-gravity interaction at the brane position. Conversely, one can always construct a bulk stress tensor such that Eq. (\ref{FRW}) holds for some $q$ \cite{ChF}; this is because the junction conditions have enough ($q$-dependent) degrees of freedom at a fixed slice in order to arrange a suitable expansion. In fact, the braneworld alone is not sufficient to fully determine the observable physics and some fundamental principle (e.g., AdS/CFT correspondence) should be advocated from the outside in order to sweep all ambiguities away \cite{NO04}. Dealing not with such elegant principles, we shall keep the following discussion on a phenomenological level.

Anyway, a patch formulation of the cosmological problem provides a compact notation for many situations. If braneworld corrections are important in the early Universe, one can follow the cosmological evolution through each energy patch in a given time interval where the patch approximation is valid, that is, far from transitions between patches; in fact, the Hubble parameter will be a more or less complicated function of the energy density, say Eq. (\ref{gabo}), with smooth transitions from an energy (and SR) regime to the other. Moreover, the case of Eqs. (\ref{higher}) and (\ref{negative}) does not lie within the braneworld picture since the only requirement is that there exists some fundamental theory which modifies general relativity at the scales of interest. In this sense the patch formalism is more powerful and unifying than expected from the very simple ansatz Eq. (\ref{FRW}), in origin devised as a \emph{high}-energy correction typical of RS and GB braneworlds.

%%%%%%%%%%%%%%%%%%%%%%%%%%%%%%%%%%%%%%%%%%%%%%%%%%%%%%%%%%%%%%%%%%%%%%%%%%%%%%%%%%%%%%%%%%%%%%%%%%%%%%%%%%%%%%%%%%%%%%%%

\subsection{Patch cosmology}\label{patsetup}

In the rest of this work, we investigate the properties of a single flat energy patch with general effective Friedmann equation (\ref{FRW}). In order to simplify the framework, we make the following assumptions:
\begin{enumerate}
\item There is a confinement mechanism such that matter lives on the brane only, while gravitons are free to propagate in the bulk. This is guaranteed as long as $\rho<m_5^4$;
\item The contribution of the Weyl tensor is neglected.
\end{enumerate}
For a perfect fluid with equation of state $p=w\rho$, assumption 1 allows the continuity equation
\be \label{conti}
\dot{\rho}+3H (\rho+p) = 0\,.
\ee
This is equivalent to the local covariant conservation of the energy-momentum tensor.\footnote{For the case of five-dimensional inflation, see \cite{HS01,HTS,HS03,MHS}.} 

Assumption 2 closes the system of equations on the brane and sets aside the nonlocal contributions from the bulk. To neglect the projected Weyl tensor implies that there is no brane-bulk exchange. The converse is not true: Given a standard continuity equation on the brane, the Friedmann equation still can get an extra dark-radiation term. In fact, in the case of brane-bulk interaction through a nondiagonal stress-energy bulk tensor, the complete continuity equation is 
\be\label{bBex}
\dot{\rho}+3 H \rho(1+w)+ r_B=0\,,
\ee
where $r_B$ is the 05-component of the bulk tensor \cite{VDBL,VDMP}. Suppose that the Friedmann equation (\ref{FRW}) acquires a time-dependent correction
\be
H^2=\rho^q+\chi_B\,,
\ee
where we have set $\beta_q=1$ and $\chi_B(t)$ is an unspecified function. Then the equation of motion for $\chi_B$ reads
\be
\dot{\chi}_B+q[3H(1+w)+r_B/\rho]\chi_B=H^2[3qH(1+w)-2H\epsilon+qr_B/\rho]\,,
\ee
where 
\be\label{epsilon}
\epsilon \equiv-\frac{d \ln H}{d \ln a}=-\frac{\dot{H}}{H^2} \,.
\ee 
We conclude that even when the bulk is empty, $r_B=0$, $\chi_B(t)$ may not vanish identically. For some studies on brane-bulk interactions, see \cite{VDBL,VDMP,LSR,KKTTZ,LMS,BBC,ApT,tet04}.

The requirement of a negligible Weyl contribution might seem too restrictive and spoiling almost all the interesting features of the model. However, bulk physics mainly affects the small-scale/late-time cosmological structure and can be consistently neglected during inflation. This is a highly nontrivial result which has been confirmed with several methods both analytically and numerically \cite{GoM,GRS,IYKOM,LCML,koy03,KLMW}. Intuitively, the RS dark radiation term, which is the simplest contribution of the Weyl tensor, scales as $a^{-4}$ and is exponentially damped during the accelerated expansion. In the following we will set $\chi_B=0$.

Differentiating Eq. (\ref{FRW}) with respect to time and using Eq. (\ref{conti}), one gets
\be
\epsilon = \case32\, q(1+w)\,. \label{dotH0}
\ee
The exponent
\be\label{theta}
\theta \equiv 2\left(1-\frac{1}{q}\right),
\ee
is shown in Fig. \ref{fig2}; it allows to rewrite Eq. (\ref{FRW}) as $H^{2-\theta}=\beta_q^{2-\theta}\rho$. Actually the parameter $\theta$ can be extended to a function $\theta(\rho)$ interpolating between the RS and GB high-energy regimes \cite{SST}. This is another (although partial) justification for keeping the discussion in the general case $\theta \in {\mathbb R}\setminus \{2\}$, and not only in the discrete set $\theta \in \{0,1,-1\}$.
\begin{figure}[ht]
\bc
\ec
\caption{\it\label{fig2} The parameter $\theta (q)$. The three cosmologies described in the text are: high-energy Gauss-Bonnet [$\theta (2/3)=-1$], standard four-dimensional [$\theta (1) =0$] and high-energy Randall-Sundrum [$\theta (2)=1$].}
\end{figure}

If one imposes the dominant energy condition, $\rho \geq |p|$ ($w \geq -1$), Eq. (\ref{dotH0}) states that the Hubble length\footnote{The Hubble length is the proper distance from the observer of an object the observer sees moving with the cosmological expansion at the speed of light.} $R_\H\equiv H^{-1}$ is monotonic during its time evolution, increasing if $q>0$ (the lower branch with $\theta <2$) and decreasing if $q<0$ ($\theta>2$). On the contrary, the particle horizon $R_c(t)\equiv a(t)\int_{t_0}^t dt'/a(t')$ \cite{rin56}, which defines the causally connected region centered in the observer, is always increasing in an expanding universe, $\dot{R}_c=1+HR_c>0$, and its comoving counterpart is always increasing also, $(R_c/a)^{^{\bm .}}>0$. We will denote with a subscript $0$ any quantity evaluated at the reference initial time $t_0$. For a constant index $w$, 
\be \label{conw}
\rho=\rho_0\left(\frac{a}{a_0}\right)^{-3(1+w)},
\ee
while the scale factor is
\be
a(t) = a_0 \left[1+\epsilon H_0(t-t_0)\right]^{1/\epsilon}.
\ee
Thus an expanding ($H>0$) solution with $q<0$ and satisfying the dominant energy condition represents a superinflationary\footnote{With standard FRW equations, $q=1$, superinflationary models are those with $w<-1$ \cite{LM1}.} ($\dot{H}>0$) expanding universe in a ``pre big bang'' era with time running from $t_0$ to eventually $\bar{t}=t_0-1/(\epsilon H_0)>t_0$, where one encounters a singularity with infinite scale factor and vanishing energy density. Since any patch should be regarded as a model with a limited time interval of validity, its long-range evolution is a true problem only for well-established regimes. Anyway, we will consider only positive $q$, which is the case of all the known realistic cosmologies, and come back to this issue in Chapter \ref{dlts}.

For $q>0$, the Hubble length is always nondecreasing; therefore, $\epsilon = \dot{R}_\H \geq 0$. Moreover, we have a precise definition for the beginning of the inflationary era, since
\be
\frac{\ddot{a}}{a}=(1-\epsilon)H^2\,.
\ee
A necessary and sufficient condition for inflation to start is $\epsilon (t)< 1$, or, for the barotropic index
\be \label{w}
w=\frac{2}{3q}\,\epsilon-1\,,
\ee
$w(t)<2/(3q)-1$. The end of inflation is set by $\epsilon(t_\text{\tiny end})=1$.

Whenever cosmological equations can be applied both in the scalar and tachyon case, the inflaton field will be generically indicated as $\psi$. Table \ref{table1} summarizes the three main cosmological regimes. For completeness, we have also shown the de Sitter (dS) solution with constant Hubble parameter, when $H=\beta_0=\text{const}$, $\ddot{a}/a=\beta_0^2>0$, and inflation is driven by a cosmological constant with equation of state $w=-1$. Also, this case can be obtained via the formal limit $q \rightarrow 0$. The de Sitter regime is the idealization of the extreme slow-roll (ESR) approximation, $\dot{\psi} \approx 0$. In this regime, the kinetic term of the scalar action is subdominant with respect to the potential itself and $H \approx \beta_qV^{q/2}$.
\begin{table}[ht]
\bc\begin{tabular}{cccccc}
Regime &   $q$   &   $\theta$  &          $\beta_q^2$           & $w_\text{\tiny max}$ \\ \hline
dS     &    0    &   $\infty$  &             $H^2$              &                 $-1$        \\
GB     &  $2/3$  &      $-1$   & $(\kappa_5^2/16\alpha)^{2/3}$  &                 $0$         \\
RS     &    2    &       $1$   &      $\kappa_4^2/6\lambda$     &             $-2/3$          \\
4D     &    1    &       $0$   &         $\kappa_4^2/3$         &             $-1/3$          \\
\end{tabular}\ec
\caption{\label{table1}The energy regimes described in the text. The de Sitter case can be seen as the asymptotic cosmology with $q \rightarrow 0$. Here, $w_\text{\tiny max}$ is the maximum value for the barotropic index (\ref{w}) allowing inflation.}
\end{table}

%%%%%%%%%%%%%%%%%%%%%%%%%%%%%%%%%%%%%%%%%%%%%%%%%%%%%%%%%%%%%%%%%%%%%%%%%%%%%%%%%%%%%%%%%%%%%%%%%%%%%%%%%%%%%%%%%%%%%%%%%%%%%%%%%%%%%%%%%%%%%%%%%%%%%%%%%%%%%%%%%%%%%%%%%%%%%%%%%%%%%%%%%%%%%%%%%%%%%%%%%%%%%%%%%%%%%%%%%%%%%%%%%%%%%%%%%%%%%%%%

\section{The inflationary setup}

%%%%%%%%%%%%%%%%%%%%%%%%%%%%%%%%%%%%%%%%%%%%%%%%%%%%%%%%%%%%%%%%%%%%%%%%%%%%%%%%%%%%%%%%%%%%%%%%%%%%%%%%%%%%%%%%%%%%%%%%

\subsection{The ordinary scalar field $\phi$}

In the following we will consider an expanding four-dimensional FRW
flat universe filled with a minimally coupled homogeneous scalar field with energy density and pressure
\ba
\rho &=& \case{1}{2} \dot{\phi}^2 + V(\phi)\,, \label{rho}\\
p &=& \case{1}{2} \dot{\phi}^2 - V(\phi)\,, \label{p}
\ea
and effective equation of motion
\be
\ddot{\phi}+3H \dot{\phi}+ V'=0\,,\label{eom}
\ee
where $V(\phi)$ is the potential of $\phi$, dots stand for synchronous-time derivatives and a prime denotes $\phi$ derivative. From Eq. (\ref{dotH0}),
\be
\dot{H} = -\case{3}{2} q \beta_q^{2-\theta} H^{\theta} \dot{\phi}^2\,;\label{dotH}
\ee
equivalently, we can regard $H$ as a function of $\phi$:
\be \label{hj2}
\frac{H'}{H^\theta} = -\case{3}{2} q \beta_q^{2-\theta} \dot{\phi}\,,
\ee
where the last passage is possible if $\phi$ varies monotonically with time.
Equations (\ref{FRW}), (\ref{conti}) and (\ref{rho}) then give
\be \label{hj3}
\frac{2}{(3q\beta_q^{2-\theta})^2}\frac{H'^2}{H^{2\theta}}-\left(\frac{H}{\beta_q}\right)^{2-\theta}+V=0\,.
\ee
Equations (\ref{FRW}), (\ref{hj2}) and (\ref{hj3}) are the cosmological equations in the Hamilton-Jacobi formulation; they are in agreement with the equations found in the low energy limit \cite{Mus90,SB}, in the Randall-Sundrum high-energy limit \cite{HaL1}, and in the Gauss-Bonnet high-energy limit \cite{MW}.

%%%%%%%%%%%%%%%%%%%%%%%%%%%%%%%%%%%%%%%%%%%%%%%%%%%%%%%%%%%%%%%%%%%%%%%%%%%%%%%%%%%%%%%%%%%%%%%%%%%%%%%%%%%%%%%%%%%%%%%%

\subsection{The cosmological tachyon $T$}
 
The deep interplay between small-scale nonperturbative string theory (especially in the effective Born-Infeld action formulation) and large-scale brane-world scenarios has raised the interest in a tachyon field as an inflationary mechanism \cite{MPP,ale02}. Subsequently, the problem has been studied in a more cosmological fashion \cite{gib02,FT,muk02,fei02,pad02,CGJP,FKS,SW,KL,ben02,sam03,SCQ,HN,PCZZ,LLH,CFM,WAS,BBS,HL,CF,gib03,GPCZ,PHZZ,AF,BSS,SV,GKMP,PS,GST,FK,GZ,KKKKL,DDK,rae04,yav04,GhM,CQ,sriv2,CPD,GST2,EQ,EJMM2}. Throughout this paper, a ``tachyon'' is by definition any scalar field $T$ with effective action $S=\int d^4x\, {\cal L}$ and Lagrangian \cite{sen4,gib02,gar00,ber00,klu00,GHY}
\be \label{ta}
{\cal L}= -V(T) \sqrt{-\det[g_{\mu\nu}-f(T) \partial_\mu T \partial_\nu T]}\,.
\ee
Here, $g_{\mu\nu}$ is the induced four-dimensional FRW metric on the brane, $T$ is a real scalar field with dimension $[T]=E^{-1}$, $f$ is a function of $T$, and $V$ is the potential, which is exact to all orders in the Regge slope $\alpha'$ but at the tree level in $g_s$. Without loss of generality we will assume that $f(T)=1$ and $T=T(t)$ is homogeneous and monotonic, say, $\dot{T}>0$. In the case of a $D-\bar{D}$ system, the field $T$ is complex due to the Chan-Paton structure, but many of the following arguments will hold in this case too. Often, the tensor $G_{\mu\nu}\equiv g_{\mu\nu}- \partial_\mu T \partial_\nu T$ is called the tachyon metric. 

We will leave the exact form of the potential unspecified, except in Sec. \ref{exact}; if $V$ were constant, the model would correspond to a brane filled with a Chaplygin gas, $p = -V^2/\rho$, which at late times behaves as an effective cosmological constant (e.g., \cite{jac00,KMP}). Otherwise, in general the potential will have a maximum at $T_0=0$ and a local minimum $V(T_*)=0$  either at finite $T_*$ or at infinity. In the latter case, there are no oscillations and a reheating mechanism appears difficult \cite{KL,CFM}.

The tachyon energy density and pressure read, respectively,
\ba
\rho &=& \frac{V(T)}{c_S}\,,\label{Trho}\\
p &=& -V(T) c_S=-\frac{V^2(T)}{\rho}\,,
\ea
where
\be \label{cS}
c_S \equiv \sqrt{-w}=\sqrt{1-\dot{T}^2}\,,
\ee
is the speed of sound. Note that when $\dot{T}^2 \rightarrow 1$, the tachyon behaves as a pressureless gas. The continuity equation (\ref{conti}) gives the equation of motion
\be \label{Teom}
\frac{\ddot{T}}{1-\dot{T}^2}+3H\dot{T}+U'=0\,,
\ee
where $U \equiv \ln V(T)$ is differentiated with respect to $T$. Equation (\ref{dotH0}) then gives
\be \label{Thj2}
\frac{H'}{H^2} = -\case{3}{2} q \dot{T}\,.
\ee
By this equation and Eqs. (\ref{FRW}) and (\ref{Trho}), we have
\be \label{Thj3}
\frac{4}{(3q\beta_q^{2-\theta})^2}\frac{H'^2}{H^{2\theta}}-\left(\frac{H}{\beta_q}\right)^{2(2-\theta)}+V^2=0\,.
\ee
Equations (\ref{FRW}), (\ref{Thj2}) and (\ref{Thj3}) are the Hamilton-Jacobi equations for the tachyon; they agree with \cite{MW,FT}.

We can give a physical interpretation of the tachyon Lagrangian (\ref{ta}), that in our case is
\be \label{Lagr}
{\cal L}= - a^3V\sqrt{1- 2\epsilon/(3q)}\,.
\ee
The first thing to note is that, if $V\neq 0$, the Lagrangian is defined only for $\epsilon<3q/2$. This implies that all the cosmologies with $q \leq 2/3$ and a tachyon on the brane with the above nonzero effective Lagrangian experience an accelerated expansion,\footnote{This may be an early-Universe inflationary phase as well as the present acceleration period.} while those with $q>2/3$ can be either accelerating or decelerating, depending on the evolution of $\epsilon$. The Gauss-Bonnet high-energy regime is the limiting case; this fact suggested a scenario with an interesting role for the tachyon \cite{LN}, which however seemed to have some problems \cite{PS}. Suppose that $q \leq 2/3$ and, at some time $t_*$, the accelerated phase stops, $\epsilon(t_*)=3q/2$: then the tachyon action vanishes. In a string-theoretical framework, when the tachyon reaches the minimum of the potential, the unstable $D$-brane on which it lives annihilates and decays into the closed string vacuum. Put into another language, in the limit $\epsilon\to 3q/2$, the tachyon metric becomes asymptotically Carrollian, $G_{\mu\nu} \sim -a^2\, \text{diag} (0,1,1,1)$. This property, called Carrollian confinement \cite{gib03,GiHY}, holds for other tachyon effective metrics. Since in the Carroll limit there is no signal propagation, again the string interpretation is that no open tachyonic modes can propagate after the condensation (see also \cite{GuS}).

On the other hand, it can be seen that the vanishing of the Lagrangian (\ref{Lagr}) is not the end of the story by reformulating the theory in the canonical formalism \cite{GHY}. Defining the conjugate comoving momentum density
\be \label{momP}
\Pi_T \equiv \frac{1}{\sqrt{-g}}\frac{\partial {\cal L}}{\partial \dot{T}} = \frac{V\dot{T}}{\sqrt{1-\dot{T}^2}}=\rho\dot{T}\,,
\ee
the density Hamiltonian ${\cal H}=\rho$ in the canonical variables is 
\be
{\cal H} = \Pi_T\dot{T}-{\cal L}= \sqrt{\Pi_T^2 + V^2}\,,
\ee
which is well defined in the condensation limit. Moreover, in string theory the absence of perturbative open modes translates to the fact that, near the minimum of the potential, the string coupling $g_s =O(1)$, and the effective action description might fail down. Possibly, in a cosmological-brane context the vanishing of the tachyon action is a fictitious effect coming from the simplified FRW equation (\ref{FRW}) and the associated dynamics. Actually, a more realistic model would have some implemented mechanism by which, and depending on the position of the minimum of the potential, the consequent cosmological evolution would experience a (pre) reheating phase, or a transition to a scalar-driven inflation, in a time interval centered in $t_*$. Similar considerations hold when $q > 2/3$ and the Hubble parameter goes through a boost of the growth rate, recovering late post-inflationary cosmology.

Soon after the first proposal by Gibbons \cite{gib02}, it became clear that the cosmology based upon a rolling tachyon 
suffers from a number of other problems, including a small number of $e$-foldings, a difficulty of reheating, and a large amplitude for density perturbations, that can be traced back to some fine-tuning requirements on the parameters of the model \cite{CGJP,FKS,KL,SCQ,PCZZ,BBS,BSS,PS,rae04}. Lately it was shown in \cite{GST} that the problem of large density perturbations is solved by considering a small warp factor in a warped metric. In addition, the problem of reheating is overcome by accounting for a negative cosmological constant which may appear by the stabilization of modulus fields \cite{KKLT}. Finally, tachyon inflation can help hybrid inflation to take place with natural initial conditions. By these reasons, it is premature to exclude the tachyon as a candidate for inflation.

%%%%%%%%%%%%%%%%%%%%%%%%%%%%%%%%%%%%%%%%%%%%%%%%%%%%%%%%%%%%%%%%%%%%%%%%%%%%%%%%%%%%%%%%%%%%%%%%%%%%%%%%%%%%%%%%%%%%%%%%%%%%%%%%%%%%%%%%%%%%%%%%%%%%%%%%%%%%%%%%%%%%%%%%%%%%%%%%%%%%%%%%%%%%%%%%%%%%%%%%%%%%%%%%%%%%%%%%%%%%%%%%%%%%%%%%%%%%%%%%

\section{The slow-roll formalism}

According to the inflationary idea, an era of accelerated expansion is driven by a scalar field slowly ``rolling'' down its potential into a local minimum. The use of the slow-roll formalism \cite{ST,LL,KV,LPB} simplifies the study of many consequences of inflation; however, it can also be considered as an effective notation for some recurrent dimensionless combinations of cosmological quantities, without imposing any condition on their magnitude. We will keep calling these parameters ``slow-roll'' in this case, too. The most commonly used SR towers rely upon two different quantities, the geometrical Hubble parameter $H$ and the dynamical inflaton potential $V$. We will name these towers H-SR and V-SR, respectively, and explore some of their properties in the general cosmology (\ref{FRW}). Other SR towers can be constructed for particular cosmological scenarios or analyses \cite{LN,STG,kin02,HaL2,RL,RL2,RL3}. We will also consider what happens in the case of a tachyonic field.

The notation we will use is the following. A subscript $V$ will denote the V-SR parameters, while the inflaton field will be indicated by its symbol as a subscript. Expressions valid for both the scalar fields will bear no subscript, save eventually $V$.

%%%%%%%%%%%%%%%%%%%%%%%%%%%%%%%%%%%%%%%%%%%%%%%%%%%%%%%%%%%%%%%%%%%%%%%%%%%%%%%%%%%%%%%%%%%%%%%%%%%%%%%%%%%%%%%%%%%%%%%%

\subsection{H-SR parameters for an ordinary scalar field}

The H-SR tower is defined as
\bs\label{hsr}
\ba
\epsilon_{\text{\tiny $\phi$},0} &\equiv& \epsilon\,,\\
\epsilon_{\text{\tiny $\phi$},n} &\equiv& \prod_{i=1}^n \left\{-\frac{d \ln \left[(H'H^{-\theta})^{(i-1)}\right]}{d \ln a}\right\}^{1/n}, \qquad n \geq 1\,,
\ea
\es
where $(n)$ is the $n$th $\phi$ derivative. For a scalar field, the first three parameters, which are those appearing in all the main expressions for cosmological observables, are
\ba
\epsilon_\S &\equiv& \epsilon_{\text{\tiny $\phi$},0} =3q \frac{\dot{\phi}^2/2}{V+\dot{\phi}^2/2}\,, \label{phepsilon}\\
\eta_\S     &\equiv& \epsilon_{\text{\tiny $\phi$},1}   = -\frac{d \ln \dot{\phi}}{d \ln a}=-\frac{\ddot{\phi}}{H\dot{\phi}} \label{eta}\,,\\
\xi^2_\S    &\equiv& \epsilon_{\text{\tiny $\phi$},2}^2 =  \frac{1}{H^2} \left(\frac{\ddot{\phi}}{\dot{\phi}}\right)^. = \frac{\dddot{\phi}}{H^2\dot{\phi}}- \eta_\S^2\,.\label{xi}
\ea
The condition $\epsilon \ll 1$ (ESR regime) permits to neglect the first term in the left-hand side of Eq. (\ref{hj3}), that can be recast as 
\be \label{hjalt}
V(\phi)= \left(1-\frac{\epsilon_\S}{3q}\right)\left(\frac{H}{\beta_q}\right)^{2-\theta},
\ee
while $|\eta_\S| \ll 1$ is equivalent to assume the attractor solution $\dot{\phi} \approx -V'/3H$ from Eq. (\ref{eom}). During our calculations, formulas containing the smallest power of any of these parameters will be referred to as ``first order SR.'' At first order SR it is possible to drop the second derivative in the equation of motion (\ref{eom}), the Hubble parameter can be considered almost constant, and the expansion becomes nearly exponential. This approximation is precisely the ESR one.

Consequently, all the dynamical information is encoded in the SR parameters. Equation (\ref{dotH}) can be rewritten in terms of $\epsilon$, giving
\be \label{useful1}
\dot{\phi}^2=\frac{2\epsilon_\S}{3q}\left(\frac{H}{\beta_q}\right)^{2-\theta}.
\ee
Thus the scalar field behaves almost like an effective cosmological constant in the SR approximation, $w \gtrsim -1$. 

Noting that $\ddot{H}=-H\dot{H}(\theta\epsilon_\S+2\eta_\S)$, we have
\ba
\dot{\epsilon}_\S &=& 2H\epsilon_\S \left(\frac{1}{q}\,\epsilon_\S-\eta_\S\right),\label{epsih'}\\
\dot{\eta}_\S     &=&  H\left(\epsilon_\S\eta_\S-\xi_\S^2\right).\label{etah'}
\ea
Differentiation with respect to the scalar field yields $\epsilon_{\text{\tiny $\phi$},n}'=\dot{\epsilon}_{\text{\tiny $\phi$},n}/\dot{\phi}\,$; by Eq. (\ref{useful1}), the resulting prefactor $H/\dot{\phi}$ can be expressed as
\be \label{usefull}
\frac{H}{\dot{\phi}}=+\left(\frac{3q\beta_q^{2-\theta}}{2} \frac{H^\theta}{\epsilon_\S}\right)^{1/2},
\ee
where the plus sign has been chosen in order to have a slow rolling down the potential with $\dot{\phi}>0$. This is always possible by a redefinition $\phi \rightarrow -\phi$.

A final comment is in order: when defined, a SR tower is dynamical (i.e., does say something about the dynamics of the Hamilton-Jacobi equations) either when constraints on the form and magnitude of the SR parameters are applied, or when distinct SR definitions are related through the Hamilton-Jacobi equations themselves. For example, the H-SR tower relies on the parameter $\epsilon$, which is its fundamental ground; as far as one does not assume any specific link between the Hubble parameter (and its derivatives) and the fields living on the brane, it is clear there will be no knowledge about the evolution of the system. However, when rewriting these H-parameters in terms of $\dot{\phi}$ through the second Hamilton-Jacobi equation, these parameters become dynamical. This ambiguity may lead to confusion in some situations; an interesting discussion on related issues can be found in \cite{lid03}.

%%%%%%%%%%%%%%%%%%%%%%%%%%%%%%%%%%%%%%%%%%%%%%%%%%%%%%%%%%%%%%%%%%%%%%%%%%%%%%%%%%%%%%%%%%%%%%%%%%%%%%%%%%%%%%%%%%%%%%%%

\subsection{V-SR parameters for an ordinary scalar field}\label{vsrph}

The H-SR hierarchy is an elegant instrument of analysis coming from the Hamilton-Jacobi formulation of the equations of motion. However, in many cases investigation starts from the inflaton potential $V$ and not from the Hubble parameter, whose shape must be determined by the Hamilton-Jacobi equations which are not always readily solvable. So, it is convenient to define another SR tower and try to relate it to the original one, namely, 
\bs \label{vsr}\ba
\epsilon_{\text{\tiny $\phi V$},0} &\equiv& \frac{q}{6\beta_q^2}\frac{V'^2}{V^{1+q}}\,,\\
\epsilon_{\text{\tiny $\phi V$},n} &\equiv& \frac{1}{3\beta_q^2}\left[\frac{V^{(n+1)}(V')^{n-1}}{V^{nq}}\right]^{1/n}, \qquad n \geq 1\,,
\ea\es
where again we have introduced the first parameter by hand. Therefore \cite{LT},
\ba
\epsilon_\sV &\equiv& \epsilon_{\text{\tiny $\phi V$},0}\,,\\
\eta_\sV     &\equiv& \epsilon_{\text{\tiny $\phi V$},1} =\frac{1}{3\beta_q^2}\frac{V''}{V^q}\,,\\
\xi_\sV^2    &\equiv& \epsilon_{\text{\tiny $\phi V$},2}^2=\frac{1}{(3\beta_q^2)^2}\frac{V'''V'}{V^{2q}}\,,
\ea
and their derivatives with respect to the scalar field are
\ba
\epsilon_\sV' &=& -q \frac{V'}{V} \left[\left(1+\frac{1}{q}\right)\epsilon_\sV-\eta_\sV\right],\label{epsiv'}\\
\eta_\sV'     &=& -\frac{q}{2\epsilon_\sV} \frac{V'}{V} \left[2\epsilon_\sV\eta_\sV-\xi_\sV^2\right],
\ea
where
\be \label{usefulV}
\frac{V'}{V} = -\left(\frac{6\beta_q^2}{q} \epsilon_\sV V^{q-1}\right)^{1/2}.
\ee
The conditions $\epsilon_\V \ll 1$ and $|\eta_\V| \ll 1$ are necessary to drop the kinetic term in Eq. (\ref{FRW}) and the acceleration term in Eq. (\ref{eom}), but they are not sufficient. In general, this SR formalism requires a further assumption, namely, $\dot{\phi} \approx -V'/3H$, which is easy enough to be satisfied. This determines the minus sign in Eq. (\ref{usefulV}), provided $\dot{\phi}>0$.

%%%%%%%%%%%%%%%%%%%%%%%%%%%%%%%%%%%%%%%%%%%%%%%%%%%%%%%%%%%%%%%%%%%%%%%%%%%%%%%%%%%%%%%%%%%%%%%%%%%%%%%%%%%%%%%%%%%%%%%%

\subsection{H-SR parameters for a tachyon}

In the tachyonic case, the first H-SR parameters are [see Eq. (\ref{epsilon})]
\ba
\epsilon_\T &=& \case{3}{2}q \dot{T}^2\,, \label{Tepsilon}\\
\tilde{\eta}_\T   &=&-\frac{\ddot{T}}{H\dot{T}}-\frac{\dot{T}}{H}\left(\frac{V'}{V}+\frac{1}{1-\dot{T}^2}\right)\,.\label{tradTeta}
\ea
Equation (\ref{Tepsilon}) shows that tachyonic inflation is similar to $k$-inflation \cite{ADM}. The condition $\epsilon_\T \ll 1$ corresponds to neglect the derivative term in Eq. (\ref{Thj3}) and set $H^2\approx \beta_q^2V^q$: using Eqs. (\ref{FRW}), (\ref{Trho}) and (\ref{Tepsilon}), one gets
\be \label{Thjalt}
V^2(T)=\left(1-\frac{2\epsilon_\T}{3q}\right)\left(\frac{H}{\beta_q}\right)^{2(2-\theta)}.
\ee
However, the expression for $\eta$ is not very precise from a dynamical point of view because the equation of motion has now a factor $1/(1-\dot{T}^2)$, attached to the second derivative, that should be taken into account when neglecting the acceleration term. This suggests to redefine the SR tower by introducing ``covariant derivation'' with respect to the tachyon metric:
\bs \label{thsr}\ba
\bar{\epsilon}_{\text{\tiny $T$},0} &\equiv& \bar{\epsilon} \equiv \epsilon\,,\\
\bar{\epsilon}_{\text{\tiny $T$},n} &\equiv& \prod_{i=1}^n \left\{-\frac{1}{1-\dot{T}^2}\frac{d \ln \left[\left(\frac{\sqrt{1-\dot{T}^2}}{V}\frac{H'}{H^\theta}\right)^{(i-1)}\right]}{d \ln a}\right\}^{1/n}\\
&=& \prod_{i=1}^n \left\{\frac{1}{w}\frac{d \ln \left[\left(\sqrt{\bar{\epsilon}_{\text{\tiny $T$},0}}\right)^{(i-1)}\right]}{d \ln a}\right\}^{1/n}, \qquad n \geq 1\,,\label{thsrc}
\ea\es
where in the last passage we have used Eqs. (\ref{Tepsilon}) and (\ref{Thj2}). From Eq. (\ref{thsr}) we have
\ba
\bar{\eta}  &\equiv& \bar{\epsilon}_{\text{\tiny $T$},1} = -\frac{1}{1-\dot{T}^2} \frac{\ddot{T}}{H\dot{T}}\,,\label{Tbareta}\\
\bar{\xi}^2 &\equiv& \bar{\epsilon}_{\text{\tiny $T$},2}^2 =  \frac{1}{1-\dot{T}^2}\frac{1}{H^2} \left(\frac{\ddot{T}}{\dot{T}}\right)^. = \frac{1}{(1-\dot{T}^2)^2}\frac{\dddot{T}}{H^2\dot{T}}- \bar{\eta}^2\,.\label{Tbarxi}
\ea
Since $\dot{T} \propto \bar{\epsilon}^{1/2}$, one can express any $T$ derivative as a time derivative with a purely geometrical factor in front. For example, $\bar{\epsilon}_{\text{\tiny $T$},n}'= \dot{\overline{\epsilon}}_{\text{\tiny $T$},n}\sqrt{3q/(2\bar{\epsilon})}$. These expressions carry an extra contribution due to the adopted SR definition, by which $\bar{\eta}\approx O[\bar{\epsilon}(1+\bar{\epsilon}+\cdots)]$. If one wants to keep the spirit of the SR expansion, and neglect by definition these next-to-lowest order terms, one may trade Eqs. (\ref{thsr}) and (\ref{hsr}) for an intermediate definition, by dropping the overall factor in Eq. (\ref{thsrc}),
\bs \label{interthsr}\ba
\epsilon_{\text{\tiny $T$},0} &\equiv& \epsilon\,,\\
\epsilon_{\text{\tiny $T$},n} &\equiv& \prod_{i=1}^n \left\{-\frac{d \ln \left[(H'H^{-2})^{(i-1)}\right]}{d \ln a}\right\}^{1/n}.
\ea\es
Applications of this SR tower will be seen in the following sections. Since $\eta_\T=-\ddot{T}/(H\dot{T})$, one gets
\ba
\dot{\epsilon}_\T &=& -2 H \epsilon_\T \eta_\T\,,\label{Tdotepsi}\\
\dot{\eta}_\T &=& H\left(\epsilon_\T\eta_\T-\xi_\T^2\right)\,.\label{Tdoteta}
\ea

%%%%%%%%%%%%%%%%%%%%%%%%%%%%%%%%%%%%%%%%%%%%%%%%%%%%%%%%%%%%%%%%%%%%%%%%%%%%%%%%%%%%%%%%%%%%%%%%%%%%%%%%%%%%%%%%%%%%%%%%

\subsection{V-SR parameters for a tachyon} \label{vsrT}

From the SR approximation $\dot{T} \approx -U'/3H$ and Eqs. (\ref{Tepsilon}) and (\ref{Tdotepsi}), we can guess the SR parameters as functions of $V$:
\ba
\epsilon_\tV &\equiv& \frac{q}{6\beta_q^2} \frac{U'^2}{V^q}\,,\\
\eta_\tV &\equiv& -\epsilon_\tV + \frac{1}{3\beta_q^2} \frac{U''}{V^q}\,.
\ea
The complete SR tower comes from the Hubble tower by substituting $\epsilon_\T$ with $\epsilon_\tV$ and putting $H=\beta_q V^{q/2}$, thus getting
\bs \label{Tvsr}\ba
\epsilon_{\text{\tiny $TV$},0} &\equiv& \epsilon_\tV\,,\\
\epsilon_{\text{\tiny $TV$},n} &\equiv& \frac{1}{3\beta_q^2}\left[\frac{(U')^{n-1}}{V^{nq/2}}\left(\frac{U'}{V^{nq/2}}\right)^{(n)}\right]^{1/n}, \qquad n \geq 1\,.
\ea\es
Different SR parameters can be found in \cite{CGJP,SV}. Note that
\ba
\epsilon_\tV' &=& q U' \eta_\tV\,,\\
\eta_\tV' &=& qU'\left(\eta_\tV+\frac{\xi_\tV^2}{2\epsilon_\tV}\right)\,.
\ea

%%%%%%%%%%%%%%%%%%%%%%%%%%%%%%%%%%%%%%%%%%%%%%%%%%%%%%%%%%%%%%%%%%%%%%%%%%%%%%%%%%%%%%%%%%%%%%%%%%%%%%%%%%%%%%%%%%%%%%%%

\subsection{SR towers and energy dependence} \label{endep}

It is possible to relate the two SR towers by some simple energy-dependent relations. Here we will restrict ourselves to the first three parameters and define $f\equiv 1/3q$. From Eq. (\ref{hjalt}) we get the exact relation
\be \label{VH1}
\epsilon_\sV= \frac{\epsilon_\S}{9}\,\frac{(3-\eta_\S)^2}{(1-f\epsilon_\S)^{1+q}}\,.
\ee
Then, noting that $V'=\dot{\phi}H(\eta_\S-3)$ and 
\be\label{v''h}
V''=H^2 [3(\epsilon_\S+\eta_\S)-\eta_\S^2-\xi_\S^2]\,,
\ee
one has
\be \label{VH2}
\eta_\sV= \frac{(\epsilon_\S+\eta_\S)-\frac{1}{3}(\eta_\S^2+\xi_\S^2)}{(1-f\epsilon_\S)^q}\,.
\ee
Finally, noting that $V'''=-3(\theta \epsilon_\S^2+3\epsilon_\S\eta_\S+\xi_\S^2)H^3+O(\epsilon_\S^3)$, we obtain, to first H-SR order,
\bs\ba
\epsilon_\S &\approx& \epsilon_\sV\,,\\
\eta_\S     &\approx& \eta_\sV-\epsilon_\sV\,,\\
\xi_\S^2    &\approx& \xi_\sV^2-3\epsilon_\sV\eta_\sV+(3-\theta)\epsilon_\sV^2\,.
\ea\es
These equations allow us to shift from one hierarchy to the other, according to the most convenient approach. Both the SR towers show an explicit dependence on the energy scale because of the definitions, Eqs. (\ref{hsr}) and (\ref{vsr}). Sometimes, this energy dependence can be hidden by proper manipulations of the definitions; however, when differentiating SR parameters, Eqs. (\ref{epsih'}) and (\ref{epsiv'}), the resulting SR combinations contain some factor $q$. 

In the tachyon case, from Eqs. (\ref{Thjalt}), (\ref{Thj2}) and (\ref{Tepsilon}) one has
\be \label{TVH1}
\epsilon_\tV= \frac{\epsilon_\T}{(1-2f\epsilon_\T)^{q/2}}\left[1-\frac{\eta_\T}{6(1-2f\epsilon_\T)}\right]^2.
\ee
Then, using
\bs\ba
\epsilon_\T' &=& -H\sqrt{\frac{2\epsilon_\T}{f}}\eta_\T\,,\\
\epsilon_\T'' &=& \frac{H^2}{f} \left(\eta_\T^2+\xi_\T^2\right),\\
\eta_\T' &=& \frac{H}{\sqrt{2f\epsilon_\T}}\left(\epsilon_\T\eta_\T-\xi_\T^2\right),\\
H' &=& -\sqrt{\frac{\epsilon_\T}{2f}}H^2\,,
\ea\es
we get
\be
\eta_\tV= \frac{1}{(1-2f\epsilon_\T)^{q/2}}\left[\eta_\T+\frac{2\epsilon_\T\eta_\T-\eta_\T^2-\xi_\T^2}{6(1-2f\epsilon_\T)}-\frac{(24f+1)\epsilon_\T\eta_\T^2}{36(1-2f\epsilon_\T)^2}\right].
\ee
Hence, to first H-SR order, 
\bs\ba
\epsilon_\T &\approx& \epsilon_\tV\,,\\
\eta_\T     &\approx& \eta_\tV\,,\\
\xi_\T^2    &\approx& \xi_\tV^2+3\epsilon_\tV\eta_\tV\,.
\ea\es

%%%%%%%%%%%%%%%%%%%%%%%%%%%%%%%%%%%%%%%%%%%%%%%%%%%%%%%%%%%%%%%%%%%%%%%%%%%%%%%%%%%%%%%%%%%%%%%%%%%%%%%%%%%%%%%%%%%%%%%%

\subsection{All in a patch}

We can treat the ordinary scalar and the tachyon on the same ground by introducing the parameter
\bs\label{tilth}\ba
\wteta&=&\theta \qquad\text{for the ordinary scalar,}\\
\wteta&=&2 \qquad\text{for the tachyon.}
\ea\es
The Hamilton-Jacobi equations (\ref{hjalt}) and (\ref{Thjalt}) are equal up to a second-SR-order term. They read
\be
V(\psi) = \left(1-\frac{\epsilon}{3q}\right)\beta_q^{\theta-2}H^{2-\theta}(\psi)+\left(\theta-\wteta\right)O(\epsilon^2)\,,\label{hjpsi}
\ee
and
\be
H'(\psi)\,a'(\psi) = -\case{3}{2}\,q\beta_q^{2-\wteta}H^{\wteta+1}(\psi)\,a(\psi)\,,\label{hj}
\ee
where $\psi=\phi, T$.

We can see that one can construct the H-SR tower of the tachyon dynamics from the scalar-field H-SR tower and vice versa.
Equation (\ref{interthsr}) is formally the same as Eq. (\ref{hsr}) when expressed as a function of the velocity field $\dot{\psi}$:
\be \label{corresp}
\epsilon_{\text{\tiny $\psi$},n}^n = \prod_{i=1}^n \frac{-d \ln [\dot{\psi}^{(i-1)}]}{d \ln a}\,.
\ee
Then
\bs\ba
\epsilon &=& \frac{3q\beta_q^{2-\wteta}}{2} \frac{\dot{\psi}^2}{H^{2-\wteta}}=\frac{2}{3q}\frac{\beta_q^{\wteta-2}}{H^\wteta}\left(\frac{H'}{H}\right)^2\,,\label{psepsi}\\
\eta     &=&  -\frac{d \ln \dot{\psi}}
{d \ln a}=-\frac{\ddot{\psi}}{H\dot{\psi}} \label{Heta}\,,\\
\xi^2    &=&   \frac{1}{H^2} 
\left(\frac{\ddot{\psi}}{\dot{\psi}}\right)^\cdot.\label{Hxi}
\ea\es
The evolution equations of the parameters with respect to synchronous time are second-SR-order expressions,
\bs \label{dotSR}
\ba
\dot{\epsilon} &=& H\epsilon \left[\left(2-\wteta\right)\,\epsilon-2\eta\right],\label{dotSRa}\\
\dot{\eta}     &=&  H\left(\epsilon\eta-\xi^2\right).
\ea
\es

%%%%%%%%%%%%%%%%%%%%%%%%%%%%%%%%%%%%%%%%%%%%%%%%%%%%%%%%%%%%%%%%%%%%%%%%%%%%%%%%%%%%%%%%%%%%%%%%%%%%%%%%%%%%%%%%%%%%%%%%

\subsection{The horizon-flow parameters} \label{Hflow}

Other definitions of the SR tower may have only implicit energy dependence through Eq. (\ref{FRW}). For example, it may be convenient to introduce the horizon-flow (HF) parameters \cite{STG,kin02}, defined by 
%%%%%%%%%%%%
\be
\epsilon_0=\frac{H_{\rm inf}}{H}\,,\qquad
\epsilon_{i+1}=\frac{ d \ln |\epsilon_i|}{dN}\,,\qquad
i \ge 0\,,
\label{hflow}
\ee
%%%%%%%%%%%%
where $H_{\rm inf}$ is the Hubble rate at some chosen time and $N \equiv \ln (a/a_i)$ is the number of $e$-folds; here $t_i$ is the time when inflation begins.\footnote{\label{foot1} Note that our definition, which counts $N$ forward in time, is in accordance with \cite{STG}, where $N(t_i)=0$ and goes up to $N(t)>0$. This is in contrast with the ``backward'' definition of \cite{kin02}, where $N=\ln (a_f/a)$ is the number of remaining $e$-folds at the time 
$t$ before the end of inflation at $t_f$. In Secs. \ref{attractor} and \ref{models} we will adopt the backward notation.} As it was shown in \cite{lid03}, these parameters (and others similarly defined) do not properly encode inflationary dynamics even if they provide a good algorithm for reconstructing the inflationary potentials. In fact, because of the absence of the $1/n$ power, the definition (\ref{hflow}) does not permit a power truncation similar to that of the traditional SR towers, unless one imposes a constraint such as $\partial^{\bar{i}+1} H=0$ and $\partial^{\bar{i}}H \neq 0$, for some maximum $\bar{i}$.

The evolution equation for the HF parameters is given by
\be 
\dot{\epsilon}_i = H\epsilon_{i}\epsilon_{i+1}\,.
\ee
The HF parameters are related to the first SR parameters, as
\bs\ba
\epsilon_1 &=& \epsilon\,,\\
\epsilon_2 &=& \left(2-\wteta\right)\epsilon-2\eta\,,\\
\epsilon_2\epsilon_3 &=& \left(2-\wteta\right)^2\epsilon^2-2\left(3-\wteta\right) \epsilon\eta+2\xi^2\,.
\ea\es

%%%%%%%%%%%%%%%%%%%%%%%%%%%%%%%%%%%%%%%%%%%%%%%%%%%%%%%%%%%%%%%%%%%%%%%%%%%%%%%%%%%%%%%%%%%%%%%%%%%%%%%%%%%%%%%%%%%%%%%%
%%%%%%%%%%%%%%%%%%%%%%%%%%%%%%%%%%%%%%%%%%%%%%%%%%%%%%%%%%%%%%%%%%%%%%%%%%%%%%%%%%%%%%%%%%%%%%%%%%%%%%%%%%%%%%%%%%%%%%%%

\section{$e$-foldings and inflationary attractor} \label{attractor}

The number of $e$-foldings, defined as $N(t)=\int^{t_*}_t H(t') dt'$, measures the amount of inflation from the time $t$, when a perturbation with comoving wave number $k(t) =a(t) H(t)$ crosses the horizon, to the end of inflation at $t_*$. A typical ``good'' number of $e$-foldings is $\approx 50-70$ and many inflationary models have quite a larger total $N$. Sometimes it is useful to perform the integral in the cosmological field; from Eq. (\ref{psepsi}) one gets
\be \label{N}
N(t)=-\frac{3q}{2}\int_{\psi(t)}^{\psi_*}d\psi \frac{H^{\wteta+1}}{H'}= \int_{\psi(t)}^{\psi_*}d\psi\left(\frac{3q}{2} \frac{H^{\wteta}}{\epsilon}\right)^{1/2},
\ee
where $\beta_q=1$. Since $k(\psi)=H(\psi)a(\psi)=a_* H(\psi) \exp[N(\psi)]$, the logarithmic scale dependence of the field is exactly
\be
\frac{d \psi}{d\ln k}=\frac{\dot{\psi}}{(1-\epsilon)H}\,.
\label{scdep}
\ee
The predictiveness of inflation depends on the behaviour of cosmological solutions with different initial conditions. If there exists an attractor behaviour such that the differences of these solutions rapidly vanish, then the inflationary (and post-inflationary) physics will generate observables which are independent of the initial conditions. Let $H_o(\psi)>0$ be a generic expanding solution (denoted with the subscript $o$) of the Hamilton-Jacobi equation (\ref{hjpsi}) and consider a linear perturbation $\delta H(\psi)$ which does not reverse the sign of $\dot{\psi}>0$. From the linearized equation of motion, exactly in the SR parameters, the perturbation is
\be
\delta H(\psi) = \delta H(\psi_o)\,\exp \left\{\left(\frac{3q}{2}\right)^2\int_{\psi_o}^\psi d\psi \left[(2-\theta)\left(1+\frac{\theta}{3}\,\epsilon\right)\right]\frac{H_o^{\wteta+1}}{H_o'}\right\}.\label{attrac}
\ee
All linear perturbations are exponentially damped when the integrand is negative definite and, since $H_o'$ and $\dot{\psi}$ have opposing signs when $q$ is positive, this occurs when the term inside square brackets is positive. There are three cases:
\begin{enumerate}
	\item $0<\theta<2\,\,(q>1)\,:$ It must be $\epsilon>-3/\theta$. This condition is always satisfied, because $\epsilon$ is positive, and it means that all linear perturbations die away at least exponentially when inflationary solutions approach one another towards the attractor;
	\item $\theta=0\,\,(q=1)\,:$ The integrand is proportional to $H_o/H_o'<0$, and any linear perturbation is suppressed;
	\item $\theta<0\,\,(0<q<1)\,:$ The damping is achieved when $\epsilon<3/|\theta|$, that is for any inflationary solution with $q>2/5$.
\end{enumerate}
All these cases enclose previous computations in literature: \cite{SB,GPCZ} for the 4D cosmology, \cite{GZZ} for the Randall-Sundrum high-energy regime, and \cite{MW} for the full Gauss-Bonnet cosmology. By Eq. (\ref{N}), assuming the slow-roll approximation $\epsilon \approx \text{const}$, the inflationary attractor translates into the condition 
\be \label{damp}
\delta H(\psi) \approx \delta H(\psi_o) \exp \left[-\left(3+\theta\epsilon\right) N\right]\,.
\ee
For a given number of $e$-foldings and $\theta>0$ (RS case), we obtain an enhanced damping with respect to the 4D case, while for $\theta<0$ (GB case) the strength of the attractor is somehow milder. In the case $\theta>2$ ($q<0$), that is when $H_o'$ and $\dot{\psi}$ have concording signs, linear perturbations are suppressed when $\epsilon>-3/\theta$; both the sides of this inequality are negative and in general this relation will not be true. When it is satisfied, we obtain an accelerating universe with both decreasing Hubble length and energy density, that is a superinflationary universe. For completeness we note that, contrary to what happens in 4D cosmology, for general $q$ it is possible to have both a contracting scale factor and perturbation damping, as it is clear from Eq. (\ref{attrac}).

We will not address the issue of how efficient inflation can be; this problem has been studied by many authors under several perspectives. For instance, in the 4D regime, a tachyonic inflationary period turns out to be too short, with a number of $e$-folding $N =O(10)$ and an early nonlinear regime, $\delta\rho/\rho \gg 1$ \cite{CGJP,FKS,KL}. This has suggested the viability of a short tachyonic inflation as a means to provide natural initial conditions for a standard scalar inflationary period,\footnote{Then, this standard inflation lasts a sufficient number of $e$-folding and dilutes the perturbation structure generated by the tachyonic phase.} similarly to what happens in fast-roll inflation \cite{lin01}. In this sense, a tachyon is not sufficient by itself; nevertheless, the study of its dynamics is worth of investigation, since in other scenarios, such as Randall-Sundrum, things can go better than in the four-dimensional case \cite{BBS,BSS}. It is important to stress again that the analysis of this section is not sufficient to explore all these topics, since we have little constrained the physics involved. This would require the knowledge of the potential and, of course, the gravity framework; perhaps, the most dramatic lack is a condition stating when the confinement of the field on the brane is reliable. These considerations are particularly true in the Gauss-Bonnet scenario, in which the damping condition is critical; see, e.g., \cite{LN,PS}.

As a final remark, we note that Eq. (\ref{damp}) roughly encodes the effects coming from extra dimensions in a term proportional to $\theta$ inside the exponential. For one noncompact extra dimension, this term contributes at most $\pm N$ extra $e$-foldings, both the sign and magnitude depending on whether the bulk physics in a given energy regime either enhances or opposes the braneworld inflationary expansion. It would be interesting to interpret this result as a general feature of braneworld models and relate the parameter $|\theta|$ to the geometrical setup of the system (number of extra dimensions and noncompact directions, number of branes and their configuration, etc.); this check would require concrete gravity models with nonstandard Friedmann equations, which is beyond the scope of the present work.

%%%%%%%%%%%%%%%%%%%%%%%%%%%%%%%%%%%%%%%%%%%%%%%%%%%%%%%%%%%%%%%%%%%%%%%%%%%%%%%%%%%%%%%%%%%%%%%%%%%%%%%%%%%%%%%%%%%%%%%%%%%%%%%%%%%%%%%%%%%%%%%%%%%%%%%%%%%%%%%%%%%%%%%%%%%%%%%%%%%%%%%%%%%%%%%%%%%%%%%%%%%%%%%%%%%%%%%%%%%%%%%%%%%%%%%%%%%%%%%%

\section{Exact solutions} \label{exact}

So far we have left undetermined the form of the potential $V(\psi)$. Investigation with a few examples shows that, in general, there exists a mapping between scalar and tachyon potentials, in the sense that, chosen a time dependence for the scale factor $a(t)$, from the Hamilton-Jacobi equations (\ref{hjalt}) and (\ref{Thjalt}) there can be found potentials that solve exactly the cosmological equations in the two cases \cite{fei02,pad02,sam03}. We are going to see this in some detail in this section. The scheme to follow is: ($i$) from $a(t)$, find $H(t)$ and the first SR parameter; ($ii$) from Eqs. (\ref{hjalt}) and (\ref{Thjalt}), find $V(t)$; ($iii$) from Eqs. (\ref{useful1}) and (\ref{Tepsilon}), find $\psi (t)$ and the other SR parameters; ($iv$) substitute $t=t(\psi)$ to find $V(\psi)$. In general, the initial time $t_0$ will not be the origin of time because each solution will be exact in a given patch and not in the entire arc of time from the big bang singularity to, say, the end of inflation. For immediate reference, we summarize the classes of solutions found for the three main energy regimes in Tables \ref{table2}, \ref{table3}, \ref{table4} and \ref{table5}. The space of parameters is chosen in order to have positive $q$, positive potentials,\footnote{Negative potentials have been studied, e.g., in \cite{lin01,FFKL}. Note that the equation of motion of the tachyon possesses a symmetry $V \rightarrow -V$.} real inflaton fields, and a strictly expanding universe; contracting cases will be discussed briefly.

\begin{table}[ht]
\bc\begin{tabular}{ccc}
Regime &     $C\,\phi(t)$    &                   $B\,V(\phi)$              \\ \hline
GB     &      $t^{-1/2}$     &                     $\phi^6$                \\
4D     &      $\ln t/t_0$    &      $\exp (-\sqrt{2\kappa_4^2/n}\,\phi)$   \\
RS2    &       $t^{1/2}$     &                    $\phi^{-2}$              \\
\end{tabular}\ec
\caption{\label{table2} Exact cosmological solutions for an expanding scale factor $a(t) =t^n$ and an ordinary scalar field. Here, $n>1/3q>0$ and $\phi_0=0$. $B$  and $C$ are proportionality coefficients depending on $q$ and $n$.}
\end{table}

\begin{table}[ht]
\bc\begin{tabular}{cccc}
Regime &        $\gamma$     &  $C\,\phi(t)$  &                            $B\,V(\phi)$                      \\ \hline
GB     &         $3/2-n$     &  $t^{n-1/2}$   & $\left[1+D\phi^{2n/(1-2n)}\right]\,\phi^{6(n-1)/(2n-1)}$    \\
GB     &           $1$       &  $\ln t/t_0$   & $[1+D\exp(-\frac{C}{2}\phi)]\,\exp(-\frac{3C}{2}\phi)$       \\
4D     &         $1-n/2$     &   $t^{n/2}$    &      $(1+D\phi^{-2})\,\phi^{4(n-1)/n}$               \\
RS2    &          $1/2$      &   $t^{1/2}$    &             $(1+D\phi^{-2n})\,\phi^{2(n-1)}$                 \\
\end{tabular}\ec
\caption{\label{table3} Exact cosmological solutions for an expanding scale factor $a(t)=\exp (p t^n)$ and an ordinary scalar field, with $\gamma=n/2+(1-n)/q$. Here, $n<1$, $\text{sgn}(p)=\text{sgn}(n)$, and $\phi_0=0$. $B$, $C$ and  $D$ are proportionality coefficients depending on $q$, $n$ and $p$.}
\end{table}

\begin{table}[ht]
\bc\begin{tabular}{ccc}
Regime &     $C\,T(t)$    &       $B\,V(T)$           \\ \hline
GB     &      $t$         &       $T^{-3}$            \\
4D     &      $t$         &       $T^{-2}$            \\
RS2    &      $t$         &       $T^{-1}$            \\
\end{tabular}\ec
\caption{\label{table4} Exact cosmological solutions for an expanding scale factor $a(t) =t^n$ and a tachyon field. Here, $n>2/3q$. $B$  and $C$ are proportionality coefficients depending on $q$ and $n$.}
\end{table}

\begin{table}[ht]
\bc\begin{tabular}{ccc}   
Regime &       $C\,T(t)$      &                             $B\,V(T)$                        \\ \hline
GB     &      $t^{1-n/2}$     & $\left[1+DT^{2n/(n-2)}\right]^{1/2}\,T^{6(n-1)/(2-n)}$     \\
4D     &      $t^{1-n/2}$     & $\left[1+DT^{2n/(n-2)}\right]^{1/2}\,T^{4(n-1)/(2-n)}$     \\
RS2    &      $t^{1-n/2}$     & $\left[1+DT^{2n/(n-2)}\right]^{1/2}\,T^{2(n-1)/(2-n)}$     \\
\end{tabular}\ec
\caption{\label{table5} Exact cosmological solutions for an expanding scale factor $a(t)=\exp (p t^n)$ and a tachyon field. Here, $n>2/3q$. $B$, $C$  and $D$ are proportionality coefficients depending on $q$ and $n$.}
\end{table}

%%%%%%%%%%%%%%%%%%%%%%%%%%%%%%%%%%%%%%%%%%%%%%%%%%%%%%%%%%%%%%%%%%%%%%%%%%%%%%%%%%%%%%%%%%%%%%%%%%%%%%%%%%%%%%%%%%%%%%%

\subsection{Ordinary scalar field models}

The two classes of models we are going to study have been widely used in literature. Let us start with a power-law scale factor,
\be \label{apl}
a(t)= t^n\,, \qquad H=\frac{n}{t}\,,\qquad n >0\,.
\ee
The SR parameters are
\be \label{plsr}
\epsilon_\S =q\,\eta_\S = \sqrt{q}\,\xi_\S=\frac{1}{n}\,,
\ee
and the potential is
\be
V(t)= \left(1-\frac{1}{3qn}\right)\left(\frac{n}{\beta_q t}\right)^{2/q},\qquad n> \frac{1}{3q}\,.
\ee
Now, since $\dot{\phi}^2=2n^{1-\theta}/(3qt^{2-\theta})$, we must discuss the case $\theta=0$ ($q=1$) separately. If $0<q \neq 1$, then
\be
\phi(t)=\frac{2}{\theta} \left(\frac{2}{3qn}\right)^{1/2}\left(\frac{n}{\beta_q}\right)^{1/q}\,t^{\theta/2}\,,
\ee
and
\be \label{v1}
V(\phi) = A_{q,n}\,\phi^{-4/(q\theta)}\,,
\ee
where $A_{q,n}$ is a coefficient depending on $q$ and $n$. Note that the potential is divergent in $\phi=0$ if $q>1$. 

If $q=1$, we obtain the 4D power-law model \cite{AbW,LM2},
\ba
\phi(t) &=& \phi_0+ \left(\frac{2n}{3\beta_1^2}\right)^{1/2} \ln \left(\frac{t}{t_0}\right),\\
V(\phi) &=& \left(1-\frac{1}{3n}\right)\left(\frac{n}{\beta_1 t_0}\right)^2\,\exp\left[-\left(\frac{6\beta_1^2}{n}\right)^{1/2}(\phi-\phi_0)\right].\label{powl}
\ea
There are no contracting solutions.

Now, consider a scale factor of the form
\be \label{aexp}
a(t)= \exp (p\, t^n)\,,\qquad H=pn\,t^{n-1}\,,\qquad \text{sgn}(p)=\text{sgn}(n)\,,
\ee
with
\be
V(t)=\left(1+\frac{n-1}{3qpn}\,t^{-n}\right)\left(\frac{pn}{\beta_q}\right)^{2/q}t^{n-2\gamma}\,,
\ee
where $2\gamma=n+2(1-n)/q$; again, $\dot{\phi} =A_{q,p,n} t^{-\gamma}$, with 
\be \label{coeff}
A_{q,p,n} =\left(\frac{2(1-n)\,(pn)^{1-\theta}}{3q\beta_q^{2/q}}\right)^{1/2},
\ee
real if $q>0$ and $n<1$. So,
\be \label{sx}
\epsilon_\S  =\frac{1-n}{pn}\,t^{-n}\,,\qquad \eta_\S =\frac{\gamma}{pn}\,t^{-n}\,,\qquad \xi_\S^2 = \frac{\gamma}{p^2n^2}\,t^{-2n}\,.
\ee
Note that the SR parameters decrease in time and inflation does not naturally end. The reality of the coefficient (\ref{coeff}) guarantees the weak energy condition ($\rho+p \geq 0$, $\rho \geq 0$) if $t_0>\sqrt[n]{(1-n)/(3qpn)}$; from Eq. (\ref{sx}) it then follows that we get inflation from the very beginning only if $q<1/3$. Same considerations are applied for the tachyonic counterpart, with an additional factor of 2 inside the root and a condition $q<2/3$. If $\gamma \neq 1$, then
\ba
\phi(t) &=& \frac{A_{q,p,n}}{1-\gamma}\,t^{1-\gamma}\,,\\
V(\phi) &=& \left[B_{q,p,n}+C_{q,p,n} \phi^{n/(\gamma-1)}\right]\,\phi^{(n-2\gamma)/(1-\gamma)}\,.
\ea
In particular, $q=2$, $\gamma=1/2$ corresponds to the solution for the Randall-Sundrum regime,  while for $q=1$, $\gamma-1=-n/2$, one recovers the 4D intermediate inflation of \cite{bar90,BS}.

If $\gamma =1$, then $0\neq\theta \neq 1$, $n=\bar{n}=\theta/(\theta-1)$ and
\ba
\phi (t) &=& \phi_0 +\bar{A}_{q,p} \ln \left(\frac{t}{t_0}\right)\,,\\
V(\phi) &=& \left\{\bar{B}_{q,p}+\bar{C}_{q,p} \exp \left[-\frac{\bar{n}}{\bar{A}_{q,p}}(\phi-\phi_0)\right]\right\} \nonumber\\
 && \times\exp \left[-\frac{2(1-\bar{n})}{q\bar{A}_{q,p}}(\phi-\phi_0)\right]\,.
\ea
This solution can be applied to just one physically known case, namely, the Gauss-Bonnet regime, with $\bar{n}=1/2$. The contracting solutions are: $p<0$, $0<n\neq1$ (the case $\gamma=1$ is possible only when $0\neq q<1$, $q>2$); $p>0$, $n<0$ (the case $\gamma=1$ is possible only when $1<q<2$). 

%%%%%%%%%%%%%%%%%%%%%%%%%%%%%%%%%%%%%%%%%%%%%%%%%%%%%%%%%%%%%%%%%%%%%%%%%%%%%%%%%%%%%%%%%%%%%%%%%%%%%%%%%%%%%%%%%%%%%%%%

\subsection{Tachyon field models}\label{lastref}

With the power law (\ref{apl}), the tachyon field is
\be
T(t) =  \left(\frac{2}{3qn}\right)^{1/2}\,t\,,\qquad \text{sgn}(n)=\text{sgn}(q)\,,
\ee
and the SR parameters read
\be
\epsilon_\T =\frac{1}{n}\,,\qquad \eta_\T = \xi_\T= \cdots=0\,;
\ee
the potential is
\ba 
V &=& \left(1-\frac{2}{3qn}\right)^{1/2}\left(\frac{n}{\beta_q t}\right)^{2/q}\\
  &=& \left(1-\frac{2}{3qn}\right)^{1/2}\left(\frac{2n}{3q\beta_q^2}\right)^{1/q} T^{-2/q}\,,\qquad n> \frac{2}{3q}>0\,.\label{TVpl}
\ea
In order to connect this cosmological solution with string theory, we must take care both of the maximum and the minimum of the potential. As regards the maximum at $T_0=0$, if $q>0$ then the potential (\ref{TVpl}) diverges; as it was shown in \cite{pad02}, it is possible to regularize $V$ and keep an approximated power-law scale factor (\ref{apl}). However, the constance of the kinetic term, $\dot{T} =\sqrt{2/3qn}<1$, which does not satisfy the conditions $\dot{T}(t_0)=0$ and $\dot{T}(t_*)=1$, suggests to regard this solution as an ``intermediate time'' model describing the rolling of the tachyon down its potential, between the very beginning and the asymptotic regime with a pressureless tachyon dust and $n=2/(3q)$. The power-law scalar model with constant SR parameters, Eq. (\ref{plsr}), suffers from the same graceful-exit problem. 

In the case of an exponential scale factor, Eq. (\ref{aexp}), the first SR parameters are
\be \label{tx}
\epsilon_\T  =\frac{1-n}{pn}\,t^{-n}\,,\qquad \eta_\T =\frac{1}{2p}\,t^{-n}\,,\qquad \xi_\T^2 = \frac{1}{2p^2n}\,t^{-2n}\,,
\ee
and the potential is
\be
V(t) = \left[1+\frac{2(n-1)}{3qpn}\,t^{-n}\right]^{1/2}\left(\frac{pn}{\beta_q}\right)^{2/q}t^{2(n-1)/q}\,.
\ee
Since $\dot{T}^2 = [2(1-n)/(3qpn)]\,t^{-n}$, one has a real expanding solution when: $p>0$ and $0<n<1$; $p<0$ and $n<0$. If $n \neq 2$, the solution is	
\ba
T(t) &=& \frac{2}{2-n}\left[\frac{2(1-n)}{3qpn}\right]^{1/2} t^{1-n/2}\,,\\
V(T) &=& \left[B_{q,p,n}+C_{q,p,n} T^{2n/(n-2)}\right]^{1/2}\,T^{4(n-1)/[q(2-n)]}\,.
\ea
If $n=2$, then $\dot{T}^2=-t^{-2}/(3qp)$ and $p<0$. The solution is
\ba
T(t) &=& T_0+\left(\frac{-1}{3qp}\right)^{1/2} \ln \left(\frac{t}{t_0}\right)\,,\\
V(T) &=& \left\{1+\frac{1}{3qpt_0^2}\,\exp [-2\sqrt{-3qp}\,(T-T_0)]\right\}^{1/2}\nonumber\\
 && \times\left(\frac{2pt_0}{\beta_q}\right)^{2/q}\exp \left[2\sqrt{\frac{-3p}{q}}\,(T-T_0)\right]\,.
\ea
In the three cosmologies of interest, this solution is contracting. Other exact models can be found in \cite{SV,GKMP}. 

It is possible to relate the solutions of the exponential model (\ref{aexp}) to those of the power-law model. In the former case, the dynamical equations are $V \propto (1+A\,t^{-n})\,t^{2(n-1)/q}$ and $\dot{\psi} \propto t^{-\lambda}$, where $\lambda=-(n/2)+(n-1)/q$ for the scalar field and $\lambda=-n/2$ for the tachyon. In the limit $n \rightarrow 0$, that is when the index of the equation of state $w \rightarrow \text{const}$, both the models formally approach the power-law solution with $V \propto t^{-2/q}$ and $\dot{\phi} \propto t^{-1/q}\,,$ $\dot{T} \propto \text{const}$.

% version:  January 19, 2005

\newpage
\thispagestyle{empty}
$\vphantom{dunno how to skip the page}$
\newpage

\chapter{Cosmological perturbations and braneworld spectra} \label{obs}

\begin{quote}
\textsl{Philolaus puts fire in the middle, around the centre, which he calls furnace of everything and abode of Zeus and mother of the gods and altar and junction and measure of nature. And then another fire at the top, surrounding the whole.}\\ --- A\"{e}tius (ed. H. Diels), \textit{Doxographi graeci}, II 7,7
\end{quote}\vspace{1cm}

The advantage of combining the cosmological patch approach with the SR formalism is to provide, at least in certain situations, a unique treatment of physical phenomena for a number of energy regimes. In this chapter we show an example of this mechanism by discussing the spectra of linear cosmological perturbations generated by an inflationary era.

Quantum fluctuations of the scalar field governing the accelerated era are inflated from Planck ($a_i\sim l_4 \approx 10^{-35}\,$m) to cosmological scales ($a_f\gtrsim l_4 e^{60}\approx 60\,$pc) because of the superluminal expansion. They constitute the seeds of both the small anisotropies observed in the microwave sky and the large-scale nonlinear structures around which gravitating matter organizes itself. Such fluctuations are coupled to those experienced by the graviton background. For an introduction of the subject in the general relativistic case, see \cite{LiL}. 

In the most common 4D situation, the metric is perturbed by a linear contribution $g_{\mu\nu} \to g_{\mu\nu}+\delta g_{\mu\nu}$ which can be decomposed as
\be
\delta g_{\mu\nu} = A_{(\mu\nu)}g_{\mu\nu}+ V^i g_{0i}+F^{i,\,j}g_{i\!j}+h_{\mu\nu}\,,
\ee
where Greek indices run from 0 to 3, Latin indices are purely spatial, a comma denotes covariant derivative, and  $\{A_{(\mu\nu)}\}$, $\{V^i,F^i\}$, and $h_{\mu\nu}$ are scalar, vector, and tensor quantities, respectively. The three types of perturbations are independent and can be treated separately. Neglecting vector perturbations, which are damped during inflation, we are left with \emph{scalar} and \emph{tensor} perturbations, describing the matter and gravitational sources of the spectrum, respectively.

The standard procedure to adopt in order to compute the perturbation spectrum is: ($i$) Write the linearly perturbed metric in terms of gauge-invariant scalar or tensor quantities; ($ii$) Compute the effective action of the scalar field fluctuation and the associated equation of motion; ($iii$) Write the perturbation amplitude as a function of an exact solution of the equation of motion with constant SR parameters; ($iv$) Perturb this solution with small variations of the parameters.

The SR formalism gives good control over the theoretical shape and amplitude of cosmological perturbations. Here we shall restrict ourselves to the linear first-order approach \cite{bar80,KS,MFB}, although it is possible to extend the discussion to second-order perturbations \cite{MMB,ABMR,MaW}, nonlinear perturbations \cite{SB,SaT}, and even to a nonperturbative setup \cite{RiS2,LyMS,LV}. The latter case will be considered in Appendix \ref{appB}.

The 5D Einstein equations for a brane with an isotropic fluid embedded in an AdS bulk are very
complicated due to both the great number of degrees of freedom for the cosmological perturbations and the nonlocal
physics coming from the possibility for Kaluza-Klein (KK) gravitational modes to propagate and interact throughout the whole
spacetime. Braneworld calculations for the perturbation spectra are much more involved because of the complicated geometrical tissue and only general formalisms or approximated approaches have been explored so far. For this reason, while the setup has been established \cite{VDBL,SMS,mar00,KoS1,lan01,LMSW,mar01,GKLR,CAW,def04,koy04,YK}, a full 5D spectrum amplitude has not been calculated yet for either scalar or tensor modes, except in the case of some particular scenarios, for example, just to mention a few possibilities, those in which the brane is de Sitter \cite{DLMS,GRS,LMW,KKT,SeT} or in the large-scale limit \cite{KoS1,lan01,GKLR,WMLL}. 

We can carry out our calculation with little effort by making a somewhat drastic simplification. In particular, in Chapter \ref{patch} we chose to neglect the projected Weyl tensor. This closes the system of equations allowing us to study brane physics without nonlocal contributions from the bulk. During inflation this is consistent with the suppression of the dark radiation term at the classical level. The good news are that this holds also for the quantum perturbations: if one chooses a conformally flat background, $\rho_{\cal E}= {\cal E} a^{-4}=0$, the fluctuation of dark radiation suffers an exponential damping during inflation, $\delta \rho_{\cal E} \propto a^{-4} \sim \exp (-4Ht)$. This oversimplified background is justified by noting that any improvement of the physics would only confirm the breaking of degeneracy between standard and braneworld consistency equations \cite{der04}, which will be one of our main points. The ``softness'' of this breaking and its evidence will strongly depend on the physics, so we cannot say hardly anything \emph{a priori} about its size in more complicated scenarios. However, as a last remark, it is important to notice that the effect of this extra physics (fluctuations in Weyl component and anisotropic stress in a high-energy regime) will be more enhanced at small scales, since in a conformally and spatially flat background it only mildly affects the density perturbations and spectrum at large scales, $k \ll aH$ \cite{GoM,GRS,IYKOM,LCML,koy03,KLMW}. Since inflationary dynamics dominates the large-scale perturbation spectrum, in our context we will expect to find results which are close to the ``true'' answer coming from a complete computation of the full Einstein equations with boundary conditions.

In order to get some general and immediate results we consider two further assumptions:
\begin{enumerate}
\item The contribution of the anisotropic stress is neglected;
\item We concentrate on the large-scale limit of the cosmological perturbations.
\end{enumerate}
The first assumption reduces the number of degrees of freedom of gauge-invariant scalar perturbations in the
longitudinal (conformal Newtonian) gauge \cite{LMSW}. As regards the second approximation, the long wavelength region of the spectrum, corresponding to the Sachs-Wolfe plateau, encodes the main physics of the inflationary era.

%%%%%%%%%%%%%%%%%%%%%%%%%%%%%%%%%%%%%%%%%%%%%%%%%%%%%%%%%%%%%%%%%%%%%%%%%%%%%%%%%%%%%%%%%%%%%%%%%%%%%%%%%%%%%%%%%%%%%%%%%%%%%%%%%%%%%%%%%%%%%%%%%%%%%%%%%%%%%%%%%%%%%%%%%%%%%%%%%%%%%%%%%%%%%%%%%%%%%%%%%%%%%%%%%%%%%%%%%%%%%%%%%%%%%%%%%%%%%%%%

\section{General spectra and observables}

By definition, the 4D spectral amplitude generated by the $k$th mode of the perturbation $\Phi$ is
\be \label{ampli}
A^2_\Phi \equiv \frac{2k^3}{25\pi^2} \left\langle |\Phi_\mathbf{k}|^2\right\rangle\!\Big|_*\,,
\ee
where angle brackets denote the vacuum expectation value of the perturbation, the subscript $\mathbf{k}$ indicates the $k$th Fourier mode of
\be\label{fk}
\Phi(x) = \int \frac{d^3{\bf k}}{(2\pi)^{3/2}} \Phi_\mathbf{k}(t) e^{i\mathbf{k} \cdot \mathbf{x}},
\ee
and the expression is evaluated at the horizon-crossing time defined by $k(t_*)=a(t_*)H(t_*)$. The quantity $\Phi$ bears no covariant indices and therefore is a scalar.

The vacuum state in which the amplitude (\ref{ampli}) is evaluated is by definition empty of particles at some initial time $t_i$ with respect to background comoving coordinates. Since this state is an attractor solution of the wave equation in de Sitter space, actually it is independent of the choice of $t_i$.

In the case of the scalar spectrum (subscript $s$), $\Phi={\mathcal R}$ is the curvature perturbation on comoving hypersurfaces, generated by the scalar field filling the early Universe. Given a field $\psi=\phi,\,T$ on the brane, this is 
\be\label{Rper}
{\cal R}=-\Psi_4-H\frac{\delta\psi}{\dot{\psi}}\,,
\ee
to linear order, where $\Psi_4 = -\delta a/a$ is the gauge-invariant potential perturbing the spatial part of the metric. Note that at large scales the curvature perturbation is conserved,
\be
\dot{\cal R} \approx \frac{H}{\rho+p}\,\delta p_\text{nad} =0\,,
\ee
since the nonadiabatic pressure perturbation $\delta p_\text{nad}\equiv \dot{p}\,[(\delta p/\dot{p})-(\delta\rho/\dot{\rho})]$ vanishes identically for a fluid with a well-defined equation of state $p=p(\rho)$, which is the case of the scalar field $\psi$.

For the gravitational spectrum (subscript $t$), $\Phi$ denotes the coefficient functions of the zero mode $h_{\mu\nu}^{(0)}(x)$ of the 4D polarization tensor.

Neglecting the contribution of the Weyl tensor and the total anisotropic stress, the system of equations closes on the brane and the number of gauge degrees of freedom conveniently reduces for longitudinal scalar perturbations; moreover, bulk effects are suppressed in the long wavelength limit, $k \ll aH$. In this case one can rely on the 4D Mukhanov equation on the brane \cite{MFB,muk85,muk89},
\be \label{muksc1}
\left(\frac{d^2}{d\eta^2}+k^2-\frac{1}{z}\frac{d^2z}{d\eta^2}\right)u_\mathbf{k}=0\,,
\ee
where derivatives are with respect to conformal time\footnote{Although the symbol $\eta$ has already been used for the second SR parameter, this one will always bear the subscript of the scalar field in any place where there might be some confusion.} 
\be \label{confor}
\eta\equiv\int \frac{dt}{a} = -\frac{1}{(1-\epsilon)aH}\,,
\ee
and $u_\mathbf{k}$ are the coefficients of the plane wave expansion of the canonical variable 
\be
u=-z\Phi\,.
\ee
The function $z$ depends on the field $\psi$, $h$ one is considering and contributes to the effective mass of the covariant perturbation induced by the original cosmological friction term in the background equation of motion. For a perfect fluid, the squared function $z$ is
\be \label{zgen}
z^2 \equiv \zeta_q \frac{(\rho+p)a^2}{H^2}= \zeta_q\frac{(1+w)a^2}{\beta_q^{2-\theta}H^\theta}\,,
\ee
where $\zeta_q$ is a proportionality coefficient related to the field $\Phi$.

The region $k^2<d_\eta^2 z/z$ characterizes the squeezing phase in which perturbations are described by stationary plane waves with fixed $k$-independent phase. At $k^2=d_\eta^2 z/z$, that is when $k \approx aH$ approximately, perturbations freeze out of the horizon and their amplitude remains constant up to the recrossing.

Note that the projection of a nonvanishing Weyl tensor only adds a source term in Eq. (\ref{muksc1}) \cite{KLMW}, which can be absorbed in the definition of $z$ \cite{der04}. Anyway, the assumptions made above are well motivated and we do not expect to find strange surprises in the dynamical and quantum behaviour of the perturbations as long as we keep the discussion at large scales.

The amplitude (\ref{ampli}) becomes
\be \label{ampli2}
A^2_\Phi = \frac{2k^3}{25\pi^2}\frac{|u_\mathbf{k}|^2}{z^2}\,.
\ee
The Mukhanov equation is exactly solvable for cosmologies with constant SR parameters. In this case we will see that $d_\eta^2 z/z \propto \eta^{-2}$. Setting
\be
\nu^2 \equiv \frac{1}{4}+\eta^2\frac{1}{z}\frac{d^2z}{d\eta^2}\,,
\ee
Eq. (\ref{muksc1}) can be rewritten as
\be \label{muksc2}
\left[\frac{d^2}{d\eta^2}+k^2-\frac{(\nu^2-1/4)}{\eta^2}\right]u_\mathbf{k}=0\,.
\ee
With constant $\nu$, the solution of this equation is $|u_\mathbf{k}| \propto (-\eta)^{1/2} H_\nu^{(1)}(-k\eta)$, where $H_\nu^{(1)}$ is the Hankel function of the first kind of order $\nu$. Since, as we shall see, $\nu \approx 3/2+O(\epsilon)$ is a combination of SR parameters, this expression describes a cosmological solution with constant SR parameters; in fact, patch power-law inflation has this feature.

In the long wavelength limit, $k/(aH)\rightarrow 0$, when the mode with comoving wave number $k$ is well outside the horizon, the appropriately normalized solution becomes
\be \label{scasol}
|u_\mathbf{k}| = \frac{2^{\nu-3/2}}{\sqrt{2k}}\frac{\Gamma(\nu)}{\Gamma(3/2)}\left(-k\eta\right)^{-\nu+1/2},
\ee
where evaluation at the horizon crossing $k=aH$ is understood. If the SR parameters are small, then they are constant to leading order because their derivatives are higher order. It is then reasonable to solve the Mukhanov equation with exactly constant SR parameters and perturb the obtained solution. Expanding the solution (\ref{scasol}) to the same SR order of $\nu$ one gets
\be
|u_\mathbf{k}| \approx [1-C(\nu-3/2)]\frac{(-k\eta)^{-\nu+1/2}}{\sqrt{2k}} \,,\qquad \nu-3/2 \ll 1\,,
\ee
where $C=\gamma+\ln 2-2 \approx -0.73$ is a numerical constant ($\gamma$ is the Euler-Mascheroni constant) coming from the expansion
\[2^x \frac{\Gamma(x+3/2)}{\Gamma(3/2)} \approx 1-Cx\,,\qquad x \ll 1\,.\]
To lowest SR order, the resulting amplitude is
\be\label{ampli3}
A^2_\Phi = \left(\frac{k}{5\pi z}\right)^2.
\ee
Finally, we define the spectral indices \cite{LL} and their runnings \cite{KT} as
\ba
n_s-1 &\equiv& \frac{d \ln A_s^2}{d \ln k}\,,\qquad n_t \equiv \frac{d \ln A_t^2}{d \ln k}\,,\label{indices}\\
\alpha_s &\equiv& \frac{d n_s}{d \ln k}\,,\qquad\quad \alpha_t \equiv \frac{d n_t}{d \ln k}\,.
\ea
At lowest SR order one can use the relation
\be \label{dotk}
\frac{d}{d \ln k} \approx \frac{d}{Hd t}\,.
\ee
The definition (\ref{indices}) allows to write down the amplitudes as powers of the comoving scale $k$ of the perturbation within some interval $\Delta k$ centered on $k$:
\be
A_s^2(k) \approx A_1 k^{n_s-1}\,,\qquad A_t^2(k)\approx A_2k^{n_t}\,,
\ee
where $A_1$ and $A_2$ are constants. The difference in the condition of scale invariance between scalar and tensor spectra ($n_s=1$ and $n_t=0$) is just a historical convention.\footnote{Scale-invariant spectra are often called Harrison-Zel'dovich \cite{har70,zel70}.}

The last observable we define is the tensor-to-scalar ratio
\be
r \equiv \frac{A_t^2}{A_s^2}\,.
\ee
Very often it is rescaled and dubbed as $R \equiv 16 r$.

%%%%%%%%%%%%%%%%%%%%%%%%%%%%%%%%%%%%%%%%%%%%%%%%%%%%%%%%%%%%%%%%%%%%%%%%%%%%%%%%%%%%%%%%%%%%%%%%%%%%%%%%%%%%%%%%%%%%%%%%%%%%%%%%%%%%%%%%%%%%%%%%%%%%%%%%%%%%%%%%%%%%%%%%%%%%%%%%%%%%%%%%%%%%%%%%%%%%%%%%%%%%%%%%%%%%%%%%%%%%%%%%%%%%%%%%%%%%%%%%

\section{Scalar perturbations}\label{scalar}

The scalar field is confined on the brane and the effective action giving the equation of motion reproduces standard four-dimensional cosmology. Quantum fluctuations of the brane field generate scalar perturbations that can be treated in the way sketched above. In this context the scalar perturbation is found to be the same as the one in the usual 4D background.

The main point we rely on is the independence of the behaviour of curvature perturbations from the gravitational part of the action at sufficiently large scales \cite{WMLL,RiS1}. It is then possible to consider a linearly perturbed effective 4D metric with the same number of gauge degrees of freedom and borrow part of the standard perturbative formalism. The perturbed 5D metric in the longitudinal gauge reads
\be
d s_5^2 \approx (1+2\Phi_5)~d t^2-a^2(t)[(1-2\Psi_5)\delta_{i\!j}+2E_{,\,i\!j}]~d x^i d x^j+2B_{,\,i}dx^i dt-dy^2,
\ee 
where $\Phi_5$, $\Psi_5$, $E$, and $B$ are gauge-invariant scalars and $y$ is the extra direction.

In the case of the ordinary scalar quantum field
\be\label{phdelph}
\varphi(t,\mathbf{ x}) = \phi(t)+\delta\phi(t,\mathbf{ x})\,,
\ee
decomposed into the homogeneous background field $\phi(t)=\langle\varphi(x)\rangle$ and its fluctuation $\delta\phi(t,\mathbf{ x})$, the 5D equations of motion in a RS braneworld with brane-bulk exchange, Eq. (\ref{bBex}), are
\ba
&&\ddot{\phi}+3H \dot{\phi}+ V'+\frac{r_B}{\dot{\phi}}=0\,,\label{eom0}\\
&&\ddot{\delta\phi}-a^{-2}\nabla^2 \delta\phi+3H \dot{\delta\phi}+ V''\delta\phi-(\dot{\Phi_5}+3\dot{\Psi_5})\dot{\phi}+2\Phi_5 V' +2 \Phi_5\frac{r_B}{\dot{\phi}}=0\,.\nonumber\\\label{eom1}
\ea
If the anisotropic stress vanishes, then $\Phi_5$ and $\Psi_5$ coincide with the gauge-invariant scalar potentials $\Phi_4$ and $\Psi_4$, respectively. By setting an empty bulk ($r_B=0$), one recovers the ordinary 4D equations. In the large-wavelength limit, $k\ll aH$, the four-dimensional perturbed metric induced on the brane is the standard one,
\be\label{permet}
d s_4^2\Big|_{\rm brane} \approx (1+2\Phi_4)\,d t^2-a^2(t)(1-2\Psi_4)\delta_{i\!j}\,d x^i d x^j\,.
\ee 
Then one can perform pure four-dimensional calculations, for example by the methods of \cite{MFB}, and obtain the Mukhanov equation (\ref{muksc1}). As one can easily check, the only new ingredient comes from the Friedmann relation (\ref{FRW}) between the Hubble parameter $H$ and the energy density $\rho$. Another derivation of the Mukhanov equation is discussed in Appendix \ref{appB}.

This equation can be solved exactly by a cosmology with constant slow-roll parameters; from Sec. \ref{exact} we know we have such solutions at our disposal. Even the full models outside the patch approximation possess exact solutions with constant or almost constant SR parameters, an important feature for constructing perturbation amplitudes. As an example, in Appendix \ref{appA} we see how the exact Randall-Sundrum solutions found in \cite{HaL1} satisfy this requirement asymptotically in time. By perturbing the exact solution with respect to small variations of the SR parameters, one gets the scalar spectral amplitude.

%%%%%%%%%%%%%%%%%%%%%%%%%%%%%%%%%%%%%%%%%%%%%%%%%%%%%%%%%%%%%%%%%%%%%%%%%%%%%%%%%%%%%%%%%%%%%%%%%%%%%%%%%%%%%%%%%%%%%%%%

\subsection{The ordinary scalar field $\phi$} \label{Apphi}

The lowest-order scalar amplitude can be directly derived from the fluctuation spectrum of a massless scalar field outside the horizon, by quantizing the classical field (\ref{phdelph}) and imposing equal-time commutation relations in curved (actually, de Sitter) spacetime \cite{BD},
\bs \label{commu}\ba
&&\left[\varphi (\mathbf{x}_1,\,t),\,\varphi(\mathbf{x}_2,\,t)\right] = 0 = \left[\Pi_\varphi (\mathbf{x}_1,\,t),\,\Pi_\varphi(\mathbf{x}_2,\,t)\right] \,,\\
&&\left[\varphi (\mathbf{x}_1,\,t),\,\Pi_\varphi(\mathbf{x}_2,\,t)\right] = i a^3 \delta^{(3)}(\mathbf{x}_1-\mathbf{x}_2)\,,
\ea\es
where $\Pi_\varphi=\dot{\varphi}$ is the conjugate momentum density. The Fourier transform of the fluctuation can be written as a combination of harmonic oscillators, 
\be \label{harmosc}
\delta\phi_\mathbf{k}(t)=w_k(t)a_\mathbf{k}+w_k^*(t)a^\dagger_{-\mathbf{k}}\,,
\ee
where $w_k$ is a complex function of $k=|\mathbf{k}|$ and the creation-annihilation operators satisfy the canonical equal-time commutation relations, $[a_{\mathbf{k}_1},\,a_{\mathbf{k}_2}^\dagger]_{t_1=t_2}=\delta_{\mathbf{k}_1\mathbf{k}_2}$, etc. Since $\delta\phi_\mathbf{k}^\dagger=\delta\phi_{-\mathbf{k}}$, the quantum operator $\varphi(x)$ is hermitian. Near horizon exit, $a = k/H$, and with negligible variation of $H$, the fluctuation amplitude turns out to be 
\bs\label{lowestS}\be
\langle|\delta\phi_\mathbf{k}|^2\rangle = |w_k|^2=\frac{H^2}{2k^3}\,.
\ee
The scalar amplitude is 
\be
A_s^2 = \frac{4}{25}\frac{H^2}{\dot{\phi}^2}\,{\cal P}_\phi\,,
\ee\es
where 
\be\label{calP}
{\cal P}_\phi \equiv \frac{k^3}{2\pi^2} \langle|\delta\phi_\mathbf{k}|^2\rangle=\left(\frac{H}{2\pi}\right)^2.
\ee
Note that this computation does not involve the Friedmann equation, Eq. (\ref{FRW}), but only the 4D equation of motion (\ref{eom}). As a function of the potential, the amplitude (\ref{lowestS}) can be written to lowest SR order as
\be \label{Vampl}
A_s^2(\phi) \approx \frac{9\beta_q^6}{25\pi^2}\frac{V^{3q}}{V'^2}\,.
\ee

%%%%%%%%%%%%%%%%%%%%%%%%%%%%%%%%%%%%%%%%%%%%%%%%%%%%%%%%%%%%%%%%%%%%%%%%%%%%%%%%%%%%%%%%%%%%%%%%%%%%%%%%%%%%%%%%%%%%%%%%

\subsection{The tachyon field $T$}

In a string-theoretical setup, the quantization of the tachyon Lagrangian (\ref{ta}) is a delicate and noncompletely explored subject; in particular, it is not clear yet if the promotion of the classical field to a quantum object correctly describes quantum string theory \cite{sen8}. Nonetheless, one may put aside high-energy motivations for the field theory (\ref{ta}) and study its quantum behaviour independently. We start from the commutation relations (\ref{commu}), with $T$ instead of $\phi$ and $\Pi_T$ given by Eq. (\ref{momP}) in the long-wavelength limit. In momentum space, the field fluctuations are quantized as in Eq. (\ref{harmosc}). Now, the equation of motion for the perturbation $\delta T_\mathbf{k}$ is not a simple Klein-Gordon equation as in the scalar case; however, one can check that, near horizon crossing, the two-point function of the fluctuation is 
\be
\langle|\delta T_\mathbf{k}|^2\rangle = |w_k|^2=\frac{H^2}{2Vk^3}\,.
\ee
The computation is performed without adopting any particular gravitation background, except for the hypothesis of quasi de Sitter expansion. In order to absorb the potential into an expression which is dependent only on the Hubble parameter, we must use Eq. (\ref{FRW}), getting
\be \label{Tspec}
{\cal P}_T \equiv \frac{k^3}{2\pi^2} \langle|\delta T_\mathbf{k}|^2\rangle = \frac{\beta_q^{2-\theta}H^\theta}{2c_S}\,,
\ee
where $c_S$ is given by Eq. (\ref{cS}). The presence of an extra $V$-term is also evident when comparing the amplitude $A_s^2(T) \propto V^{3q+1}/V'^2$ with that in the scalar scenario, Eq. (\ref{Vampl}). Slow-roll corrections to this result can be computed by a more refined treatment including back-reaction from the effective 4D metric. The Mukhanov equation for the tachyon case has been derived in a $k$-inflationary context; it reads \cite{FKS,HN,MFB,GM}
\be\label{tacmuk}
\left(\frac{d^2}{d\eta^2}+c_S^2k^2-\frac{1}{z}\frac{d^2z}{d\eta^2}\right)u_\mathbf{k}=0\,.
\ee
To lowest SR order, it is the same amplitude as that in the nontachyonic case, if expressed as a function of $H$ and its time derivative:
\be \label{Sdeg}
A_s^2 \approx \frac{3q\beta_q^{2-\theta}}{25\pi^2}\frac{H^{2+\theta}}{2\epsilon}\,.
\ee
This is not surprising, since in the ESR regime the dynamics are almost the same, as explained in Sec. \ref{vs}. 

%%%%%%%%%%%%%%%%%%%%%%%%%%%%%%%%%%%%%%%%%%%%%%%%%%%%%%%%%%%%%%%%%%%%%%%%%%%%%%%%%%%%%%%%%%%%%%%%%%%%%%%%%%%%%%%%%%%%%%%%%%%%%%%%%%%%%%%%%%%%%%%%%%%%%%%%%%%%%%%%%%%%%%%%%%%%%%%%%%%%%%%%%%%%%%%%%%%%%%%%%%%%%%%%%%%%%%%%%%%%%%%%%%%%%%%%%%%%%%%%

\section{Tensor perturbations} \label{Tpert}

Tensor perturbations are more difficult to deal with than the scalar ones since gravity is free to propagate in the whole five-dimensional spacetime. In general, the graviton zero-mode, localized on the brane, interacts with KK massive modes generating an infinite tower of coupled differential equations. The key point is the knowledge of the behaviour of gravity modes on the brane, that is, how Kaluza-Klein modes couple with the zero-mode. Many of the problems arise because of the extra degree of freedom provided by the radion and, in general, a complete solution of the Einstein equations with boundary conditions is difficult to achieve. 

In some limits the analysis simplifies considerably, for instance in a de Sitter brane ($\rho+p=0$) \cite{DLMS,LMW}. In this case it is possible to decouple the graviton zero-mode from the massive tower of Kaluza-Klein gravitons, since the maximal symmetry of the dS brane permits a variable separation of the wave equation for the Kaluza-Klein modes, $h_{\mu\nu}(x,y) \rightarrow h_{\mu\nu}^{(m)}(x)\xi_m(y)$.  The normalization of the bulk-dependent part of the zero-mode ($m=0$), calculated on the brane position $y_b$, determines the mapping function $F\equiv\xi_0(y_b) \kappa_5/\kappa_4$. It turns out that $F$ is a complicated function of the couplings of the theory, the Hubble parameter $H$, and $\chi$, the inverse of the bulk curvature scale. The effective Newton constant on a GB brane is $\kappa_4^2 = \kappa_5^2\chi/(1+4\alpha\chi^2)$, which in RS ($\alpha=0$) reduces to $\chi^2=\lambda\kappa_4^2/6$. Given the 4D amplitude 
\be\label{8dec}
A_{t(\text{4D})}^2 = \frac{\kappa_4^2}{50 \pi^2} H^2\,,
\ee
which is Eq. (\ref{ampli3}) with $z_{\text{4D}}=\sqrt{2}a/\kappa_4$, the braneworld tensor amplitude is
\be\label{tensor}
A_t=A_{t(\text{4D})} F (H/\chi)\,,
\ee
with \[F^{-2}(x)=\sqrt{1+x^2}-\left(\frac{1-4\alpha\chi^2}{1+4\alpha\chi^2}\right)x^2\sinh^{-2}\frac{1}{x}\,.\] 
In the limit $x \gg 1$ (GB high-energy regime), 
\be
F^2(x) \approx \frac{1+4\alpha\chi^2}{8\alpha\chi^2} \frac{1}{x}\,;
\ee
in the RS case and in the limit $\rho \gg \lambda$, 
\be
F^2(x) \approx \case{3}{2} x\,,
\ee
while in the low-energy limit one recovers the standard spectrum, $F^2(x) \approx 1$. 

One way to state the result (\ref{tensor}) is that, to lowest SR order, the brane-world tensor amplitude is given by the 4D expression (\ref{8dec}) under the mapping
\be \label{hmap1}
h_0:\quad H \mapsto H F(H/\chi)\,. 
\ee
In \cite{KKT} a perturbed RS de Sitter brane whose Hubble constant is experiencing a discontinuous variation $\delta H = H_1-H_2$ is studied. A 5D spectral amplitude $A_{t(\text{5D})}$, obtained as an expansion in $\delta H/H \ll 1$, is provided at mildly large scales, $k \ll H^2/\delta H$, and is compared with the 4D braneworld amplitude $A_{t,\eff}$ derived from a generalization of the map (\ref{hmap1}) acting on the standard amplitude $A_{t(\text{4D})} \propto \tilde{H}$, namely,
\be \label{hmap2}
h_1:\quad \tilde{H} \mapsto \tilde{H} F(\tilde{H}/\chi)\,.
\ee
Here, $\tilde{H}=H[1+O(\delta H/H)]$ includes the correction to the lowest-order result coming from the 4D zero-mode--zero-mode Bogoliubov coefficients. It turns out that the lowest-order [$O(1)$, superscript (lo)] and next-to-lowest order [$O(\delta H/H)$, superscript (ntlo)] effective amplitudes match the five-dimensional result, 
\be \label{eff}
A_{t,\eff}^{(\text{lo})}=A_{t(\text{5D})}^{(\text{lo})}\,,\qquad A_{t,\eff}^{(\text{ntlo})}=A_{t(\text{5D})}^{(\text{ntlo})}\,;
\ee
outside the horizon $\left|\left(A_{t(\text{5D})}-A_{t,\eff}\right)/A_{t(\text{5D})}^{(\text{lo})}\right| \sim O[(\delta H/H)^2]$, while inside the horizon there can be significant discrepancies.

The case of a continuous smooth variation of the Hubble parameter, which is typical during the inflationary regime long before the reheating, has to be treated separately because of the time dependence of $H$ in the equations of motion of the KK modes. However, one expects the slow-roll generalization of the previous result to display similar features \cite{KKT}, since $\epsilon=O(\delta H/H)$. In fact, in order the variation $\delta H$ not to be damped by the accelerated expansion, it must occur in a time interval $\delta t \sim H^{-1}$, that is, $\dot{H} \sim -\delta H / \delta t \sim -\delta H H$, giving the heuristic correspondence $\epsilon \sim \delta H / H$. Thus we can conjecture an analogous mapping (\ref{hmap2}) in the SR expansion such that (\ref{eff}) holds, with $\tilde{H}=H[1+O(\epsilon)]$. We schematically represent it in figure \ref{fig3}.
\begin{figure}[ht]
\bc
\ec
\caption{\it\label{fig3} The $h$-mappings (\ref{hmap1}) and (\ref{hmap2}) between 4D and effective 5D tensor amplitudes.}
\end{figure}
Here $f_0$ maps any next-to-lowest-order function to its lowest-SR-order form.\footnote{This is equivalent to setting $\epsilon=\eta=0$ only in the amplitude $A_t$, because, for instance, $f_0(A_s) \propto \epsilon^{-1/2}$.} We remark once again that, while $f_0$ is trivial and $g_0 = 1$ was first demonstrated in \cite{LMW}, it would be a nontrivial goal to show that $g_1 = 1$ for a non-de Sitter inflationary brane or, more generally, to find the region in the space of the effective amplitudes where $f_0$ is one-to-one.

There is however another possibility. According to Eq. (\ref{ampli3}), the mapping (\ref{hmap1}) can be regarded as acting on $z$ rather than on the Hubble parameter \cite{cal2}:
\be \label{map}
h:\quad z_{\text{4D}} \mapsto z=\frac{z_{\text{4D}}}{F(H/\chi)}\,,
\ee
for the tensor perturbation. This more elegant formulation has two advantages: first, it encodes all the needed SR information at any order in one single relation, so $h_1$ and $h_2$ collapse to each other; secondly, it is reasonably valid in more general braneworld scenarios. 

From the above discussion it emerges that another effective prescription is viable for a general braneworld tensor amplitude: 
\be\label{ansatz}
A_t = A_{t(\text{4D})} \frac{z_{\text{4D}}}{z}\,,
\ee
evaluated at the horizon crossing. To be consistent with the patch solution (\ref{FRW}), we must consider the approximated version $F_q$ of $F$ in the proper energy limits. With Eq. (\ref{ansatz}), one may encode the phenomenology of the transverse direction into the map (\ref{map}) acting on the function $z$, $z_{4D} \mapsto z=z_{4D}/F_q$.

We can find the patch version of $F$ with a trick, by noting that in four dimensions the graviton background can be formally described by Eq. (\ref{zgen}) with $\zeta_1(h)=1$ and a perfect fluid $p_h= -\rho_h/3$ which does not contribute to the cosmic acceleration, since $\ddot{a}\propto(1-\epsilon)$ and $\epsilon=1$. Generalizing this stationary solution one has $w_h=2/(3q)-1$ and \cite{cal4}
\bs\label{zgrav}\ba
z(h)  &=& \frac{\sqrt{2}a}{\kappa_4 F_q}\,,\\
F^2_q &\equiv& \frac{3q\beta_q^{2-\theta}H^\theta}{\zeta_q(h)\kappa_4^2}\,.
\ea\es
This is equivalent to take the 4D tensor amplitude and substitute the gravitational coupling with $\kappa_4^2 \sim (H^2/\rho)_{\text{4D}} \rightarrow H^2/\rho$. Although these arguments do not completely justify Eq. (\ref{zgrav}) as the general patch solution for the tensor amplitude, the proposed ingredients do match the results coming from both the 4D and full 5D calculations in Randall-Sundrum and Gauss-Bonnet scenarios. We suggest this picture to be valid in other cases, too; direct contact with explicit gravity models is reduced to a minimum through the coefficients $\beta_q$ and $\zeta_q$, but only the latter is indispensable for the consistency relations. 

From now on we shall drop the argument $h$ of $\zeta_q$, $\zeta_q=\zeta_q(h)$. In the Randall-Sundrum case $\zeta_2=2/3$ \cite{LMW}, while in the Gauss-Bonnet case $\zeta_{2/3}=1$ \cite{DLMS}.

%%%%%%%%%%%%%%%%%%%%%%%%%%%%%%%%%%%%%%%%%%%%%%%%%%%%%%%%%%%%%%%%%%%%%%%%%%%%%%%%%%%%%%%%%%%%%%%%%%%%%%%%%%%%%%%%%%%%%%%%%%%%%%%%%%%%%%%%%%%%%%%%%%%%%%%%%%%%%%%%%%%%%%%%%%%%%%%%%%%%%%%%%%%%%%%%%%%%%%%%%%%%%%%%%%%%%%%%%%%%%%%%%%%%%%%%%%%%%%%%

\section[Braneworld spectra and consistency relations]{Braneworld spectra and\\ consistency relations}

Let us list the results for the cosmological spectra to next-to-leading SR order. In all the calculations we use Eq. (\ref{confor}) together with 
\be
d_\eta^2=a^2(Hd_t +d_t^2)\,,
\ee
and drop $O(\epsilon^2)$ terms in $z$. In fact, higher-order contributions might be too faint to play an important role in the determination and discrimination of the spectra. Then we do not push the patch SR expansion up to the second order, although some authors have gone in this direction \cite{KM,KM2,KLM,wei04}.

Note that we do not need to specify $\beta_q$ since it will not appear in the slow-roll expressions for the cosmological observables, except in the amplitudes. In this case, CMB data can constrain the parameters of the gravitational model encoded in $\beta_q$.

%%%%%%%%%%%%%%%%%%%%%%%%%%%%%%%%%%%%%%%%%%%%%%%%%%%%%%%%%%%%%%%%%%%%%%%%%%%%%%%%%%%%%%%%%%%%%%%%%%%%%%%%%%%%%%%%%%%%%%%

\subsection{Graviton field}

For the tensor amplitude we have
\ba
\frac{1}{z}\frac{d^2z}{d\eta^2} &=& (aH)^2\left[2+\left(\frac{3\theta}{2}-1\right)\epsilon\right]\,,\\
\nu_h &=& \frac{3}{2}+\left(1+\frac{\theta}{2}\right)\epsilon\,,
\ea
and
\bs \label{At}\ba
A_t^2 &=& \left\{1-\left[(2+\theta)C+2\right]\epsilon\right\}\frac{3q\beta_q^{2-\theta}}{25\pi^2}\frac{H^{2+\theta}}{2\zeta_q}\,,\label{Ah}\\
n_t &=& -(2+\theta)\epsilon\left[1+\omega_t\epsilon-2\left(C+\frac{2}{2+\theta}\right)\eta\right]\,,\\
\alpha_t &=& (2+\theta)\epsilon\left[2\eta-\left(2-\wteta\right)\epsilon\right]\,,
\ea
where 
\be
\omega_t \equiv \left(2-\wteta\right)C+\frac{6-\wteta}{2+\wteta}\,.
\ee\es
The $O(\epsilon^2)$ part of the tensor index and its running depend on the assumed scalar field model through Eq. (\ref{dotSRa}). In the tachyon case, $\omega_t= 1$. 

%%%%%%%%%%%%%%%%%%%%%%%%%%%%%%%%%%%%%%%%%%%%%%%%%%%%%%%%%%%%%%%%%%%%%%%%%%%%%%%%%%%%%%%%%%%%%%%%%%%%%%%%%%%%%%%%%%%%%%%

\subsection{Ordinary scalar field $\phi$}

For the ordinary scalar field on the brane,
\be
z(\phi) = \frac{a\dot{\phi}}{H}\,,
\ee
with $\zeta_q(\phi)=1$ in Eq. (\ref{zgen}).  Then
\ba
\frac{1}{z}\frac{d^2z}{d\eta^2} &=&  (aH)^2\left(2+2\epsilon_\S-3\eta_\S\right)\,,\\
\nu_\S  &=& \case{3}{2}+2\epsilon_\S-\eta_\S\,,
\ea
and
\bs\label{AsS} \ba 
A_s^2(\phi) &=& [1-2(2C+1)\epsilon_\S+2C\eta_\S]\frac{3q\beta_q^{2-\theta}}{25\pi^2}\frac{H^{2+\theta}}{2\epsilon_\S}\,,\label{phiS}\\
n_s-1 &=& \left(2\eta_\S-4\epsilon_\S\right)+2(5C+3)\epsilon_\S\eta_\S- 2C\xi_\S^2-2[(4-\theta)+ 2(2-\theta)C]\, \epsilon_\S^2\,,\nonumber\\\\
\alpha_s &=& 2\left[2(\theta-2)\, \epsilon_\S^2+5\epsilon_\S\eta_\S-\xi_\S^2\right],\\
r &=&  \frac{\epsilon_\S}{\zeta_q}\left[1-(\theta-2)C\epsilon_\S-2C\eta_\S\right].
\ea\es
The expression (\ref{phiS}) has the asymptotic form at large scales, $k \ll aH$, but is written in terms of quantities evaluated at the horizon crossing of the perturbation. By fixing the term in square brackets equal to 1, one gets the lowest-order expression (\ref{lowestS}) obtained via a de Sitter calculation of the correlation function of the fluctuation $\delta\phi \approx u/a$ outside the horizon. With $\theta=0$, one recovers the four-dimensional results \cite{LS,SL,CKLL}.

%%%%%%%%%%%%%%%%%%%%%%%%%%%%%%%%%%%%%%%%%%%%%%%%%%%%%%%%%%%%%%%%%%%%%%%%%%%%%%%%%%%%%%%%%%%%%%%%%%%%%%%%%%%%%%%%%%%%%%%

\subsection{Tachyon field $T$}

In the Mukhanov equation (\ref{tacmuk}),
\be
z(T) = \frac{(\rho+p)^{1/2}a}{c_S H}= \frac{1}{\beta_q^{1/q}} \frac{a \dot{T}}{c_S H^{\theta/2}}\,,
\ee
where we have used $\rho+p=\rho \dot{T}^2$.  Despite of what happens in the scalar field case, $\theta$ appears explicitly in the definition of $z$. In the extreme SR approximation we can set $\zeta_q(T)= 1/c^2_S\approx 1$ in Eq. (\ref{zgen}).

The solution of (\ref{tacmuk}) is again of Hankel type but with a rescaled wave number, $k \rightarrow c_Sk$. For the tachyon one gets
\ba
\frac{1}{z}\frac{d^2z}{d\eta^2} &=& (aH)^2\left[2+\left(\frac{3\theta}{2}-1\right)\epsilon_\T-3\eta_\T\right]\,,\label{tac1}\\
\nu_\T  &=& \frac{3}{2}+\left(1+\frac{\theta}{2}\right)\epsilon_\T-\eta_\T\,,\label{tac2}
\ea
and
\bs \label{AsT} \ba 
A_s^2(T) &=& \left(1-2\omega_s\epsilon_\T+2C\eta_\T\right)\frac{3q\beta_q^{2-\theta}}{25\pi^2}\frac{H^{2+\theta}}{2\epsilon_\T}\,,\label{tacS}\\
n_s-1 &=& \left[2\eta_\T-(2+\theta)\,\epsilon_\T\right]+2\left(C+1+2\omega_s\right)\epsilon_\T\eta_\T-2C\xi_\T^2-(2+\theta)\epsilon_\T^2\,,\nonumber\\\\
\alpha_s &=& 2\left[(3+\theta)\,\epsilon_\T\eta_\T-\xi_\T^2\right],\label{Talp}\\
r &=& \frac{\epsilon_\T}{\zeta_q}\left[1-(2-\theta)\frac{\epsilon_\T}{6}-2C\eta_\T\right].
\ea
Here,
\be
\omega_s \equiv \left(C+\frac{5}{6}\right)+\frac{\theta}{2}\left(C+\frac{1}{6}\right).\label{tac3} 
\ee\es
In the four-dimensional case $\theta=0$, we recover the results of \cite{SV,GM}.

%%%%%%%%%%%%%%%%%%%%%%%%%%%%%%%%%%%%%%%%%%%%%%%%%%%%%%%%%%%%%%%%%%%%%%%%%%%%%%%%%%%%%%%%%%%%%%%%%%%%%%%%%%%%%%%%%%%%%%%%

\subsection{The consistency relations}

One might find interesting to rewrite the SR expressions (\ref{AsS}) and (\ref{AsT}) for the scalar perturbation in a minimal fashion, by substituting $\theta$ with 
\be\label{oteta}
2\oteta=2+\theta-\wteta\,,
\ee
in Eqs. (\ref{tac1}), (\ref{tac2}) and (\ref{tac3}); in the ordinary scalar case, $\omega_s= 2C+1$. Then,
\bs\label{SRexpr}\ba
A_s^2(\psi) &=& \left(1-2\omega_s\epsilon+2C\eta\right)\frac{3q\beta_q^{2-\theta}}{25\pi^2}\frac{H^{2+\theta}}{2\epsilon}\,,\\
n_s-1 &=& \left[2\eta-2\left(1+\oteta\right)\,\epsilon\right]+2(C+1+2\omega_s)\epsilon\eta\nonumber\\&&-2\left[\left(2-\wteta\right)\omega_s +1+\oteta\right]\epsilon^2-2C\xi^2\,,\\
\alpha_s &=& 2\left[2\left(\wteta-2\right)\epsilon^2 +\left(3+2\oteta\right)\epsilon\eta-\xi^2\right],\\
r &=& \frac{\epsilon}{\zeta_q}\left\{1-\left[(2+\theta)C+2(1-\omega_s)\right]\epsilon-2C\eta\right\}.
\ea\es
We can collect Eqs. (\ref{At}), (\ref{AsS}) and (\ref{AsT}) in the set of consistency equations:
\bs\label{ce}\ba
\alpha_s(\phi) &=& \zeta_q r [4(3+\theta)\zeta_qr+5(n_s-1)]-\xi_\S^2\,, \label{cealpsS}\\
\alpha_s(T) &=& (3+\theta)\zeta_q r [(2+\theta)\zeta_qr+(n_s-1)]-\xi_\T^2\,, \label{cealpsT}\\
n_t(\phi) &=& \zeta_q r [-(2+\theta)+(2+\theta)\zeta_qr+2(n_s-1)]\,,\label{cenS}\\
n_t(T) &=& \zeta_q r \left[-\vphantom{\frac{1}{1}}(2+\theta)+2(n_s-1)+\frac{(2+\theta)(4+\theta)}{6}\zeta_qr\right],\label{cenT}\\
\alpha_t &=& (2+\theta)\zeta_qr[(2+\theta)\zeta_qr+(n_s-1)]\,.\label{cealpt}
\ea\es
If
\be\label{prior}
|\xi| \ll \min(\epsilon,|\eta|)\,,
\ee
the set closes and the scalar running depends only on the observables.
In fact, this approximation can be put into a milder form when considering different patches, $\theta$ and $\theta'$, with the same inflaton field: namely, that the parameter $\xi$ is almost constant in the energy regime, $\xi(\theta) \approx \xi(\theta')$; this will be sufficient in order to compare theoretical results with observations. In the case of confrontation between a scalar patch and a tachyon patch, it should be $\xi_\S(\theta) \approx \xi_\T(\theta')$. We will see in Sec. \ref{vs} that SR parameters of the two scenarios, with the same scale factor, approach one another order by order when $q$ increases; therefore, the last approximation is valid at a certain confidence level if $q,q' \gg 1$. Whichever assumption is chosen to neglect the parameter $\xi$ and write the scalar running in terms of observables, it is important to keep in mind that in general $\xi$ cannot be fairly eliminated, even if this is indeed the case in many reasonable situations. One has $\xi = 0$ when either $\psi \propto t$ (e.g., tachyon power-law inflation) or $\psi \propto e^t$; however, in many simple models of inflation like the ordinary power-law ($a = t^n$), the dynamical features of the system are such that $|\xi| =O(\epsilon,|\eta|)$. We will come back to this issue in Sec. \ref{like1}.

The consistency equation for the tensor index depends on the chosen scalar action, as first pointed out in \cite{SV}. Conversely, Eq. (\ref{cealpt}) is valid both for the scalar and tachyon field, but in general it is model-dependent \cite{SV}. The next-to-lowest-order equations (\ref{cenS}) and (\ref{cenT}) generalize the lowest-order equation
\be\label{ce1}
n_t = -(2+\theta)\zeta_q r\,,
\ee
which is insensitive to the type of scalar field. Since at second order there appear only quantities present in first-order expressions, it is reasonable to consider Eqs. (\ref{cenS}) and (\ref{cenT}) not as extensions of (\ref{ce1}) but as \emph{the} consistency equations.

The consistency equations relate cosmological observables in a way typical of inflationary scenarios, in which the scalar and gravitational spectra are originated by a unique mechanism. When considering them in the braneworld case, they can give different (and testable) signatures of the early-Universe inflationary expansion.

%%%%%%%%%%%%%%%%%%%%%%%%%%%%%%%%%%%%%%%%%%%%%%%%%%%%%%%%%%%%%%%%%%%%%%%%%%%%%%%%%%%%%%%%%%%%%%%%%%%%%%%%%%%%%%%%%%%%%%%%
%%%%%%%%%%%%%%%%%%%%%%%%%%%%%%%%%%%%%%%%%%%%%%%%%%%%%%%%%%%%%%%%%%%%%%%%%%%%%%%%%%%%%%%%%%%%%%%%%%%%%%%%%%%%%%%%%%%%%%%%

\section[Degeneracy of consistency equations: testing the braneworld]{Degeneracy of consistency equations:\\ testing the braneworld}

In this section we address the issue of possible theoretical degeneracies between next-to-leading-SR-order consistency relations of different inflaton/braneworld models, to be distinguished from observational degeneracies coming from particular values of the observables, for example when the spectrum is nearly scale invariant in the extreme SR approximation. This problem arose for the first time when the degeneracy of the 4D and RS relation between the tensor spectral index and the tensor-to-scalar ratio was discovered \cite{HuL1,HuL2}; such a property also holds in induced-gravity braneworld inflation \cite{BMW} and in generalized Einstein theories including four-dimensional dilaton gravity and scalar-tensor theories \cite{TG04}. Several works then showed that this degeneracy is most likely incidental (e.g., \cite{cal2} and references therein). We are going to confirm this result in quite a general manner and pave the way to the classification of eventual future braneworld scenarios. 

It is useful to stress that even in the standard general relativistic case the consistency relations are violated in some simple situations, for example, in multi-field inflation \cite{SaSt} (see, e.g., \cite{WBMR} and references therein). In this sense, a deviation from the standard equations would not provide the smoking gun for the existence of extra dimensions.

We claim and next show that, while observational degeneracy of the consistency relations is achievable within the range of cosmological parameters determined by recent experiments, the theoretical structure is unstable against even long-wavelength 5D contributions, via the effective Friedmann equation (\ref{FRW}). 

%%%%%%%%%%%%%%%%%%%%%%%%%%%%%%%%%%%%%%%%%%%%%%%%%%%%%%%%%%%%%%%%%%%%%%%%%%%%%%%%%%%%%%%%%%%%%%%%%%%%%%%%%%%%%%%%%%%%%%%%

\subsection{Theoretical degeneracy}

To leading order, the consistency relation for a scalar-driven inflation is the same in 4D and RS scenarios and a discrimination between them, at least by this method, is not possible. However, quasi de Sitter computations show a break of the degeneracy and a possible nonclosed structure \cite{KKT,SeT}. In the GB case even the lowest-order tensor index consistency relation is no longer degenerate \cite{DLMS}. Some evidence of the degeneracy breaking in the case of smoothly varying Hubble parameter is provided by showing that departures from the standard form of the scalar amplitude would spoil the 4D structure of the consistency relations.

\subsubsection{4D vs braneworld}

One may wonder if there can exist SR perturbation amplitudes that give the same consistency equations for the standard four-dimensional cosmology, thus ruling out the possibility to discriminate between the two scenarios. However, to construct mathematical expressions without any physical content would be of little use since they would not provide a physical explanation why consistency equations should be still degenerate; so the consistency equation approach would still be worth investigation. 

After this preamble, it should be noted that it is not possible to construct simple perturbation amplitudes that give degenerate consistency equations and reproduce the extreme SR limit (\ref{tensor}). The reason is the following. We will concentrate on the ordinary scalar model and the consistency equation involving the scalar running $\alpha_s$, since it is the most relevant relation from an observational point of view. It is an equation coming from the lowest-SR-order part of the scalar index, 
\be\label{noteta}
n_s-1 = 2\eta-2\left(1+\oteta\right)\epsilon\,,
\ee
which is generated by the lowest-order part of the scalar amplitude, Eq. (\ref{Sdeg}). We recall that each time derivative raises the SR order by one at any step: the time derivative of the functional part of the amplitude gives the linear part of the spectral index, while another derivation gives the running, through Eq. (\ref{dotSR}). Therefore one should change the lowest-order part of the amplitude $A_s$ in order to impose 
\be\label{run1}
\alpha_s=\alpha_{s(\text{4D})} \equiv\alpha_s^\text{\tiny $(\phi,0)$} = r \left[12r+5 (n_s-1)\right],
\ee
regardless of the next-to-leading-order SR structure. Since the bulk source term of (\ref{eom0}) has not been taken into consideration, this could be feasible \emph{a priori}, even if Eq. (\ref{Sdeg}) is well supported in many respects.\footnote{We stress once again that the lowest-order scalar amplitude can be calculated, in the long wavelength part of the spectrum, by a number of different background-independent techniques.} An example of a generalized scalar amplitude is
\be
A_s^2 = [1+f_s(\epsilon_\S, \eta_\S)]B_s\epsilon^{-\gamma}\frac{H^c}{\dot{\phi}^b}\,,\label{Sgen}
\ee
where $b$, $c$ and $\gamma$ are constants, $B_s$ is a normalization prefactor, and $f_s$ is a linear function of the SR parameters and of any dimensionless combination of cosmological quantities such as $m_4$, $\lambda$, $H$ and its time derivatives. By imposing the amplitude to be dimensionless, if $d_s$ is the dimension of $B_s$ then it must be $d_s+c-2b=0$. 

As has been said in Sec. \ref{Tpert}, the $h$-mappings are completely well motivated only in the quasi de Sitter case, so in principle a more general high-energy structure for the tensor amplitude than that of Eqs. (\ref{hmap2}) and (\ref{map}) is possible, keeping the functional part constrained by the zero-mode 5D calculation:
\ba \label{Tgen}
A_t^2 = [1+f_t(\epsilon_\S, \eta_\S)] B_t H^{2+\theta}\,,
\ea
where $f_t$ ($B_t$) is the tensor counterpart of $f_s$ ($B_s$). For reasons of simplicity, we have dropped the $H$ dependence in the SR functions since it gives rise to a polynomial structure for $A_s$ which does not change the main argument \cite{cal2}. Also, for the above considerations we can ignore these SR functions.

Equation (\ref{Sgen}) generates the spectral index
\be \label{ngen}
n_s-1 = (b +2\gamma) \eta_\S-(c +\gamma)\epsilon_\S\,,
\ee
and the running
\be \label{alphaq}
\alpha_s = \epsilon_\S \left[\left(1+\frac{2c+2\gamma}{b+2\gamma}\right)(n_s-1)+\left(\theta-1+\frac{2c+2\gamma}{b+2\gamma}\right)(c+\gamma)\epsilon_\S\right].
\ee
The case $\gamma=b=0\neq c$ ($\gamma=c=0\neq b$) is discarded because it is not possible to absorb the $\eta_\S$ ($\epsilon_\S$) dependence of the running into a cosmological observable. The standard case $c=2b=4$, $\gamma=0$ is trivial. 

Using  $\dot{\phi}^2 \propto \rho\epsilon_\S$, the tensor-to-scalar ratio reads, to lowest SR order,
\be\label{T/Sgen}
r \propto \epsilon_\S^{\gamma+ b/2} \rho^{[2(2q-1)+b-qc]/2}\,;
\ee
the constant $\gamma$ ha been introduced for trying to compensate by hand for the $\epsilon$ factor of Eq. (\ref{T/Sgen}). Now, assuming
\be \label{r}
r \approx a_r \epsilon_\S\,,
\ee
$a_r$ being a constant, in order to satisfy Eqs. (\ref{run1}) and (\ref{r}) we must solve the system
\ba
&&\gamma+ \frac{b}{2} = 1\,,\nonumber\\
&&2(2q-1) + b -qc   = 0\,,\nonumber\\
&&2\frac{c+\gamma}{b+2\gamma}   = 5a_r-1\,,\nonumber\\
&&\left(\theta-2+5a_r \right)(c+\gamma) = 12a_r^2\,.\nonumber
\ea
For general $q$ there do exist nontrivial solutions, but in GB ($a_r=1$) and RS ($a_r=3/2$) only the standard case is allowed. These considerations imply that a high-energy relation such that Eq. (\ref{r}) is possible only in the standard case $c=2b=4$, $\gamma=0$. In general, a traditional consistency relation will not be obtained in a braneworld context.

%%%%%%%%%%%%%%%%%%%%%%%%%%%%%%%%%%%%%%%%%%%%%%%%%%%%%%%%%%%%%%%%%%%%%%%%%%%%%%%%%%%%%%%%%%%%%%%%%%%%%%%%%%%%%%%%%%%%%%%%

\subsubsection{Patch vs patch} \label{pavspa}

Let us now seek what are the necessary conditions for obtaining the same set of consistency equations in two models $(\psi,\,\theta)$ and $(\psi',\,\theta')$. In the discussion on degeneracy we will not restrict ourselves to the RS and GB scenarios, since other gravity models can generate patches (i.e., effective Friedmann equations with $\theta$) different from $\theta = 0,\,\pm 1$.

There are several possible degeneracies which arise particular attention. The first one is \emph{exact}, that is $\alpha_s=\alpha_s'$, $\alpha_t=\alpha_t'$, and $n_t=n_t'$ to next-to-leading SR order; this model correspondence would open up many compelling possibilities, for example to construct a complicated braneworld scenario starting from a simple one. A second, more operative degeneracy is \emph{effective}, namely, one considers only the scalar running and the lowest-SR-order tensor index. Differences in next-to-leading-order tensor indices and in tensor runnings are neglected since the observational uncertainty on these quantities would blur any theoretical mutual deviation, at least for near-future experiments. When neither exact nor effective degeneracy are achieved, we will say that the two classes of models are \emph{definitely} nondegenerate. Another choice could be to consider \emph{tensor} degeneracy, of either lowest or next-to-lowest order, when the tensor index and its running are degenerate; tensor-degenerate models give the same gravitational wave spectrum. This degeneracy is useful when reducing the space of parameters in numerical analyses via the tensor-index consistency equation. 

The first degeneracy we investigate is between ordinary-scalar and tachyon-field scenarios. Let a prime denote the tachyon model; in order to get $(\phi,\,\theta)=(T,\,\theta')$, we match Eqs. (\ref{cealpsS}) and (\ref{cealpsT}), giving
\bs \label{deg1}
\ba
\theta &=& \frac{14+13\theta'}{4(3+\theta')}\,,\qquad \theta'\neq -3\,,\\
\zeta_q &=& \frac{3+\theta'}{5}\zeta_{q'}\,.
\ea
\es
From Eqs. (\ref{cenS}) and (\ref{cenT}) one gets either $\theta=2=\theta'$ or $\theta=-2=\theta'$; Eq. (\ref{cealpt}) is automatically degenerate for all $\theta$. Therefore, exact degeneracy is not allowed for finite $q$. For the effective degeneracy it is sufficient that 
\be \label{lSRce}
(2+\theta)\zeta_q=(2+\theta')\zeta_{q'}\,,
\ee
from the lowest-order tensor indices; coupling this condition with Eq. (\ref{deg1}) again gives $\theta=2=\theta'$. Therefore, $\phi$- and $T$-models are definitely nondegenerate for finite $q$. The quantity $\zeta_q$ is determined by the specific gravitational-geometric configuration one is considering, although from our point of view it plays the role of a purely numerical coefficient; it contributes to the normalization of the tensor amplitude through its general definition, Eq. (\ref{zgen}). By comparing Gauss-Bonnet and 4D scenarios, it is clear that the case $\zeta_q=\zeta_{q'}$ is possible even when $\theta \neq \theta'$. 

Scalar models in different braneworlds are definitely nondegenerate, $(\phi,\,\theta) \neq (\phi,\,\theta')$, since it must be $\theta=\theta'$ in the scalar running. The same conclusion holds for tachyon models, $(T,\,\theta)\neq(T,\,\theta')$ if $\theta\neq\theta'$.

Tensor degeneracy is straightforward: all the previous models are tensor degenerate to lowest SR order when Eq. (\ref{lSRce}) holds. In particular, ($i$) scalar and tachyon scenarios in a given patch and ($ii$) 4D and RS models  are tensor-degenerate at lowest order. Models with the same inflaton field $\psi$ and $\zeta_q=\zeta_{q'}$ are not tensor-degenerate; obviously, 4D and GB scenarios are not tensor-degenerate. Next-to-leading-order tensor degeneracy is possible only between $(\phi,\,-2)$ and $(T,\,-2)$, when
\bs\ba
\alpha_s(\phi) &\approx& \zeta_{1/2} r [4\zeta_{1/2}r+5(n_s-1)]\,, \label{cealpsS-2}\\
\alpha_s(T) &\approx& \zeta_{1/2} r (n_s-1)\,, \label{cealpsT-2}\\
n_t &=& 2\zeta_{1/2} r(n_s-1)\,,\label{cen-2}\\
\alpha_t &=&0\,.\label{cealpt-2}
\ea\es
As far as the author knows, no gravity model giving an effective Friedmann equation with $q=1/2$ has been developed so far. Note that this cosmology gives a scale-invariant spectrum ($n_s=1$) in the case of power-law inflation, $a=t^n$, irrespective of $n$ \cite{KM}. Table \ref{table6} summarizes the various degeneracies for finite $q$.
\begin{table}[ht]
\bc\begin{tabular}{l|cc}
Degeneracy               & $(\phi,\,\theta)-(T,\,\theta')$ & $(\psi,\,\theta)-(\psi,\,\theta')$ \\\hline
Exact                    &                $-$              &                $-$                 \\
Effective                &                $-$              &                $-$                 \\
Tensor n.t.l. SR &         $\theta=\theta'=-2$     &                $-$                 \\
Tensor lowest SR         & Eq. (\ref{lSRce})  &  Eq. (\ref{lSRce}) with $\theta \neq \theta'$\\ 
\end{tabular}\ec
\caption{\label{table6} Patch degeneracies for finite $q$. ``n.t.l.'' stands for next-to-lowest order.}
\end{table}

%%%%%%%%%%%%%%%%%%%%%%%%%%%%%%%%%%%%%%%%%%%%%%%%%%%%%%%%%%%%%%%%%%%%%%%%%%%%%%%%%%%%%%%%%%%%%%%%%%%%%%%%%%%%%%%%%%%%%%%%

\subsection{Observational degeneracy: theory vs data}\label{pert}

A new era of high-precision cosmological observations can now constrain the high-energy physics of braneworld models predicting significant deviations from the standard four-dimensional big bang scenario \cite{TSM,LCML,koy03,LT,LeL,BFM,LiS,TL}. The first-year data of WMAP \cite{ben03,kom03,spe03,pei03,bri03} confirm the standard scenario of a flat, adiabatic universe with Gaussian,\footnote{See Sec. \ref{gausspri}.} almost scale-invariant anisotropies, with a scalar amplitude $A_s^2 \approx 3.5 \cdot 10^{-10}$ at large scales \cite{pei03}. Bennett \emph{et al.} \cite{ben03} put a bound on the tensor-to-scalar ratio, 
\be\label{upr}
r<r_\text{max} = 0.06\,,
\ee
while the best fit for the scalar index is
\be\label{bfit}
n_s \approx 0.95\,. 
\ee 
Data analyses carried out for WMAP make use of the consistency equation (\ref{ce1}) in order to fix the tensor index and its running in the space of parameters, so a direct confrontation between these quantities and an experimental result with an associate error is not possible. The next-to-lowest order consistency relations (\ref{cenS}) and (\ref{cenT})  indeed break the degeneracy between scalar-field and tachyon-field models. Taking $n_t \approx -0.1$, one has $\big|n_t^{(\theta,\psi)}-n_t^{(\theta',\psi')}/n_t\big| \sim O(r^2)$; testing this effect would require an experimental uncertainty less than 1\% for the tensor index, a very difficult goal to hit for the missions of this and next generation.

A more useful quantity might be the scalar running (the consistency equation for the running of the tensor index is degenerate). In terms of SR parameters, this is a second-order quantity but it comes from the lowest-order part of the scalar amplitude. To quantify the effect of the extra dimension, we can use the recent CMB data coming from WMAP. With the upper bound (\ref{upr}) and the best fit (\ref{bfit}), the relative scalar running in two different patches is
\ba 
\Delta\alpha_s^\text{\tiny $(\psi_q \psi'_{q'})$} &\equiv&\alpha_s^{(\theta,\psi)}-\alpha_s^{(\theta',\psi')}\label{deltalp} \\
&\sim& O(10^{-2})\,, \nonumber
\ea
which is comparable both with the error in the estimate of Bridle \emph{et al.} \cite{bri03}, $\alpha_s=-0.04 \pm 0.03$, coming from the combination of WMAP and 2dFGRS (\textit{2 Degree Field Galaxy Redshift Survey}) data, and with the uncertainty estimate of the Planck mission \cite{planc,CGL}.

This estimate will be highly improved by both the updated WMAP data set and near-future experiments, including the European Planck satellite, for which the forecast precision should be ameliorated by one order of magnitude, $\Delta\alpha_s \sim O(10^{-3})$ \cite{BCLP}. 
 
For lower tensor-to-scalar ratios, the effect quits the window of detectability. For instance, scalar-driven and tachyon-driven scenarios lead to different predictions already in the standard 4D model when $r\approx r_\text{max}$,
\ba 
\Delta\alpha_s^\text{\tiny $(\phi_0 T_0)$} &=& 2r\left[3r+ (n_s-1)\right]\label{enfin}\\ 
&\approx& 0.01\,;\nonumber
\ea
however, with a lower tensor-to-scalar ratio $r$ and a scalar spectrum closer to scale invariance, discrimination between the two scenarios becomes harder to carry out via consistency equations. For a scale-invariant spectrum and taking the ratio $r = r_\text{max}/2=0.03$, which is within the $2\sigma$ likelihood bound of \cite{TL}, from Eq. (\ref{enfin}) we have $\Delta\alpha_s^\text{\tiny $(\phi_0 T_0)$} \approx 0.005$, while for $r =  r_\text{max}/3=0.02$ we get $\Delta\alpha_s^\text{\tiny $(\phi_0 T_0)$} \approx 0.002$, one order of magnitude smaller than the most optimistic high-ratio case. 

Therefore the most direct way to obtain experimental degeneracy is to consider functionally different
consistency equations, whatever they are, in the two regimes and small tensor-to-scalar ratios as well as a
nearly invariant spectrum, an eventuality which is quite possible in the range of the currently available data. A more refined analysis will be performed in the next chapter.

As a last comment we note that the comparison of the observable quantities $n_t$, $\alpha_t$, and $\alpha_s$ must be done with the consistency equations (\ref{ce}) and not through the SR parameters expressions, Eq. (\ref{SRexpr}). This is because both we are dealing with independent expressions and there is an evident ambiguity in relations between SR parameters and observables when considering energy-scale finite differences of the quantities of interest.

% version:  January 19, 2005

\newpage
\thispagestyle{empty}
$\vphantom{dunno how to skip the page}$
\newpage

\chapter{Noncommutative inflation} \label{noncom}

\begin{quote}
\textsl{What shall we use to fill the empty spaces?} \\--- Roger Waters (Pink Floyd), \textit{The wall}
\end{quote}\vspace{1cm}

%%%%%%%%%%%%%%%%%%%%%%%%%%%%%%%%%%%%%%%%%%%%%%%%%%%%%%%%%%%%%%%%%%%%%%%%%%%%%%%%%%%%%%%%%%%%%%%%%%%%%%%%%%%%%%%%%%%%%%%%
%%%%%%%%%%%%%%%%%%%%%%%%%%%%%%%%%%%%%%%%%%%%%%%%%%%%%%%%%%%%%%%%%%%%%%%%%%%%%%%%%%%%%%%%%%%%%%%%%%%%%%%%%%%%%%%%%%%%%%%%

\section{Introduction}

The idea that the early Universe experienced a phase of accelerated expansion has come to a crucial point. Born as a panacea for some problems of the standard big bang scenario, the inflationary paradigm has been developed and refined during these years, always successfully explaining the available observational data. The upcoming generation of high-precision cosmological experiments such as WMAP and Planck might definitely operate a selection on the great amount of inflation-inspired models. On the other hand, new theoretical scenarios in which the high-energy physics grows more and more in importance have produced a set of interesting research fields implementing the traditional 4D cosmology: therefore we have string cosmology, braneworld cosmology, noncommutative cosmology, and so on.

In their seminal paper \cite{BH}, Brandenberger and Ho presented a model of large-scale perturbation spectra, in which a noncommutative geometrical structure is generated by the stringy spacetime uncertainty relation (SSUR)
\be \label{SSURph}
\Delta t \Delta x_p \geq l_s^2\,,
\ee
where $l_s$ is the string length scale and $x_p=a(t)x$ is a physical space coordinate. It has been argued that this is a universal property for string and brane theory \cite{yon87,LY,yon00}. This picture (henceforth ``BH'') has then been further explored in \cite{cal5,cal4,CT,HL1,TMB,HL2,HL3,LLi1,KLM1,KLLM,cai04,myu04,LLi2,BY} and presents many common features with trans-Planckian scenarios with a modified dispersion relation \cite{dan02,EGKS,MB,cre03,ACT,KoKS,SrS,SP}.\footnote{A noncommutative spacetime structure may arise also by taking into account an holographic bound on the information contained in frozen perturbation modes per comoving volume \cite{hog04}.}

In this chapter we construct another noncommutative model based on the same philosophy of \cite{BH} and confront it with  BH in its two versions, the first one with the FRW 2-sphere factored out in the action measure and the second one with a unique effective scale factor. Scalar and tensor amplitudes and indices as well as consistency equations are obtained through the slow-roll formalism, both for an ordinary scalar field and a Born-Infeld cosmological tachyon. All the observables turn out to be functions of a noncommutative parameter, called $\mu$, measuring the magnitude of the Hubble energy $H$ at the time of horizon crossing in comparison with the fundamental string mass $M_s\equiv l_s^{-1}$.

Some works have studied the inflationary perturbations treating $\mu$ on either almost \cite{HL3,cai04} or exactly \cite{KLM1,KLLM} the same ground of the SR parameters, computing ultraviolet amplitudes and indices via a double or SR expansion for small parameters, respectively. Here we will follow a different approach and consider $\mu$ as a distinct object with respect to the SR tower; we will keep only the lowest-SR-order part of the observables and regard any $\mu$-term as pertaining these leading-SR-order quantities. We stress that, while the parameter $\mu$ accounts for nonlocal effects coming from the string scale $l_s$, the SR tower is determined by the dynamics of the cosmological inflationary expansion. Therefore, they describe quite distinct physical phenomena. In fact, there is no connection between $\mu$ and the recursively-defined SR tower, although even the first SR parameter is introduced by hand; the elements of the tower are built up of time derivatives of $H$ and they all vanish in a dS background, while $\mu$, which contains only the Hubble parameter and the string scale, does not. In particular, the lowest-SR-order spectral amplitudes, equivalent to those obtained in a quasi de Sitter model, will depend on $\mu$.
Beside this motivation, such a procedure has additional advantages. For example, we can study regimes with not-so-small $\mu$ within the SR approximation; secondly, if one keeps the magnitude of $\mu$ unconstrained, one can also explore the infrared region of the spectrum, $\mu \gg 1$, through appropriate techniques. 

These effective noncommutative models can be extended to braneworld scenarios in which the 3-brane experiences a cosmological expansion governed by an effective Friedmann equation. The precise theoretical setup is highly nontrivial even in the commutative case, because of the number of requirements to impose on the background forms and spacetime geometry in order to have a cosmological four-dimensional variety. We will phenomenologically assume to have a 3-brane in which the SSUR (\ref{SSURph}) holds for all the braneworld coordinates $\{x^\nu\}$, $\nu = 0,1,2,3$, while the extra dimension $y$ along the bulk remains decoupled from the associated *-algebra. 

%%%%%%%%%%%%%%%%%%%%%%%%%%%%%%%%%%%%%%%%%%%%%%%%%%%%%%%%%%%%%%%%%%%%%%%%%%%%%%%%%%%%%%%%%%%%%%%%%%%%%%%%%%%%%%%%%%%%%%%%
%%%%%%%%%%%%%%%%%%%%%%%%%%%%%%%%%%%%%%%%%%%%%%%%%%%%%%%%%%%%%%%%%%%%%%%%%%%%%%%%%%%%%%%%%%%%%%%%%%%%%%%%%%%%%%%%%%%%%%%%

\section{General setup}

We will keep the general framework of a noncommutative 3-brane in which, either in a limited time interval during its evolution or in a given energy patch, the cosmological expansion satisfies the effective Friedmann equation (\ref{FRW}). To diagonalize the noncommutative algebra and induce a pure 4D SSUR on the brane one might fix the expectation values of the 11D background fields such that the extra direction commutes, $[y,x^\nu]=0$. Some other subtleties to deal with are discussed in Sec. \ref{see}.

We identify the noncommutative string mass as the fundamental energy scale of the full theory. Best-fit analyses of BH noncommutative models give estimates for the string scale $M_s \sim 10^{11} - 10^{17}$ GeV \cite{HL1,TMB}. In typical Ho\v{r}ava-Witten scenarios, the fundamental scale is of order of the GUT scale, $M_s \sim 10^{16}$ GeV, consistently with our hypothesis.

%%%%%%%%%%%%%%%%%%%%%%%%%%%%%%%%%%%%%%%%%%%%%%%%%%%%%%%%%%%%%%%%%%%%%%%%%%%%%%%%%%%%%%%%%%%%%%%%%%%%%%%%%%%%%%%%%%%%%%%%

\subsection{Leading-order noncommutative observables}\label{ncsetup}

Let $A_\Phi$ denote a lowest-order perturbation amplitude, $A_\Phi \in \{A_t,\,A_s\}$; in general, it can be written as
\be \label{Anoncom}
A_\Phi(\mu_*,\,H,\,\psi) = A^{(c)}_\Phi (H,\,\psi)\,\Sigma (\mu_*)\,,
\ee
where $\mu_*$ is a noncommutative parameter to be defined later, $A^{(c)}_\Phi=A_\Phi(\Sigma\!\!=\!\!1)$ is the amplitude in the commutative limit, and $\Sigma(\mu_*)$ is a function encoding leading-SR-order noncommutative effects. The commutative observables derived in the previous chapter will be denoted by a superscript $(c)$.

It will turn out that, up to $O(\epsilon^2)$ terms,
\be \label{dotsig}
\frac{d \ln \Sigma^2}{d \ln k} = \sigma \epsilon \,,
\ee
where $\sigma = \sigma(\mu_*)$ is a function of $\mu_*$ such that $\dot{\sigma}=O(\epsilon)$. The spectral index is
\be \label{spindex}
n \equiv \frac{d \ln A_\Phi^2}{d \ln k} = n^{(c)}+\sigma\epsilon\,;
\ee
for the scalar spectrum, $n=n_s-1$. The index running is
\be
\alpha \equiv \frac{d n}{d \ln k} = \alpha^{(c)}+\frac{d^2 \ln \Sigma^2}{d \ln k^2}\,.
\ee
The last term can be written as
\be
\frac{d^2 \ln \Sigma^2}{d \ln k^2} = \sigma\epsilon \left[\left(2-\wteta-\bar{\sigma}\right)\epsilon-2\eta\right],
\ee
with $\bar{\sigma}\equiv -\dot{\sigma}/(\sigma H\epsilon)$ to first SR order. Because of Eq. (\ref{Anoncom}), the tensor-to-scalar ratio is $r=r^{(c)}$ and the consistency equations for the scalar runnings read
\ba 
\alpha_s(\phi) &=& r \zeta_q \left\{(5-\sigma)(n_s-1)+\left[4(3+\theta)-\sigma (7+\theta+\bar{\sigma}-\sigma)\right]r \zeta_q\right\}-\xi_\S^2,\nonumber\\\label{salph}\\
\alpha_s(T) &=& r \zeta_q \left\{(3+\theta-\sigma)(n_s-1)+\left[(2+\theta)(3+\theta)\right.\right.\nonumber\\
&&\qquad\qquad\qquad-\left.\left.\sigma (5+2\theta+\bar{\sigma}-\sigma)\right]r \zeta_q\right\}-\xi_\T^2.\label{talph}
\ea
The lowest-SR-order consistency equation for the tensor index is
\be \label{ntconeq}
n_t = [\sigma-(2+\theta)] \zeta_q r\,,
\ee
and its running is
\be
\alpha_t = r \zeta_q \left\{(2+\theta-\sigma)(n_s-1)+
\left[(2+\theta-\sigma)^2-\sigma \bar{\sigma} \right]r \zeta_q\right\}.\label{tenalph}
\ee
There is also a next-to-leading order version of Eq. (\ref{ntconeq}), which we will not consider here.

%%%%%%%%%%%%%%%%%%%%%%%%%%%%%%%%%%%%%%%%%%%%%%%%%%%%%%%%%%%%%%%%%%%%%%%%%%%%%%%%%%%%%%%%%%%%%%%%%%%%%%%%%%%%%%%%%%%%%%%%

\section{Noncommutative models}

Let us introduce the new time variable $\tau \in \mathbb{R}^+$, $\tau \equiv\int a\,dt=\int da/H$. With a constant SR parameter $\epsilon$, an integration by parts with respect to $a$ gives
\be \label{tauusef}
\tau = \frac{a}{(1+\epsilon)H} \approx \frac{a}{H}\,.
\ee
Inequality (\ref{SSURph}) can be rewritten in terms of comoving coordinates as
\be \label{SSUR}
\Delta \tau \Delta x \geq l_s^2\,,
\ee
and the corresponding algebra of noncommutative spacetime is time independent, 
\be \label{alg}
[\tau,x]=il_s^2\,.
\ee
The *-product realizing Eq. (\ref{alg}) is defined as
\be \label{*}
(f*g)(x,\tau)=e^{-(il_s^2/2)(\partial_x\partial_{\tau'}-\partial_{\tau}\partial_{x'})}f(x,\tau)g(x',\tau')\big|_{\text{\tiny \begin{tabular}{l} $x'=x$ \\ $\tau'=\tau$ \end{tabular}}}\,.
\ee
This realization of noncommutativity is in contrast with
\be \label{alt*}
[x_\mu,x_\nu]=i\theta_{\mu\nu}\,,
\ee
where $\theta_{\mu\nu}$ is a noncommutative parameter. This type of noncommutative cosmology, which does not preserve the FRW symmetries, has been studied in \cite{ChGS,LMMP,BG}. Other implementations can be found in \cite{LMMS,LMM,ABM,FKM,FKM2,GOR,AN,BPN,bar04,PM}.

%%%%%%%%%%%%%%%%%%%%%%%%%%%%%%%%%%%%%%%%%%%%%%%%%%%%%%%%%%%%%%%%%%%%%%%%%%%%%%%%%%%%%%%%%%%%%%%%%%%%%%%%%%%%%%%%%%%%%%%%

\subsection{BH models}

In the following we will adopt the short notation 
\be
a=a(\tau)\,,\qquad a_\pm\equiv a(\tau\pm kl_s^2)\,.
\ee
Consider now the action of a free scalar field $\Phi(\tau,x)$ living in a (1+1)-dimensional FRW space:
\be
S=\int d\tau dx\, \frac{1}{2}\left(a^2\partial_\tau \Phi^\dagger\partial_\tau \Phi-a^{-2}\partial_x\Phi^\dagger\partial_x\Phi\right).
\ee
In the noncommutative models we will study, each conventional product is replaced by the *-product (\ref{*}); thus, the gravitational sector of the theory is not a completely passive spectator but is involved via the *-coupling of the metric with the matter content. The new 2D action reads, noting that $a^2=a*a$ \cite{BH},
\be \label{bhaction}
S_\text{BH}=\int d\tau dx\, \frac{1}{2}\left(\partial_\tau \Phi^\dagger*a^2*\partial_\tau \Phi-\partial_x\Phi^\dagger*a^{-2}*\partial_x\Phi\right).
\ee
In the comoving momentum space,
\be
\Phi(\tau,x)= \int_{k<k_0} \frac{dk}{\sqrt{2\pi}}\,\Phi_k(\tau) e^{ikx}\,,
\ee
where $k_0$ is a cutoff realizing the stringy uncertainty relation. The most convenient way to recast the action is to write the scale factor as a Fourier integral, $a^2(\tau)=\int d\omega\, a_\tau^2(\omega) e^{i\omega\tau}$, and perform the *-products of the complex exponentials in the integrand, removing the cutoff in the limit $k_0 \rightarrow \infty$ when absorbing the $\delta^{(3)}$-integrals in momentum spaces. The result is
\be\label{ncac}
S \approx \int_{k<k_0} d\tau dk\, \frac{1}{2}\left(\beta_k^+ \partial_\tau \Phi_{-k}\partial_\tau \Phi_k-\beta_k^-k^2\Phi_{-k}\Phi_k\right)\,,
\ee
where
\be \label{BHbeta}
\beta_k^\pm = \case{1}{2} \left(a_+^{\pm 2}+a_-^{\pm 2}\right).
\ee
Defining two new objects
\ba
a^2_\eff &\equiv& \sqrt{\frac{\beta^+_k}{\beta^-_k}}=a_+ a_-\,,\label{aeffbh}\\
y^2 &\equiv& \sqrt{\beta^+_k \beta^-_k}=\frac{a^2_+ + a^2_-}{2a_+a_-}\,,
\ea
and the effective conformal time coordinate
\be \label{teta}
\teta \equiv \int\frac{d\tau}{a^2_\eff}\,,
\ee
the scalar action becomes
\be \label{1+1}
S \approx \int_{k<k_0} d\teta dk\, \frac{1}{2}y^2\left(\partial_\teta\Phi_{-k}\partial_\teta\Phi_k-k^2\Phi_{-k}\Phi_k\right)\,.
\ee
To estimate the cutoff $k_0$, we note that the energy for the mode $k$ with respect to the time variable $\tau$ is, by the action (\ref{ncac}), $E_k=ka_\eff^{-2}$. Then, the saturated SSUR (\ref{SSUR}) with $\Delta x \sim k^{-1}$ and $\Delta \tau \sim E_k^{-1}$ yields 
\be \label{cutoff}
k_0 \equiv M_s a_\eff\,.
\ee

%%%%%%%%%%%%%%%%%%%%%%%%%%%%%%%%%%%%%%%%%%%%%%%%%%%%%%%%%%%%%%%%%%%%%%%%%%%%%%%%%%%%%%%%%%%%%%%%%%%%%%%%%%%%%%%%%%%%%%%%

\subsection{A new prescription for noncommutativity}

Cyclic permutations of the *-product inside the integral (\ref{bhaction}) leave the action invariant. Therefore, it is natural to see whether a different noncyclic ordering of the factors gives a theory with interestingly new predictions. The other nontrivial noncommutative action one can obtain is
\be
S_\text{new} = \int d\tau dx\, \frac{1}{2}\left(\partial_\tau \Phi^\dagger*a*\partial_\tau \Phi*a-\partial_x\Phi^\dagger*a^{-1}*\partial_x\Phi*a^{-1}\right).
\ee
The same computational pattern of the previous section leads to Eq. (\ref{1+1}) with $\beta_k^\pm$ given by
\be
\beta_k^\pm = \frac{a^{\pm 1}}{2} \left(a_+^{\pm 1}+a_-^{\pm 1}\right)\,,
\ee
and
\ba
a_\eff^2 &=& a \sqrt{a_+ a_-}\,, \label{aeffnew}\\
y^2 &=& \frac{a_+ + a_-}{2\sqrt{a_+a_-}}\,.
\ea
In this case there is only a partial smearing of the product of scale factors and one might guess that the resulting noncommutative phenomenology would be less pronounced than that of BH model. In the UV limit it will turn out that, within a given variation of the noncommutative parameter and in \emph{some} region in the space of parameters, the range of $\Delta\alpha_s^\text{\tiny $(\psi_q \psi'_{q'})$}$ is slightly smaller than in the BH model but always of the same order of magnitude. In the infrared region, however, the two models are almost undistinguishable; see Sec. \ref{disc}.

%%%%%%%%%%%%%%%%%%%%%%%%%%%%%%%%%%%%%%%%%%%%%%%%%%%%%%%%%%%%%%%%%%%%%%%%%%%%%%%%%%%%%%%%%%%%%%%%%%%%%%%%%%%%%%%%%%%%%%%%

\subsection{Four-dimensional effective actions and amplitudes} \label{see}

When going to 3+1 dimensions, the measure $z_k^2$ of the integral will contain the nonlocal effect coming from the SSUR:
\be \label{action}
S \approx \int_{k<k_0} d\teta d^3\mathbf{k}\, \frac{1}{2}z_k^2\left(\partial_\teta\Phi_{-\mathbf{k}}\partial_\teta\Phi_\mathbf{k}-k^2\Phi_{-\mathbf{k}}\Phi_\mathbf{k}\right)\,.
\ee
Here we will consider two classes of models. In the first one, we suppose the total measure to be given by the product of the noncommutative (1+1)-measure and the commutative one:
\be \label{mod1}
z_k = z y\,;
\ee
then, as we are going to show in a moment,
\be \label{SigmaI}
\Sigma = \frac{a^2_\eff}{a^2 y} \qquad \mbox{(class 1)}\,.
\ee
These models, in which the FRW 2-sphere is factored out, will be dubbed as ``1.'' Another interesting prescription consists in replacing the commutative scale factor in the measure with the effective one; then, $ay \rightarrow a_\eff$, 
\be \label{mod2}
z_k=z \frac{a_\eff}{a}\,,
\ee
and
\be\label{SigmaII}
\Sigma = \frac{a_\eff}{a}  \qquad \mbox{(class 2)}\,;
\ee
models with this $\Sigma$ will be named ``2.''

Let us now look at cosmological perturbations coming from an inflationary era and assume, as it is the case, that $\Phi$ is a generic perturbation satisfying the action (\ref{action}). The spectral amplitude coming from the $k$th mode of the perturbation is given by Eq. (\ref{ampli}), where the expression is evaluated at the reference time $\teta_*$ to be discussed in a while. Via a change of variable, 
\be
u_\mathbf{k}=-z_k\Phi_\mathbf{k}\,,
\ee
the action (\ref{action}) gives the Mukhanov equation
\be \label{muk}
\left(\frac{d^2}{d\teta^2}+k^2-\frac{1}{z_k}\frac{d^2z_k}{d\teta^2}\right)u_\mathbf{k}=0\,.
\ee
Noting that $d \teta /d\eta =(a/a_\eff)^2$, we get the useful relation
\be \label{useful}
\teta \approx \frac{-1}{aH}\left(\frac{a}{a_\eff}\right)^2,
\ee
in the lowest SR approximation. If the SR parameters are small, then they are constant to leading order because their derivatives are higher order. It is then possible to solve the Mukhanov equation with exactly constant SR parameters and perturb the obtained solution. Such cosmological solutions do exist and can be constructed in a variety of situations [see the discussion below Eq. (\ref{efold2})]; among them, a particularly important one is power-law inflation, which we will use when considering the infrared region of the spectrum. Therefore,
\be
\frac{1}{z_k}\frac{d^2z_k}{d\teta^2} \approx \left(\frac{a_\eff}{a}\right)^4 \frac{1}{z}\frac{d^2z}{d\eta^2}=\left(\frac{a_\eff}{a}\right)^4\frac{\nu^2-1/4}{\eta^2}\approx\frac{\nu^2-1/4}{\teta^2}\,,
\ee
where $\nu = 3/2+O(\epsilon)$. With constant $\nu$, the solution of this equation is the same as that of the commutative case, namely $|u_\mathbf{k}| \propto (-\teta)^{1/2} H_\nu^{(1)}(-k\teta)$. In the long wavelength limit $k/(aH)\rightarrow 0$, when the mode with comoving wave number $k$ is well outside the horizon, the appropriately normalized solution becomes, from Eq. (\ref{useful}),\footnote{For the tachyon the comoving momentum is rescaled as $k\to kc_S$, but at lowest SR order $c_S\approx 1$.}
\be\label{ncdSsol}
|u_\mathbf{k}|^2=\frac{1}{2k}\left(\frac{-1}{k\teta}\right)^2=\frac{1}{2k}\left(\frac{aH}{k}\right)^2\left(\frac{a_\eff}{a}\right)^4;
\ee
finally, one gets Eq. (\ref{Anoncom}) by inserting either definition (\ref{mod1}) or (\ref{mod2}) in Eq. (\ref{ampli}).

Given a noncommutative brane in a commutative bulk, the nonlocal smearing will only affect the pure four-dimensional part of the graviton-zero-mode action, while leaving the pure transversal normalization unchanged; from the discussion in Sec. \ref{Tpert}, it is then clear that the noncommutative tensor spectral amplitude will be $A_t^2 =A_t^{(c)2}\Sigma^2\propto \xi_0^2(y_b) A_{t(\text{4D})}^2$. Therefore, for the gravitational spectrum, $\Phi$ denotes the coefficient functions of the noncommutative 4D polarization tensor $h_{\mu\nu}^{(0)}(*x)$ and $z$ is given by Eq. (\ref{zgrav}).

The action and Mukhanov equation for a perturbation generated by a tachyon field has an additional factor in front of $k^2$ in Eqs. (\ref{action}) and (\ref{muk}), namely the speed of sound for the perturbation: $k^2 \rightarrow c_S^2k^2$. Since the SSUR does not affect products of homogeneous quantities, the noncommutative generalization of the tachyonic scalar amplitude is straightforward \cite{LLi1}. Now, one may ask how the inhomogeneous version of the original Born-Infeld Lagrangian (\ref{ta}) is modified when inserting the *-products. Let us recall that noncommutativity naturally arises in string theory when a Neveu-Schwarz--Neveu-Schwarz (NS-NS) $B$-field is switched on in the low-energy tree-level action. However, this results in a linearization of the tachyonic action and, on the other hand, a large noncommutative parameter may trigger brane decay processes \cite{DRM}; therefore, the simple noncommutative version of the cosmological tachyon might seem too na\"{i}ve.

Anyhow, tachyon scenarios are not new to counterintuitive behaviours. In the slow-roll approximation, $\epsilon \propto \dot{T}^2 \ll 1$, the Lagrangian (\ref{ta}) can be linearized and the rescaled field $\phi=\sqrt{V}T$ behaves like an ordinary scalar; nevertheless, the theoretical prediction encoded in the consistency relations is different with respect to that of the genuine scalar scenario, see Eqs. (\ref{salph}) and (\ref{talph}). Here, there happens something similar, imagining to turn on and increase the $B$-field smoothly, and the final result differs from the scalar case indeed. 

Moreover, the stringy linearization is a feature of realization (\ref{alt*}) rather than (\ref{alg}) and the former may give rise to a different cosmological model in which FRW isotropy is not preserved \cite{LMMP}; also, \textit{a priori} it would be highly nontrivial to construct a Lorentz-violating cosmological brane model (in fact, in the case of a dS brane, maximal symmetry is crucial for coordinate-separating the graviton wave equation \cite{DLMS,LMW}).

To further understand the difficulties lying in a full implementation of noncommutative string theory in cosmology, it is important to stress that all that has been said about the algebra (\ref{alt*}) (i.e. instability and cosmological scenarios) is true only in a purely spatial *-product, $\theta_{0i}=0$. When trying to introduce noncommutativity in both space and time, as is the case of realization (\ref{alg}), it may be difficult to achieve a coherent, well-defined theory. In fact, in the Seiberg-Witten limit reproducing the noncommutative geometry, $\theta_{\mu\nu}$ and $\alpha'B_{\mu\nu}$ are kept fixed while $B_{\mu\nu}\rightarrow \infty$ and the Regge slope $\alpha' \rightarrow 0$. Let $E_i=B_{0i}$ be the electric part of the NS 2-form and assume $|E_i|\neq 0$. Then, while the $B$-field goes to infinity and approaches the critical value $E_{cr}=(2\alpha')^{-1}$, a classical instability develops and the rate of open string pair production diverges \cite{BP}; heuristically, the string is tore apart by the increasing electric field strength. For these reasons we regard algebra (\ref{alg}) as the starting point of the cosmological setup rather than the ultimate product of some high-energy theory, for the moment leaving the details of the latter aside. 

%%%%%%%%%%%%%%%%%%%%%%%%%%%%%%%%%%%%%%%%%%%%%%%%%%%%%%%%%%%%%%%%%%%%%%%%%%%%%%%%%%%%%%%%%%%%%%%%%%%%%%%%%%%%%%%%%%%%%%%%

\subsection{The UV region}

In order to correctly evaluate the perturbation spectra, one must determine the time $\teta_0$ when the $k$th mode is generated and, later, when it crosses the Hubble horizon. Because of the momentum cutoff (\ref{cutoff}), the analysis for the noncommutative case must be conducted separately in the mildly and strongly noncommutative regions. 

From the very beginning, one can define the time $\teta_*$ when a perturbation with wave number $k$ crosses the horizon by the formula $k_* \equiv k(\teta_*) = a(\teta_*) H(\teta_*)$. This relation provides an operative definition of the number of $e$-foldings ($k \propto H e^N$) and the time variation of $k$, Eq. (\ref{dotk}). Of course, this is valid for any cosmology in which time definitions have zero uncertainty, that is for commutative cosmologies and noncommutative cosmologies in the range far from the upper bound (\ref{cutoff}), in the so-called ultraviolet (UV) region, where $k_* \ll k_0$. In fact, the time of horizon crossing is different from its commutative counterpart $\teta_c$, since $\teta_c < \teta_*$ and the crossing mode is delayed \cite{BH}. In \cite{KLLM} this effect is quantified as $k_c/k_* = \exp (\teta_c-\teta_*)$.

On the contrary, one might define the horizon crossing through the $z_k$ function as
\be \label{optpivot}
k^2_* = \frac{1}{z_k}\frac{d^2z_k}{d\teta^2} \approx 2(aH)^2\,,
\ee
and get an extra factor of 2; due to the structure of the Mukhanov equation, this approach would be valid in any case, let it be the commutative or the noncommutative one. 

In the UV region, the cosmological energy scale when the perturbation is generated is much smaller than the stringy scale, $H(\teta\! >\!\!\teta_0) \leq H(\teta_0) \ll M_s$, and noncommutative effects are soft; thus, the smeared versions $a_\pm$ of $a$ can be approximated by $a$ since
\be \label{UV}
kl_s^2 \ll \tau_* \qquad \mbox{(UV region)}\,,
\ee
from Eq. (\ref{tauusef}). It is convenient to define the noncommutative parameter
\be \label{mu}
\mu \equiv \left(\frac{kH}{a M_s^2}\right)^2,
\ee
whose time derivative is
\be \label{dotmu}
\dot{\mu} = -4H\mu\epsilon\,.
\ee
Note that this relation states that $\mu$ is almost constant in a rapidly accelerating background, regardless of its magnitude. The analogy with the evolution equations of the SR tower, e.g., Eq. (\ref{dotSR}), suggested the authors of \cite{KLM1,KLLM} treat $\mu$ as a sort of SR parameter, keeping all the parameters at the same truncation level in the expressions of the UV observables.

At horizon crossing,
\be \label{mu*}
\mu_*=\mu |_{k=\sqrt{2}aH}=2\left(\frac{H}{M_s}\right)^4,
\ee
and Eq. (\ref{dotmu}) is valid for $\mu_*$, too. The ultraviolet region is by definition the region in which $H/M_s \ll 1$;\footnote{Without risk of confusion, we will continue to use the symbol $\mu$ to indicate the ratio $H/M_s$ when discussing the UV limit ($\mu \ll 1$) of spectral quantities.} it is characterized by long wavelength perturbations generated inside the Hubble radius and, in a CMB spectrum, this would correspond to the portion of the Sachs-Wolfe (inflationary) plateau with not-too-small spherical modes, $10 \lesssim l \lesssim 100$.

In the commutative case, to use one pivot scale instead of the other amounts to different next-to-lowest-order expansions in the SR parameters; the 4D consistency equations are thus unaffected, since the introduction of the optimized pivot scale (\ref{optpivot}) results in a rescaled coefficient $C \rightarrow C+\ln \sqrt{2}$ which is not present in them (see, e.g., \cite{KLLM} and references therein for details). This is also true in the RS scenario \cite{cal2} as well as in general patch cosmology \cite{cal5}.

In the noncommutative case, the change of the pivot scale doubles the magnitude of the parameter (\ref{mu*}). The resulting models will display the same theoretical features of the $k=aH$ models, but shifted backward along the energy scale determined by the ratio $H/M_s$. Observational constraints should take the rescaling of the string mass into account, when changing the pivot scale.

In the limit (\ref{UV}), we can Taylor expand the scale factors $a_\pm$ around $\tau$ for small $k$. To first order in the SR parameters and to all orders in $\mu$, the nonlocal dependence of the scale factor is
\ba
a(\tau \pm kl_s^2) &=& a(\tau)\,\left\{1 \pm \sqrt{\mu}+\left[\pm\sqrt{\mu}-(1 \pm\sqrt{\mu}) \ln (1\pm\sqrt{\mu})\right]\epsilon\right\}+O(\epsilon^2)\,,\nonumber\\
\ea
where the factor in front of $\epsilon$ comes from a series whose radius of convergence is $\mu \leq 1$. More precisely, when $\mu_* \leq 1$ then $H/M_s \lesssim 0.8\,.$ Since we are interested in lowest-SR-order amplitudes, we can neglect the SR tower and find
\be \label{aUV}
a_\pm \approx \left(1 \pm \sqrt{\mu}\right)a\,.
\ee
The concrete procedure to compute the spectral amplitudes will be to use the horizon-crossing formula (\ref{optpivot}) at $\teta_*$ in the UV region, and the saturation time $\teta_0$ in the infrared (IR) region. In \cite{BH} and other papers these instants are dubbed $\teta_k$ and $\teta_k^0$, respectively, to highlight the dependence on the wave number.

%%%%%%%%%%%%%%%%%%%%%%%%%%%%%%%%%%%%%%%%%%%%%%%%%%%%%%%%%%%%%%%%%%%%%%%%%%%%%%%%%%%%%%%%%%%%%%%%%%%%%%%%%%%%%%%%%%%%%%%%

\subsection{BH model IR region}

In the IR region things are quite different: the wave modes are generated outside the horizon and, since they are frozen until they cross the horizon, their magnitude depends on the time when they were generated. This corresponds to the ($k$-dependent) time $\teta_0$ when the SSUR is saturated, $k(\teta_0)=k_0(\teta_0)$, and quantum fluctuations start out with their vacuum amplitude. The effective and smeared scale factors must be evaluated at this instant; the expansion (\ref{aUV}) is no longer valid since $H \gg M_s$ in the infrared. To proceed one can explicitly use the exact solution around which the equation of motion for the perturbation has been expanded. The power-law solution $a=a_0t^n$ corresponds to a constant index $w$, when the scale factor is $a(\tau)=\alpha_0 \tau^{n/(n+1)}$, and $H= n\alpha_0\tau^{-1/(n+1)}/(n+1)$, where $\alpha_0=(n+1)^{n/(n+1)}a_0^{1/(n+1)}$. For an exponential scale factor (de Sitter expansion, $n \rightarrow \infty$), $a(\tau)= H\tau$, in accordance with Eq. (\ref{tauusef}). From Eqs. (\ref{cutoff}) and (\ref{aeffbh}),
\be
\tau_0 = kl_s^2 \sqrt{1+\delta}\,,
\ee
where $\tau_0=\tau(\teta_0)$ and
\be
\delta \equiv \left(\frac{2}{\mu_*}\right)^{1/2} = \left(\frac{M_s}{H}\right)^2.
\ee
In the infrared region, 
\be \label{IR}
kl_s^2 \approx \tau_0 \qquad \mbox{(IR region)}\,,
\ee
and
\ba
a     &=& Hkl_s^2 \sqrt{1+\delta}\,,\label{adelta}\\
a_\pm &=& Hkl_s^2 \left(\sqrt{1+\delta} \pm 1\right),\label{apmdelta}
\ea
where evaluation at $\tau_0$ is understood. When $\delta \gg 1$, we recover the UV or quasicommutative region since $kl_s^2 \ll \tau_0 \leq \tau_*$. Actually, the UV and IR spectra may be joined together in an intermediate region, as it was shown in \cite{TMB}; in particular, see their Eq. (12), corresponding in the de Sitter limit to $\Sigma^2 \sim \delta (1-3\sqrt{\mu_*/2})$. We will not be able to recover this spectrum within our formalism; however, we will describe other hybrid regimes by using the methods adopted in the IR region ($2 \leq l \lesssim 10$) for $\delta \gg 1$. For future reference, note that
\be \label{dotdelta}
\dot{\delta} =2\delta H\epsilon\,.
\ee

%%%%%%%%%%%%%%%%%%%%%%%%%%%%%%%%%%%%%%%%%%%%%%%%%%%%%%%%%%%%%%%%%%%%%%%%%%%%%%%%%%%%%%%%%%%%%%%%%%%%%%%%%%%%%%%%%%%%%%%%

\subsection{New model IR region}

In the ``New'' model, the effective scale factor is given by Eq. (\ref{aeffnew}). From Eq. (\ref{cutoff}),
\be
\tau_0 = kl_s^2 \sqrt{1+\gamma}\,,
\ee
where
\be
\gamma \equiv \frac{1}{2} \left(\sqrt{1+4\delta^2}-1\right).
\ee
With this definition, the new expressions for $a$ and $a_\pm$ are identical to Eqs. (\ref{adelta}) and (\ref{apmdelta}), with $\delta$ replaced by $\gamma$. Equation (\ref{dotdelta}) is replaced by
\be
\dot{\gamma} = \frac{4\gamma(\gamma+1)}{1+2\gamma}H\epsilon\,.
\ee
In the far IR region, $\gamma \approx \delta^2 \ll 1$, while in the UV limit $\gamma \approx \delta \gg 1$.

Without further justifications, the IR region of the spectrum, $H \gg M_s$, may be not very satisfactory from a string-theoretical point of view, either because we are above the fundamental energy scale\footnote{However, the space-momentum stringy uncertainty relation, implying $\Delta x_p \geq l_s$, is not a universal property of the theory.} and due to the above-mentioned classical instabilities. As it is done in many other occasions in early-Universe cosmology, we will turn a blind eye to this point and seek what are the observational consequences of the extreme regime of the present noncommutative models.

%%%%%%%%%%%%%%%%%%%%%%%%%%%%%%%%%%%%%%%%%%%%%%%%%%%%%%%%%%%%%%%%%%%%%%%%%%%%%%%%%%%%%%%%%%%%%%%%%%%%%%%%%%%%%%%%%%%%%%%%
%%%%%%%%%%%%%%%%%%%%%%%%%%%%%%%%%%%%%%%%%%%%%%%%%%%%%%%%%%%%%%%%%%%%%%%%%%%%%%%%%%%%%%%%%%%%%%%%%%%%%%%%%%%%%%%%%%%%%%%%

\section{Noncommutative zoology} \label{zoo}

We are ready to collect all the machineries developed so far and inspect the noncommutative models at hand. 

%%%%%%%%%%%%%%%%%%%%%%%%%%%%%%%%%%%%%%%%%%%%%%%%%%%%%%%%%%%%%%%%%%%%%%%%%%%%%%%%%%%%%%%%%%%%%%%%%%%%%%%%%%%%%%%%%%%%%%%%

\subsection{BH1}

In the BH1 case, 
\be
\Sigma^2 = \frac{2(a_+a_-)^3}{a^4 (a^2_+ + a^2_-)}\,.
\ee
In the UV region,
\be
\Sigma^2 = \frac{(1-\mu_*)^3}{1+\mu_*}\,,\qquad \sigma = \frac{8\mu_*(2+\mu_*)}{1-\mu_*^2}\,,\qquad 
\bar{\sigma} = \frac{8(\mu_*^2+\mu_*+1)}{(2+\mu_*)(1-\mu_*^2)}\,.
\ee 
For $\mu \ll 1$ one recovers the nearly commutative, $\mu$-expanded behaviour\footnote{Throughout the paper we will keep only the leading-order term in the approximated $\bar{\sigma}$ since there is a $\sigma$ factor in front of it in Eqs. (\ref{salph}) and (\ref{talph}).} 
\be
\Sigma^2 \approx 1-4\mu_*\,,\qquad \sigma \approx 16\mu_*\,,\qquad \bar{\sigma} \approx  4\,.
\ee
In the IR region,
\be
\Sigma^2 = \frac{\delta^3}{(2+\delta)(1+\delta)^2},\quad
\sigma = \frac{4(2\delta+3)}{(2+\delta)(1+\delta)},\quad
\bar{\sigma} = \frac{2\delta(2\delta^2+6\delta+5)}{(3+2\delta)(2+\delta)(1+\delta)}.
\ee
In the commutative limit ($\delta \gg 1$), $\Sigma^2 \approx 1$, while in the strongly noncommutative regime ($\delta \ll 1$), $\Sigma^2 \approx \delta^3/2$ and $\sigma \approx 6-5\delta$, in agreement with \cite{TMB}.\footnote{Eqs. (44)--(47) of \cite{BH} are not correct, due to a missing power of $y$ in the inserted $z_k^2$; in Eqs. (23)--(25) of \cite{TMB} the correct amplitude is recovered.}

%%%%%%%%%%%%%%%%%%%%%%%%%%%%%%%%%%%%%%%%%%%%%%%%%%%%%%%%%%%%%%%%%%%%%%%%%%%%%%%%%%%%%%%%%%%%%%%%%%%%%%%%%%%%%%%%%%%%%%%%

\subsection{BH2}

From Eqs. (\ref{aeffbh}) and (\ref{SigmaII}),
\be
\Sigma^2 = \frac{a_+a_-}{a^2}\,.
\ee
In the UV,
\be
\Sigma^2 = 1-\mu_*\,,\qquad \sigma = \frac{4\mu_*}{1-\mu_*}\,,\qquad \bar{\sigma} = \frac{4}{1-\mu_*}\,.
\ee 
For $\mu \ll 1$, $\sigma \approx 4\mu_*$ and $\bar{\sigma} \approx 4$. In the IR,
\be
\Sigma^2 = \frac{\delta}{\delta+1}\,,\qquad \sigma = \frac{2}{\delta+1}\,,\qquad \bar{\sigma} = \frac{2\delta}{\delta+1}\,.
\ee 
When $\delta \ll 1$, $\sigma \approx 2(1-\delta)$.

%%%%%%%%%%%%%%%%%%%%%%%%%%%%%%%%%%%%%%%%%%%%%%%%%%%%%%%%%%%%%%%%%%%%%%%%%%%%%%%%%%%%%%%%%%%%%%%%%%%%%%%%%%%%%%%%%%%%%%%%

\subsection{New1}

The correction to the commutative amplitude reads
\be
\Sigma^2 = \frac{2(a_+a_-)^{3/2}}{a^2 (a_+ + a_-)}\,.
\ee
In the UV region,
\be
\Sigma^2 = (1-\mu_*)^{3/2},\qquad \sigma = \frac{6\mu_*}{1-\mu_*}\,,\qquad \bar{\sigma} = \frac{4}{1-\mu_*}\,.
\ee 
In the IR limit,
\be
\Sigma^2 = \left(\frac{\gamma}{1+\gamma}\right)^{3/2},\qquad \sigma = \frac{6}{1+2\gamma}\,,\qquad \bar{\sigma} = \frac{8\gamma(\gamma+1)}{(1+2\gamma)^2}\,.
\ee 
In the strongly noncommutative limit ($\gamma \ll 1$), $\Sigma^2=\gamma^{3/2}$ and $\sigma = 6 +O(\delta^2)$.

%%%%%%%%%%%%%%%%%%%%%%%%%%%%%%%%%%%%%%%%%%%%%%%%%%%%%%%%%%%%%%%%%%%%%%%%%%%%%%%%%%%%%%%%%%%%%%%%%%%%%%%%%%%%%%%%%%%%%%%%

\subsection{New2}

From Eqs. (\ref{aeffnew}) and (\ref{SigmaII}),
\be
\Sigma^2 = \frac{\sqrt{a_+a_-}}{a}\,.
\ee
The UV limit gives
\be
\Sigma^2 = \sqrt{1-\mu_*}\,,\qquad \sigma = \frac{2\mu_*}{1-\mu_*}\,,\qquad \bar{\sigma} = \frac{4}{1-\mu_*}\,.
\ee
In the IR region,
\be
\Sigma^2 = \left(\frac{\gamma}{\gamma+1}\right)^{1/2},\qquad \sigma = \frac{2}{1+2\gamma}\,,\qquad
\bar{\sigma} = \frac{8\gamma(\gamma+1)}{(1+2\gamma)^2}\,.
\ee
For $\gamma \ll 1$, $\sigma = 2 +O(\delta^2)$.

%%%%%%%%%%%%%%%%%%%%%%%%%%%%%%%%%%%%%%%%%%%%%%%%%%%%%%%%%%%%%%%%%%%%%%%%%%%%%%%%%%%%%%%%%%%%%%%%%%%%%%%%%%%%%%%%%%%%%%%%

\subsection{Discussion} \label{disc}

To summarize, we can compare the considered models in the perturbative limits, that is, the UV commutative limit ($\mu \ll 1$) and the IR noncommutative limit ($\delta \approx \sqrt{\gamma} \ll 1$). Trivially, in the nonperturbative or commutative IR region ($\delta\approx\gamma \gg 1$), $a \approx a_\pm$ and one recovers the standard spectrum, $\Sigma^2=1$ and $\sigma=0$; also, by construction, the noncommutative UV region is ill-defined. 

In general, we can write the UV commutative limit of the relevant quantities as
\bs \label{deepuv}
\ba
\Sigma^2     &\approx& 1-b\mu_*\,,\\
\sigma       &\approx& 4b\mu_*\,,\label{sigapp}\\
\bar{\sigma} &\approx& 4\,,
\ea
\es
where $b$ is a constant. As anticipated, the structure of the IR amplitudes also permits a perturbative expansion around $1/\delta$; in this case, spectral amplitudes are evaluated at $k \lesssim k_0$ via the power-law solution. The IR commutative limit is then
\bs \label{uvir}
\ba
\Sigma^2     &\approx& 1-b\sqrt{\mu_*/2}\,,\\
\sigma       &\approx& 2b\sqrt{\mu_*/2}\,,\\
\bar{\sigma} &\approx& 2\,;
\ea
\es
from the previous discussions, it is natural to interpret this as an intermediate momentum region at the edge of the UV regime, around $\mu \lesssim 1$ where Eq. (\ref{aUV}) ceases to be valid, and corresponding to perturbations generated across the Hubble horizon. In fact, what one does is hit this region starting from the low-momentum IR side. The above-mentioned junction spectrum of \cite{TMB} is located somewhere closer to the infrared. 

Table \ref{table7} shows that all the models display similar asymptotic limits towards different numerical coefficients, the BH ones being larger than the New ones; the coefficient of BH1 is 4 times that of model 2 within each region (UV or IR), while this ratio is reduced to $b_1/b_2=3$ in the New model. Thus, there is less difference between model New1 and model New2 with respect to that occurring between BH1 and BH2, further confirming that the ``half-smearing'' of the new scenario somehow softens noncommutative effects.
\begin{table}[ht]
\begin{center}
\begin{tabular}{l|c||cc}
        &   UV    &  \multicolumn{2}{c}{IR}\\
        &   $b$   &    $\Sigma^2$ & $\sigma$ \\ \hline
BH1     &    4    &  $\delta^3/2$ &     6    \\
New1    &   3/2   &  $\delta^3$   &     6    \\
BH2     &    1    &   $\delta$    &     2    \\
New2    &   1/2   &   $\delta$    &     2    \\
\end{tabular}\end{center}
\caption{\label{table7} Noncommutative perturbation amplitudes in the UV (first order in $\mu \ll 1$) and IR (first order in $\delta \ll 1$) limits.}\end{table}

The intermediate spectrum (\ref{uvir}) breaks down when $\Sigma^2<0$, that is when $H/M_s >$ 0.5 (BH1), 0.8 (New1), 1 (BH2) and 1.4 (New2); therefore Eq. (\ref{uvir}) well describes class 2 models at the UV boundary $\mu \lesssim 1$ while it is not particularly reliable for class 1 models.

In the deep UV or commutative limit, the linear approximation (\ref{deepuv}) properly encodes all the phenomenology of the models; however, the exact noncommutative amplitude better describes the behaviour of the cosmological observables in the full span of the UV region. To see this, let us compare the function $\sigma$, governing the energy dependence of the spectral index (\ref{spindex}), with its approximated version $\sigma_\text{\tiny appr}$ given by Eq. (\ref{sigapp}); we plot the quantity $(\sigma-\sigma_\text{\tiny appr})/\sigma$ for the UV models in Fig. \ref{fig4}. The BH2, New1 and New2 models display the same linear trend in $\mu_*$, while the BH1 curve is a little below the bisector; the approximation error is up to 50\% for $\mu_* \lesssim 0.5$, corresponding to $H/M_s \lesssim 0.7$, and goes below 10\% when $H/M_s \lesssim 0.5$.
\begin{figure}[ht]
\bc
\ec
\caption{\it\label{fig4} The relative approximation error $(\sigma-\sigma_\text{\tiny appr})/\sigma$ vs $\mu_*$ in the UV sector. The thin line is for BH1, the thick line is a superposition of BH2, New1 and New2.}
\end{figure}
An analogous treatment of Eqs. (\ref{salph}) and (\ref{talph}) shows that the difference between the $\mu$-exact and the approximated scalar running may be even greater than the WMAP experimental error for this observable, $\alpha_s-\alpha_{s,\text{\tiny appr}} \gtrsim 10^{-2}$, for any $\theta$ and suitable values for $n_s$ and $r$ in the allowed range. Therefore, the following analysis has been conducted with the full nonlinear amplitude.

Table \ref{table7} reports the noncommutative high-energy limit in the IR region. In particular, the spectral amplitude of New1 is twice the amplitude of BH1; however, within each class (1 and 2) a unique set of consistency relations is generated. In the perturbative noncommutative limit, $\delta \ll 1$, the IR version of $(\sigma-\sigma_\text{\tiny appr})/\sigma$ is shown in Fig. \ref{fig5}. The relative approximation error is up to 20\% for the BH models and $\delta \lesssim 0.5$, while it is up to 40\% for the New models. The curves of New1 and New2 models coincide.

\begin{figure}
\bc
\ec
\caption{\it\label{fig5} The relative approximation error $(\sigma-\sigma_\text{\tiny appr})/\sigma$ vs $\delta$ in the IR sector. The thin solid line is for BH1, the thin dashed line is for BH2 and the thick line is a superposition of New1 and New2.}
\end{figure}

In standard cosmology, the consistency equation relating the tensor index $n_t$ and $r$ is adopted in order to reduce the space of parameters. The function $\sigma$ in Eq. (\ref{ntconeq}) contains a new theoretical parameter, the string energy scale $M_s$, which enlarges the standard space of cosmological variables. In principle, this might pose some problems if one wanted a reasonably stringent constraint on the observables, facing an uncertainty similar to that one gets when keeping $n_t$ unfixed \cite{efs02}. In the UV commutative region $\sigma \ll 1$, however, one can use the known results for the 4D, RS and GB likelihood analyses in order to compare the consistency equations in the allowed range.

The IR noncommutative limit is easier to deal with since the asymptotic form of Eq. (\ref{ntconeq}) is independent of the string scale, as it is shown in Table \ref{table8}. Some features are particularly interesting: ($i$) The infrared RS2 models are the only ones with a negative tensor tilt, other noncommutative realizations giving a tilt sign opposite to that of the commutative case; ($ii$) 4D class 2 models predict an exactly scale-invariant tensor spectrum to lowest order in SR, setting $n_t \sim O(\epsilon^2)$; ($iii$) The highest proportionality coefficient is provided by GB class 1 models, allowing a greater tilt given the same tensor-to-scalar ratio.
\begin{table}
\bc\begin{tabular}{c|ccc}
   (Non)commutative         &\multicolumn{3}{c}{$n_t/r$}\\
      models                &       GB     &           RS    &   4D   \\ \hline
Commutative UV ($\sigma=0$) &      $-1$    &         $-2$    &   $-2$ \\
Class 1 IR ($\sigma=6$)     &       5      &           2     &    4   \\ 
Class 2 IR ($\sigma=2$)     &       1      & $-\frac{2}{3}$  &    0   \\
\end{tabular}\ec
\caption{\label{table8}The consistency equation (\ref{ntconeq}) in the commutative UV and noncommutative IR limit.}
\end{table}
Then the perturbation spectra tend to be blue tilted in the IR region relative to the UV commutative case.\footnote{It is interesting to note that the sign of the correction to the scalar index ($+$) and its running ($-$) agrees with the results coming from a pure spatial realization of the noncommutative algebra \cite{LMMP,AN}.} 

Although there are $3\cdot2^4=48$ models at hand and a great amount of information to deal with, some preliminary considerations will permit us to simplify such an intricate taxonomy and draw theoretical curves in a reasonable region in the $n_s$-$r$ plane. Let us first compare the BH scenario with the New one and define $|\sigma|\equiv(\sigma_\text{\tiny BH}+\sigma_\text{\tiny New})/2$ and $\Delta \equiv (\sigma_\text{\tiny BH}-\sigma_\text{\tiny New})/|\sigma|$. Figure \ref{fig6}(\textit{a}) shows that in the commutative region BH and New models are considerably different, being $\Delta_1^\text{\tiny UV}=2(\mu_*+5)/(7\mu_*+5)\sim 10/11$ when $\mu_* \rightarrow 0$ and $\Delta_2^\text{\tiny UV}= 2/3$. In the limit $\mu_* \rightarrow 1$, $\Delta_1^\text{\tiny UV} \rightarrow \Delta_2^\text{\tiny UV} $; this is a spurious effect due to the breaking of the Taylor expansion (\ref{aUV}), as one can see by considering the commutative limit of the IR spectra in Fig. \ref{fig6}(\textit{b}). In fact, $\Delta_1^\text{\tiny IR} \neq \Delta_2^\text{\tiny IR}$ when $\delta \rightarrow \sqrt{2}$ and, as expected, $\Delta_1^\text{\tiny IR} \rightarrow 10/11$ and $\Delta_2^\text{\tiny IR} \rightarrow 2/3$ when $\delta \rightarrow \infty$. All this is in accordance with Table \ref{table7}. However, in the IR noncommutative limit there is little difference between BH and New models, being $\Delta \lesssim 10\%$. Therefore, we will only show the results of New in the infrared and skip the almost identical counterparts in BH.
\begin{figure}
%\bc\includegraphics[width=8.6cm]{fig6.eps}\ec
\caption{\it\label{fig6}Comparison of BH and New models in the ultraviolet (\textit{a}) and infrared (\textit{b}) region (thick line: 1-models; dashed line: 2-models).}
\end{figure}

A similar inspection shows that class-1 and class-2 models are quantitatively nondegenerate, getting $\sigma_1 =3\sigma_2$ for New and BH-IR, and $\sigma_1=4\sigma_2$ for BH-UV, in agreement with Table \ref{table7}. Note that these results are independent of the bulk physics.

The versatility of the patch formalism allows coupling it to a noncommutative background in a great number of ways. For example, a realistic picture of the cosmological evolution would be to adopt one particular patch regime in a time interval when a given region of the (non)commutative spectrum is generated; one may then associate the IR region of  extra-horizon-generated perturbations with the early-Universe high-energy period, when the extra dimension opens up and the Friedmann equation suffers either GB or RS modifications. The consequent evolution is GB-IR $\rightarrow$ RS-IR/UV $\rightarrow$ 4D-UV. Another possibility is to consider pure energy patches and study the noncommutative spectrum in GB, RS, and 4D separately.

%%%%%%%%%%%%%%%%%%%%%%%%%%%%%%%%%%%%%%%%%%%%%%%%%%%%%%%%%%%%%%%%%%%%%%%%%%%%%%%%%%%%%%%%%%%%%%%%%%%%%%%%%%%%%%%%%%%%%%%%%%%%%%%%%%%%%%%%%%%%%%%%%%%%%%%%%%%%%%%%%%%%%%%%%%%%%%%%%%%%%%%%%%%%%%%%%%%%%%%%%%%%%%%%%%%%%%%%%%%%%%%%%%%%%%%%%%%%%%%%

\section{Consistency equations and observations}

The introduction of the new degree of freedom provided by the noncommutative parameter does not complicate the analysis of Sec. \ref{pavspa}, nor triggers further degeneracies in any of the concrete (non)commutative braneworlds. Let us consider the consistency relations (\ref{salph}), (\ref{talph}), (\ref{ntconeq}) and (\ref{tenalph}) in their closed form, assuming the prior (\ref{prior}). Since theoretical degeneracy of consistency relations should be independent of the particular value of horizon-crossing quantities, we investigate only the IR case with constant $\sigma$, that is in the IR limit $\delta\ll 1$ ($\bar{\sigma}=0$) .

We do not care for exact degeneracy and consider only effective and tensor degeneracy between two models $(\psi,\,\theta,\,\sigma)$ and $(\psi',\,\theta',\,\sigma')$, always discarding the commutative case $\sigma =\sigma'=0$. Patch models display lowest-order tensor degeneracy when 
\be \label{SRcenc}
(2+\theta-\sigma)\zeta_q=(2+\theta'-\sigma')\zeta_{q'}\,.
\ee
If $\sigma\neq\sigma'$, this translates into $\sigma_{4D}-\sigma_{GB}=1$, $\sigma_{4D}=2\sigma_{RS}/3$, or $2\sigma_{RS}/3-\sigma_{GB}=1$. As regards effective degeneracy:
\begin{itemize}
\item $\phi \leftrightarrow T$: Scalar and tachyon models are never degenerate;
\item $\phi \leftrightarrow \phi$: Degeneracy is possible only when $\sigma \neq \sigma'$ and at least one of the two models is not 4D, RS or GB. In general, it must be $\theta=(3\sigma'+10\theta'-2\sigma'\theta'-15)/(2-\theta')$, $\sigma=(\sigma'-5\theta'+5)/(2-\theta')$ and $\zeta_q=\zeta_{q'}(2-\theta')$;
\item $T \leftrightarrow T$: The degeneracy conditions read $\theta-\sigma=\theta'-\sigma'$ and $\zeta_q=\zeta_{q'}$. Therefore, $\sigma_{4D}-\sigma_{GB}=1$.
\end{itemize}
To summarize, among the known braneworld commutative models there are two lowest-SR-order tensor degeneracies, one between scalar and tachyon cosmologies and one between the four-dimensional scenario and the Randall-Sundrum braneworld; when noncommutativity is turned on, these braneworld models can be degenerate with suitable values of the noncommutative parameter, but not in the classes investigated above. This result holds under the standard assumption (\ref{prior}), which permits to close the expression for the scalar running. If the inflaton potential does not satisfy such a dynamical constraint, as in the case of power-law ordinary inflation, then the consistency relations (\ref{salph}) and (\ref{talph}) are modified. For example, we will see that it may be convenient to perform numerical analyses via the horizon-flow formalism; when one neglects the third flow parameter $\epsilon_3$ with respect to $\epsilon_1=\epsilon$, it turns out that the scalar and tachyon scenarios are always effectively degenerate if Eq. (\ref{SRcenc}) holds, since the scalar running is then given by Eq. (\ref{tenalph}) in both cases \cite{CT}. The discussion for the lowest-order tensor degeneracy would thus also apply to the effective degeneracy. Should this be the most realistic scenario, the adoption of one field instead of the other would be important only at second order in this model-independent context, in which nothing about the shape of the potential is said; anyway, the universal equation (\ref{SRcenc}) would continue to determine the condition for degeneracy, excluding coincident predictions from the braneworld (non)commutative setups we have considered.

%%%%%%%%%%%%%%%%%%%%%%%%%%%%%%%%%%%%%%%%%%%%%%%%%%%%%%%%%%%%%%%%%%%%%%%%%%%%%%%%%%%%%%%%%%%%%%%%%%%%%%%%%%%%%%%%%%%%%%%%

\subsection{A first estimate of noncommutative effects}\label{pert2}

Let us compare the running of the scalar index of ordinary-inflaton and tachyon-inflaton fields, 
\be \label{da}
\Delta\alpha_s \equiv \alpha_s(\phi)-\alpha_s(T)\,.
\ee
Since the graphic material is very abundant, we give just a selection of it; the full set of bi- and three-dimensional figures of this and other combined analyses are available upon request to the author. Different analyses would point out other important aspects of the models; one may set his/her fancy free by looking at cross comparisons with general relative running Eq. (\ref{deltalp}). It should be noted that observations might not distinguish between the ordinary scalar and the tachyon, $\Delta\alpha_s=0$ (see Sec. \ref{like1}). However, the example provided by Eq. (\ref{da}) still gives an idea of the order of magnitude of noncommutative effects that can discriminate between one patch and another.

In Fig. \ref{fig7} the relative running $\Delta\alpha_s(n_s\!=\!1,r,\mu_*)$ is presented for 4D noncommutative models in the ultraviolet. Two-dimensional slices are then displayed in Figs. \ref{fig8} and \ref{fig9}. Figure \ref{fig8} shows that the relative running in Randall-Sundrum is rather modest; on the contrary, in GB and 4D noncommutativity may conspire to bias Eq. (\ref{da}) and, in particular, the scalar running above the current WMAP uncertainty estimates, $O(10^{-2})$. Braneworld effects, if any, should become more apparent in Planck data, for which the forecasted error is one order of magnitude smaller, $\Delta\alpha_s \sim O(10^{-3})$ \cite{BCLP}. In each 2D plot we keep the commutative model as a reference. Note that to increase either $n_s$ or $\delta$ ($\mu_*^{-1}$) pushes $\Delta\alpha_s$ towards positive values. Finally, Figs. \ref{fig10} and \ref{fig11} show some features of the New scenarios in the infrared region.

\begin{figure}
%\bc\includegraphics[width=\textwidth]{fig7.eps}\ec
\caption{\it\label{fig7}$\Delta\alpha_s$ as a function of $r$ and $\mu_*$ in the 4D ultraviolet region with $n_s=1$. The noncommutative models are BH1 (\textit{a}), BH2 (\textit{b}), New1 (\textit{c}) and New2 (\textit{d}).}
%\end{figure}

%\begin{figure}
%\bc\includegraphics[width=8.6cm]{fig8.eps}\ec
\caption{\it\label{fig8}$\Delta\alpha_s$ in Randall-Sundrum BH1 ultraviolet as a function of $r$. The values for the scalar index are $n_s=0.9$ (dashed lines), 1 (solid lines), 1.1 (dot-dashed lines); the values for $\mu_*$ are 0 (thin lines), 0.2 (thick lines) and 0.4 (very thick lines).}
\end{figure}

\begin{figure}
%\bc\includegraphics[width=\textwidth]{fig9.eps}\ec
\caption{\it\label{fig9}$\Delta\alpha_s$ as a function of $r$ for ultraviolet GB-BH1 (\textit{a}), 4D-BH1 (\textit{b}), GB-BH2 (\textit{c}), 4D-BH2 (\textit{d}), GB-New1 (\textit{e}), 4D-New1 (\textit{f}), GB-New2 (\textit{g}), 4D-New2 (\textit{h}). The values for the scalar index are $n_s=0.9$ (dashed lines), 1 (solid lines), 1.1 (dot-dashed lines); the values for $\mu_*$ are 0 (thin lines), 0.2 (thick lines) and 0.4 (very thick lines).}
\end{figure}

\begin{figure}
%\bc\includegraphics[width=\textwidth]{fig10.eps}\ec
\caption{\it\label{fig10}$\Delta\alpha_s$ as a function of $r$ and $\delta$ in the GB infrared region with $n_s=1$. The noncommutative models are New1 (\textit{a}) and New2 (\textit{b}).}
\end{figure}

\begin{figure}
%\bc\includegraphics[width=\textwidth]{fig11.eps}\ec
\caption{\it\label{fig11}$\Delta\alpha_s$ as a function of $r$ for infrared GB-New1 (\textit{a}), 4D-New1 (\textit{b}), GB-New2 (\textit{c}), 4D-New2 (\textit{d}). The values for the scalar index are $n_s=0.9$ (dashed lines), 1 (solid lines), 1.1 (dot-dashed lines); the values for $\delta$ are 0.2 (thin lines), 0.4 (thick lines) and 0.6 (very thick lines).}
\end{figure}

%%%%%%%%%%%%%%%%%%%%%%%%%%%%%%%%%%%%%%%%%%%%%%%%%%%%%%%%%%%%%%%%%%%%%%%%%%%%%%%%%%%%%%%%%%%%%%%%%%%%%%%%%%%%%%%%%%%%%%%%%%%%%%%%%%%%%%%%%%%%%%%%%%%%%%%%%%%%%%%%%%%%%%%%%%%%%%%%%%%%%%%%%%%%%%%%%%%%%%%%%%%%%%%%%%%%%%%%%%%%%%%%%%%%%%%%%%%%%%%%

\section{Large-field noncommutative models} \label{models}

Even if the dynamical conditions are slightly more precise within the Hubble SR formalism, the V-SR towers of Secs. \ref{vsrph} and \ref{vsrT} fit better for numerical analyses like that we are going to carry out in Sec. \ref{likeli}. Using the relations of Sec. \ref{endep}, the inflationary observables $A_s^2$, $n_s$, and $R=16r$  read, to lowest SR order,
%%%%%%%%%%%%
\bs\ba
A_s^2(\phi) &=& \frac{9\beta_q^6}{25\pi^2}\frac{V^{3q}}{V'^2}\Sigma^2\,,\\
n_s-1  &=& 2\eta_\S-(4-\sigma)\,\epsilon_\S\\
&=&\frac{1}{3\beta_q^2V^q}\left[2V''+(\sigma-6)\frac{q}{2}\frac{V'^2}{V}\right],\\
R &=& \frac{16q}{6\beta_q^2\zeta_q}\frac{V'^2}{V^{q+1}}\,.
\ea\es
For the tachyon, the inflationary observables are
\bs\ba 
A_s^2(T) &=& \frac{9\beta_q^6}{25\pi^2}\frac{V^{3q+1}}{V'^2}\Sigma^2\,,\\
n_s-1    &=& 2\eta_\T-(2+\theta-\sigma)\,\epsilon_\T\\
         &=& \frac{1}{3\beta_q^2V^q}\left[2U''-(4+\theta-\sigma)\frac{q}{2}U'^2\right],\\
R        &=& \frac{16q}{6\beta_q^2\zeta_q} \frac{V'^2}{V^{q+2}}\,.
\ea\es
In this section we consider an important class of inflaton potentials, namely, the large-field models
\be \label{power}
V(\psi)=V_0 \psi^p\,,\qquad \psi=\phi,\,T\,,
\ee
in which the inflaton field starts with a large initial value and rolls down towards the potential minimum at smaller $\psi$. The linear potential  with $p=1$ corresponds to the border of large-field and small-field models. The exponential potential
\be \label{expo}
V=V_0\exp\left(-\psi/\psi_0\right)\,,
\ee
characterizes the border of large-field and hybrid models. This case can be regarded as the $p \to \infty$ limit of 
the polynomial potential (\ref{power}).

Making use of the ESR approximation we obtain
\be\label{efold1}
N(\phi) \approx -3\beta_q^2 \int_\phi^{\phi_f}\frac{V^q}{V'}\,d\phi\,,
\ee
for the scalar field $\phi$, and
\be\label{efold2}
N(T) \approx -3\beta_q^2 \int_T^{T_f}\frac{V^{q+1}}{V'}\,dT\,,
\ee
for the tachyon field $T$. Here we make use of the backward definition $N=\ln (a_f/a)$ of the number of $e$-foldings.

The potentials (\ref{power}) and (\ref{expo}) cover a number of exact solutions either exactly or approximately. In fact, the commutative solutions of Sec. \ref{exact} are perfectly viable in the noncommutative case too, since the nonlocal physics does not affect the homogeneous background. The unique apparently subtle point is that in the IR region one explicitly uses the exponential solution to construct the perturbation amplitudes, contrary to the UV case in which it is implicitly assumed in the approximation of constant SR parameters. However, the subtended philosophy is 
quite the same, that is to find a general solution with constant nonzero SR parameters and then to perturb it with small time variations. Despite these simple considerations, the predictions of these homogeneous models definitely change when spacetime becomes noncommutative.

%%%%%%%%%%%%%%%%%%%%%%%%%%%%%%%%%%%%%%%%%%%%%%%%%%%%%%%%%%%%%%%%%%%%%%%%%%%%%%%%%%%%%%%%%%%%%%%%%%%%%%%%%%%%%%%%%%%%%%%%

\subsection{The ordinary scalar field $\phi$}

For the scalar potential (\ref{power}) with the ordinary scalar field $\phi$, we have 
\ba
n_s-1 &=& -\frac{pV_0^{1-q}}{6\beta_q^2}\frac{p(6q-\sigma q-4)+4}{\phi^{2+(q-1)p}}\,,\label{npow}\\
R &=& \frac{16qp^2}{6\beta_q^2\zeta_qV_0^{q-1}}\phi^{(1-q)p-2}\,.
\ea
We can estimate the field value at the end of inflation by setting $\epsilon_\phi(t_f)=1$, which yields 
$\phi_f^{p(q-1)+2} \approx qp^2/(6\beta_q^2V_0^{q-1})$.\footnote{One may adopt the criterion $\eta_\sV(t_f)=1$ to estimate the value $\phi_f$, but the difference is small as long as $p/N \ll 1$.}
Then the number of $e$-foldings (\ref{efold1}) is
\ba \label{efolds}
N =\frac{3\beta_q^2V_0^{q-1}}{p[p(q-1)+2]}\,\phi^{p(q-1)+2}-
\frac{qp}{2[p(q-1)+2]}\,,
\ea
which is valid for $p \ne 2/(1-q)$. The scalar index and the tensor-to-scalar ratio are
\ba
n_s-1 &=& -\frac{p(6q-\sigma q-4)+4}{2N(pq-p+2)+pq}\,,\label{nspower}\\
R &=& \frac{16qp}{\zeta_q}\frac{1-n_s}{p(6q-\sigma q-4)+4}\,.\label{Rpower}
\ea
As discussed in Sec. \ref{ncsetup}, the tensor-to-scalar ratio $R$ does not involve the parameter $\sigma$, since this quantity is invariant by taking the noncommutative effect into account; this is evident when expressing Eq. (\ref{Rpower}) in terms of $N$. The main change due to spacetime noncommutativity appears for the spectral index $n_s$.

For the commutative spacetime ($\sigma=0$) one can easily verify that the above results reduce to what was derived in \cite{LiS} for the 4D and RS scenarios. In these cases scalar perturbations are red tilted ($n_s<1$). The spectrum can be blue tilted when noncommutativity is switched on. For example, let us consider the noncommutative limit $\sigma \to 6$. In this case we have
\ba \label{nSlimit}
n_s-1=\frac{4(p-1)}{2N(pq-p+2)+pq}\qquad \text{for}~~\sigma \to 6\,,
\ea
which means $n_s>1$ for $p > 1$. Therefore it is possible to explain the loss of power in the spectrum at large scales,
as we shall see in Sec. \ref{cmbsuppr}.

The exponential potential (\ref{expo}) corresponds to the limit $p \to \infty$ in Eqs. (\ref{nspower}) and (\ref{Rpower}), thereby yielding 
\ba
n_s-1 &=& \frac{4-(6-\sigma)q}{2N(q-1)+q}\,,\\
R &=& \frac{16q}{\zeta_q[2N(q-1)+q]}\,,
\ea
which is valid for $q\neq 1$.\footnote{The power-law inflation does not end for the 4D case unless the slope of 
the exponential potential changes.} This gives the border between large-field and hybrid models
\be \label{border1s}
R = -\frac{16q}{\zeta_q(6q-\sigma q-4)}(n_s-1)\,.
\ee
In the case of 4D ($q=1$) we find that the border of large-field and hybrid models extends to the region of $n_s>1$ for $\sigma>2$. Thus, in the regime where the noncommutative effect becomes important ($2 \leq \sigma \leq 6$), one can obtain a blue-tilted spectrum even in the large-field models, which is not possible in the commutative case. 

Note that the border of large-field and small-field models corresponds to $p=1$, giving
\be
R = -\frac{16}{\zeta_q(6-\sigma)}(n_s-1)\,.
\ee
This border does not extend to the region $n_s>1$ for $\sigma<6$.

%%%%%%%%%%%%%%%%%%%%%%%%%%%%%%%%%%%%%%%%%%%%%%%%%%%%%%%%%%%%%%%%%%%%%%%%%%%%%%%%%%%%%%%%%%%%%%%%%%%%%%%%%%%%%%%%%%%%%%%%

\subsection{The tachyon field $T$}

For the scalar potential (\ref{power}) with the tachyon field $T$, we have 
\ba
n_s-1 &=& -\frac{p}{6\beta_q^2V_0^q}\frac{pq(4+\theta-\sigma)+4}{T^{2+qp}}\,,\\
R &=& \frac{16qp^2}{6\beta_q^2\zeta_qV_0^q}T^{-2-qp}\,.
\ea
Since inflation ends at $T_f^{qp+2}\approx qp^2/(6\beta_q^2V_0^q)$, the number of $e$-foldings is estimated as 

\be
N =\frac{3\beta_q^2V_0^q}{p(pq+2)}\,T^{pq+2}-\frac{qp}{2(pq+2)}\,,\qquad p \neq -\frac{2}{q}\,.
\ee
Then we get
\ba
\label{nsT}
n_s-1 &=& -\frac{p(6q-\sigma q-2)+4}{2N(pq+2)+pq}\,,\\
R &=& \frac{16qp}{\zeta_q}\frac{1-n_s}{p(6q-\sigma q-2)+4}\,,\label{RT}
\ea
where we used Eq. (\ref{theta}). For the 4D0 spacetime, these results reproduce what was obtained in \cite{SV}. The tensor-to-scalar ratio is smaller relative to the case of the ordinary scalar field $\phi$, thus preferred observationally \cite{GST}. The effect of noncommutativity can lead to a blue-tilted spectrum ($n_s>1$) as is similar to the case of the field $\phi$.

For the exponential potential (\ref{expo}) one gets
\ba
n_s-1 &=& -\frac{4+\theta-\sigma}{2N+1}\,,\\
R &=& \frac{16}{\zeta_q(2N+1)}\,,
\ea
which is obtained by taking the limit $p \to \infty$ in Eqs. (\ref{nsT}) and (\ref{RT}). This gives the border of large-/hybrid-field models
\be
R=-\frac{16}{\zeta_q(4+\theta-\sigma)} (n_s-1)\,.
\ee
In the 4D case this border belongs to the region $n_s>1$ for $\sigma>4$. 

The border of large-/small-field models is 
\be \label{border1t}
R = -\frac{16q}{\zeta_q(6q-\sigma q+2)}(n_s-1)\,,
\ee
which does not extend to the region $n_s>1$ for $\sigma<6$.

%%%%%%%%%%%%%%%%%%%%%%%%%%%%%%%%%%%%%%%%%%%%%%%%%%%%%%%%%%%%%%%%%%%%%%%%%%%%%%%%%%%%%%%%%%%%%%%%%%%%%%%%%%%%%%%%%%%%%%%%

\subsection{The difference between $\phi$ and $T$} 

By Eqs. (\ref{nspower}) and (\ref{nsT}) we find that the spectral index $n_s$ of the ordinary field and the tachyon 
field differs both in the denominator and the numerator. By Eqs. (\ref{Rpower}) and (\ref{RT}) the difference for the ratio $R$ only appears in the denominator
\be 
\label{ratiouni}
R = \frac{16qp}{\zeta_q[2N(qp+2-b)+qp]}\,,
\ee
where $b=p$ for $\psi=\phi$, and $b=0$ for $\psi=T$. Therefore the tensor-to-scalar ratio in the tachyon case is smaller than in the ordinary scalar field case when $p>0$. This property implies that tachyon inflation is less affected by observational pressure as was pointed out in the 4D commutative case \cite{GST}. In the next section we shall study this issue in detail in the context of noncommutative inflation.

%%%%%%%%%%%%%%%%%%%%%%%%%%%%%%%%%%%%%%%%%%%%%%%%%%%%%%%%%%%%%%%%%%%%%%%%%%%%%%%%%%%%%%%%%%%%%%%%%%%%%%%%%%%%%%%%%%%%%%%%

\subsection{Theoretical structure of the $n_s$-$R$ plane}

Before considering each noncommutative case, it is important to understand the behaviour of the theoretical curves on the $n_s$-$R$ plane. The effect of the noncommutative parameter $\sigma$ has a straightforward geometrical interpretation. Let us define $x\equiv n_s-1$ and $y \equiv R$, together with the polar coordinates $\varrho\equiv\sqrt{x^2+y^2}$ and $\sin\vartheta \equiv y/\varrho$ centered at $(1,0)$ in the $n_s$-$R$ plane. From the last section, we know that
\ba
y &=& \gamma(q,p,\sigma) x\,,\\
\gamma(q,p,\sigma) &=& -\frac{16qp}{\zeta_q[p(6q-\sigma q-c)+4]}\,,
\ea
where $c=4$ for $\psi=\phi$, and $c=2$ for $\psi=T$. Then, $\varrho^2=(1+\gamma^2)x^2$ and $\tan \vartheta = \gamma$. 
Since $pq>0$ in the cases we consider, $\vartheta$ is a decreasing function in terms of $\sigma$. Therefore, as $\sigma$ increases, the theoretical points are rotated clockwise in the $\varrho$-$\vartheta$ plane. This rotation is mainly governed by $\sigma$ rather than $p$ when $p$ is large, which can be seen from the computation of the logarithmic variation of $\gamma$:
\be \label{rotat}
\frac{d \ln \gamma}{d p} =\frac{4}{qp^2}\frac{d \ln \gamma}{d \sigma}\,.
\ee
This also implies that, for a given $\sigma$, the three models $p=2,4,+\infty$ lie on a wider range of radii for smaller values of $q$. As we shall see later, this effect is particularly evident in the Gauss-Bonnet case with 
respect to the 4D and RS cases in the same (non)commutative class.

The divergence of $\gamma$ at the asymptote $\vartheta=\pi/2$ identifies those models generating a scale-invariant scalar spectrum $n_s=1$. They are listed in Table \ref{table9} for fixed $\sigma$ and $q$. In particular, ordinary-scalar class 2 models cannot give $n_s=1$ if one imposes the condition $p(q-1)+2\neq 0$ for inflation to have a natural end. The tachyonic counterparts are those with $pq+2 \neq 0$, and only the 4D2 case is excluded.

Note that class 1 patch models admit only one scale-invariant potential for each inflaton, that is, the linear potential for $\phi$ and the quadratic one for $T$. Another frequent case is $p=-2$ (scalar 4D0, tachyon 4D2 and tachyon GB2), which however does not match with the exact power-law solutions of Sec. \ref{exact} (RS scalar and 4D tachyon). Anyway our interest in this paper are the models with positive $p$ ($p \ge 2$) that lead to natural reheating.
\begin{table}[ht]
\bc\begin{tabular}{cc|rrr}
       &    &  \multicolumn{3}{c}{$\sigma$} \\
       &    &   0  (Class 0)      &  6  (Class 1)   &   2 (Class 2)   \\ 
\hline
       & 4D &   $-2$     &  1 &  $\infty$   \\
$\phi$ & RS &   $-1/2$   &  1    &   $-1$   \\
       & GB &   $\infty$ &  1      &   3   \\ \hline
       & 4D &   $-1$     &  2     &   $-2$   \\
$T$    & RS &   $-2/5$   &  2   &   $-2/3$   \\
       & GB &   $-2$     &  2     &   $-6$   \\
\end{tabular}\ec
\caption{\label{table9}Values of $p$ for scale-invariant models.}
\end{table}

%%%%%%%%%%%%%%%%%%%%%%%%%%%%%%%%%%%%%%%%%%%%%%%%%%%%%%%%%%%%%%%%%%%%%%%%%%%%%%%%%%%%%%%%%%%%%%%%%%%%%%%%%%%%%%%%%%%%%%%%%%%%%%%%%%%%%%%%%%%%%%%%%%%%%%%%%%%%%%%%%%%%%%%%%%%%%%%%%%%%%%%%%%%%%%%%%%%%%%%%%%%%%%%%%%%%%%%%%%%%%%%%%%%%%%%%%%%%%%%%

\section[Likelihood analysis: noncommutative inflation]{Likelihood analysis: noncommutative\\ inflation} \label{likeli}

In this section we study constraints on a number of patch inflationary models in noncommutative spacetime using a compilation of recent high-precision observational data. In particular, we perform likelihood analyses\footnote{The likelihood analysis is one of the so-called ``top-down'' approaches: one asks what is the probability that a theory predicting a given set of observables would realize the observed experimental data \cite{AgT}.} in terms of inflationary observables using the new consistency relation (\ref{ntconeq}) and confront them with large-field inflationary models with potential $V \propto \psi^p$ in the two classes of IR noncommutative scenarios. In \cite{LiS,TL} it was shown that the 4D/RS quartic chaotic potential ($V \propto \phi^4$) is under a strong observational pressure and that steep inflation driven by an exponential potential is ruled out. This situation changes if we account for the Gauss-Bonnet  curvature invariant in five dimensions. One effect of the GB term is to break the degeneracy of the standard consistency relation \cite{DLMS}. Although this does not lead to a significant change for the likelihood results of the inflationary observables, the quartic potential is rescued from marginal rejection for a wide range of energy scales \cite{TSM}. Even steep inflation exhibits marginal compatibility for a sufficient number of $e$-folds ($N \gtrsim 55$).

Here we implement both braneworld and noncommutative frameworks as well as the standard 4D commutative/noncommutative paradigm. Our numerical analysis based on recent observational data will show that the general shape of likelihood contours in the $n_s$-$R$ plane is deformed independently by braneworld and noncommutative effects. The major modification to the 4D/RS commutative cosmology appears for the upper bound of $R=16r$, roughly setting it in a $2\sigma$ interval with $0.5 \lesssim R_\text{max} \lesssim 0.7$. The scalar index always ranges $0.9<n_s<1.1$ at the $2\sigma$ level. 

For later convenience we dub the commutative spacetime ($\sigma=0$) as ``class 0.'' Also, in the figures the 4D case is indicated as ``GR'' (general relativistic).

%%%%%%%%%%%%%%%%%%%%%%%%%%%%%%%%%%%%%%%%%%%%%%%%%%%%%%%%%%%%%%%%%%%%%%%%%%%%%%%%%%%%%%%%%%%%%%%%%%%%%%%%%%%%%%%%%%%%%%%%

\subsection{HF consistency equations}

At first order in the HF parameters defined in Sec. \ref{Hflow}, the spectral indices of scalar and tensor perturbations are given by
\ba \label{ns}
n_s-1&=& -(2+\theta-\sigma)\epsilon_1-\epsilon_2\,, \\
n_t &=& -(2+\theta-\sigma)\epsilon_1\,.\label{nt}
\ea
The lowest-order ratio of tensor-to-scalar perturbations is
\be \label{ratio}
R =\frac{16\epsilon_1}{\zeta_q}\,,
\ee
while the runnings of the spectral indices are given by
\ba
\alpha_s &=& -(2+\theta-\sigma)\epsilon_1\epsilon_2-\sigma\bar{\sigma}\epsilon_1^2-\epsilon_2\epsilon_3\,,\\
\alpha_t &=& -(2+\theta-\sigma)\epsilon_1\epsilon_2-\sigma\bar{\sigma}\epsilon_1^2\,.
\ea
The resulting (nonclosed) set of consistency equations is
\ba
n_t &=& -(2+\theta-\sigma) \frac{R\zeta_q}{16}\,,\label{ntconeq2}\\
\alpha_s &=& \frac{R\zeta_q}{16} \left\{(2+\theta-\sigma)(n_s-1)+\left[(2+\theta-\sigma)^2-\sigma \bar{\sigma}\right]\frac{R\zeta_q}{16}\right\}-\epsilon_2\epsilon_3\,,\nonumber\\ \label{alps}\\
\alpha_t &=& \frac{R\zeta_q}{16} \left\{(2+\theta-\sigma)(n_s-1)+\left[(2+\theta-\sigma)^2-\sigma \bar{\sigma} 
\right]\frac{R\zeta_q}{16} \right\}\,.
\ea
Notably these relations do not depend on which inflaton field one is assuming on the brane, except for the term $\epsilon_2\epsilon_3$. This means that the likelihood analysis for the field $\phi$ in terms of the variables $n_s$, $R$, and $\epsilon_3$ with given values of $q$, $\sigma$, and $\bar{\sigma}$ is identical to the one for the tachyon $T$.

With the SR parameters the running of the scalar perturbations splits into Eqs. (\ref{salph}) and (\ref{talph}); then one would need to perform two separate likelihood analyses. In this sense, the HF parameters are a more convenient choice for numerical purposes than the SR parameters, while the adoption of the SR parameters better highlights the difference between the scalar and tachyon dynamics already to first order.

%%%%%%%%%%%%%%%%%%%%%%%%%%%%%%%%%%%%%%%%%%%%%%%%%%%%%%%%%%%%%%%%%%%%%%%%%%%%%%%%%%%%%%%%%%%%%%%%%%%%%%%%%%%%%%%%%%%%%%%%

\subsection{Likelihood contour bounds} \label{like1}

In \cite{TMB} the CMB spectrum of power-law inflation was divided into three main regions, ultraviolet, intermediate, and infrared. If the IR spectrum corresponds to the scale around $1<l\lesssim 10$, then the cosmologically relevant modes with $10\lesssim l\lesssim 1000$ also belong to the same IR spectrum. This is because the characteristic scale of this spectrum is $k_{s3}=10^{-5}k_{s2}$, where $k_{s2}$ is an intermediate scale around which the IR description becomes invalid [see Eq.~(12) of \cite{TMB}]. However, it was assumed that it is the intermediate spectrum that dominates at large scales ($1<l\lesssim 10$), following the approach of Ref.~\cite{HL1}.

In this section we will adopt another perspective, that is, to consider the far IR regime as a dominant contribution to 
the large-scale spectrum. So we shall assume that the IR spectra $\Sigma^2 \approx \delta^3$ [class 1, which is Eq.~(23) of \cite{TMB} in the de Sitter limit] and $\Sigma^2 \approx \delta$ (class 2) correctly describe the large-scale sector with $1<l \lesssim 10$. Since one generally has $k_{s2} \gg k_{s3}$, it is natural to use the IR power spectrum over the cosmologically relevant scales with $1<l\lesssim 2000$.

We have run the Cosmological Monte Carlo \textsc{cosmomc} code together with the \textsc{camb} program \cite{LCS,LB,camb}, applied to the latest observational data coming from the data set of WMAP \cite{wmap}, 2dF 
\cite{per01}, and SDSS \cite{spe03,teg04}. We implement the band powers on small scales ($800 \lesssim l \lesssim 
2000$) coming from CBI \cite{CBI}, VSA \cite{VSA}, and ACBAR \cite{ACBAR} experiments.

The set of inflationary observables is $\{A_s^2,R,n_s,n_t,\alpha_s,\alpha_t,\sigma\}$. The tensor index is absorbed via the consistency equation (\ref{ntconeq2}) while $\alpha_t$ is ignored since its cosmological impact is too small to be detected in current observations. The actual set of parameters is $\{A_s^2,\epsilon_1,\epsilon_2,\epsilon_3,\sigma\}$ or equivalently $\{A_s^2, n_s, R, \epsilon_3,\sigma\}$. For several fixed values of $\sigma$ ($\sigma=0, 2, 6$) we have numerically found that $\epsilon_3$ is poorly constrained and is consistent to be set to zero according to the standard assumption
\be
|\epsilon_3| \ll \epsilon_1\,.
\ee
We also ran the numerical code when the SR parameters $\epsilon$, $\eta$, and $\xi$ are varied. Since the running $\alpha_s$ is constrained to be $|\alpha_s| \lesssim 0.03$ in order for the Taylor expansion of the power spectrum to be valid \cite {LeL}, one cannot put large values of the prior on $\xi^2$. Making use of the fact that $\xi^2$ is of the same order as $\epsilon_1 \epsilon_2$ and the two HF parameters are constrained to be $\epsilon_1 \lesssim 0.03$ and $|\epsilon_2| \lesssim 0.1$ \cite{LeL}, we should put the prior around $\xi^2<0.003$. In this case our likelihood analysis shows that $\xi^2$ vanishes consistently, Eq. (\ref{prior}). We found that the likelihood values of inflationary and cosmological parameters are very similar to the case in which the HF parameters are used.

In what follows we shall show the numerical results obtained by using HF parameters, since this is more convenient 
because of the degeneracy between the ordinary field $\phi$ and the tachyon field $T$.

We ran the numerical code for the 4D case by varying $A_s^2$, $n_s$, $R$, and $\sigma$ with $\epsilon_3=0$. We chose the parameter range $0 \le \sigma \le 6$ and found that $\sigma$ does not show a good convergence. This means that current observations do not choose a preferred commutative or noncommutative model. Then the analysis was performed for three fixed values of $\sigma$, including the commutative limit ($\sigma=0$) and two classes of noncommutative IR limits ($\sigma=6, 2$).

In the 4D case the consistency relation (\ref{ntconeq2}) reads $n_t=-(2-\sigma)R/16$, which means that the ratio $n_t/R$ ranges $-1/8 \le n_t/R \le 1/4$. As found in  Fig. \ref{fig12}, $\sigma$ does not select a preferred value, since the $R=0$ case is not ruled out anyway. Therefore noncommutative inflation is allowed observationally as well as the case of commutative spacetime.

%%%%%%%%%%
\begin{figure}
%\bc\includegraphics[height=3.5in,width=3.5in]{fig12.eps}\ec
\caption{\it\label{fig12}
Marginalized probability distributions of inflationary parameters ($n_s$, $R$, $A_s^2$, $\sigma$) for the 4D case with prior $0 \le \sigma \le 6$ and $\epsilon_3=0$. The likelihood analysis does not choose a preferred value of the
noncommutative parameter $\sigma$ \cite{CT}.}
\end{figure}
%%%%%%%%%%

Since the UV commutative case has already been investigated in literature, we will concentrate ourselves to the IR noncommutative region. This choice is also dictated by a technical reason. In the commutative limit $\sigma$ depends upon both the Hubble parameter, evaluated at the horizon crossing, and the string mass $M_s$. As we have seen, even if the tensor spectral index is fixed by Eq.~(\ref{ntconeq2}), the introduction of the extra degree of freedom $\sigma$ results in a poor constraint on the parameter itself. On the contrary, in the far IR region the function $\sigma$ approaches nonzero constant values as shown in Table \ref{table7}. This allows us to impose the consistency equation (\ref{ntconeq2}) and concretely reduce the space of parameters, setting a meaningful scheme of analysis for the noncommutative models. Moreover, the amplitude of gravitational waves is strongly damped for angular scales with 
$l \gtrsim 10$ and the relations (\ref{ratio}) and (\ref{ntconeq2}) only affect the large scales with $l \lesssim 10$, corresponding to the IR region. In this sense, using a constant $\sigma$ is a good approximation.

In Fig. \ref{fig13} we plot the $1\sigma$ and $2\sigma$ observational contour bounds for the 4D case with $\sigma=0$ (GR0), $\sigma=6$ (GR1) and $\sigma=2$ (GR2). Figures \ref{fig14} and \ref{fig15} correspond to the likelihood contours for the RS and GB cases, respectively.\footnote{The likelihood contour for the Gauss-Bonnet case is slightly different from what was obtained in \cite{TSM}. This is because in that paper the authors considered the exact GB scenario and assumed that the running of the spectral indices is zero, since the expressions of the exact RS and GB regimes are very complicated. This resulted in a lower upper bound for $R$.} These results hold for both the scalar field $\phi$ and 
the tachyon field $T$ because of the use of the HF parameters.

%%%%%%%%%%
\begin{figure}
%\bc\includegraphics[height=3.5in,width=3.5in]{fig13.eps}\ec
\caption{\it\label{fig13}
The $1\sigma$ and $2\sigma$ observational contour bounds for the 4D case. Each contour curve corresponds to (\textit{a}) GR0 ($\sigma=0$), solid line; (\textit{b}) GR1 ($\sigma=6$), dashed line; (\textit{c}) GR2 ($\sigma=2$), dotted line. We also show the border of large-field and hybrid inflationary models for (\textit{a}) GR0, (\textit{b}) GR1 and (\textit{c}) GR2 cases. The region on the left of each border corresponds to the parameter space in large-field models. Noncommutative spacetime allows the border extending to the region $n_s>1$ \cite{CT}.}
\end{figure}
%%%%%%%%%%

%%%%%%%%%%
\begin{figure}
%\bc\includegraphics[height=3.5in,width=3.5in]{fig14.eps}\ec
\caption{\it\label{fig14}
The $1\sigma$ and $2\sigma$ observational contour bounds for the RS case. The meaning of the curves and the borders are 
the same as in Fig.~\ref{fig13} \cite{CT}.}
\end{figure}
%%%%%%%%%%

%%%%%%%%%%
\begin{figure}
%\bc\includegraphics[height=3.5in,width=3.5in]{fig15.eps}\ec
\caption{\it\label{fig15}
The $1\sigma$ and $2\sigma$ observational contour bounds for the GB case. The meaning of the curves and the borders are 
the same in Fig.~\ref{fig13} \cite{CT}.}
\end{figure}
%%%%%%%%%%

In the 4D case, the class 2 ($\sigma=2$) is rather special since $R$ and $n_t$ vanish. The class 2 contour extends to higher values of $R$ relative to the commutative plot, while the class 1 contour allows larger values of $|n_s-1|$ but with a smaller $R_\text{max}$. Thus the noncommutativity of a model is not monotonically measured by $\sigma$ (with greater $\sigma$ corresponding to larger effects) and nonlocal features make their appearance in a nontrivial way.

We can do similar considerations for the RS case (where the maximal elongation is achieved for $\sigma=6$) and for the GB one (where the class 1 behaves in a totally different manner); see Figs. \ref{fig14} and \ref{fig15}. Note that the degeneracy between 4D and RS is removed for $\sigma>0$, both from a theoretical and observational point of view.

%%%%%%%%%%%%%%%%%%%%%%%%%%%%%%%%%%%%%%%%%%%%%%%%%%%%%%%%%%%%%%%%%%%%%%%%%%%%%%%%%%%%%%%%%%%%%%%%%%%%%%%%%%%%%%%%%%%%%%%%
%%%%%%%%%%%%%%%%%%%%%%%%%%%%%%%%%%%%%%%%%%%%%%%%%%%%%%%%%%%%%%%%%%%%%%%%%%%%%%%%%%%%%%%%%%%%%%%%%%%%%%%%%%%%%%%%%%%%%%%%

\section[Likelihood analysis: constraints on large-field models]{Likelihood analysis: constraints on\\ large-field models} \label{like2}

We are ready to place constraints on large-field noncommutative inflationary models using the observational contour bounds obtained in Sec. \ref{like1}. We plot the theoretical values of $n_s$ and $R$ for $N=45,50,55,60$ on the likelihood contours. Typically one can restrict the number of $e$-folds to $N \lesssim 65$ \cite{LL2}, but it is sufficient to show the values up to $N=60$ to judge whether the models we consider are ruled out or not.

%%%%%%%%%%%%%%%%%%%%%%%%%%%%%%%%%%%%%%%%%%%%%%%%%%%%%%%%%%%%%%%%%%%%%%%%%%%%%%%%%%%%%%%%%%%%%%%%%%%%%%%%%%%%%%%%%%%%%%%%

\subsection{The ordinary scalar field $\phi$}

Let us first study the observational constraints on the large-field models for the ordinary field $\phi$. In Figs. \ref{fig16}--\ref{fig18} the theoretical values (\ref{nspower}) and (\ref{Rpower}) for the potential (\ref{power}) are plotted in the 4D, RS, and GB cases together with $1\sigma$ and $2\sigma$ contour bounds. Hereafter we shall consider each case separately in order to clarify the situation.

%%%%%%%%%%%%%%
\subsubsection{4D case}
%%%%%%%%%%%%%%

It is well known that the commutative 4D case ($\sigma=0$) is observationally disfavoured for the quartic potential ($p=4$). In this case the theoretical points are outside of the $2\sigma$ contour bound for a number of $e$-folds
$N<60$.

In the noncommutative class 1 case ($\sigma=6$) the spectral index $n_s$ is larger than 1 by Eq.~(\ref{nSlimit}).
The tensor-to-scalar ratio $R$ is independent of  $\sigma$, so this value is the same as the one in the class 0.
As one can see in Fig. \ref{fig16}, the quartic potential is outside of the $2\sigma$ bound for $N<55$. Therefore this case is also marginal as in the class 0 case.

The noncommutative class 2 case ($\sigma=2$) corresponds to a scalar spectral index smaller than 1, but it is closer to a scale-invariant spectrum relative to the class 0 case. This shifts the theoretical points inside of the $2\sigma$ bound and allows the quartic potential even for $N=45$. Then a ``mild'' spacetime noncommutativity in which $\sigma$ is close to 2 is favoured for the observational compatibility of the quartic potential. Note that the quadratic potential ($p=2$) is allowed in all three models, irrespective of the degree of noncommutativity, as clearly shown in Fig. \ref{fig16}. 

%%%%%%%%%%
\begin{figure}
%\bc\includegraphics[height=8.5in,width=3.2in]{fig16.eps}\ec
\caption{\it\label{fig16}
Observational constraints on large-field models for the 4D ordinary field $\phi$ together with the $1\sigma$ and $2\sigma$ contour bounds for three classes of (non)commutative scenarios. The theoretical values correspond to 
(\textit{a}) $p=2$ (dots) and (\textit{b}) $p=4$ (triangles), respectively, with the number of $e$-folds
$N=45, 50, 55, 60$ (from top to bottom in each panel and for each $p$) \cite{CT}.}
\end{figure}
%%%%%%%%%%

%%%%%%%%%%%%%%
\subsubsection{RS case}
%%%%%%%%%%%%%%

In commutative RS spacetime, the quartic potential is under a strong observational pressure as is similar to the 4D0 case, and the steep inflation driven by an exponential potential ($p \to \infty$) is ruled out \cite{LiS,TL}.

This situation is improved in the class 1 noncommutative scenario. Since the spectral index $n_s$ takes a value which is slightly larger than 1 and the $2\sigma$ contour bounds extend to the region with $R>0.6$, even the steep inflation is allowed (see Fig. \ref{fig17}).

Meanwhile in the class 2 case the exponential potential is outside of the $2\sigma$ bound unless the number of 
$e$-folds $N$ is larger than 60. The quartic potential moves inside of the $2\sigma$ bound relative to the class 0 case, thus becoming compatible with observations.

In the RS case ``strong'' noncommutativity close to $\sigma=6$ is favoured observationally rather than ``mild'' noncommutativity like $\sigma=2$, in contrast with the 4D case.

%%%%%%%%%%
\begin{figure}
%\bc\includegraphics[height=8.5in,width=3.2in]{fig17.eps}\ec
\caption{\it\label{fig17}
Observational constraints on large-field models for the RS ordinary field $\phi$. Each case corresponds to (\textit{a}) $p=2$ (dots), (\textit{b}) $p=4$ (triangles) and (\textit{c}) exponential potential with $p \to \infty$ (squares), respectively, with the number of $e$-folds $N=45, 50, 55, 60$ (from top to bottom in each panel and for each $p$) \cite{CT}.}
\end{figure}
%%%%%%%%%%

%%%%%%%%%%
\begin{figure}
%\bc\includegraphics[height=8.5in,width=4.5in]{fig18.eps}\ec
\caption{\it\label{fig18}
Observational constraints on large-field models for the GB ordinary field $\phi$. Each case corresponds to 
(\textit{a}) $p=2$ (dots) and (\textit{b}) $p=4$ (triangles), respectively, with the number of $e$-folds
$N=45, 50, 55, 60$ (from top to bottom in each panel and for each $p$). In the GB1 case the quadratic potential is far outside of the $2\sigma$ bound \cite{CT}.}
\end{figure}
%%%%%%%%%%

%%%%%%%%%%%%%%
\subsubsection{GB case}
%%%%%%%%%%%%%%

In the Gauss-Bonnet braneworld cosmology the GB dominant stage with $q=2/3$ is followed by the RS stage with $q=2$.
In Ref.~\cite{TSM} theoretical values of $n_s$ and $R$ were derived for the case where inflation ends in the RS regime.
Here we study a situation in which the end of inflation corresponds to the GB regime. In this case we do not have a sufficient amount of $e$-folds for $p>6$, so it is not meaningful to consider steep inflation.

In commutative spacetime the quartic potential is ruled out observationally, while the quadratic potential is inside of the $1\sigma$ bound, see Fig. \ref{fig18}. 

In the class 1 case the spectral index for the quartic model is larger than 1.1 for a number of $e$-folds $N<65$, thus far outside of the $2\sigma$ bound. In this sense the effect of ``strong'' noncommutativity close to $\sigma=6$ is not welcome to save the quartic potential. On the other hand, the quadratic potential is not ruled out due to a little 
departure from scale invariance.

The class 2 noncommutative scenario exhibits an interesting feature to have $n_s$ close to 1 even for the quartic potential. As seen in Fig. \ref{fig18} the quartic potential is within the $2\sigma$ bound for $N>50$, thereby compatible with observations. This situation is similar to the 4D case.

%%%%%%%%%%%%%%%%%%%%%%%%%%%%%%%%%%%%%%%%%%%%%%%%%%%%%%%%%%%%%%%%%%%%%%%%%%%%%%%%%%%%%%%%%%%%%%%%%%%%%%%%%%%%%%%%%%%%%%%%

\subsection{The tachyon field $T$}

Let us next consider the observational constraint on the tachyonic large-field models. 

%%%%%%%%%%%%%%
\subsubsection{4D case}
%%%%%%%%%%%%%%

The 4D commutative case was already investigated in \cite{SV,GST}. Since the tensor-to-scalar ratio is smaller relative to the normal scalar field case, this leads to the compatibility with observations. Even steep inflation is 
deep within the $2\sigma$ contour bound.

Because of this small value of $R$, the class 1 and class 2 noncommutative scenarios are also allowed as shown in Fig. \ref{fig19}. The class 1 scenario corresponds to a spectral index $n_s$ larger than 1, but this does not deviate from a
scale-invariant spectrum. All cases with $p=2$, $p=4$ and $p=\infty$ are inside of the $2\sigma$ contour bound.

%%%%%%%%%%
\begin{figure}
%\bc\includegraphics[height=8.3in,width=3in]{fig19.eps}\ec
\caption{\it\label{fig19}
Observational constraints on large-field models for the 4D tachyon field $T$ together with the $1\sigma$ and $2\sigma$ 
contour bounds for three classes of (non)commutative scenarios. Each case corresponds to (\textit{a}) $p=2$ (dots), (\textit{b}) $p=4$ (triangles) and (\textit{c}) exponential potential with $p \to \infty$ (squares), respectively, with the number of $e$-folds $N=45, 50, 55, 60$ (from top to bottom in each panel and for each $p$) \cite{CT}.}
\end{figure}
%%%%%%%%%%

%%%%%%%%%%%%%%
\subsubsection{RS case}
%%%%%%%%%%%%%%

The RS case exhibits larger values of the tensor-to-scalar ratio compared to the 4D case. However, the quadratic and 
quartic potentials are always within the $2\sigma$ bound. The exponential potential is also allowed for the $e$-folds with $N \gtrsim 50$. See Fig. \ref{fig20}.

%%%%%%%%%%
\begin{figure}
%\bc\includegraphics[height=8.5in,width=3.2in]{fig20.eps}\ec
\caption{\it\label{fig20}
Observational constraints on large-field models for the RS tachyon field $T$. Each case corresponds to (\textit{a}) $p=2$ (dots), (\textit{b}) $p=4$ (triangles) and (\textit{c}) exponential potential with $p \to \infty$ (squares), respectively, with the number of $e$-folds $N=45, 50, 55, 60$ (from top to bottom in each panel and for each $p$) \cite{CT}.}
\end{figure}
%%%%%%%%%%

%%%%%%%%%%%%%%%%
\subsubsection{GB case}
%%%%%%%%%%%%%%%%

By Eq. (\ref{rotat}) each inflationary model ($p=2, 4, \infty$) in the GB case ($q=2/3$) lies on a wider range of radii $\varrho$ relative to the 4D and RS cases. In spite of this property, even steep inflation is compatible with observations in both commutative and noncommutative spacetimes. In summary, tachyon inflation is allowed irrespective of the slope of the potential due to a small tensor-to-scalar ratio in all patch cosmologies we have considered. See Fig. \ref{fig21}.

%%%%%%%%%%
\begin{figure}
%\bc\includegraphics[height=8.5in,width=3.2in]{fig21.eps}\ec
\caption{\it\label{fig21}
Observational constraints on large-field models for the GB tachyon field $T$. Each case corresponds to 
(\textit{a}) $p=2$ (dots), (\textit{b}) $p=4$ (triangles) and (\textit{c}) exponential potential with $p \to \infty$ 
(squares), respectively, with the number of $e$-folds $N=45, 50, 55, 60$ (from top to bottom in each panel and for each $p$) \cite{CT}.}
\end{figure}
%%%%%%%%%%

%%%%%%%%%%%%%%%%%%%%%%%%%%%%%%%%%%%%%%%%%%%%%%%%%%%%%%%%%%%%%%%%%%%%%%%%%%%%%%%%%%%%%%%%%%%%%%%%%%%%%%%%%%%%%%%%%%%%%%%%

\subsection{Suppression of CMB low multipoles}\label{cmbsuppr}

In \cite{HL1,TMB,HL2} it was shown that it is possible to explain the loss of power at low multipoles at least partially using the modified spectrum in the UV regime ($\tau \gg kl_s^2$). Here we will consider the situation in which the spectrum on cosmologically relevant scales is generated in the IR noncommutative regime.

As seen in the likelihood contours, the best-fit value of $n_s$ is smaller than 1 and is insensitive to which prescription we adopt for the spacetime (non)commutative structure. The loss of power on large scales is difficult to 
be explained in the standard concordance scenario. If we take the effect of spacetime noncommutativity into account, it is possible to have a suppression of power due to a blue-tilted spectrum. For example, the potential (\ref{power}) gives rise to the blue spectrum for the 4D1 noncommutative case. Of course, the large spectral index $n_s \gtrsim 1.05$
is ruled out as seen in Fig. \ref{fig16}, but the quadratic potential ($p=2$) gives the observationally allowed value around $n_s \sim 1.02$. The quartic potential ($p=4$) corresponds to a marginal compatibility with observations,
but it is welcome to explain the loss of power on the largest scales.

In Fig. \ref{fig22} we plot the CMB angular power spectra for several different cases. The spectrum exhibits some suppression around $1<l\lesssim 10$ in noncommutative spacetime relative to the commutative one.
The quartic potential leads to a stronger suppression compared to the quadratic one, but the smaller-scale spectrum tends to show some disagreement with observations for larger $n_s$. Anyway, it is intriguing that single-field noncommutative inflation leads to a blue-tilted spectrum suitable for explaining the suppression of low multipoles, since this is difficult to be achieved in commutative spacetime, even with a very blue-tilted spectrum $n_s \gtrsim 3$ \cite{PTZ}, unless we introduce another scalar field as in the case of hybrid inflation (see also \cite{LMMR}). Another mechanism generating a blue spectrum at large scales can be found in \cite{YY}.

%%%%%%%%%%
\begin{figure}
%\bc\includegraphics[height=3.5in,width=3.5in] {fig22.eps}\ec
\caption{\it\label{fig22}
The CMB angular power spectrum showing the effects of suppression of power at low multipoles. Curve (\textit{a}) is the 4D commutative model with $(n_s, R)=(0.967, 0.132)$ corresponding to the quadratic potential. Curves (\textit{b}) and (\textit{c}) are the 4D1 noncommutative scenario with $(n_s, R)=(1.018, 0.144)$ and $(n_s, R)=(1.049, 0.263)$, respectively. Note that these values are achieved for the quadratic and quartic potentials in the 4D1 case, respectively \cite{CT}.}
\end{figure}
%%%%%%%%%%

%version: January 20, 2005

\newpage
\thispagestyle{empty}
$\vphantom{dunno how to skip the page}$
\newpage

\chapter{Dualities in patch cosmology and other issues} \label{dlts}

\begin{quote}
\textsl{Tirem-me daqui a metaf\'{\i}sica!}\\ --- Fernando Pessoa, \textit{Lisbon revisited (1923)\bigskip\\
\textsl{Spare me metaphysics!}}
\end{quote}\vspace{1cm}

The ultimate theory of everything, if any, is a long-living mirage that physicists and mathematicians have been pursuing for years in the attempt to solve many fundamental problems rooted in our modern view of the Universe. One of the open issues is how to reconcile general relativity and quantum physics, two separate branches that experiments and observations have widely accepted as meaningful descriptions of natural phenomena, at least each in its own range of influence. The marriage between the two would require deep modifications of both and, although great progress has been made in this direction thanks to string theory, a happy ending to the story is still missing. In particular, the  mostly successful big bang model of cosmological evolution, which manages to glue gravitation and microphysics together in a very nontrivial way, sits on the paradox of the initial singularity: the original point from which all came decrees the failure of general relativity as a self-contained framework, since the relevant cosmological quantities diverge by definition when going back to the first instant of the past. At present we know that quantum effects can resolve such a point into a finite speck and smooth out the worried infinities \cite{str1,str2,str3,str4,str5,str6,str7,str8,str9,str10,str11,str12,str13,str14}.

From a philosophical perspective, the big bang has raised many questions about the nature of time and its birth, leading to the (indeed not new\footnote{For instance, as regards the Theravada Buddhist tradition see Buddhaghosa, \textit{Visuddhimagga}, 13.404-409.}) hypothesis that the Universe may experience a cyclic succession of expansions and contractions in which the big bang singularity is just a transitory phase (a bounce) in a wider process of evolution; see \cite{old1,old2,old3,old4,old5,old6,old7} for old attempts to implement this idea. At a semiclassical level, the structure of the perturbations generated through the bounce can be more complicated than the standard one in a monotonically expanding universe; for example, vector modes cannot be neglected during the contracting phase in contrast to their decaying behaviour in the post big bang phase \cite{BB}. General phenomenology of cyclic models and bouncing cosmological perturbations have been studied, e.g., in \cite{PP1,CDC,PPG,KST1,EWST,AW,pin04,CNZ}; the case of a bouncing closed universe has been investigated in \cite{clos1,clos2,clos3,clos4,clos5,clos6}. 

Moreover, bouncing flat cosmologies may require a violation of the null energy condition, according to which light rays are focused by matter:
\be\label{nec}
\rho+p \geq 0\,,
\ee
where $\rho$ and $p$ are the energy density and pressure of a perfect fluid describing the matter content of the early Universe \cite{PP1,PPG,AW,BM1,BM2,PP2}. Until now, we have adopted the standard lore of well-established energy conditions. What about abandoning the old path in favour of more speculative scenarios? In particular, can some of the most popular objections against embarrassing forms of matter be circumvented? Recently, many people have been considering scenarios in which the dark energy content of the observable Universe is of a nonconventional nature, namely, violating the null energy condition (\ref{nec}). The field associated to an equation of state $p=w\rho$ with $w<-1$ is called ``phantom'' \cite{pha1} and its properties can give rise to a new sort of singularity as well as to an explanation of current observations of dark energy \cite{pha2,pha3,pha5,pha4,pha6,pha7,pha16,pha8,pha9,pha10,pha11,pha14,pha12,pha15,pha17,pha18,pha13,pha19,pha20,pha23,ph23a,pha25,pha24,pha26,pha27,pha28,ph24a,ph25a,ph26a,ACL,ph27a,ph28a,pha29,pha30,pha31,pha32,pha33,pha34,pha35,pha36,pha37,pha38,pha40,pha39,pha41,pha42,pha43,pha44,pha45,pha46,pha47,ph47a,pha48,pha49,pha50,pha51,pha52}. For such a scalar field, the violation of the null energy condition is achieved by a kinetic term with negative sign and this may lead to unitarity problems when quantizing the field (also: particles with negative energy propagate forward in time). However, negative kinetic energies arise in supersymmetric models and higher-derivative-gravity theories \cite{nil84,pol88}, while string models can describe brane physics in which the effective 4D null energy condition is not preserved \cite{CM}; also, anti de Sitter configurations do violate the dominant energy condition.

An interesting singularity-free setup, alternative to inflation and motivated by string theory, is the ekpyrotic scenario, which explains the large-scale small anisotropies of the cosmic microwave background via a collision between wrinkling branes \cite{ekpy1,ekpy2,ekpy3,KOSST,ekpy4,ekpy9,ekpy5,ekpy6,ekpy7,ekpy8,ekp10,ekp12,ekp11,ekp13,ekp14}. A general-relativistic treatment of ekpyrotic/cyclic scenarios predicts a scale-invariant scalar spectrum (with scalar index $n_s-1\approx 0$) and a blue-tilted tensor spectrum $n_t\approx 2$, while standard inflation generates almost scale-invariant spectra. In the latter case, this is a consequence of the SR approximation, stating that both the parameter (\ref{epsilon}) and its time derivative must be sufficiently small.\footnote{In the following we will refer to $\epsilon$ as the ``SR parameter'' even when the slow-roll approximation $\epsilon \ll 1$ is not applied.} Conversely, the cyclic model achieves scale invariance when $\bar{\epsilon}\equiv \epsilon_\text{cyclic} \gg 1$.

The search of viable bouncing mechanisms has led to explore several possibilities that involve, for instance, varying couplings \cite{BKM}, noncommutative geometry \cite{MMMZ}, quantum gravity and cosmology \cite{quan1,quan2,quan3,quan4} (see also \cite{sri04}). In particular, a Randall-Sundrum modification of the Friedmann equations has been considered \cite{SS}, in which a phantom component may help to tear apart black holes during the bounce \cite{BFK,PZ2}. Also, the big bang singularity can be avoided by the combined effect of Gauss-Bonnet and induced gravity terms \cite{KoMP}.

Recently, two remarkable dualities were discovered in flat cosmology, one relating inflationary to ekpyrotic/cyclic spectra \cite{GKST,KST2,BST,PZ1,pia04,lid04} and the other connecting inflationary to phantom spectra  \cite{ACL,lid04,chi02,ACJL,CL,DSS}. More precisely, given an inflationary model there exist both cyclic and phantom cosmologies with the same spectra and such that
\bs\label{4Dduality}\ba 
\bar{\epsilon} &=& 1/\epsilon\,,\label{4Ddual}\\
\hat{\epsilon} &=& -\epsilon\,,\label{4Dphdual}\\
\bar{\epsilon} &=& -1/\hat{\epsilon}\,,
\ea\es
where $\hat{\epsilon}\equiv \epsilon_\text{phantom}$. Other dualities can be found in \cite{BGV,wan99}. In four dimensions, these dualities (inflation-contraction, inflation-phantom, and contraction-phantom) are exact for arbitrary (even varying) $\epsilon$, in the sense that not only the dynamics but also the cosmological spectrum of scalar perturbations is preserved. The issue has then been generalized to the braneworld context \cite{cal9,CLLM}. Since braneworld spectra are broken under duality, and mapped into a quantitatively different contractinglike or phantomlike spectra, these transformations are not symmetries in the strict meaning of the word.

In this chapter we shall investigate the above-mentioned triality for a general commutative patch and show that Eq. (\ref{4Dduality}) no longer realizes exact correspondences between cyclic, inflationary, and phantom patches. According to the new relations we will establish, any expanding universe is mapped to either a contracting or phantom universe which no longer display exactly the same scalar perturbations. In addition, it will turn out that the generalized version of the 4D contracting (phantom) mapping gives rise to a phantom (contracting) dual solution when flipping the sign of $q$.

By considering the general invariance of the patch Hamilton-Jacobi equations we will be able to construct dual solutions with regular behaviour, that is not suffering sudden future singularities, between different patches, which we will call cross dualities (Sec. \ref{patdua}). Then, we shall clarify the relation between patch cosmologies dominated by an ordinary scalar field and the tachyonic ones (Sec. \ref{vs}).

Another interesting insight comes from the comparison of $(q>0,w<-1)$ and $(q<0,w>-1)$ patches. To see this, take the case of a normal scalar field, Eq. (\ref{rho}); an ESR expansion of the energy density yields $\rho^q \propto q \dot{\phi}^2_\text{eff}/2+V_\text{eff}$, where the effective theory includes the dimensional contribution of $\beta_q$. If $q<0$, the kinetic term has the same wrong sign of phantom models. From a mathematical point of view, the phantom universe displays interesting properties such as the presence of a finite-time singularity when $w$ is constant [big smash or big rip \cite{pha1,pha5,pha10}, see Eq. (\ref{conw}) with $w<-1$] and, as said before, a correspondence resembling the scale-factor duality of pre-big-bang cosmology which is a symmetry of the low-energy string effective action and is obtained with the mapping $a(t)\rightarrow a^{-1}(-t)$ \cite{ven91,GV1} (for some reviews on string and pre-big-bang cosmology, see \cite{gas99,LWC,GV2}). Even expanding patch cosmologies with negative $q$ have a finite-time singularity with divergent scale factor, although the density evolution shows the opposite trend. This fact, together with the noncanonical effective theory which seems to characterize such models, invites to investigate if there is some relation between $q<0$ patch cosmologies and scenarios with phantom fluids, which we will do in Secs. \ref{phaqdu} and \ref{qbounce}. At last, inspired by a modified version of the phantom duality we shall outline some proposals for ($i$) a new bouncing scenario, ($ii$) the generation of features in the power spectrum breaking scale invariance, and ($iii$) an alternative to standard inflation.

%%%%%%%%%%%%%%%%%%%%%%%%%%%%%%%%%%%%%%%%%%%%%%%%%%%%%%%%%%%%%%%%%%%%%%%%%%%%%%%%%%%%%%%%%%%%%%%%%%%%%%%%%%%%%%%%%%%%%%%%
%%%%%%%%%%%%%%%%%%%%%%%%%%%%%%%%%%%%%%%%%%%%%%%%%%%%%%%%%%%%%%%%%%%%%%%%%%%%%%%%%%%%%%%%%%%%%%%%%%%%%%%%%%%%%%%%%%%%%%%%

\section{Preliminary remarks}

%%%%%%%%%%%%%%%%%%%%%%%%%%%%%%%%%%%%%%%%%%%%%%%%%%%%%%%%%%%%%%%%%%%%%%%%%%%%%%%%%%%%%%%%%%%%%%%%%%%%%%%%%%%%%%%%%%%%%%%%

\subsection{Broken dualities}

We can see that the dualities (\ref{4Ddual}) and (\ref{4Dphdual}) are broken in their simplest form when considering nontrivial patches. A first evidence comes from the equations of motion for scalar perturbations, which in four dimensions are invariant under the mapping $\epsilon \rightarrow \epsilon^{-1}$ for dominant and subdominant modes, separately \cite{GKST,KST2,BST}. The effective 4D equations of motion of the (Fourier-transformed) scalar-perturbation modes in the longitudinal gauge are the Mukhanov equation (\ref{muksc1}) and
\be\label{mukha}
\left(\frac{d^2}{d\eta^2}+k^2-\frac{1}{\vartheta}\frac{d^2\vartheta}{d\eta^2}\right)v_\mathbf{k} = 0\,,\\
\ee
where $v\equiv-\Phi_4/\dot{\phi}$ is a gauge-invariant variable [$\Phi_4$ is the Newtonian potential that appears in Eq. (\ref{permet})] and $\vartheta \equiv z^{-1}$. We can express $\vartheta$ in terms of the slow-roll parameter (\ref{epsilon}) and its variation $\gamma \equiv d\ln \epsilon/d{\cal N}$ with respect to 
\be
{\cal N} \equiv \ln\frac{a_fH_f}{aH}\,,
\ee
where the subscript $f$ denotes evaluation at the end of the inflationary or ekpyrotic phase; the standard forward definition of the number of $e$-foldings, $N=\ln (a/a_i)$, is related to this quantity by $d{\cal N}=(\epsilon-1)dN$. Neglecting $O(\gamma^2)$ and $O(d\gamma/d{\cal N})$ terms, one finds
\be
\frac{1}{\vartheta}\frac{d^2\vartheta}{d\eta^2} \approx \left(1+\frac{\theta}{2}\right) \left(1+\frac{\theta}{2}\epsilon\right)\frac{\epsilon}{(\epsilon-1)^2}+\frac{\epsilon^2-1}{2(\epsilon-1)^2}\gamma\,.\label{exam}
\ee
In general relativity, a first step towards the duality $\epsilon \leftrightarrow \epsilon^{-1}$ is to note that the equation of motion for $v$ is invariant under the mapping (\ref{4Ddual}). However, when $\theta \neq 0$ this duality is explicitly broken by the term inside the second round brackets, which by the way contributes to the only piece surviving for a constant $\epsilon$ ($\gamma=0$). In the case of a standard tachyonic field, Eq. (\ref{exam}) has an extra term proportional to $\gamma(3+\theta\epsilon)(\epsilon-1)\epsilon/(3q-2\epsilon)$, which breaks the invariance even in four dimensions.

%%%%%%%%%%%%%%%%%%%%%%%%%%%%%%%%%%%%%%%%%%%%%%%%%%%%%%%%%%%%%%%%%%%%%%%%%%%%%%%%%%%%%%%%%%%%%%%%%%%%%%%%%%%%%%%%%%%%%%%%

\subsection{Contracting and phantom patches}

We can make the previous argument more rigorous by means of the Hamilton-Jacobi formulation of the cosmological dynamics. First we extend the setup of Chapter \ref{patch} to the case of phantom fields. For an homogeneous scalar field $\phi$ with potential $V$,
\be
\rho(\phi)=\frac{\ell}{2}\dot{\phi}^2+V(\phi)\,,
\ee
while for a DBI tachyon
\be
\rho(T) = \frac{V(T)}{\sqrt{1-\ell\dot{T}^2}}\,.
\ee
Here $\ell=1$ for ordinary causal fields and $\ell=-1$ for phantoms. To avoid confusion, we will call ``standard (phantom) ordinary scalar'' the $\phi$ field with $\ell=1$ ($\ell=-1$), ``standard (phantom) tachyon'' the $T$ field with $\ell=1$ ($\ell=-1$) and ``scalar'' (or, sometimes, ``inflaton'') the field satisfying the continuity equation (\ref{conti}), regardless of its action.

So far we considered expanding cosmologies, $H>0$. It is now time to extend the equations and discussion to the general case $H\in \mathbb{R}$. The Hamilton-Jacobi equations (\ref{hjpsi}) and (\ref{hj}) become
\ba \label{hjphi}
V(\phi) &=& \left(1-\frac{\epsilon_\S}{3q}\right)|H|^{2-\theta},\label{Vphi}\\
V^2(T)  &=& \left(1-\frac{2\epsilon_\T}{3q}\right)H^{2(2-\theta)},\label{Vtac}\\
H'(\psi)a'(\psi)    &=& -\frac{3q}{2}\ell |H(\psi)|^{\wteta}H(\psi)a(\psi)\,,\label{hjgen}
\ea
where we have set $\beta_q=1$; the absolute value of $H$ is necessary and sufficient to preserve the invariance under time reversal of the original equations of motion. The Hamilton-Jacobi equation (\ref{hjgen}) can be recast as
\be
y'(\psi)a'(\psi)=- y(\psi)a(\psi)\,,\label{hjy}
\ee
where the variable $y(\psi)$ is
\ba 
y(\phi) &\equiv& H^{2\ell/3}(\phi),\qquad\qquad\qquad \theta=0\,,\label{4Dy}\\
y(\psi) &\equiv& \exp \left[\alpha |H(\psi)|^{-\wteta}\right],\qquad \wteta \neq 0\,,\label{y}
\ea
and the coefficient $\alpha \equiv -2\ell/(3q\wteta)$ is $\alpha=-1/3$ and $\alpha=1$ for a RS and GB braneworld without phantoms, respectively. Then, $\text{sgn}(dy)=\text{sgn}(dH)$ when $q>0$. Figure \ref{fig23} shows the function $y(H)$ for the RS and GB case.
\begin{figure}
%\bc\includegraphics[width=8.6cm]{fig23.eps}\ec
\caption{\it\label{fig23}
The function $y(H)$ in the GB (upper curve) and RS (lower curve) expanding braneworlds. The image of $H$ is $y\geq 1$ in the first case and $0\leq y< 1$ in the second one.}
\end{figure}

It is convenient to define the new parameter
\be\label{ve}
\ve \equiv \frac{\epsilon}{q|H|^{\wteta}}= \frac{3}{2}\ell\left(\frac{a}{a'}\right)^2;
\ee
then one can express the spectral amplitudes as $A_s^2(\phi)\propto H^2/\ve$, $A_s^2(T)\propto H^\theta/\ve$, and get the spectral indices from the evolution equation $\dot{\ve}=2H\ve(\epsilon-\eta)$, which reproduces the 4D one when $\ve=\epsilon$. From now on we will set $\wteta=\theta$ for lighter notation. 

Equation (\ref{epsilon}) defines the time variation of the Hubble radius $R_\H= H^{-1}$; this parameter can also be expressed either via derivatives of the inflaton field or through the continuity and Friedmann equations:
\ba 
\epsilon &=& -\frac{a}{a'}\frac{H'}{H}\label{epsi1}\\
         &=& -\frac{(\ln y)'^2}{\theta \ln y}\,,\\
				 &=& \frac32\,q(1+w)-\frac{\dot{q}}{q}\frac{\ln H^2}{2H}\,.\label{epsiy}
\ea
In the last formula, we have considered the general case of time-dependent $q(t)$. When $q=\text{const}$ (which we shall assume throughout the chapter except when stated otherwise), then $\epsilon>0$ when $\text{sgn}(q)=\text{sgn}(w+1)$, while phantom matter with $q>0$ (or ordinary matter with $q<0$) reverses the sign of $\epsilon$.

%%%%%%%%%%%%%%%%%%%%%%%%%%%%%%%%%%%%%%%%%%%%%%%%%%%%%%%%%%%%%%%%%%%%%%%%%%%%%%%%%%%%%%%%%%%%%%%%%%%%%%%%%%%%%%%%%%%%%%%%
%%%%%%%%%%%%%%%%%%%%%%%%%%%%%%%%%%%%%%%%%%%%%%%%%%%%%%%%%%%%%%%%%%%%%%%%%%%%%%%%%%%%%%%%%%%%%%%%%%%%%%%%%%%%%%%%%%%%%%%%

\section{Patch dualities} \label{patdua}

The Hamilton-Jacobi equations encode all the dynamical information for the cosmological evolution. If two different models ($\psi$,$\theta$) and ($\psi'$,$\theta'$) display the same set of equations, then we will say there is a duality between them. Let us now consider what transformations are symmetries of Eq. (\ref{hjy}). In general, a symmetry transformation can be written as
\bs\label{map1}\ba
\da(\psi)&=&f_1(\psi)\,,\\
\dy(\psi)&=&f_2(\psi)\,,\label{map1b}
\ea\es
provided that $[\ln f_1(\psi)]'[\ln f_2(\psi)]'=-1$. In Eq. (\ref{map1b}) all the elements of $\dy$, including $\theta$, are evaluated in the dual patch. Since in principle it is not possible to set $\beta=1=\bar{\beta}$ consistently, one should restore the dimensional factors in the previous and following expressions, noting that $[\beta]=E^{(\theta+2)/(\theta-2)}$.

%%%%%%%%%%%%%%%%%%%%%%%%%%%%%%%%%%%%%%%%%%%%%%%%%%%%%%%%%%%%%%%%%%%%%%%%%%%%%%%%%%%%%%%%%%%%%%%%%%%%%%%%%%%%%%%%%%%%%%%%

\subsection{Singular dualities}

A simple realization of Eq. (\ref{map1}) is
\bs\ba
\da(\psi)&=&[y(\psi)]^{p(\psi)}\,,\\
\dy(\psi)&=&[a(\psi)]^{1/p(\psi)}\,.
\ea\es
In order to satisfy the above integrability condition, the function $p(\psi)$ must be either a constant or
\be \label{pfun}
p(\psi)=p_0\frac{\ln a(\psi)}{\ln y(\psi)}\,,
\ee
where $p_0$ is an arbitrary real constant. For constant $p=p_0$,
\bs\label{map2}\ba
\da(\psi)&=&[y(\psi)]^{p_0},\label{map2a}\\
\dy(\psi)&=&[a(\psi)]^{1/p_0},
\ea\es
also considered in \cite{CLLM}.\footnote{Their results are in agreement with ours when the duality transformations act on scaling solutions of a single patch with $L=H^{\theta/2}$.} The set of equations describing the dual solution can be obtained from Eqs. (\ref{map2a}), (\ref{y}) and (\ref{ve}):
\ba
\da(\psi) &=& \exp \left(-p_0\int^\psi d\psi \frac{a}{a'}\right),\\
|\dH(\psi)| &=& \left[\frac{\dalp p_0}{\ln a(\psi)}\right]^{1/\dth},\label{Hdual}\\
\bar{\ve}(\psi)\,\ve(\psi) &=& \frac{9\ell\bar{\ell}}{4p_0^2}\,.\label{epsidual}
\ea
The right-hand side of Eq. (\ref{Hdual}) is positive when $\text{sgn}(\theta)=\text{sgn}(1-a)$ and $q>0$. Then $a<1$ for RS and tachyon scenarios and $a>1$ for the GB braneworld. Equation (\ref{epsidual}) reproduces Eq. (\ref{4Ddual}) in the ordinary scalar case with $\theta=0$ and $p_0=3/2$, although the 4D auxiliary variable $y$, Eq. (\ref{4Dy}), is constructed in a different way. In the case of cyclic duality ($\bar{\ell}=\ell=1$), the mapping (\ref{map2}) relates a standard  accelerating ($\epsilon<1$) expanding universe with a standard decelerating ($\bar{\epsilon}>1$) contracting phase with the typical properties of cyclic cosmology.

The transformation (\ref{map2}) connects the scale factor of the expanding cosmology to that of a dual cosmology when expressed in terms of the scalar field. In the dual model, the scalar field acquires a different time dependence relative to its expanding counterpart. The time variable can be written as an integral over $\psi$,
\be\label{tnorm}
t=\int^\psi \frac{d\psi}{H} \frac{a'}{a}\,;
\ee
the time variable $\bar{t}$ of the dual solution is then
\be
\bar{t} = \frac{2\ell p_0}{3}\int^\psi \frac{d\psi}{a^{3\bar{\ell}/(2p_0)}} \frac{H'}{H}\,,\label{4Dt}
\ee
in the 4D$\to$4D case,
\be
\bar{t} = \frac{2\ell p_0}{3}\int^\psi d\psi\, (\ln a)^{1/\dth}(\ln H)'\,,
\ee
for the pure braneworld dual of the 4D scenario, while for a general cross duality with $\dth\neq0\neq\theta$, using Eqs. (\ref{y}) and (\ref{hjy}),
\be
\bar{t} = -\frac{p_0}{(\dalp p_0)^{1/\dth}}\int^\psi d\psi \frac{(\ln a)^{1/\dth}}{(\ln a)'}\,.\label{t}
\ee
Everywhere we have omitted $\text{sgn}(\bar{H})$ which is implicit in the time-reversal symmetry of the dual solution.
The dual evolution of the scalar field will be denoted as $\bar{\psi}(t)\equiv \psi(\bar{t})$. For $\bar{q}=q=1$ and $p_0=3/2$, these relations reproduce the already known four-dimensional standard triality.

The exact inversion of the SR parameter $\bar{\epsilon}\epsilon=1$ is achieved in any dimension by the stationary cosmology $a(t)=t$. Otherwise, the fixed points of the transformation (\ref{map2}) are those with 
\be \label{self}
\ve_\text{self-dual} \equiv \frac{3}{2p_0}\,.
\ee
In general, we define a \emph{self-dual solution} as the set of roots of Eq. (\ref{self}). In four dimensions with $p_0=3/2$, Eq. (\ref{self}) reduces to the self-dual condition $\epsilon=1$. From the dual of the SR parameter as given by Eq. (\ref{ve}), it is clear that dual cosmologies superaccelerate either in the phantom case with $\dq>0$ or in the ordinary one for $\dq<0$.

Let us discuss what is the structure of the cyclic duality in a patch framework with positive $q$ and $p_0$. For clarity, we compare the cases $\theta=0,\pm 1$. By definition, standard inflation is characterized by a monotonically varying scalar field which can be assumed to be increasing with time, $\dot{\psi}=Ha/a'>0$. A parity transformation $\psi\to-\psi$ always achieves this condition. Therefore $a'>0$ (since $H>0$) and $H'<0$. On the contrary, the dual scale factor is a decreasing function of $\bar{\psi}$ since $\bar{a}'/\bar{a}\propto -a/a'<0$. 

In the four-dimensional ordinary scalar case, the expanding ($\bar{H}>0$) dual solution has $\dot{\bar{\phi}}=aH/H'<0$, $\bar{a}'<0$, and $\bar{H}'>0$; also, from Eq. (\ref{epsi1}) $\bar{\epsilon}>0$. Under the time reversal
\begin{eqnarray*}
t \in [0,+\infty[  &\to& t \in\,\,]-\!\!\infty,0]\,,\\
\dot{\bar{\psi}}(t)&\to& -\dot{\bar{\psi}}(-t)\,,\\
\bar{a}(t)        &\to& \bar{a}(-t)\,,\\
\bar{H}(t)        &\to& -\bar{H}(-t)\,,\\
\bar{\epsilon}(t) &\to& \bar{\epsilon}(-t)\,,
\end{eqnarray*}
the dual cosmology becomes contracting while keeping the condition $\dot{\bar{\phi}}>0$ and $\bar{\epsilon}>0$ (i.e., it does not superaccelerate). In a general expanding patch, the dual time evolution of the scalar field is $\dot{\bar{\psi}} \propto (\ln a)^{-1/\theta}$, which shows that in the RS, GB, and tachyon scenarios the evolution of the dual cosmology is not regular because of the factor $\ln a$.

To be consistent with the image of $y$ and Eqs. (\ref{map2}) and (\ref{Hdual}), we require $a<1$ in the RS scenario and $a>1$ in the GB one. In this case the above considerations hold with the same signs as in 4D and we get contracting solutions after a time reversal. Let $t_*$ be the time when $a(t_*)=1$; then $\infty>H>H_*=H(t_*)$ and the dual RS scale factor $\bar{a}$ ranges from $\bar{a}_*=\exp [-1/(2H_*)]$ to 1. In the GB case, $\infty>H_*>H>0$ and $\infty>\bar{a}>\exp(3H_*/2)$. As a matter of fact, in the example below the range of the GB power-law dual solution is modified according to the sign of the scalar field (negative for an expanding cosmology) but the underlying message in unchanged: Because of the different range of the variables involved in the mapping (\ref{map2}), the dual cosmology is only a portion of a contracting cosmology evolving from the infinite past to the origin. For this reason one might consider Eq. (\ref{map2}) as an ``incomplete'' mapping; rather, the restriction on the range of $a$ makes these solutions ``complete'' although very peculiar, since the dual Hubble parameter indeed goes from infinity to zero but in a finite time interval.

Note that these features are \emph{not} an effect of the patch approximation we have used for simplifying the cosmological evolution. For $\theta=1$, Eq. (\ref{y}) is a good approximation of the exact Randall-Sundrum case, where \cite{lid04}
\be \label{yRS}
y^2_\text{RS} \equiv \frac{\rho}{\rho+2\lambda}\,.
\ee
In order to make it manifest, we restore the dimensional factors and temporarily redefine the variable $y_\text{temp} \equiv y^{p_0}$ with $p_0=3/2$; then, in the patch approximation $\delta\equiv \lambda/\rho \ll 1$,
\be
y_\text{temp}= \exp \left(\frac{-\kappa_4^2}{6\beta_2H}\right) \approx 1-\sqrt{\frac{\lambda\kappa_4^2}{6H^2}}=1-\delta\,,
\ee
which reproduces Eq. (\ref{yRS}) in the high-energy RS limit. Note that even in the exact RS scenario $\epsilon$ is not exactly inverted under the transformation (\ref{map2}), since $\epsilon \propto [(1+y^2_\text{RS})/(1-y^2_\text{RS})](y'_\text{RS}/y_\text{RS})^2$. The dual Hubble parameter through $\da=y_\text{RS}$ is
\be\label{HRS}
\bar{H}(\phi)=\frac{\sqrt{2\lambda} a(\phi)}{1-a^2(\phi)}\,,
\ee
which is positive as far as $a<1$. It agrees with Eq. (\ref{Hdual}) in the above limit $a =\bar{y}_\text{temp}\approx 1-\bar{\delta}$, as $\bar{H}\approx (2\bar{\delta})^{-1}$. Also,
\be\label{vpRS}
\dot{\bar{\phi}}=-\ell\frac{\sqrt{8\lambda} a'}{1-a^2}\,.
\ee
Equations (\ref{yRS}), (\ref{HRS}) and (\ref{vpRS}) fully confirm what we have said about the structure of the dual solution, since the image of $y_\text{RS}$ is the same as that of $y$ for $q=2$.

\subsubsection{Self-dual solutions and power-law expansion: ordinary scalar case}

The self-dual solutions of the three scenarios with an ordinary scalar field are
\ba
a(t) &=& \exp\left[-p_0\exp \left(-t/p_0\right)\right]\,,\quad\qquad \theta=-1\,,\label{selfGB}\\
a(t) &=& \exp (\sqrt{2p_0t/3})\,,\qquad\qquad\qquad \theta=1\,,\label{selfRS}\\
a(t) &=& t^{2p_0/3} \,,\qquad\qquad\qquad\qquad\qquad \theta=0\,.\label{self4D}
\ea
As an example of the duality, let us consider the power-law inflation,
\be
a(t)=t^n\,,\qquad \epsilon = 1/n\,.
\ee
The ordinary scalar field associated with this expansion is such that $\dot{\phi}^2\propto t^{\theta-2}$. In four dimensions, the exact cosmological solution corresponding to this scale factor is
\be
\phi(t) = \phi_0 \ln t\,,\qquad V(\phi)=V_0e^{-2\phi/\phi_0},
\ee
where $\phi_0=\sqrt{2n/3}$. The scale factor and Hubble parameter read
\be\label{pls}
a(\phi)=e^{n\phi/\phi_0},\qquad H(\phi)=ne^{-\phi/\phi_0},
\ee
respectively. From Eq. (\ref{4Dt}) with $p_0=3/2$, the (time reversed) cyclic-dual solution is
\be
\bar{a}(t)=(-t)^{1/n},\qquad \bar{\epsilon} = n\,, \qquad \phi(t) =-\frac{2}{3\phi_0}\ln (-t)\,,
\ee
after a redefinition $n\bar{t}\rightarrow t$. The dual of the potential can be obtained by taking the dual of Eq. (\ref{Vphi}).

In the Randall-Sundrum scenario, the power-law expansion is realized by
\be
\phi(t)=\phi_0t^{1/2},\qquad V(\phi)=V_0\phi^{-2},
\ee
where $\phi_0=\sqrt{4/3}$. Then, after a redefinition $\phi/\phi_0 \to \phi$,
\be\label{rs}
a(\phi)=\phi^{2n},\qquad H(\phi)=n\phi^{-2},
\ee
and Eq. (\ref{t}) gives $t=1+\phi^2(\ln \phi^2-1)$; the dual RS cosmology under the mapping (\ref{map2}) has 
\bs\label{RSdual}\ba
\da &=& \exp[p_0(1-\phi^2)/(3n)]\,,\\
|\dH| &=& -p_0[3n\ln \phi^2]^{-1},\\
\de&=& -3n(p_0\phi^2 \ln \phi^2)^{-1},
\ea\es
where we have chosen the normalization of the scale factor such that $\da(1)=1$. A changing in the sign of $p_0$ results in different dual solutions. Figure \ref{fig24} shows the time behaviour of $\phi$ in the two separate regions $0<\phi<1$ and $\phi>1$; the quantities of Eq. (\ref{RSdual}) with $p_0=3/2$ are depicted in the left side ($\phi<1$) of Fig. \ref{fig25}, while the cyclic duals with $p_0=-3/2$ are in the right portion ($\phi>1$). Time flows from $\phi=1$, where the vertical line in each panel separates the two dual solutions. 

In the allowed region $\phi<1$ with $p_0>0$ (no phantoms, $\de>0$), the dual scale factor $\da(\phi)$ increases from $\da(1)$ to $\bar{a}(0)$ in a finite time interval, while the dual Hubble parameter goes from infinity to zero in the meanwhile. Solutions with $\phi>1$ and $p_0<0$ behave much better, since they extend not only up to the infinite future, but also are nonsingular at the origin, a very promising feature in classical bouncing models.

One gets the RS contracting solution simply by reversing the time direction [so that the dual scale factor $\bar{a}(\phi)$ decreases from $\bar{a}(0)$ or $\bar{a}(\infty)$ to $\bar{a}(1)$] and flipping the sign of $\bar{H}$. The dual slow-roll parameter does not change under time reversal and keeps being positive. 
%%%%%%%%
\begin{figure}
%\bc\includegraphics[width=8.6cm]{fig24.eps}\ec
\caption{\it\label{fig24}
The normalized RS scalar field $\phi$ as a function of time. The solid horizontal line divides the solutions of the duality (\ref{map2}) with $p_0=3/2$ (region $\phi<1$) and $p_0=-3/2$ (region $\phi>1$).}
\end{figure}
%%%%%%%%%%%%%%
\begin{figure}
%\bc\includegraphics[width=8.6cm]{fig25.eps}\ec
\caption{\it\label{fig25}
The Randall-Sundrum solutions dual to RS power-law inflation for $n=5,10,30$ (increasing thickness). From top to bottom, each panel corresponds to the $\phi$ behaviour of the dual scale factor, the Hubble parameter and SR parameter under the duality (\ref{map2}) with $p_0=3/2$ (region $\phi<1$) and $p_0=-3/2$ (region $\phi>1$).}
\end{figure}
%%%%%%%%%%%%%%

By inverting Eq. (\ref{t}) in the region $\phi>1$, one gets the time dependence of the scale factor as $\bar{a}(t) \propto \exp[t/W(t/e)]$, where $W(x)$ is the product log function solving the nonlinear equation $x=W e^W$.

In the Gauss-Bonnet case we have
\be
\phi(t)=-2nt^{-1/2},\qquad V(\phi)=V_0\phi^6,
\ee
together with
\be
a(\phi) = \phi^{-2n},\qquad H(\phi)=n\phi^2,
\ee
where $-\phi/2n\to \phi$. Equation (\ref{t}) gives $\bar{t}$ in terms of $\phi$: it turns out that $t =4\bar{t}/9=\int d\phi\, \phi (\ln \phi)^{-1}=-\text{Ei}[\ln \phi^2]$, where Ei is the exponential integral function plotted in Fig. \ref{fig26}. From Eqs. (\ref{map2a}), (\ref{Hdual}) and (\ref{epsidual}), the dual GB cosmology is
\bs\label{GBdual}\ba
\da &=& \exp[p_0 n(\phi^2-1)]\,,\\
|\dH| &=& -(n/p_0)\ln \phi^2,\\
\de &=& -(p_0n\phi^2\ln \phi^2)^{-1},
\ea\es
and again the cyclic solution with ordinary matter evolves with $\da<\infty$ for all $t$ and $p_0>0$. On the contrary, in the branch with $p_0<0$ the dual scale factor $\da$ does not collapse to zero at the origin and diverges in the infinite future (see Fig. \ref{fig27}). Under time reversal the cyclic solution evolves from $\phi=1$ to $\phi=0$.
%%%%%%%%%%%%%%
\begin{figure}
%\bc\includegraphics[width=8.6cm]{fig26.eps}\ec
\caption{\it\label{fig26}
The normalized GB scalar field $\phi$ as a function of time. The solid horizontal line divides the solutions of the duality (\ref{map2}) with $\text{sgn}(p_0)= \pm 1$ at the infinite future $\phi=1$.}
\end{figure}
%%%%%%%%%%%%%%
\begin{figure}
%\bc\includegraphics[width=8.6cm]{fig27.eps}\ec
\caption{\it\label{fig27}
The Gauss-Bonnet solutions dual to GB power-law inflation for $n=5,10,30$ (increasing thickness). From top to bottom, each panel corresponds to the $\phi$ behaviour of $\ln\bar{a}(\phi)$, $\bar{H}(\phi)$, and $\bar{\epsilon}(\phi)$ under the duality (\ref{map2}) with $p_0=3/2$ (region $\phi<1$) and $p_0=-3/2$ (region $\phi>1$).}
\end{figure}
%%%%%%%%%%%%%%

Things do not change when exploring cross dualities. We can try to see what happens, say, for the GB dual of a RS cosmology ($\theta=1$, $\dth=-1$). Starting from Eq. (\ref{rs}), one gets Eq. (\ref{GBdual}) with $p_0 \to -p_0$, modulo an irrelevant positive constant. The image of the function $\phi(\bar{t})$ is either $\{\phi<1\}$ or $\{\phi>1\}$.

Dual potentials can be obtained via the dual of Eq. (\ref{Vphi}) or (\ref{Vtac}). Figure \ref{fig28} shows the potential corresponding to the cosmology Eq. (\ref{GBdual}). Depending on the choice of the parameters $n$ and $p_0$, the function $\bar{V}(\phi)$ has a number of local minima and maxima, can assume negative values, and also be unbounded from below.
%%%%%%%%%%%%%%
\begin{figure}
%\bc\includegraphics[width=8.6cm]{fig28.eps}\ec
\caption{\it\label{fig28}
Gauss-Bonnet potential dual to GB power-law inflation under the mapping (\ref{map2}), for some values of $n$ and $p_0$. The region with $\phi<1$ ($\phi>1$) corresponds to duals with $p_0>0$ ($p_0<0$).}
\end{figure}
%%%%%%%%%%%%%%

The properties of cosmological potentials change very interestingly when going from the 4D picture to the braneworld. Take as examples fast-roll inflation with a standard scalar field and negative potentials \cite{lin01,FFKL}. Fast-roll inflation occurs by definition when the kinetic energy of the scalar field is small with respect to the potential energy, $\dot{\phi}^2 \gg V(\phi)$. In this case one obtains a stiff equation of state ($p=\rho$) and a regime described by
\be
a \sim t^{1/3},\qquad \dot{\phi}^2 \sim t^{-2},\qquad \phi \sim \ln t\,,
\ee
from the Klein-Gordon and Friedmann equations. This implies that at early times the kinetic term dominates over any monomial potential energy $V=\phi^m$. In particular, under these conditions the behaviour of the singularity will depend on the kinetic energy regardless of the choice of the potential. In a generic patch with $\theta \neq 0$, the fast-roll regime is given by
\be
a \sim t^{1/(3q)},\qquad \dot{\phi}^2 \sim t^{\theta-2},\qquad \phi \sim t^{\theta/2}.
\ee
Thus the scalar field evolves quite differently in the RS ($\theta=1$) and GB ($\theta=-1$) case. Near the origin, $t\sim 0$, the fast-roll regime is achieved for any $\theta \neq 0$ when $m=2$ and for $\theta>4/(2-m)$ when $m>2$. Therefore the behaviour of the singularity may depend nontrivially on both the contributions of the energy density for suitable (and still simple) potentials on a brane (see also \cite{SST}).

Another result in four dimensions is that potentials with a negative global minimum do not lead to an AdS spacetime. According to the Friedmann equation $H^2=\rho$, the energy density cannot assume negative values; therefore at the minimum $V_\text{min}<0$ the scalar field does not oscillate and stop but increases its kinetic energy until this dominates over the potential contribution. Then one can describe the instability at the minimum in the fast-roll approximation through the only kinetic term; the Hubble parameter vanishes and becomes negative [so that $(\dot{\phi}^2/2)^\cdot>0$], and the Universe undergoes a bounce.

In a braneworld scenario this might not be the case. In fact, in the RS brane the Friedmann equation is $H^2=\rho [1+\rho/(2\lambda)]$. If the negative minimum is larger than the brane tension, $|V_\text{min}| \gtrsim \lambda$, then, after an eventual fast-roll transition, the quadratic correction dominates near the minimum and $H^2 \approx \rho^2$. The scalar field can relax without spoiling the constraints from the equations of motion.

All that we have said can be investigated in greater detail by means of phase portraits in the three-dimensional space $(\phi,\dot{\phi},H)$. Here we shall not explore the subject further and limit ourselves to the above qualitative comments, whose aim was to stress that complicated dual potentials cannot be discarded by general classical or semiclassical considerations. Rather, from one side they should be studied case by case; from the other side, one or more local features encountered by the scalar field during its evolution could induce interesting phenomena at the quantum level, for instance triggering premature reheating or a series of quantum tunnelings.

\subsubsection{Self-dual solutions and power-law expansion: tachyon case}

In the tachyonic case, from Eq. (\ref{Hdual}) we have $\bar{H}^2=(-2q\ln a)^{-1}$, with $p_0=3/2$ for convenience; in order to have a real Hubble parameter with positive $q$, the dual solution corresponds to the time region with $a<1$. The tachyon solution ($\theta=2$) to Eq. (\ref{self}) is $a(t) = \exp \sqrt[3]{9t^2/(8q)}$.

Power-law inflation is achieved with a tachyon profile
\be
T(t)=T_0 t\,,\qquad V(T)=V_0 T^{\theta-2},
\ee
and
\be
a(T)=(T/T_0)^n,\qquad H(T)=nT_0/T,
\ee
for all $q$, where $T_0=\sqrt{2/(3qn)}$. Defining $z\equiv (T/T_0)^2$, Eq. (\ref{t}) gives $\dot{z}\propto -(-\ln z)^{-1/2}$, and we get a real dual solution provided $0<z<1$. Since $z$ is a monotonic function of time (see Fig. \ref{fig29} for $z<1$), we express the dual quantities in terms of $z$ itself:
\bs\ba
\bar{a} &=& \exp[-z/(2qn^2)]\,,\\
|\bar{H}| &=& (-\bar{q}n\ln z)^{-1/2},\\
\bar{\epsilon} &=& qn(-z\ln z)^{-1}.
\ea\es
Consistently with Eq. (\ref{epsidual}), the dual Hubble radius decreases with time and in fact the dual cosmology decelerates ($\bar{\epsilon}>1$). Figure \ref{fig30} shows the behaviour of the found solution, together with the dual with $p_0=-3/2$.
%%%%%%%%%%%%%%
\begin{figure}[ht]
%\bc\includegraphics[width=8.6cm]{fig29.eps}\ec
\caption{\it\label{fig29}
Numerical plot of the function $z(\bar{t})$ describing the tachyonic cosmologies dual to power-law tachyon inflation. The solid line divides the solutions of the two dualities at $z=1$.}
\end{figure}
%%%%%%%%%%%%%%
%%%%%%%%%%%%%%
\begin{figure}
%\bc\includegraphics[width=8.6cm]{fig30.eps}\ec
\caption{\it\label{fig30}
Tachyon cosmology dual to power-law tachyon inflation for arbitrary positive values of $q$ and $n$. Each panel corresponds to the behaviour of $\bar{a}(z)$, $\bar{H}(z)$, and $\bar{\epsilon}(z)$ (from top to bottom).}
\end{figure}
%%%%%%%%%%%%%%

\subsubsection{Phantom and $q$-duality} \label{phaqdu}

Another duality relates standard solutions to phantom ($\hat{l}=-1$) superinflationary ($\hat{\epsilon}<0$) cosmologies through Eq. (\ref{map2}). For $p_0=-3/2$, one has
\bs\label{mapph}\ba
\hat{a}(\psi)&=&y^{-1}(\psi)\,,\\
\hat{y}(\psi)&=&a(\psi)\,.
\ea\es
The mapping (\ref{mapph}) together with Eq. (\ref{tnorm}) gives $\hat{t}=-\bar{t}$, and we can get the phantom dual solution from the cyclic-dual one:
\bs\ba
\hat{\psi}(t) 			&=& \bar{\psi}(-t)\,,\\
\hat{a}(t) 				&=& \bar{a}^{-1}(-t),\\
\hat{H}(t)        &=& \bar{H}(-t)\,,\\
\hat{\epsilon}(t) &=& -\bar{\epsilon}(-t)\,.
\ea\es
One may realize a similar evolution with superaccelerating scale factor by preserving the null energy condition ($\ell\equiv 1$) and flipping the sign of $q$. The mapping we impose is then
\bs\ba
q_*&=&-q\,,\\
\theta_*&=&4-\theta\,,\\
\epsilon_*&=&-\epsilon\,.
\ea\es
The effect of this correspondence is also clear from Eqs. (\ref{ve}) and (\ref{Hdual}). Actually, the choice of the sign of $q$ determines whether the dual solution is superaccelerating or not. Since the region with $q<0$ generates a phantom cosmology, the name ``cyclic'' often adopted for the transformation (\ref{map2}) with $p_0>0$ is therefore misleading in a braneworld scenario with $q<0$. Same considerations hold for the ``phantom'' mapping, which in this case would generate a solution without phantoms.

Sometimes we will say that cosmologies with $q<0$ mimic scenarios with phantom matter; by this we refer to the above matching of the Hamilton-Jacobi equations and do not mean that there is an effective equivalence between the two, since in the first case the energy density decreases when the scale factor expands, while in the phantom case the energy density increases with $a$.

%%%%%%%%%%%%%%%%%%%%%%%%%%%%%%%%%%%%%%%%%%%%%%%%%%%%%%%%%%%%%%%%%%%%%%%%%%%%%%%%%%%%%%%%%%%%%%%%%%%%%%%%%%%%%%%%%%%%%%%%

\subsection{Regular dualities}

It is not yet clear whether the dual solutions constructed so far, especially those with $p_0>0$, describe reasonable (not to mention viable) scenarios. At this point there are two possibilities. The first one is to accept these non-superaccelerating cosmologies and try to explain them by means of some deeper and still missing theoretical ingredient. The second one is to consider their exotic behaviour as a signal that we cannot impose $p_0>0$ (or even $p=$ const) consistently in pure high-energy braneworlds (at least in the RS and GB cases), while the 4D cosmology can be dual to another 4D cosmology. This is due to the fact that the functions $a(\psi)$ and $y(\psi)$ live in different real image sets. Then a new path to follow is to find some mechanism which ``regularizes'' the dual solutions at the asymptotic past and future. The only degree of freedom we could exploit is given by the parameter $p$, which by this line of reasoning must depend on $\psi$. Therefore we are forced to assume Eq. (\ref{pfun}), which generates the transformation
\bs\label{map3}\ba
\da(\psi)&=&[a(\psi)]^{p_0}\,,\label{map3a}\\
\dy(\psi)&=&[y(\psi)]^{1/p_0}\,.
\ea\es
For $\dth\neq0\neq\theta$, the other dual quantities read
\bs \label{altdu1}\ba
\dH(\psi) &=& \left(\frac{p_0 \dalp}{\alpha}\right)^{1/\dth} H^{\theta/\dth}(\psi)\,,\\
\bar{\ve}(\psi) &=& \frac{\ell\bar{\ell}}{p_0^2}\,\ve(\psi) \quad\Rightarrow\\
\de(\psi) &=& \frac{\theta}{\dth p_0}\,\epsilon(\psi)>0\,,
\ea\es
while
\be \label{altdu2}
\bar{t} = p_0 \left(\frac{\alpha}{p_0\dalp}\right)^{1/\dth}\int^\psi d\psi \frac{(\ln a)'}{H^{\theta/\dth}}\,,%
\ee
so that $\bar{\psi}(t)=\psi(t)$ when $\dth=\theta$. 

In 4D ($\dth=\theta=0$),
\bs\ba
\dH(\psi) &=& H^{\ell\bar{\ell}/p_0}(\psi)\,,\\
\bar{t}   &=& p_0\int^\psi d\psi (\ln a)'/H^{\ell\bar{\ell}/p_0}\,,\\
\de       &=& \ell\bar{\ell}\epsilon/p_0^2.
\ea\es
The cross duality between the general-relativistic framework ($\theta=0$) and a high-energy braneworld ($\dth\neq 0$) is, after a time redefinition,
\bs\ba
\dH(\psi) &=& [\ln H(\psi)]^{-1/\dth},\\
\de(\psi) &=& -\epsilon(\psi)[p_0\dth\ln H(\psi)]^{-1},\\
\bar{t}  &=& p_0 \int^\psi d\psi\, (\ln H)^{1/\dth}(\ln a)'\,.
\ea\es
Clearly, the effect of Eqs. (\ref{map3}), (\ref{altdu1}) and (\ref{altdu2}) results in a rescaling of time when $\dth=\theta$, as one can verify by making the substitution
\be
p_0 \to p_0 \frac{\ln \phi^2}{\phi^2}\,,
\ee
in the RS and GB power-law duals, Eqs. (\ref{RSdual}) and (\ref{GBdual}). In this case (which includes tachyon-tachyon dualities) duals without phantoms are achieved as long as $p_0>0$. 

When $\dth\neq\theta$, this transformation relates the dynamics of different braneworld scenarios. According to the cross duality between RS and GB standard inflation, the dual solution does not superaccelerate if, and only if, $p_0<0$. The power-law case is trivial since the dual GB solution is 
\be
\da = \phi^{-2\bar{n}}\,,\qquad \dH = \phi^2\,,\qquad \de = \bar{n}^{-1}\,,
\ee
where $\bar{n}\equiv -np_0$ and $\phi(t) \propto t^{-1/2}$. 

In the power-law case the mapping (\ref{map3}) can be realized also by
\bs\label{map4}\ba
\da(\psi) &=& [\ln y(\psi)]^s\,,\\
\dy(\psi) &=& \exp\left(\frac{1}{s\theta}\int^\psi d\psi \frac{H}{H'}\right),
\ea\es
where $s$ is a real constant, giving a power-law dual $\da = t^{|s|}$. Note the domain range of the dual scale factor. The dual parameter $\de$ is
\be
\de = \frac{\alpha}{\dalp\dth\theta s^2}\frac{|\dH|^{\dth}}{|H|^\theta}\frac{1}{\epsilon}\,,
\ee
which shows how in general the mapping (\ref{map4}) is not equivalent to Eq. (\ref{map3}). This can be seen also by considering the action of the former in four dimensions, where the dual Hubble parameter reads
\be
\dH = \exp\left[-\frac{3\bar{\ell}}{2 s}\int^\phi d\phi\frac{\ln H}{(\ln H)'}\right].
\ee
The 4D dual of the power-law solution (\ref{pls}) is $\da =\phi^s$, $\dH =\exp (-\phi^2/s)$, and $\de=2\phi^2/s^2$, with potential $\bar{V}=(1-\de/3)\exp(-s\de)$. If $s<0$, there is an instability as $\phi \to \infty$, while for positive $s$ the potential has a local minimum at $\de_*=3+1/s$ [being $V''(\de_*) \propto s$] and vanishes at large $\phi$. 

One can devise other transformations of the Hamilton-Jacobi equation than Eqs. (\ref{map2}), (\ref{map3}) and (\ref{map4}). The last example we give is the following:
\bs\label{map5}\ba
\da(\psi) &=& \exp\left(-\frac{1}{r}\int^\psi \frac{d\psi}{a'}\right)\,,\\
\dy(\psi) &=& \exp [r a(\psi)]\,,
\ea\es
where $r$ is a real constant. For $\dth \neq 0\neq \theta$, the basic equations are
\be
|\dH|=\left(\frac{\dalp}{ra}\right)^{1/\dth},\qquad \de= -\frac{r}{\dth}\frac{a'^2}{a}\,,\qquad \dot{\bar{\psi}}=-ra'\left(\frac{\dalp}{ra}\right)^{1/\dth}.
\ee
The RS$\to$RS dual ($r<0$) has 
\bs\ba
\da &\sim& \exp t^{1-n},\\
\dH &\sim& t^{-n},\\
\de &\sim& t^{n-1}.
\ea\es
The RS$\to$GB dual ($r>0$) has 
\bs\ba
\da &\sim& \exp t^{(1-n)/(1-2n)},\\
\dH &\sim& t^{n/(1-2n)},\\
\de &\sim& t^{(1-n)/(1-2n)}.
\ea\es
The GB$\to$GB dual ($r>0$) has 
\bs\ba
\da &\sim& \exp t^{(1+n)/(1+2n)},\\
\dH &\sim& t^{-n/(1+2n)},\\
\de &\sim& t^{-(1+n)/(1+2n)}. 
\ea\es
In the limit $n \to \infty$, the GB dual of both RS and GB cosmology is $\da \sim \exp \sqrt{t}$, that is the Randall-Sundrum self-dual solution with respect to Eq. (\ref{map2}) and $p_0=3/2$.

%%%%%%%%%%%%%%%%%%%%%%%%%%%%%%%%%%%%%%%%%%%%%%%%%%%%%%%%%%%%%%%%%%%%%%%%%%%%%%%%%%%%%%%%%%%%%%%%%%%%%%%%%%%%%%%%%%%%%%%%%%%%%%%%%%%%%%%%%%%%%%%%%%%%%%%%%%%%%%%%%%%%%%%%%%%%%%%%%%%%%%%%%%%%%%%%%%%%%%%%%%%%%%%%%%%%%%%%%%%%%%%%%%%%%%%%%%%%%%%%

\section{Relations between $\phi$ and $T$ cosmologies}\label{vs}

%%%%%%%%%%%%%%%%%%%%%%%%%%%%%%%%%%%%%%%%%%%%%%%%%%%%%%%%%%%%%%%%%%%%%%%%%%%%%%%%%%%%%%%%%%%%%%%%%%%%%%%%%%%%%%%%%%%%%%%%

\subsection{Slow-roll correspondence}

Because of the nonstandard kinetic term in the equation of motion for the tachyon, there is no field redefinition connecting Eqs. (\ref{eom0}) and (\ref{Teom}); in other words, the scalar and tachyon fields are dynamically inequivalent. However, we have seen that in the extreme slow-roll approximation the two descriptions are not distinguishable to lowest SR order, since, near a local extremum $V\approx \text{const}$, one can rescale $T$ such that $\phi=\sqrt{V} T$ and $V(\phi) \approx V(T)$ [see Eq. (\ref{hjpsi})]. For this reason, any cosmological observable generated by an inflationary mechanism with sufficiently slow rolling will be rather insensitive to which of the equations of motion is governing the dynamics. As another example, we consider inflationary non-Gaussianity in Appendix \ref{appB}. 

In general, this first-order correspondence between scalar-filled and tachyon-filled backgrounds allows to relate cosmologies with different index $q$ \cite{muk02}. Thus, one might expect similar predictions for first-order quantities when there is no brane-bulk exchange; however, second-order effects may not be irrelevant when comparing the theory with observations, as it has been seen in Secs. \ref{pert} and \ref{pert2}. Outside the SR regime, the tachyon dynamics may lead to qualitatively different scenarios \cite{GKMP}.

Also, there is a sort of triality among the Mukhanov equations for the scalar, tachyon, and tensor amplitudes: in fact, $\nu_\S=\lim_{\theta\rightarrow 2}\nu_\T$ and $\nu_h=\lim_{\eta\rightarrow 0}\nu_\T$. The first condition is a consequence of the definitions of the SR towers; the second one states that, when $\epsilon_\T \propto \dot{T}^2 \approx \text{const}$, the quantum field $u_\mathbf{k}(\delta T)$ evolves like its gravitational counterpart $u_\mathbf{k}(h)$. 

It is worth noting that in higher-derivative theories the scalar and the tachyon may behave in a radically different way. The expression (\ref{FRW}) is not the most general outcome from alternative gravitational theories. An example outside the braneworld framework is given by the four-dimensional gravitational action (\ref{higher}) with $f(R)=\ln R$ \cite{ABF1}. In the high-energy limit the Friedmann evolution reads
\be\label{alle}
H^2 \approx e^\rho\,,
\ee
where we have absorbed some dimensional positive factors. The SR parameter (\ref{epsilon}) is
\be
\epsilon= -\frac{\dot{\rho}}{2H}=\frac32 (1+w) \ln H^2\,.
\ee
Notably, the exponential behaviour of Eq. (\ref{alle}) does not completely spoil the patch equations we constructed above. Indeed, for an ordinary scalar field we have
\be
\dot{\phi}=-\frac{2H'}{3H^2\ln H^2}\,,
\ee
while the Hamilton-Jacobi equations read
\bs\ba
V(\phi) &=& \ln H^2-\frac{\epsilon_\S}{3\ln H^2}\,,\label{hjalle1}\\
H'(\phi)a'(\phi) &=& -\case{3}{2}\ln H^2(\phi)H^3(\phi)a(\phi)\,.\label{hjalle2}
\ea\es
Therefore Eq. (\ref{hjalle2}) reproduces Eq. (\ref{hj}) when $q=q(t)=\ln H^2$ and $\wteta=2$ are independent quantities;
the extra logarithmic term makes the dynamics deviate from the general patch. For the tachyon, the Hamilton-Jacobi system is
\bs\ba
V^2(T) &=& (\ln H^2)^2-\case{2}{3}\epsilon_\T\,,\label{Thjalle1}\\
H'(T)a'(T) &=& -\case{3}{2}[\ln H^2(T)]^2 H^3(T)a(T)\,.\label{Thjalle2}
\ea\es
Equation (\ref{Thjalle1}) does not reproduce the normal-scalar potential (\ref{hjalle1}) even at first order in $\epsilon$.

%%%%%%%%%%%%%%%%%%%%%%%%%%%%%%%%%%%%%%%%%%%%%%%%%%%%%%%%%%%%%%%%%%%%%%%%%%%%%%%%%%%%%%%%%%%%%%%%%%%%%%%%%%%%%%%%%%%%%%%%

\subsection{$q$-correspondence}

Another useful correspondence appears when taking the limit $q \to \infty$ ($\theta \to 2$), which is another way to look at the parameter (\ref{tilth}). Then, the SR towers (\ref{hsr}) and (\ref{interthsr}) acquire the same dependence on the Hubble parameter; this fact, together with Eq. (\ref{corresp}), tells us that the inflaton field formally tends to an evolution equation $\dot{\psi} \sim H'/(qH^2)$. 

Here ``formally'' means that, from a dynamical point of view, this limit is trivial because it forces the inflaton field to a static background $\psi\approx\text{const}$ [Eq. (\ref{attrac}) guarantees that perturbations are frozen]. Nonetheless, if one keeps nonvanishing slow-roll parameters, it can help to derive and check tachyon H-SR tower and formulas from those of the scalar case; see Eqs. (\ref{epsih'}) and (\ref{Tdotepsi}). In fact, general SR combinations will contain $\theta$ factors and remain asymptotically finite, a fact which we have translated into the adoption of $\wteta$; a cross comparison of the slow-roll equations in Chapters \ref{patch} and \ref{obs} nicely shows this feature. In \cite{KM} it was noted that the asymptotic cosmology $\theta=2$ gives the largest scalar spectrum and smallest scalar spectral index for a power-law inflationary expansion.

Looking at the exact solutions of Sec. \ref{exact}, when going to the limit $q \rightarrow \infty$ in the kinetic term $\dot{\psi}$, scalar solutions approach the tachyonic ones within a given background scale factor; in particular, the parameter defined in the end of Sec. \ref{lastref} $\lambda \rightarrow -n/2$, and Eq. (\ref{sx}) matches Eq. (\ref{tx}). Using this trick, dynamically inequivalent setups are connected when considering the formal time evolution of the inflaton field with respect to the asymptotic gravitational background. In the holographic language \cite{LVL,hal02,van04}, this is equivalent to consider the static solution as the common fixed point of the scalar and tachyon theories, with $\beta$-function given by $\beta \sim \dot{\psi}$ and in the limit in which the horizon-flow tower of the scalar theory approaches the H-SR tower and becomes dynamical.

We conclude with an interesting remark. The above dualities connect not only different braneworlds with the same type of scalar field but also patches with different scalars. If one wishes to construct cosmologies with a DBI tachyon, it is sufficient to start from a generic scenario $(\psi,q,\wteta)$ and hit the dual $(T,\bar{q},2)$ via either Eq. (\ref{map2}), (\ref{map3}), (\ref{map4}) or (\ref{map5}). In particular, with Eq. (\ref{map3}) 
\ba
\dH(T) &=& [H(\psi \to T)]^{\theta/2},\qquad\qquad \theta\neq 0\,,\\
\dH(T) &=& [\ln H(\psi \to T)]^{-1/2},\qquad \theta= 0\,,
\ea
in agreement with the previous results on power-law standard and tachyon inflation.

%%%%%%%%%%%%%%%%%%%%%%%%%%%%%%%%%%%%%%%%%%%%%%%%%%%%%%%%%%%%%%%%%%%%%%%%%%%%%%%%%%%%%%%%%%%%%%%%%%%%%%%%%%%%%%%%%%%%%%%%%%%%%%%%%%%%%%%%%%%%%%%%%%%%%%%%%%%%%%%%%%%%%%%%%%%%%%%%%%%%%%%%%%%%%%%%%%%%%%%%%%%%%%%%%%%%%%%%%%%%%%%%%%%%%%%%%%%%%%%%

\section{Remarks on cosmologies with $q<0$} \label{qbounce}

Let us come back to nonstandard cosmologies with negative $q$ and make some considerations on their features. If the bulk moduli vary with time, the resulting Friedmann evolution on the brane changes accordingly and can be written as in Eq. (\ref{FRW}) but with a time-dependent exponent $q(t)$, at least in a small time interval and under particular energy approximations. The SR parameter $\epsilon$ would not be constant even in the case of constant index of state $w$, see Eq. (\ref{epsiy}). We stress once again that it is left to see whether such a moduli evolution can be consistently implemented in string theory. A sensible treatment of the moduli sector is crucial for a clear understanding of string cosmology; concrete examples have been constructed, e.g., in \cite{KKLT,KKMM,IT,KLi,buc04,MO2}. Nonetheless, a few preliminary remarks might trigger some research in this direction.

%%%%%%%%%%%%%%%%%%%%%%%%%%%%%%%%%%%%%%%%%%%%%%%%%%%%%%%%%%%%%%%%%%%%%%%%%%%%%%%%%%%%%%%%%%%%%%%%%%%%%%%%%%%%%%%%%%%%%%%%

\subsection{$q$-bounce?}

In the end of Sec. \ref{patdua} we have seen that there is a formal duality, similar to the ``phantom'' duality, relating standard expanding solutions with $q>0$ to superaccelerating cosmologies with $q<0$. Now it would be interesting to see what are the properties of these solutions and whether they can play some role in bouncing scenarios, as true phantom components may do. For this reason, let us assume that ($i$) the moduli variation is such that a contracting period with $q<0$ is smoothly followed by a standard $q>0$ expansion, and ($ii$) some stabilization mechanism is effective after the shift in the moduli space, so that the cosmological expansion on the brane can be described by one of the previous models (4D, RS, GB) at sufficiently late times. In the simplest toy model, we can consider a sharp transition from $-q$ to $q$ at the big bang, with $0<q=\text{const}\ll 1$ around the bounce.

The contracting phase is deflationary since $\ddot{a}>0$ and is actually superaccelerating if the brane content is not phantomlike. The absolute value of the Hubble rate decreases to zero while the energy density $\rho(t)$ approaches the singularity at $\rho(0)=\infty$; in a standard contracting phase (Fig. \ref{fig31} for $t<0$) it is the Hubble radius that decreases.
%%%%%%%%%%%%%%
\begin{figure}
%\bc\includegraphics[width=8.6cm]{fig31.eps}\ec
\caption{\it\label{fig31}
Inflationary expansion with a short $q\to -q \to q$ transition. The physical Hubble length $|R_\H|$ (upper panel) and the comoving one $|R_\H|/a$ (lower panel) are plotted in arbitrary units of time. $k$ denotes the comoving wave number of a perturbation exiting the horizon during the $q$-transition. For $t<0$ a standard contracting behaviour is represented.}
\end{figure}
%%%%%%%%%%%%%%

Note that at neither this nor any other stage we are saying anything about ``the creation of the Universe,'' since all these considerations regard the cosmological evolution from a brane-observer point of view rather than the global spacetime structure. Although the back-reaction on the brane is governed by the moduli evolution, the braneworld as a geometrical object does not undergo any dramatical transition and is considered to be present at any time in order to make sense of the modified Friedmann equation before, during, and after the bounce. Genuine braneworld creation has been considered in \cite{KoS2,AM}.

There are several advantages in constructing a model of bounce with varying $q$. First, it avoids the reversal problem due to the monotonicity of the Hubble parameter in general relativity \cite{KOSST}. Second, one does not encounter the classical instabilities of background contracting solutions with $w=\text{const}$ found in \cite{EWST,GKST}. In this case, from the continuity equation the energy density scales as $\rho =a^{-3(1+w)}$, up to some constant factor. In a contracting universe, if $w \lesssim -1$ the energy density of the scalar field is nonincreasing, while an extra matter or radiation component increases with time. Therefore solutions with $w \lesssim -1$ are not attractors as regards the isotropic cosmological evolution, while solutions with $w > 1$ ($\gg 1$ in cyclic or ekpyrotic scenarios) are stable. 

Put into another way, for constant $w$ one has $R_\H \sim t$ and $a \sim t^{1/\epsilon}$. When $0<\epsilon<1$, $a$ grows more rapidly than the Hubble radius and quantum fluctuations can leave the horizon; for $q>0$, $\epsilon>1$, and $H<0$, a necessary condition for getting a scale-invariant spectrum is that $R_\H$ shrinks more rapidly than $a$, that is $\epsilon >1$. When $q<0$, the scale factor shrinks as the Hubble radius decreases and vice versa, and no apparent critical index of state is required.

Since there is no concrete model motivating a patch transition, this scenario is not less arbitrary than those invoking an \emph{ad hoc} phantom matter. Matter with $w<-1$ has been advocated both in the context of bouncing cosmologies and for explaining modern data on cosmic acceleration. Although it has been criticized in many respects \cite{pha9,pha28,crit1,crit2} and is not strictly necessary to bring current observations to account \cite{OW1,OW2,HoN,LuS}, a phantom component still can be embedded in string theory \cite{CM,fra02} and has attractive features; for instance, in a cyclic phantom universe black holes are tore apart and are prevented to cannibalize the cosmological horizon during one of the contracting phases \cite{BFK,BDE,GoD}. Of course this is not the case for $q$-cosmologies in which the null energy condition, determining the evolution of the black hole mass, is preserved.

Another clear shortcoming is that there is no apparent reason why the scale factor should reverse its evolution exactly during the $q$-bounce. Therefore there is no immediate relation between solutions with negative $q$ and bouncing models of the early Universe. Anyway Eq. (\ref{FRW}) is only a particular case of a wider and more realistic class of cosmological evolutions, to which the RS scenario itself does belong. If the nonstandard behaviour of the 4D Friedmann equation arises as a correction to the linear term, then it is natural to write it down as a polynomial (rather than a monomial) in $\rho$:
\be
H^2=b_1\rho^{q_1}-b_2\rho^{q_2}\,,
\ee
where $q_1,q_2,b_1,b_2$ are constants; one can always set one of the $b_i$'s to 1 in appropriate units. In the RS two-brane case $q_1=1$, $b_1=1$, and $q_2=2$, while $b_2=-(2\lambda)^{-1}$ in the type 2 model (matter on the brane with positive tension) and $b_2=(2|\lambda|)^{-1}$ in the type 1 model (matter on the brane with negative tension). If $b_1,b_2>0$, then a bounce occurs at
\be \label{rhob}
\rho_b\equiv(b_2/b_1)^{1/(q_1-q_2)}\,.
\ee
Under the additional assumption that $\text{sgn}(q_1)\neq \text{sgn}(q_2)$, a period of nonphantom superacceleration may dominate at some point of the evolution, according to the sign of the coefficients. But about this we will say no more.

%%%%%%%%%%%%%%%%%%%%%%%%%%%%%%%%%%%%%%%%%%%%%%%%%%%%%%%%%%%%%%%%%%%%%%%%%%%%%%%%%%%%%%%%%%%%%%%%%%%%%%%%%%%%%%%%%%%%%%%%

\subsection{$q$-bump?}

Another possibility arises when the evolution of the moduli in the bulk is such that $q$ changes from positive to negative to again positive values in some interval $\Delta t=t_e-t_i$. In the case the transition $q \rightarrow -q \rightarrow q$ happens during the inflationary period, some interesting features in the power spectrum may be generated. A bump in the power spectrum would occur for those perturbations crossing the horizon during the patch transition. In the toy model, $q(t)$ is a step function with sharp transitions and the characteristic time of the event is small with respect to the total duration $\Delta t_\text{inf}$ of the accelerated expansion, $\Delta t/\Delta t_\text{inf}\ll 1$; in Fig. \ref{fig31} the interval $\Delta t$ is exaggerated. Perturbations leaving the horizon during this period will break scale invariance in a comoving wave number interval $\Delta k=k(t_e)-k(t_i)$. 

To get some idea of the properties of the arising feature it is more convenient to consider a smoothly varying $q$, for example with a Gaussian profile centered at some time $t_0$,
\be 
q(t)=q_0-q_1 e^{-(t-t_0)^2/\sigma^2},
\ee
where $q_1>q_0>0$ and $\Delta t \sim \sigma$ is the region of validity of the approximation.

Let us recall that the expressions for the squared scalar and tensor amplitudes and their ratio are ($\beta_q=1$)
\be\label{qspec}
A_s^2(\psi) \propto qH^{2+\theta}/\epsilon\,,\qquad
A_t^2      \propto |q|H^{2+\theta}/\zeta_q\,,\qquad
r          = |\epsilon|/\zeta_q\,,
\ee
where $\zeta_q$ is a $O(1)$ coefficient depending on the concrete gravity model and it has been assumed to be positive without loss of generality. Equation (\ref{qspec}) is valid to lowest SR order; in fact, around $q\sim 0$ the parameter $\epsilon \sim 0$ and the SR approximation still holds. The scalar and tensor indices are, near $k_0=k(t_0)$,
\ba
n_s-1 &\approx& -(2+\theta)\epsilon+(1-\epsilon)\gamma\,,\\
n_t   &\approx& -(2+\theta)\epsilon\,.
\ea
A negative $q \approx q_0-q_1$ corresponds to $\theta > 2$ and a very blue-tilted gravitational wave spectrum, an effect that has been found in ekpyrotic models also \cite{ekp11,KST2,BST,PZ1,pia04}. However, models with $\gamma > 4\epsilon$ have even a blue-tilted scalar spectrum; if $k_0 \ll 10$, that is at long wavelengths, this might fit with the loss of power in the CMB quadrupole region found in recent data. Note that the divergence $\theta \to \infty$ at the bounce is typical of purely adiabatic perturbations. In general relativity, the consistent introduction of entropy perturbations, generated by the mixing modes of a multicomponent fluid, compensates the curvature divergence \cite{PPG} and a similar mechanism might operate in this case, too.

%%%%%%%%%%%%%%%%%%%%%%%%%%%%%%%%%%%%%%%%%%%%%%%%%%%%%%%%%%%%%%%%%%%%%%%%%%%%%%%%%%%%%%%%%%%%%%%%%%%%%%%%%%%%%%%%%%%%%%%%

\subsection{$q$-inflation?}

Because of its features one might think to regard an expanding $q<0$ era as a substitute of standard inflation. For example, we can devise a superaccelerating universe filled by a not-slow-rolling scalar field with a generic potential. The expansion inflates the fluctuations of the field (thus explaining the large-scale anisotropies) until the moduli evolution changes the sign of $q$ and gracefully exits to a normal, decelerating expansion. A few properties of expanding $q$-models were already outlined in Chapter \ref{patch}.

One of the most important strongholds of inflation is its capability to select a de Sitter vacuum from a non fine-tuned set of initial conditions. This property is encoded in the definition of the inflationary attractor of Sec. \ref{attractor}. If there exists an attractor behaviour such that cosmological solutions with different initial conditions (i.c.) rapidly converge, then the (post-)inflationary physics will generate observables which are independent of such conditions. When $\theta>2$ ($q<0$) in Eq. (\ref{attrac}), $H_o'$ and $\dot{\psi}$ have concording signs. In this case, linear perturbations are suppressed when $|\epsilon|<3/\theta$; in the large $\theta$ limit, that is when $q$ is close to vanish in the realistic case of smoothly varying moduli, this condition leads to a trivial de Sitter expansion $H=\beta_0$ insensitive of the matter content. Solutions with a greater SR parameter would depend on the initial conditions in an unpleasant way. 

The condition $|\epsilon|<3/2$, though more stringent than those of standard inflationary scenarios with positive $\epsilon$ (4D and RS: $\forall \epsilon$; GB: $\epsilon < 3$), does not severely constrain the dynamics of the scalar field in order to have a sufficiently flat potential, provided not a too negative $q$. However, it is important to stress that this new picture might not replace inflation because of this possible fine tuning, $|q|\ll 1$. Therefore it is not clear whether the dependence on i.c. would survive or not after the bump, although a sufficient amount of $q$-inflation might have erased any memory of the i.c. at this time.

%version: January 20, 2005

\chapter{Discussion and conclusions} \label{concl}

\begin{quote}
\textsl{And I said to my spirit, When we become the enfolders of those orbs and the pleasure and knowledge of every thing in them, shall we be filled and satisfied then? \\ And my spirit said No, we level that lift to pass and continue beyond.} --- Walt Whitman, \textit{Leaves of grass (1855 edition)}
\end{quote}\vspace{1cm}

%%%%%%%%%%%%%%%%%%%%%%%%%%%%%%%%%%%%%%%%%%%%%%%%%%%%%%%%%%%%%%%%%%%%%%%%%%%%%%%%%%%%%%%%%%%%%%%%%%%%%%%%%%%%%%%%%%%%%%%%%%%%%%%%%%%%%%%%%%%%%%%%%%%%%%%%%%%%%%%%%%%%%%%%%%%%%%%%%%%%%%%%%%%%%%%%%%%%%%%%%%%%%%%%%%%%%%%%%%%%%%%%%%%%%%%%%%%%%%%%

\section{Summary of the results}

In this work we have considered an inflationary period started by a single scalar field, with either an ordinary or Born-Infeld action, slowly ``rolling'' down its potential and driving an early-Universe period of accelerated expansion. Quantum fluctuations of this scalar field generate the perturbation structure explaining the small anisotropies of the cosmic microwave background. By means of the slow-roll formalism, several consistency relations have been derived and used to compare theoretical predictions and modern experimental data. Cosmological models with a variety of different high-energy ingredients have also been confronted, using a modified effective Friedmann equation (describing the cosmological evolution on the brane) and/or a maximally symmetric realization of noncommutative spacetime. In the latter case, spacetime is assumed to have a ``fine-grained'' structure at quantum scale, which rather surprisingly modifies the \emph{large}-scale spectrum of primordial perturbations. Modifications of the Friedmann equation were assumed to be valid within finite time intervals or, equivalently, in particular energy regimes (or ``patches'') experienced by the inflaton field during the early cosmological evolution. Thanks to the patch approach, we have obtained a Hamilton-Jacobi and SR formulation of the cosmological evolution which is valid for many known gravitational theories either in a particular energy limit or time interval. 

Despite all the shortcomings of this approximated treatment of extra-dimensional physics, it gives several important first-impact informations. Different braneworld models are treated in a simple, unified way. The four-dimensional scenario as well as tachyon inflation are automatically included, without performing separate analyses of the cosmological dynamics. We have achieved the following results for commutative inflation:
\begin{itemize}
\item[--]	Previous assessments on tachyon and normal scalar inflation have been extended to the patch context, in particular regarding the inflationary attractor, exact solutions, perturbation spectra, non-Gaussianities, Hamilton-Jacobi formulation, and dualities.
\item[--]	The consistency relations describing the inflationary spectra are definitely broken in the presence of extra dimensions and can discriminate between standard four-dimensional and braneworld scenarios. We have also provided many elements useful for probing the viability of braneworld models through the latest observational data of the cosmic microwave background.
\item[--] We have generalized the four-dimensional triality between inflationary, cyclic, and phantom cosmologies to the patch case. The simple 4D relations between the SR parameters of models with an ordinary scalar field are broken and extended consequently. The self-dual solutions and the duals of power-law inflation have been provided in the presence of either a normal scalar field or a Born-Infeld tachyon. The structure of the triality is deeply modified: The cosmologies dual to inflation either display singularities within finite time intervals or are not singular at the origin. This last feature is appealing as regards the construction of nonsingular bounces. Finally, starting from a new version of the ``phantom'' duality, we have set some remarks on cosmologies with $q<0$.
\item[--] Under certain assumptions, it can be shown that the linear cosmological spectrum comes from the first term of a gradient perturbative expansion of a nonlinear curvature perturbation satisfying a generalized Mukhanov equation of motion. The bispectrum of this quantity, which involves it at second order, governs the non-Gaussian signature eventually detectable in the CMB. By neglecting the projected Weyl tensor on the brane, we have found that the pure inflationary contribution to the nonlinearity parameter $f_{\rm NL}$ is proportional to the braneworld scalar spectral index and therefore unobservable, in agreement with past 4D calculations. 
\end{itemize}
In Chapter \ref{noncom} we have considered several classes of noncommutative inflationary models within an extended version of patch cosmological braneworlds, starting from a maximally invariant *-generalization of the action for scalar and tensor perturbations. The noncommutative cosmological model by Brandenberger and Ho has been developed from both the theoretical and experimental point of view, showing that new compelling features arise when considering the presence of a noncommutative scale. A full analysis of these models and their observational consequences have been interpreted in the light of WMAP data (in collaboration with Shinji Tsujikawa). The main results are:
\begin{itemize}
\item[--] Class 1 and class 2 models are appreciably distinct from each other in the full span of the spectrum.
\item[--] BH and New models give almost the same predictions in the IR region of the spectrum.
\item[--] The relative running (\ref{da}) is generally more pronounced in the GB scenario than in 4D, while in RS the effect is less evident. Either increasing $n_s$ or going to the commutative limit, $H/M_s\rightarrow 0$, the relative running $\Delta\alpha_s$ tends towards positive values.
\item[--] The consistency relation $n_t \propto R$, Eq. (\ref{ntconeq2}), greatly differs from one noncommutative model to another. The perturbations are always blue-tilted for the class 1 scenario, thus giving positive values of $n_t/R$.
This unusual property comes from the fact that the mechanism for generating fluctuations is different from the standard case due to the stringy uncertainty relation in momentum space.
\item[--] Expressing the inflationary observables \{$A_s^2$, $R$, $n_s$, $n_t$, $\alpha_s$, $\alpha_t$\} in terms of the horizon-flow parameters, the likelihood analysis of these quantities is the same for both types of scalar fields. One can find some difference in the $n_s$-$R$ plane by the noncommutative modification of consistency relations. The main change appears in the maximum value of $R$ ($=R_\text{max}$) and it ranges in the region $0.5 \lesssim R_\text{max} \lesssim 0.7$.
\item[--] We have also placed constraints on the large-field monomial potentials $V=V_0 \psi^p$ (including the exponential potential $V=V_0e^{-\psi/\psi_0}$ by taking the limit $p \to \infty$) in the 4D, RS, and GB cases in (non)commutative spacetime. For the ordinary scalar field $\phi$:
\begin{itemize}
\item[.] The quartic potential is rescued from the marginal rejection in the noncommutative class 2 4D case ($\sigma=2$).
\item[.] Steep inflation driven by an exponential potential is excluded in the commutative RS scenario, but is allowed in the noncommutative class 1 RS case ($\sigma=6$).
The quartic potential is compatible with observations both in the class 1 and class 2 RS cases, but it is not so in the RS commutative case.
\item[.] The quartic potential exhibits a compatibility with observations for the class 2 GB case, while it does not in the other two cases (GB0 and GB1).
\end{itemize}
For the tachyon field $T$:
\begin{itemize}
\item[.] A scale-invariant spectrum ($n_s=1$) is generated for $p=2$ in the noncommutative class 1 case irrespective of the kind of patch cosmologies.
\item[.] Even steep inflation is allowed due to small values of the tensor-to-scalar ratio in the three patch classes.
\end{itemize}
All these properties have been investigated both analytically and numerically.
\item[--]  We have also pointed out a possibility to explain the suppression of CMB low multipoles using a blue-tilted spectrum generated in the IR regime. Although noncommutativity can provide a better fit of the spectrum for low multipoles, it is not easy to fully explain the loss of power. Anyway, this spectral region chiefly suffers from cosmic variance and the experimental data at large scales ($l=3,4$) are not determined with sufficient accuracy.
\item[--] Inflationary non-Gaussianity does not change significantly when considered in a noncommutative framework.
\end{itemize}

%%%%%%%%%%%%%%%%%%%%%%%%%%%%%%%%%%%%%%%%%%%%%%%%%%%%%%%%%%%%%%%%%%%%%%%%%%%%%%%%%%%%%%%%%%%%%%%%%%%%%%%%%%%%%%%%%%%%%%%%%%%%%%%%%%%%%%%%%%%%%%%%%%%%%%%%%%%%%%%%%%%%%%%%%%%%%%%%%%%%%%%%%%%%%%%%%%%%%%%%%%%%%%%%%%%%%%%%%%%%%%%%%%%%%%%%%%%%%%%%

\section{Open questions}

The patch formalism will prove adequate for developing new scenarios with a modified Friedmann equation, different from those constructed so far. However, it is important to stress what are the assumptions and eventual shortcomings of this approach in order to push forward our knowledge and walk through a path that -- for myself, I do not even say has been taken yet -- people have begun to catch a glimpse of.

%%%%%%%%%%%%%%%%%%%%%%%%%%%%%%%%%%%%%%%%%%%%%%%%%%%%%%%%%%%%%%%%%%%%%%%%%%%%%%%%%%%%%%%%%%%%%%%%%%%%%%%%%%%%%%%%%%%%%%%%

\subsection{Bulk physics}

With no reference to the gravitational sector, two important assumptions, intimately connected with the evolution of the matter content, emerge; namely, to consider an empty bulk and neglect the Weyl tensor contribution. In particular, there is no source term in the continuity equation (\ref{conti}). In the Randall-Sundrum model, several works have shown that bulk physics mainly affects the small-scale or late-time cosmological structures, i.e., that part of the spectrum which is dominated by post-inflationary physics \cite{GoM,GRS,IYKOM,LCML,koy03,KLMW}. However, it is possible that a nonzero brane-bulk flux would modify the inflationary spectra. For instance, production of particles when the inflaton does not lie in its vacuum state can generate a non-Gaussianity signature during the accelerated expansion \cite{LPS,MRS,GMS}. CMB observations strongly constrain the maximum number density of these particles and the $n$-point correlation functions of the resulting perturbations; with a brane-bulk exchange mechanism and interactions at the KK energy scale, this number density, as well as the predicted non-Gaussianity, may vary nontrivially. Thus, the adoption of a modified continuity equation may lead to a richer scenario. See, e.g., \cite{VDMP,LSR,KKTTZ,LMS,BBC,ApT,tet04} for Randall-Sundrum cosmologies with nondiagonal bulk stress-energy tensor and \cite{LeT} for a six-dimensional example (but see also \cite{kof04}).

Future studies of patch cosmology with implemented bulk contributions will be crucial for several reasons. For example, one should consider the contribution of the nonlocal physics of the bulk in order to set a truly consistent picture of braneworld cosmologies and dualities. 

In the typical inflationary context, Eq. (\ref{FRW}) encodes the most part of the braneworld effective evolution; in fact, the simplest contribution of the projected Weyl tensor is $\propto a^{-4}$ and is damped away during the accelerated expansion. However, the dark radiation term is no longer negligible in a shrinking universe and should be taken into account when relating an inflationary evolution to its contracting dual.

Also, a way to generate a stronger non-Gaussian signal in the braneworld context might be to include the Weyl contribution, but as we have seen in Sec. \ref{patsetup} we should expect it to play a negligible role in the long wavelength limit. Nevertheless, this issue will deserve further attention for at least two good reasons. The first is that Weyl damping was considered and shown only in the case of linear perturbations, while the stochastic Langevin equation for the curvature invariant holds at all orders. The second is that bulk physics intrinsically provides a noise source to the Mukhanov equation through an infinite tower of Kaluza-Klein scalar modes dominating at short wavelengths \cite{KLMW}. Therefore, while Eq. (\ref{Qeom}) would keep being valid, an important contribution to the stochastic noise term (\ref{lange2}) might be lacking in the present analysis. In this case the consistency equations, included Eq. (\ref{fns}) even in the squeezed limit, would be spoiled anyway \cite{der04}.

%%%%%%%%%%%%%%%%%%%%%%%%%%%%%%%%%%%%%%%%%%%%%%%%%%%%%%%%%%%%%%%%%%%%%%%%%%%%%%%%%%%%%%%%%%%%%%%%%%%%%%%%%%%%%%%%%%%%%%%%

\subsection{More on noncommutativity}

Noncommutative models are far from being completely explored. For instance, one could impose also the extra dimension(s) to be noncommutative and extend the algebra (\ref{alg}) or other realizations to the transverse direction(s). A brane with finite thickness would emerge because of the minimum length scale $l_s$; in this case our analysis could be thought as performed on mean-valued quantities along the brane thickness. For example, $\rho \rightarrow \langle\rho\rangle \sim \int_\text{brane} \rho\, dy$, $p \rightarrow \langle p\rangle$, and so on. The subject requires further investigation and a good starting point might be the cosmological thick brane setup \cite{KKOP1,KKOP2,CEHS,KKS,ML,wan02,BGS,BMe,GY,KaS1}.

An interesting possibility is to choose another vacuum state rather than the adiabatic vacuum with which the perturbation spectrum is usually calculated. This scheme has been outlined in \cite{dan02} and developed in \cite{cai04,ACT}. Another important aspect is the extension of SR calculations to next-to-leading order; the use of the gravitational version of the function $z(\teta)$, Eq. (\ref{zgrav}), would permit to compute higher-order expressions for both the tensor amplitude and the consistency equation for the tensor index. 

%%%%%%%%%%%%%%%%%%%%%%%%%%%%%%%%%%%%%%%%%%%%%%%%%%%%%%%%%%%%%%%%%%%%%%%%%%%%%%%%%%%%%%%%%%%%%%%%%%%%%%%%%%%%%%%%%%%%%%%%

\subsection{More on patches and beyond}

Beside Weyl physics, other possible ingredients have been left aside in this work. One of them is the influence of quantum correction to the braneworld model, embodied in an induced gravity term in the 4D brane action \cite{BMW,KoMP,kof01,def01,def02,KTT,MMT,PaZ,ZC,BW}. 

In parallel, it would be interesting to explore two other directions. The first, most important issue should be to
find new cosmological scenarios with $\theta\neq 0,\pm 1, 4$ and exploit the compact formalism provided by the patch formulation of the cosmological dynamics. Certainly there could be a lot of work for M/string theorists in this sense.
In particular, we would like to motivate $q<0$ scenarios within string theory, super or quantum gravity, since at this stage they are rather speculative. 

\emph{A priori}, it would be useful to investigate whether some regions of the line of patches $\theta$ are excluded or not by observations. A clear answer in this respect would constrain any new braneworld scenario with a nonstandard Friedmann equation with $\theta\neq 0, \pm 1$. In the 4D case ($\theta=0$), we have addressed a similar question for the noncommutative quantity $\sigma$ and performed a likelihood analysis with a very large prior ($|\sigma|<100$) \cite{CT}. The parameter did not show a good convergence, as the tensor index $n_t$ can be made smaller by choosing a smaller $R$ in Eq. (\ref{ntconeq2}). Since the same result holds when varying $\theta$ and assuming the set $\{n_t, \zeta_qR\}$ to be constrained by Eq. (\ref{ntconeq2}), one has to consider \emph{fixed} values of any extra parameter which modifies the four-dimensional scenario. Then we have not been able to say anything about the viability of a general patch cosmology.

The second direction is related to the dual picture and goes towards a study of the cosmological perturbations through the bounce, by further modeling the too simple step-function transitions we presented. A more concrete model would try to provide a smooth big crunch/big bang phase and allow a nonsharp transition in $q$.
 
In order to fully resolve the singular bounce we should rely on a description more general than classical gravity. To find reasonable solutions of the big bang singularity and embed a bouncing picture in a well-established (stringy) theoretical framework will perhaps be one of the most promising lines of research in the following years, not only for the immediate cosmological implications (observability of pre-inflationary physics and comprehension of the  high-energy early Universe) but also because it might lead to a better understanding of the still controversial but intriguing \emph{landscape} of vacua \cite{carg,ban04}.

Let us conclude with a fundamental question which lay hidden in these lines and nevertheless should be answered: What about cosmic confusion? Can we rely on the consistency equations and CMB observations as a smoking gun for both braneworld and noncommutative scenarios? In the context of the patch formalism the answer, presumably, is no. As it typically happens in cosmology, other completely different frameworks could mimic the features we have exploited, and even simple 4D multifield configurations produce a nonstandard set of consistency relations \cite{WBMR,BMR,TPB}. Some general relativistic models may predict a set of values for the observables $\{n_t,R,n_s,\alpha_s,\dots\}$  close to that of a braneworld within the experimental sensitivity. Even noncommutativity may not escape this ``cosmic degeneracy'' since, for example, a blue-tilted spectrum can be achieved by 4D hybrid inflation. So we can talk about clues but not proofs for high-energy cosmologies when examining the experimental data. 

The subject has to be further explored in a more precise way than that provided by the patch formalism in order to find out more characteristic and sophisticated predictions, extending the discussion also to the small-scale region of the spectrum. Nonambiguous physical evidences for extra dimensions or other aspects of string/M theory would open up a new season (we would daresay ``era'') for our modern view of the high-energy and geometrical structure of spacetime and would dramatically boost the theoretical research for a viable, completely consistent theory. Therefore I do believe it will be important to capitalize at least part of our efforts in the inspection of models giving reliable predictions to be tested in the near future. Whatever the final answer turns out to be, there is hope of tracking down braneworld signatures through the inflationary physics and related experiments.

\appendix

% version:  December 15, 2004

\newpage
\thispagestyle{empty}
$\vphantom{dunno how to skip the page}$
\newpage

\chappendix{Exact solutions in the RS braneworld} \label{appA}

Hawkins and Lidsey \cite{HaL1} have found several exact solutions for the Randall-Sundrum single brane inflation. Here we will consider just two of them and show that at sufficiently late times they approach the 4D power-law solution with constant SR parameters, thus providing a reasonable background around which to construct lowest-order perturbation amplitudes \cite{cal1}. The first model has
\ba
a(\tau) &=& \left(\tau+\sqrt{\tau^2-1}\right)^p\,,\\
\phi(\tau) &=& \frac{1}{\gamma} \ln \left(\tau+\sqrt{\tau^2-1}\right)\,,
\ea
where $p>1/3$, $\gamma=\sqrt{4\pi/(p m_4^2)}$, and $\tau=\sqrt{4\pi\lambda/(3 p^2 m_4^2)}~(t-t_0)$ is the rescaled time, with $t_0$ being an arbitrary integration constant. Using the notation $v_\tau=d v/d\tau$ for the variable $v$, we have
\be
\tilde{H} \equiv \frac{a_\tau}{a}= \frac{p}{\sqrt{\tau^2-1}}\,.
\ee
The slow-roll parameters are
\ba
\epsilon =\eta &=& \frac{1}{p}\frac{\tau}{\sqrt{\tau^2-1}}\,,\\
\xi^2 &=& \frac{1}{p^2}\frac{\tau^2+1}{\tau^2-1}\,.
\ea
In the second model,
\ba
a(\tau) &=& (4\tau^2-1)^{p/2}\,,\\
\cosh [\gamma\phi(\tau)] &=& 2\tau\,,\\
\tilde{H} &=& \frac{4p\tau}{4\tau^2-1}\,,
\ea
and
\ba
\epsilon &=& \frac{1}{p}\left(1+\frac{1}{4\tau^2}\right)\,, \label{epsi2}\\
\eta &=& \frac{1}{p}\frac{4\tau^2-1}{4\tau^2+1}\,, \label{eta2}\\
\xi^2 &=& \frac{1}{p^2}\left(1+\frac{1}{4\tau^2}\right)\left(\frac{4\tau^2-1}{4\tau^2+1}\right)^2\,.\label{xi2}
\ea
In the limit $\tau \rightarrow \infty$, these solutions tend to power-law inflation \cite{LM2}, with $H=p/t$ and $\epsilon=\eta=\xi=1/p$. This occurs both when the brane tension is very large and at late times. Actually one can fix the integration constant $t_0$ such that $\phi(t=0)=0$, whence $t_0 \sim -m_4/\sqrt{\lambda}$; with this choice,
 one can show that the approximation with constant SR parameters is valid at large times, for instance when
$t \gtrsim 10^3\,t_4$, where $t_4 = m_4^{-1}\approx 5.4 \cdot 10^{-44}$s is the Planck time. Thus we have shown that there exist exact solutions with late time constant SR parameters. Anyway, the reader can convince oneself that the approximation is good by checking the behaviour of the scale factor $a$ near the origin of time (first model: $\tau\approx 1$; second model: $\tau\approx 1/2$). In both cases, again, one obtains a power law ($a\approx \tau^p$ and $a\sim t^{p/2}$, respectively) which, combined with the asymptotic behaviour of $\phi(\tau)$, generates constant SR parameters.

%version: December 22, 2004

\chappendix{Inflationary non-Gaussianity} \label{appB}

After a brief introduction to the issue of Gaussianity, in this section we achieve three goals. The first is to demonstrate the validity of the Mukhanov equation for the braneworld, at least at large scales (Sec. \ref{nlmukeq}); not only the method we shall adopt gives independent support to other proofs (e.g., \cite{KLMW,der04}), but is valid beyond the linear order in perturbation theory. The second outcome is the bispectrum of perturbations generated by either a cosmological tachyon or scenarios, such as high-energy braneworlds, with a modified Friedmann equation (Sec. \ref{bispectra}). Finally, in Sec. \ref{ncbisp} these results are extended to the case of a maximally symmetric noncommutative spacetime.

%%%%%%%%%%%%%%%%%%%%%%%%%%%%%%%%%%%%%%%%%%%%%%%%%%%%%%%%%%%%%%%%%%%%%%%%%%%%%%%%%%%%%%%%%%%%%%%%%%%%%%%%%%%%%%%%%%%%%%%%%%%%%%%%%%%%%%%%%%%%%%%%%%%%%%%%%%%%%%%%%%%%%%%%%%%%%%%%%%%%%%%%%%%%%%%%%%%%%%%%%%%%%%%%%%%%%%%%%%%%%%%%%%%%%%%%%%%%%%%%

\section{Linear perturbations are Gaussian} \label{gausspri}

Once one knows the spectrum $A^2_\Phi$ (also denoted ${\cal P}_\Phi$, depending on the numerical coefficient in the definition) for a given scalar or tensor perturbation $\Phi$, one can ask what are the probability distributions $P[\Phi]$ and $P[\Phi_\mathbf{k}]$ of the fluctuations in real and Fourier space, respectively. In the latter, 
\be\label{phk}
\Phi(x) = \int \frac{d^3\mathbf{k}}{(2\pi)^{3/2}} \Phi_\mathbf{k}(t) e^{i\mathbf{k} \cdot \mathbf{x}},
\ee
where the Fourier coefficients describe the fluctuation $\hat{\Phi}_\mathbf{k} \propto \delta\hat{\psi}_\mathbf{k}$ for each wave number (hats indicate quantum operators).

In the vacuum state $|0_\mathbf{k}\rangle$, the components of the field $\hat{\Phi}_\mathbf{k}$ do not have definite values. Rather, we can expand the vacuum state into a combination of eigenstates of the fluctuation:
\ba
|0_\mathbf{k}\rangle &=& \sum_u c_u |0_\mathbf{k}\rangle_u\,,\\
\hat{\Phi}_\mathbf{k}|0_\mathbf{k}\rangle_u &=& \Phi^u_\mathbf{k}                                           |0_\mathbf{k}\rangle_u\,, \qquad \forall\,\mathbf{k}\,,
\ea
where the sum is over the ensemble of all the possible universes $u$. The probability to find a given distribution $P[\Phi_\mathbf{k}^u]$ of fluctuations will be given by the squared coefficient $|c_u|^2$  in the vacuum expansion. As we can observe only our Universe, in order to proceed we must assume that the Universe we live in corresponds to a particular choice $\bar{u}$ in the ensemble. Since in this case, at variance with the usual quantum experiments in laboratory, to choose an eigenstate is not equivalent to perform a measure, there arises the problem to understand what mechanism has forced the vacuum state into one of the eigenstates it is made of. The issue of the cosmic choice, intimately related with the quantum to classical transition of cosmological perturbations, is still unclear in many respects; some authors have considered it as a decoherence process, that is, a destruction of interference terms in density matrices \cite{sak88,hal89,pad89,mos96}.

Let us admit to have a set of well-defined Fourier eigenvalues $\Phi^{\bar{u}}_\mathbf{k}$ for a fixed universe choice; ignoring henceforth the superscript $\bar{u}$, they can be written as
\be\label{reim}
\Phi_\mathbf{k} = |\Phi_\mathbf{k}| e^{i \vartheta_\mathbf{k}}= \Re (\Phi_\mathbf{k}) + i\,\Im (\Phi_\mathbf{k})\,.
\ee
In order to describe the statistical behaviour of cosmological perturbations, we recall some results on random fields \cite{adl81,BBKS}:
\begin{enumerate}
\item If a random field $\Phi(x)$ can be expressed as a Fourier superposition of coefficients $\Phi_\mathbf{k}$
 such that $\Re(\Phi_\mathbf{k})$ and $\Im(\Phi_\mathbf{k})$ are statistically independent and with the same distribution for all $\mathbf{k}$, then the probability distribution $P[\Phi]$ of the field is Gaussian. This is equivalent to state that the phases $\vartheta_\mathbf{k}$ are randomly distributed.
\item The statistical properties of homogeneous and isotropic Gaussian random fields with zero mean, $\langle \Phi(x)\rangle=0$, are completely described by the two-point correlation function 
\ba
{\cal P}_\Phi(\mathbf{x},\mathbf{x}') &\equiv& \langle \Phi(x)\Phi(x')\rangle_{t=t'}\\
&=&{\cal P}_\Phi(\varrho)\,,
\ea
where $\varrho=|\mathbf{x}-\mathbf{x}'|$, or, equivalently, by the power spectrum 
\be\label{powsp}
{\cal P}_\Phi(k)\propto k^3\langle \Phi_\mathbf{k}\Phi_{-\mathbf{k}}\rangle_t\,,
\ee
where $\langle fg\rangle_t\equiv\langle f(t)g(t)\rangle$. Here, angle brackets indicate the mean on a collection of universes or, in the case of a quantum operator, its mean value in the vacuum state. For Gaussian random fields, the ($2n+1$)-point correlation functions\footnote{And the connected part of the $2n$-point correlation functions \cite{pee80}.} vanish identically, while the $2n$-point correlation functions can be expressed through the only power spectrum.
\item A Gaussian field is ergodic, that is its spatial means in a given realization $\bar{u}$ are equal to the expectation values on the entire ensemble, if, and only if, its spectrum is continuous.
\end{enumerate}
When applying these theorems to the above cosmological quantum fluctuations, one gets the following results:
\begin{enumerate}
\item The real and imaginary part of each coefficient, Eq. (\ref{reim}), behave like two independent harmonic oscillators for each $\mathbf{k}$, as shown by the equation of motion for the fluctuation. This is clearer for the gauge-invariant fluctuation $u_\mathbf{k}$, which is an oscillator in conformal time with squared mass $(k^2-d^2_\eta z/z)$. In the vacuum state they have the same probability distribution, given by the ground-state wave eigenfunction of an harmonic oscillator: a Gaussian. Another way to see this is to note that under this condition the phases $\vartheta_\mathbf{k}$ are mutually independent, randomly distributed in the interval $0\leq \vartheta_\mathbf{k} < 2\pi$, and such that $\vartheta_{-\mathbf{k}}=-\vartheta_\mathbf{k}$. If the phase of each mode is random, then the central limit theorem guarantees that the superposition (\ref{phk}) is Gaussian if the number of modes is large. That is, one has $\langle\Phi_{-\mathbf{k}}^*\Phi_\mathbf{k}\rangle_t=\langle\Phi_\mathbf{k}^2\rangle_t=0$, and the probability to have a fluctuation $\Phi(\mathbf{x},t)$ at the point $\mathbf{x}$ is
\be
P[\Phi]=\frac{e^{-\Phi^2/(2\sigma_\Phi^2)}}{\sqrt{2\pi\sigma_\Phi^2}}\,,
\ee
where
\be
\sigma_\Phi^2\equiv \langle\Phi^2(\mathbf{x})\rangle_t=\int_0^\infty \frac{dk}{k}{\cal P}_\Phi(k)\,,
\ee
is the variance of the distribution. \emph{In the approximation of linear theory, cosmological fluctuations have a Gaussian probability distribution}.
\item By the cosmological principle, the resulting distribution is homogeneous and isotropic, implying that the coefficients $\Phi_\mathbf{k}$ as well as the spectrum ${\cal P}_\Phi$ depend only upon the absolute value $k$. The two-point correlation function depend only on $\varrho$ (Wiener-Khintchine theorem):
\be
{\cal P}_\Phi (\varrho)=\int_0^\infty \frac{dk}{k}{\cal P}_\Phi(k)\frac{\sin (k\varrho)}{k\varrho}\,.
\ee
Also, \emph{cosmological fluctuations are completely described by the power spectrum}.
\item The definition of the perturbation amplitude (\ref{powsp}) has the ergodic property, therefore being consistent with the initial assumption for the cosmic choice for any continuous transfer function describing the time evolution of the perturbation. Then, \emph{the statistical properties of the perturbations are evaluated in the ensemble of spatial points in the sky vault}.
\end{enumerate}
The Gaussianity of the statistical distribution for the perturbations is a direct consequence of ($i$) neglecting second-order terms in the equation of motion and ($ii$) taking the cosmological principle for granted. Actually both these are approximations, although very good according to experiments, of the theoretical setup and the real world, respectively. When going beyond the linear theory and accepting some deviation from perfect isotropy, as CMB probes do indicate, small departures from the Gaussian distribution will appear and provide new interesting features to explore.

%%%%%%%%%%%%%%%%%%%%%%%%%%%%%%%%%%%%%%%%%%%%%%%%%%%%%%%%%%%%%%%%%%%%%%%%%%%%%%%%%%%%%%%%%%%%%%%%%%%%%%%%%%%%%%%%%%%%%%%%%%%%%%%%%%%%%%%%%%%%%%%%%%%%%%%%%%%%%%%%%%%%%%%%%%%%%%%%%%%%%%%%%%%%%%%%%%%%%%%%%%%%%%%%%%%%%%%%%%%%%%%%%%%%%%%%%%%%%%%%

\section{Bispectrum and non-Gaussianity}

According to the inflationary paradigm, small quantum fluctuations of the inflaton field are amplified to cosmological scales by the accelerated expansion. These perturbations then leave their imprint into the cosmic microwave background as thermal anisotropies. Two main physical observables are generated by this mechanism, namely, the scalar spectrum, which is (the Fourier transform of) the two-point correlation function of scalar perturbations, and the bispectrum, coming from the three-point function \cite{GaM,WK,KSp}. For the curvature perturbation on comoving hypersurfaces $\cal R$, this reads 
\ba
\langle{\cal R}({\mathbf k}_1){\cal R}({\mathbf k}_2){\cal R}({\mathbf k}_3)\rangle &=& -(2\pi)^3\delta^{(3)}\left({\mathbf k}_1+{\mathbf k}_2+{\mathbf k}_3\right)\nonumber\\
&& \times \sum_{i<j} 2f_\text{NL}({\mathbf k}_i, {\mathbf k}_j)\langle|{\cal R}_\text{L}(k_i)|^2\rangle \langle|{\cal R}_\text{L}(k_j)|^2\rangle\,,\nonumber\\\label{fnl}
\ea
where $\langle|{\cal R}_\text{L}(k)|^2\rangle$ is the power spectrum of the linear Gaussian part ${\cal R}_\text{L}$ of the curvature perturbation with comoving wave number $k$, sum indices run from 1 to 3, and $f_\text{NL}$ is the \emph{nonlinearity parameter}.\footnote{Our definition of the nonlinearity parameter (sometimes dubbed $f_\text{NL}^{\cal R}$ in literature) is $- 2\cdot 3/5$ that of \cite{GLMM,gan94} (there denoted $\Phi_3$), $- 3/5$ times that of \cite{KSp}, and $3/5$ times that of \cite{ABMR,mal03}. The factor of 3/5 comes from the definition of $f_\text{NL}$ through the peculiar gravitational potential $\Psi_4$, which is $\Psi_4=-3{\cal R}/5$ during matter domination.}  If $f_\text{NL}$ is momentum independent, one can write the gravitational potential in terms of ${\cal R}_\text{L}$: in real space,
\be\label{nonlinear}
{\cal R}(\mathbf{x})={\cal R}_\text{L}(\mathbf{x}) -f_\text{NL}\left[{\cal R}^2_\text{L}(\mathbf{x})- \left\langle {\cal R}^2_\text{L}(\mathbf{x})\right\rangle \right],
\ee
which gives Eq. (\ref{fnl}) with $f_\text{NL}$ shifted outside the summation over the $\mathbf{k}_i$'s. When the statistical distribution is Gaussian, $f_\text{NL}=0$, the three-point function vanishes. In terms of the CMB temperature fluctuation $\Delta T(\mathbf{\hat{e}})/T$, measured along the direction $\mathbf{\hat{e}}$, the limit of the bispectrum at zero angular separation is the skewness, ${\cal S}_3(\mathbf{\hat{e}})\equiv \langle(\Delta T/T)^3\rangle$. For practical purposes, this is a less sensitive probe for non-Gaussianity than the bispectrum \cite{KSp}.

So far in this work we have restricted the discussion of cosmological perturbations to the power spectrum, implicitly assuming to deal with all the relevant informations that can be extrapolated from the sky. In many respects, the measured two-point correlation function is able, all by itself, to both describe the microwave sky in great detail and place observational constraints on the features of early-Universe models such as inflation.

However, the experiments of the last generation have radically changed the general attitude towards cosmology and made possible what is now recognized as a ``precision era.'' The physical scenarios explaining the large-scale structure of the Universe can be refined by more and more accurate observational inspections. Therefore it is natural to consider the bispectrum, too, and ask what signatures of non-Gaussianity we might expect from a given theoretical model (see \cite{BKMR} for a comprehensive review).

A non-Gaussian spectrum can arise according to a number of different mechanisms: just to mention some, late-time nonlinear evolution of cosmic structures \cite{MVH}, multifield inflation or inflation with scalar spectators \cite{AGW,KP,MMOL,LiM,BU1,BU2,zal04,EV}, curvaton scenario \cite{LyW,curv1,curv2}, nonvacuum inflation \cite{LPS,MRS,GMS}, higher-dimension operators in the inflaton Lagrangian \cite{pcr03}, DBI-like inflaton \cite{AST},\footnote{This model is quite distinct from the DBI cosmological tachyon. While the former predicts a strong non-Gaussian signature, the latter is more similar to standard inflation.} ghost inflaton \cite{ACMZ}, and, more commonly, self-interacting inflaton \cite{OLM,HBKP,FRS,YV2,YV3}. Non-Gaussianity was also considered in generalized Brans-Dicke gravity \cite{KoKS2} and $D$-$\bar{D}$ brane inflation \cite{EJMM2}.

In four dimensions, it turns out that the inflationary contribution to the nonlinear parameter is $4f_\text{NL}\approx n_s-1 =O(\epsilon)$ for a single ordinary scalar field \cite{ABMR,mal03}. It would be interesting to see how this result is modified for braneworld cosmologies and tachyon-driven scenarios. Intuitively, we do not expect a dramatic quantitative change in the effect since the patch formalism does not interfere with SR expansions except for the value of the coefficients in front of the SR parameters themselves. We have verified this guess in two ways. 

A preliminar confirmation comes from the stochastic approach of Gangui \textit{et al.} \cite{GLMM}. This approach permits to estimate the order of magnitude of the effect by just considering second-order fluctuations of a self-interacting inflaton field and no gravitational fluctuations. This might seem too crude an approximation, since one should go up to second order in perturbation theory in order to fully take gravitational back-reaction into account and treat the bispectrum consistently. Surprisingly, the inflaton perturbation really encodes the main feature of the model apart from the resulting incorrect combination of SR parameters. 

Stochastic inflation is an approximated method according to which the solutions of the equation of motion for the scalar perturbation in the long wavelength limit $k \ll aH$ are connected to those in the $k \gg aH$ limit at the Hubble horizon \cite{sta86,rey87,GLM,HMN,SB2,YV1}. The scalar field (or other derived quantities) is separated into a ``classical'' or coarse-grained contribution $\psi^{(c)}$, encoding all the modes larger than the Hubble horizon, and a quantum or fine-grained part $\psi^{(q)}$ taking into account the in-horizon modes. Therefore, the classical part is the average of the scalar field on a comoving volume with radius $\sim (aH)^{-1}$. With this decomposition, the equation of motion for $\psi$ becomes a Langevin equation with a stochastic noise source generated by the fine-grained contribution of the quantum fluctuations (see \cite{MMR} for an extension to stochastic inflation with colored noise).

Actually, we have performed the calculation of \cite{GLMM} with the general FRW equation (\ref{FRW}) and the V-SR tower both for the scalar field and the tachyon. So, $f_\text{NL} = O(\epsilon)$ for a generic braneworld filled with a scalar field $\psi$. As we shall see, the key assumption giving rise to this behaviour is the validity of the standard continuity equation, Eq. (\ref{conti}). The presence of a brane-bulk flow would introduce genuinely novel features and considerably complicate the set of dynamical equations. Also, in the presence of extra dimensions the gravitational contribution may lead to a nontrivial behaviour of second-order perturbations, since to this order the interplay between extra-horizon scales and small scales may become quite delicate. This would impose a more rigorous treatment and a full second-order calculation in order to carefully evaluate non-Gaussianity produced during inflation, e.g., like that performed in \cite{ABMR} for the four-dimensional case.

A very powerful approach in this sense is the space-gradient formalism of \cite{RiS2,LyMS,RiS1,RiS3}, a development of the separate universe method \cite{WMLL} (see also the earlier works \cite{SaT,EB,SaS,PSS,NT}). We shall follow the notation of the cited papers closely, and skip details that can be found there unaltered by brane or tachyon physics. To begin with, it is convenient to work in the induced metric \cite{ArDM}
\be
d s_4^2\Big|_{\rm brane} = N^2(t,\mathbf{x})dt^2-a^2(t,\mathbf{x})dx_idx^i,
\ee
where $N$ is the lapse function and $a(t,\mathbf{x})$ is a locally defined scale factor; in the synchronous gauge we used so far, $N=1$. The Hubble parameter reads $H=\dot{a}/(Na)$, where dots will denote derivatives with respect to $t$. Also, we define $\Pi\equiv\dot{\psi}/N$. The SR parameters (\ref{psepsi}) and (\ref{Heta}) are rewritten as ($\beta_q$ units)
\ba
\epsilon &=& -\frac{\dot{H}}{NH^2}=\frac{3q}{2}H^{\wteta} \frac{\Pi^2}{H^2}\,,\label{Npsepsi}\\
\eta     &=&  -\frac{\dot{\Pi}}{NH\Pi}\,,\label{NHeta}
\ea
while their evolution equations are
\ba
\dot{\epsilon} &=& NH\epsilon \left[\left(2-\wteta\right)\epsilon-2\eta\right],\\
\dot{\eta}     &=&  NH\left(\epsilon\eta-\xi^2\right),\\
\xi^2    &=& \frac{1}{NH^2} \left(\frac{\dot{\Pi}}{N\Pi}\right)^\cdot.
\ea
In the separate universe approach, the physical quantities such as $H(t,{\mathbf x})$, the scalar field $\psi(t,{\mathbf x})$, cosmological perturbations and observables are defined on an inhomogeneous background and evolve separately through the dynamical equations at each point once the initial conditions have been specified. Then we can transform time derivatives into spatial gradients like
\be\label{spgrad}
\frac{\partial_i H}{H}=-\epsilon\frac{H}{\Pi}\partial_i\psi\,,\qquad \partial_i\epsilon=2\epsilon\left[\left(\oteta-\frac{\theta}{2}\right)\epsilon-\eta\right]\frac{H}{\Pi}\partial_i\psi\,,
\ee
where the index $i$ runs from 1 to 3, $\oteta=1$ for the ordinary scalar field, and $\oteta=\theta/2$ for the tachyon.

At large scales, second-order gradient terms can be neglected and the gradient generalization of the Bardeen potential \cite{BaST} is conserved \cite{RiS1}:
\ba
\zeta_i &\equiv& X_i-NH\frac{\partial_i\rho}{\dot{\rho}}\,,\\
\dot{\zeta}_i &=& 0\,,
\ea
where $X_i\equiv\partial_i \ln a$. In the braneworld case, the long wavelength limit was advocated for consistently neglecting the projected Weyl tensor; this in turn is deeply intertwined with the other fundamental constraint, that is a standard continuity equation
\be \label{Nconti}
\dot{\rho}+3HN\rho (1+w) = 0\,.
\ee
For a scalar field, the curvature perturbation $\zeta$ on hypersurfaces with constant energy density coincides with the curvature perturbation ${\cal R}$ on comoving hypersurfaces, which as a vector quantity reads 
\be\label{Ri}
{\cal R}_i \equiv X_i-\frac{H}{\Pi}\partial_i\psi\,.
\ee
In the linear approximation, Eq. (\ref{Ri}) is the spatial gradient of ${\cal R}$, Eq. (\ref{Rper}). The spatial gradient of the Mukhanov-Sasaki variable $u$ corresponds to the linear limit of
\be
{\cal Q}_i \equiv -z\zeta_i = -z{\cal R}_i= \tilde{a}\left(\partial_i\psi-\frac{\Pi}{H}X_i\right), \label{Q}
\ee
where $z$ is given by Eq. (\ref{zgen}) and $\tilde{a}\equiv zH/\Pi$. In the generalization to the nonsynchronous gauge $N\neq 1$,
\be\label{zPi}
z=\frac{a\Pi}{c_S H^{\oteta}}= a \left(\frac{2\epsilon}{3qc_S^2 H^\theta}\right)^{1/2},
\ee
where the speed of sound $c_S$ is $c_S=1$ for the ordinary scalar field and $c_S=\sqrt{1-2\epsilon_\T/(3q)}$ for the tachyon. For the ordinary scalar field, $\tilde{a}=a$. From Eqs. (\ref{spgrad}), (\ref{Q}) and (\ref{zPi}) we get
\be\label{partz}
\partial_i z = -{\cal Q}_i+\tilde{a}\partial_i\psi [1+\oteta\epsilon-\eta+(\oteta-1)\varsigma^2]\,,
\ee
where the quantity
\be
\varsigma^2 \equiv \frac{q}{NH} \frac{\dot{c}_S}{c_S}= \frac{4q\epsilon_\T\eta_\T}{3q-2\epsilon_\T}\,,
\ee
is second order in the SR parameters.

%%%%%%%%%%%%%%%%%%%%%%%%%%%%%%%%%%%%%%%%%%%%%%%%%%%%%%%%%%%%%%%%%%%%%%%%%%%%%%%%%%%%%%%%%%%%%%%%%%%%%%%%%%%%%%%%%%%%%%%%%%%%%%%%%%%%%%%%%%%%%%%%%%%%%%%%%%%%%%%%%%%%%%%%%%%%%%%%%%%%%%%%%%%%%%%%%%%%%%%%%%%%%%%%%%%%%%%%%%%%%%%%%%%%%%%%%%%%%%%%

\section[Generalized Mukhanov equation and stochastic inflation]{Generalized Mukhanov equation and\\ stochastic inflation}\label{nlmukeq}

The evolution equation for ${\cal Q}_i$ can be computed within the multifield framework of \cite{RiS2}. Here we consider the same calculation for a single field $\psi$ in a patch given by Eq. (\ref{FRW}); the results will match each other in the 4D case. The first time derivative of Eq. (\ref{Q}) is
\be
\dot{\cal Q}_i=\frac{\dot{z}}{z}{\cal Q}_i=NH{\cal Q}_i \left[1+\oteta\epsilon-\eta+(\oteta-1)\varsigma^2\right].\label{dotQ}
\ee
Another derivation gives
\be\label{Qmuktemp}
\ddot{\cal Q}_i-\frac{\ddot{z}}{z}{\cal Q}_i=0\,.
\ee
However, it is more convenient to keep the decaying mode implicitly dropped in the recursion of Eq. (\ref{dotQ}); then the equation of motion can be recast as
\be\label{Qeom}
\ddot{\cal Q}_i-F\dot{\cal Q}_i+\Omega{\cal Q}_i=0\,,
\ee
where
\ba
F &\equiv& \frac{\dot{N}}{N}-NH\,,\\
\Omega &\equiv& F\frac{\dot{z}}{z}-\frac{\ddot{z}}{z}\\
&=&2+\left(3\oteta-1\right)\epsilon-3\eta-4\oteta\epsilon\eta+\left(1+\oteta-\wteta\right)\oteta\epsilon^2+\eta^2+\xi^2+(\oteta-1)g_\varsigma\varsigma^2.\nonumber\\
\ea
The extra tachyonic term is
\be
g_\varsigma \equiv 3+(\theta-1)\epsilon_\T-2\eta_\T+\left(\frac{\theta}{2}-1\right)\varsigma^2+\frac{(\varsigma^2)^{^{\bm .}}}{\varsigma^2}\,.
\ee
The expression for $\Omega$ is in accordance with the computation in the linear theory \cite{cal3} and, as that, is \emph{exact} in the SR parameters. The expression found in \cite{RiS2,RiS3} ($\theta=0$,$\,\oteta=1$) is recovered via Eq. (\ref{v''h}).

Equation (\ref{Qeom}) is equivalent to a generalized Mukhanov equation (with the Laplacian term dropped) when expressed via conformal time $d\eta=Ndt/a$; since $d_t^2=(N/a)^2d_\eta^2+(N/a)Fd_\eta$ and $\Omega=-(N/a)^2d^2_\eta z/z$, one has
\be\label{Qmuk}
\left(\frac{d^2}{d\eta^2}-\frac{1}{z}\frac{d^2z}{d\eta^2}\right){\cal Q}_i=0\,.
\ee
In the linear approximation and in momentum space, Eqs. (\ref{Qeom}) and (\ref{Qmuk}) hold for ${\cal Q}_i \approx \partial_i u\to ik_iu_\mathbf{k}$. From the de Sitter calculations of Sec. \ref{scalar}, the mean value of the quantum field $u_\mathbf{k}$ is
\be\label{dSsol}
\sqrt{\langle|u_\mathbf{k}|^2\rangle}=\frac{aH}{\sqrt{2(kc_s)^3}}\,.
\ee

The equation of motion for $u_\mathbf{k}$ can be written as an equation for the coarse-grained part of $u_\mathbf{k}$ sourced by a stochastic noise term. The coarse-grained part of the Mukhanov variable is $u_\mathbf{k}^{(c)}=u_\mathbf{k}{\cal W}(kR)$, where ${\cal W}$ is the Fourier transform of a window function $W(|\mathbf{x}-\mathbf{x}'|/R)$ falling off at distances larger than $R$. The scale $R$ is of order of the comoving Hubble radius, $R=h(aH)^{-1}$, with $h>1$ so as to encompass the whole horizon. With $h$ sufficiently larger than 1 we can safely discard the $k^2$ term in the Mukhanov equation.

The final result is then extended to the nonlinear gradient variable ${\cal Q}_i$ at large scales, getting
\bs\label{lange}\ba
&& \ddot{\cal Q}_i-F\dot{\cal Q}_i+\Omega{\cal Q}_i=\xi_i(t,\mathbf{x})\,,\\
&& \xi_i(t,\mathbf{x})=\int \frac{d^3\mathbf{k}}{(2\pi)^{3/2}}ik_ie^{i\mathbf{k}\cdot\mathbf{x}}\left[\ddot{\cal W}+\dot{\cal W}(2d_t-F)\right]u_\mathbf{k}\alpha(\mathbf{k})+\text{c.c.}\,,\nonumber\\\label{lange2}
\ea\es
where the superscript $(c)$ in ${\cal Q}_i$ is understood and  $\alpha(\mathbf{k})$ is a complex stochastic quantity such that the ensemble average $\langle\alpha(\mathbf{k})\alpha^*(\mathbf{k}')\rangle= \delta^{(3)}(\mathbf{k}-\mathbf{k}')$; it simulates the continuous crossing of modes outside the horizon and their addition to the coarse-grained part.
As it stands, Eq. (\ref{lange}) properly encodes the full stochastic contribution. This would not be the case if we started from Eq. (\ref{Qmuktemp}), where the velocity degree of freedom associated with the decaying mode has been absorbed (see the discussion in \cite{RiS2} for more details).

Equation (\ref{lange}) is the nonlinear extension of the Langevin-type equation we used in the first-SR-order heuristic computation in synchronous gauge:
\be
\dot{\psi}^{(c)}(t)=-\frac{U'}{3H}-\frac{\ddot{\psi}^{(c)}(t)}{3Hc_S^2}+\xi(\mathbf{x},t)\approx -\frac{U'}{3H}+\xi(\mathbf{x},t)\,,
\ee
where $U(\phi)=V(\phi)$ and $U(T)=\ln V(T)$. In this and the other equation one can see that there are basically two sources of nonlinearity. The first one is the back-reaction of the field fluctuations on the background encoded in the noise term, the second one is the self-interaction of the scalar field represented by the potential contribution $\Omega$ [or $-U'/(3H)$]. Therefore, \emph{a priori} the statistical distribution of ${\cal Q}_i^{(c)}$ ($\psi^{(c)}$) will be non-Gaussian, even if quantum fluctuations are random.

%%%%%%%%%%%%%%%%%%%%%%%%%%%%%%%%%%%%%%%%%%%%%%%%%%%%%%%%%%%%%%%%%%%%%%%%%%%%%%%%%%%%%%%%%%%%%%%%%%%%%%%%%%%%%%%%%%%%%%%%%%%%%%%%%%%%%%%%%%%%%%%%%%%%%%%%%%%%%%%%%%%%%%%%%%%%%%%%%%%%%%%%%%%%%%%%%%%%%%%%%%%%%%%%%%%%%%%%%%%%%%%%%%%%%%%%%%%%%%%%

\section{Braneworld and tachyon bispectrum}\label{bispectra}

In order to compute the scalar spectrum and bispectrum, we fix the gauge to the time slicing with respect to which the $k$th mode crosses the horizon simultaneously for all spatial points \cite{SB2}. Then $t=\ln (aH)$, $NH=(1-\epsilon)^{-1}$, and $R=he^{-t}$. In this gauge, the gradient curvature perturbation and $\partial_i z$ are, respectively,
\ba
{\cal Q}_i &=& \tilde{a}\partial_i\psi (1-\epsilon)\,,\\
\partial_i z &\approx& \left[\left(1+\oteta\right)\epsilon-\eta\right]{\cal Q}_i\\
             &=& -\case12 (n_s-1){\cal Q}_i\,,\label{zgfix}
\ea
where in the last passage we have used Eq. (\ref{noteta}). At first SR order,
\be
F \approx -1\,, \qquad \Omega \approx 2+3\left(\oteta+1\right)\epsilon-3\eta\,.
\ee
Equation (\ref{lange}) can be expressed as a Langevin differential equation in the curvature perturbation:
\be
\ddot{\cal R}_i+\left(2\frac{\dot{z}}{z}-F\right)\dot{\cal R}_i=-\frac{1}{z}\,\xi_i(t,\mathbf{x})\,.
\ee
To lowest SR order and neglecting the $\ddot{\cal R}_i$ term,
\be\label{finR}
\dot{\cal R}_i \approx -\frac{1}{3z}\xi_i(t,\mathbf{x})\,,
\ee
where $c_S \approx 1$ inside $\xi_i$ and $z$. The power spectrum is given by the solution of the linearized equation (\ref{finR}): at first order in a perturbative expansion,
\be
\dot{\cal R}_i^{(1)} \approx -\frac{\xi_i^{(1)}}{3z^{(0)}}\,,
\ee
where $z^{(0)}$ is $z$ defined on the homogeneous background. The time integration of $\xi/z$ from the initial time $t_i$ to $t$ is proportional to $A_s(k) B(kR)$, where $B(kR)=\int_{t_i}^{t} dt' (aH)^{-1}[\ddot{\cal W}+\dot{\cal W}(2d_{t'}+1)](aH)=[1+(kR)^2/3]{\cal W}(kR)-[1+(kR_i)^2/3]{\cal W}(kR_i)$ to lowest SR order. Here we have used a Gaussian window function 
\be
{\cal W}=\exp(-k^2R^2/2)\,,\qquad \dot{\cal W}=(kR)^2{\cal W}\,,\qquad \ddot{\cal W}=[(kR)^2-2]\dot{\cal W}\,,
\ee
and the lowest-SR-order eigenvalue equation $d_t u_\mathbf{k}=u_\mathbf{k}$. In the limit of asymptotic past ($t_i \to -\infty$) and future ($t\to +\infty$), $B(kR)\to 1$. The integration over $k$ yields ${\cal R}_\text{L}\equiv{\cal R}^{(1)}=\partial^{-2}\partial^i {\cal R}_i^{(1)}$ and the lowest-order amplitude (\ref{Sdeg}), after a computation almost identical to that of \cite{RiS3}.

At second order in the perturbation, 
$-3\dot{\cal R}_i^{(2)} =\xi_i^{(1)}(z^{-1})^{(1)}+\xi_i^{(2)}/z^{(0)}$. Since $\xi_i^{(2)}=O(\epsilon^2)$, the only surviving term at lowest SR order is, by the first-order version of Eq. (\ref{zgfix}),
\ba
\dot{\cal R}_i^{(2)} &\approx&  -(z^{-1})^{(1)}\frac{\xi_i^{(1)}}{3} =- \left(-\frac{\partial^{-2}\partial^j\partial_j z^{(1)}}{z^{(0)}}\right)\frac{\xi_i^{(1)}}{3z^{(0)}}\\
                     &=& -\case12 (n_s-1){\cal R}^{(1)}\dot{\cal R}_i^{(1)}\,,\label{R2}
\ea
After computing the nonlinear term ${\cal R}^{(2)}=-\case12 (n_s-1)[{\cal R}^{(1)}]^2$, one is ready to write down the curvature perturbation at second order,\footnote{In \cite{LyR} various definitions of the second-order curvature perturbation are reviewed.} 
\be
{\cal R} \approx {\cal R}^{(1)}+\case12{\cal R}^{(2)}\,,
\ee
and the bispectrum (\ref{fnl}). Actually, the integration of the window function inside the noise term corresponds to a nonlinearity parameter with nontrivial momentum dependence, for which Eq. (\ref{nonlinear}) does not hold. When one of the momenta is negligible relative to the others, that is in the squeezed limit $k_3 \ll k_1,k_2$, by Eqs. (\ref{R2}) and (\ref{fnl}) we get
\be\label{fns}
4f_\text{NL} = (n_s-1)+4f(\mathbf{k}_1,\mathbf{k}_2,\mathbf{k}_3) \approx n_s-1\,.
\ee
Although we have not written explicitly the momentum structure $f(k)$, we can draw some important conclusions.

$(i)$ Tachyon and ordinary inflation generate the same non-Gaussian signature in the limit of collapsing momentum dependence and at first SR order. Outside this approximation the nonlinearity parameter $f_\text{NL}$ acquires explicit dependence on the type of scalar field and braneworld through the parameter $z$: $4f_\text{NL}=(n_s-1)+4f^{(\theta,\psi)}(\epsilon,\eta;\mathbf{k}_1,\mathbf{k}_2,\mathbf{k}_3)$. This is in agreement with the correspondence between lowest-SR-order tachyon and ordinary observables.

$(ii)$ Written in terms of $n_s$, braneworld non-Gaussianity does not differ from the 4D picture, except perhaps in higher-order contributions. What changes is the inflationary model one has to impose in order to predict a given scalar spectral index. In this sense, we could regard Eq. (\ref{fns}) as a first-SR-order consistency equation joining the traditional set we explored until now, since both $n_s$ and $f_\text{NL}$ (through the bispectrum) are observables \cite{gru04,CZ}.\footnote{A small non-Gaussian component comes also from the 3-point functions involving the graviton zero-mode. Using the $z$ function (\ref{zgrav}) for braneworld tensor perturbations, one finds a contribution proportional to the tensor amplitude and spectral index $n_t$, in accordance with the 4D result \cite{mal03}. However, since the tensor amplitude is much smaller than the scalar one, we can neglect this term with respect to the scalar bispectrum.}

$(iii)$ On the other hand, one should note that the post-inflationary era greatly enhances non-Gaussianity, up to $f_\text{NL}^{\text{post}} =O(1)$ \cite{BMR1,BMR2,BMR3}\footnote{This is basically due to the fact that at second order the longitudinal gauge condition $\Phi_4^{(1)}-\Psi_4^{(1)}=0$ is modified as $\Phi_4^{(2)}-\Psi_4^{(2)}= 4\left(\Psi_4^{(1)}\right)^2$ at large scales, thus providing a nontrivial second-order correction to the Sachs-Wolfe effect \cite{ABMR}.} or even $f_\text{NL}^{\text{post}} \sim 50$ from a suitable preheating phase \cite{EJMMV}. As explained in \cite{BKMR}, the true observed nonlinearity parameter is not the bare inflationary result (\ref{fns}). In addition to the post-inflationary contribution, one must consider angular averaging. The total observed $f_\text{NL}$ is in fact, and at least, $f_\text{NL}^{\rm obs}=O(1)+f_\text{NL}$. Therefore the nonlinear effect of braneworld or 4D inflation, if the SR approximation holds as we required, is always subdominant.

Although these features can be obvious when inspecting patch cosmology, here we have derived them quantitatively.\footnote{The authors of \cite{CZ} drew similar conclusions, claiming that Eq. (\ref{fns}) is a model-independent consistency equation under the assumption of single-field inflation and in the squeezed limit. However, a proper treatment of the second-order nonlinear Bardeen potential seems missing there, as already emphasized in \cite{BKMR} (Sec. 8.4.1). Reference \cite{gru04} deals with the de Sitter case only.} Moreover, the gradient+stochastic approach can be the basis for next-to-leading-order SR and perturbation calculations as well as for numerical simulations \cite{RiS2}. 

According to the 1st-year WMAP analysis, the power spectrum fully characterizes the statistical properties of CMB anisotropies: that is, $f_\text{NL}$ vanishes consistently. More precisely, the constrain on the nonlinearity parameter is $-58<f_\text{NL}<134$ \cite{kom03}, which does not discard not only inflationary non-Gaussianity, but also other models predicting a more robust effect, $f_\text{NL} \gg 1$. See also \cite{GaW,cab04} for other analyses. The next-year WMAP data and the Planck satellite should significantly improve the accuracy of the measure, with the inclusion of polarization anisotropies: $f_\text{NL}^\text{min}(\text{WMAP})\sim 11-15$, $f_\text{NL}^\text{min}(\text{Planck})\sim 3-5$ \cite{BZ}.

%%%%%%%%%%%%%%%%%%%%%%%%%%%%%%%%%%%%%%%%%%%%%%%%%%%%%%%%%%%%%%%%%%%%%%%%%%%%%%%%%%%%%%%%%%%%%%%%%%%%%%%%%%%%%%%%%%%%%%%%%%%%%%%%%%%%%%%%%%%%%%%%%%%%%%%%%%%%%%%%%%%%%%%%%%%%%%%%%%%%%%%%%%%%%%%%%%%%%%%%%%%%%%%%%%%%%%%%%%%%%%%%%%%%%%%%%%%%%%%%

\section{Noncommutative bispectrum}\label{ncbisp}

Until now we have considered a commutative background throughout the whole spacetime. We can make a step further and phenomenologically assume to have a 3-brane in which the stringy spacetime uncertainty relation (\ref{SSUR}) is realized, where now $\tau =\int dt Na$. The separate universe approach does not contrast with a noncommutative background. To understand this point, we can use the linear picture of \cite{WMLL}, and in particular their Fig. 1. The basic idea is that a comoving large-scale perturbation is independently specified in two comoving locally homogeneous regions separated by a distance $\lambda$, if the size $\lambda_s\gtrsim H^{-1}$ of these regions is small with respect to $\lambda$. Perturbations are defined on a given background, that is a region of scale $\lambda_0$ much larger than our present horizon. Then the required hierarchy of scales is $\lambda_0\gg\lambda\gg\lambda_s\gtrsim H^{-1}$. In the presence of a nonlocal algebra, the string scale can play the role of natural marker in the hierarchy. For instance, setting $\lambda_s\sim l_s$ we just consider the IR region of *-models. If $\lambda_s\gtrsim H^{-1}>l_s$, the previous argument is unchanged. 

Since the *-product (\ref{*}) does not involve homogeneous quantities (i.e., it preserves the FRW maximal symmetry), the Mukhanov equation for a noncommutative 4D \cite{BH} or braneworld \cite{cal4} scenario is, at linear order and large scales, Eq. (\ref{muk}). In the separate universe approach, $a_\pm$ acquires a spatial dependence like the other quantities, $a_\pm(\tau)\to a_\pm(\mathbf{x},\tau)$.\footnote{Note that the correct procedure is first to smear the scale factor in a sufficiently small homogeneous neighbourhood, $a(\tau)\to a_\pm (\tau)$, and then to extend it to very large scales $a_\pm(\tau)\to a_\pm(\mathbf{x},\tau)$. The ``top-down'' smearing $a(\mathbf{x},\tau)\to a_\pm(\mathbf{x},\tau)$ does not lead to Eq. (\ref{muk}).} The measure $z_k$ is given by the product of $z$ and a correction factor $f_z$ depending on the particular noncommutative model one is assuming. The de Sitter solution of Eq. (\ref{muk}) is Eq. (\ref{ncdSsol}), i.e. the commutative solution (\ref{dSsol}) multiplied by $f_a^2\equiv (a_\eff/a)^2$, which is the relative rescaling of commutative to noncommutative conformal time.

The above discussion on conserved nonlinear perturbations is not modified by the introduction of a fundamental length scale. Therefore we are tempted to directly generalize Eq. (\ref{muk}) with the gradient variable ${\cal Q}_i$.
Since the mass term $d_\teta^2 z_k/z_k$ now depends on $k$, we define the Fourier mode of ${\cal Q}_i$ as
\be
{\cal Q}(\mathbf{k}) = \int \frac{d^3 \mathbf{x}}{(2\pi)^{3/2}}\, e^{-i \mathbf{k}\cdot\mathbf{x}}\partial^{-2}\partial^i{\cal Q}_i(\mathbf{x})\equiv -z_k{\cal R}(\mathbf{k})\,.
\ee
Then Eq. (\ref{muk}) holds for $u_\mathbf{k}\to {\cal Q}(\mathbf{k})$. One gets the $t$-version Eq. (\ref{Qeom}) by using the effective time $d\tilde{t}=d\teta a_\eff/N=d t/f_a$ and setting $\tilde{N}=N$ without loss of generality:
\be
d^2_{\tilde{t}}{\cal Q}(\mathbf{k})-\tilde{F}d_{\tilde{t}}{\cal Q}(\mathbf{k})+\tilde{\Omega}{\cal Q}(\mathbf{k})=0\,,
\ee
where $F$ and $\Omega$ are
\ba
\tilde{F}     &=& f_aF-\dot{f}_a\,,\\
\tilde{\Omega} &=& \tilde{F}\frac{d_{\tilde{t}}z_k}{z_k}-\frac{d^2_{\tilde{t}}z_k}{z_k}=-\left(\frac{N}{a_\eff}\right)^2\frac{d^2_\teta z_k}{z_k}=f_a^2(\Omega-\omega)\,,\\
\omega        &\equiv& \left[2\frac{\dot{z}}{z}\left(\frac{\dot{f}_a}{f_a}+\frac{\dot{f}_z}{f_z}\right)-F\frac{\dot{f}_z}{f_z}\right]+ \left(\frac{\ddot{f}_z}{f_z}+2\frac{\dot{f}_a}{f_a}\frac{\dot{f}_z}{f_z}\right)\,.\label{ncomeg}
\ea
In the commutative limit, $\tilde{F}\to F$, $\omega\to 0$, and $\tilde{\Omega}\to\Omega$. The term in square brackets in Eq. (\ref{ncomeg}) contains the $O(\epsilon)$ contribution of $\tilde{\Omega}$, since $\dot{f}_a/f_a=O(\epsilon)=\dot{f}_z/f_z$. In the infrared limit (strongly noncommutative regime), $\omega$ and its components loose their momentum dependence. The Langevin equation for the curvature perturbation reads
\be\label{langRnc}
d^2_{\tilde{t}}{\cal R}(\mathbf{k})+\left(2f_a\frac{\dot{z}_k}{z_k}-\tilde{F}\right)d_{\tilde{t}}{\cal R}(\mathbf{k})=-\frac{1}{z_k}\,\xi(\tilde{t},\mathbf{k})\,.
\ee
The noncommutative version of Eq. (\ref{partz}) gets an extra term from the redefinition of $z$; in momentum space and at first SR order, $z_k \approx -{\cal Q}(\mathbf{k})+\tilde{a}_\mathbf{k}\psi_\mathbf{k}[1+\oteta\epsilon-\eta+\dot{f}_z/(NHf_z)]$. 

With the gauge choice $\teta\approx -(aHf_a^2)^{-1}=-e^{-\tilde{t}}$, by definition one has $\partial_i \tilde{t}=0$ on surfaces of constant noncommutative time. Then 
\be
NHf_a=\frac{1-2\dot{f}_a}{1-\epsilon}= 1+O(\epsilon)\,,\qquad \tilde{F}=-1+O(\epsilon^2)\,,\qquad d_{\tilde{t}}u_\mathbf{k}=u_\mathbf{k}\,.
\ee
Neglecting the second-derivative term in Eq. (\ref{langRnc}), one has
\be
d_{\tilde{t}}{\cal R}(\mathbf{k}) \approx -\frac{1}{3z_k}\xi(\tilde{t},\mathbf{k})\,.
\ee
With the procedure of the last section, at lowest SR order and first perturbative order, one obtains the noncommutative power spectrum (\ref{Anoncom}), which is the commutative amplitude corrected by a factor $\Sigma^2=(f_a^2/f_z)^2$. Actually, if one wanted to go to coordinate space the integration over momenta should be performed up to the UV cutoff $k_0$, while the characteristic scale $R$ should be pushed towards the asymptotic limit $R \sim k_0^{-1}$ at the infinite future. However, the approximation $k_0\to \infty$ fits well at this stage of accuracy.

The gauge-fixed variables ${\cal Q}$ and $z_k$ are
\ba
{\cal Q}(\mathbf{k}) &=& \tilde{a}_\mathbf{k}\psi_\mathbf{k} (1-\epsilon+2\dot{f}_a)\,,\\
z_k &\approx& \left[\left(1+\oteta\right)\epsilon-\eta+f_a\frac{\dot{f}_z}{f_z}-2\dot{f}_a\right]{\cal Q}(\mathbf{k})= -\frac12 (n_s-1){\cal Q}(\mathbf{k}).\qquad\label{zkgfx}
\ea
At first SR order, the spectral index (\ref{noteta}) has acquired an extra term
\be\label{signga}
\sigma\epsilon=\frac{d\ln\Sigma^2}{d\ln k}=\frac{2}{NH}\left(2\frac{\dot{f}_a}{f_a}-\frac{\dot{f}_z}{f_z}\right)\approx 2f_a\left(2\frac{\dot{f}_a}{f_a}-\frac{\dot{f}_z}{f_z}\right).
\ee
A computation of the second-order curvature perturbation, by Eq. (\ref{zkgfx}), gives again Eq. (\ref{fns}), with the spectral index now depending on the noncommutative parameter $\sigma$.

%\printindex

%%%%%%%%%%%%%%%%%%%%%%%%%%%%%%%%%%%%%%%%%%%%%%%%%%%%%%%%%%%%%%%%%%%%%%%%%%%%%%%%%%%%%%%%%%%%%%%%%%%%%%%%%%%%%%%%%%%%%%%%%%%%%%%%%%%%%%%%%%%%%%%%%%%%%%%%%%%%%%%%%%%%%%%%%%%%%%%%%%%%%%%%%%%%%%%%%%%%%%%%%%%%%%%%%%%%%%%%%%%%%%%%%%%%%%%%%%%%%%%%

%\newpage
%\addcontentsline{toc}{chapter}{Index} \printindex

\end{document}